\documentstyle[amssymb,fullpage,11pt,subfigure,epsfig]{article}

\newif\iffigure

\figuretrue

\def\agt{\gtrsim}
\def\alt{\lesssim}

\title{\bf {\boldmath$d_{x^2-y^2}$}-wave superconductivity 
and the Hubbard model}
\author{N. BULUT\\
Department of Physics, Ko\c{c} University \\
Sariyer, 80910 Istanbul, Turkey\\ }

\date{} 

\begin{document}

\maketitle

\centerline{to appear in {\it Advances in Physics}, {\bf 51}, no. 6 (2002)}

\bigskip

\begin{abstract}

The numerical studies of $d_{x^2-y^2}$-wave pairing in the 
two-dimensional (2D) and the 2-leg Hubbard models
are reviewed.
For this purpose, the results obtained from the 
determinantal Quantum Monte Carlo 
and the density-matrix renormalization-group 
calculations are presented.
These are calculations which 
were motivated by the discovery of the 
high-$T_c$ cuprates.
In this review, the emphasis is placed on 
the microscopic many-body processes which are responsible for 
the $d_{x^2-y^2}$-wave pairing correlations observed in the 2D and 
the 2-leg Hubbard models.
In order to gain insight into these processes, 
the results on the effective pairing interaction
as well as the magnetic, 
density and the single-particle excitations will be reviewed.
In addition,
comparisons will be made with the other numerical approaches
to the Hubbard model and the numerical results on the
$t$-$J$ model.
The results reviewed here 
indicate that an effective pairing
interaction which is repulsive at $(\pi,\pi)$ momentum transfer
and enhanced single-particle spectral weight near the 
$(\pi,0)$ and $(0,\pi)$ points of the Brillouin zone 
create optimum conditions for $d_{x^2-y^2}$-wave pairing.
These are two effects which act to enhance the $d_{x^2-y^2}$-wave 
pairing correlations in the Hubbard model.
Finding additional ways is an active research problem.

\end{abstract}


\tableofcontents

\newpage
\setcounter{equation}{0}\setcounter{figure}{0}
\section{Introduction}

Since the discovery of high-temperature superconductivity
in the layered cuprates in 1986
[Bednorz and M\"uller 1986],
it has been established that 
the superconducting order parameter has the $d_{x^2-y^2}$-wave 
symmetry in a number of these materials
[Schrieffer 1994, Scalapino 1995, van Harlingen 1995,
Tsuei and Kirtley 2000]. 
This is important, because 
the $d_{x^2-y^2}$-wave symmetry of the order parameter
suggests the possibility of 
an electronically mediated pairing mechanism.
Perhaps, the simplest model used for modelling the low-energy
electronic correlations of the layered cuprates is 
the two-dimensional (2D) Hubbard model.
Within this context, the nature of the pairing correlations 
in the Hubbard model as well as the nature of its low-lying electronic
excitations has received considerable attention. 

In 1987 Anderson suggested that the 2D Hubbard model 
is relevant to the cuprates 
[Anderson 1987].
However, 
even today questions remain about this model.
In this article, 
what has been learned about the physical properties 
of the 2D and the 2-leg Hubbard models from the numerical studies 
will be reviewed.
The emphasis will be placed on the $d_{x^2-y^2}$ pairing 
correlations seen in these models and their microscopic origin. 
The implications of these calculations 
for the $d_{x^2-y^2}$-wave pairing 
in the high-$T_c$ cuprates is of current interest. 
In particular, 
one is interested in knowing whether the 
$d_{x^2-y^2}$-wave superconducting order 
could exist in the ground
state of the 2D Hubbard model, and, if it does, whether
it would have sufficient strength to explain the 
superconducting transition temperatures as high as
those seen in the cuprates.
There is also much interest in finding ways of enhancing the 
strength of the $d_{x^2-y^2}$-wave pairing correlations observed
in the Hubbard model. 
Within the past ten to fifteen years, 
the determinantal Quantum Monte Carlo (QMC) and 
the density-matrix renormalization-group (DMRG) 
techniques have been used to address these questions. 
These numerical studies are the subject of this review article.

For the 2D Hubbard model, the QMC data 
on the magnetic, charge and the single-particle excitations 
will be presented.
In order to investigate the pairing correlations, 
the QMC data on the irreducible particle-particle 
interaction and the solutions 
of the particle-particle Bethe-Salpeter equation will be shown. 
Using the DMRG method, the equal-time pair-field correlation function
has been calculated in the ground state of the 2-leg 
Hubbard ladder. 
These DMRG data along with the QMC results 
on the 2-leg ladder will be shown 
and compared with each other. 
In addition, 
these results will be compared with the numerical studies of 
the $t$-$J$ model.

Figure 1.1 shows a schematic drawing of the CuO$_2$ plane
of the cuprates consisting of the one-electron Cu($3d_{x^2-y^2}$)
and O($p_x,p_y$) orbitals that give rise to the 
band in which the $d_{x^2-y^2}$-wave superconducting pairs form.
In an electronic model of the CuO$_2$ plane, 
there would be an onsite Coulomb repulsion at the 
Cu($3d_{x^2-y^2}$) and the O($p_x,p_y$) orbitals and one-electron 
hopping matrix elements between the neighbouring 
Cu-O and the O-O orbitals
as well as longer-range hoppings and Coulomb interactions.
A simplified approach is to use the single-band Hubbard model
on the 2D square lattice which has an onsite Coulomb repulsion
and one-electron hopping matrix elements between the 
nearest-neighbour sites, as illustrated in Fig.~1.2(a).
To some extent, 
this is motivated by the notion that the 
nonperturbative effects due to the onsite Coulomb repulsion
will dominate at low temperatures and low energies.
Extensive numerical calculations have been 
carried out for studying the properties of the 
one-band 2D Hubbard model.
Another model which has proven useful for studying 
$d_{x^2-y^2}$-wave pairing is the 2-leg Hubbard ladder
illustrated in Fig.~1.2(b).
This model has an onsite Coulomb repulsion $U$ and 
the intrachain and interchain hopping matrix elements 
$t$ and $t_{\perp}$, respectively.
The 2-leg Hubbard ladder is an important model 
where the pairing correlations can be studied in detail 
in the ground state for systems with up to 32 rungs.
Here, the numerical results on the 2D and the 2-leg 
Hubbard models will be used to discuss the nature of the 
$d_{x^2-y^2}$-wave pairing found in the CuO$_2$ layers of the 
high-$T_c$ cuprates.

\begin{figure}
\centering
\iffigure
\epsfig{file=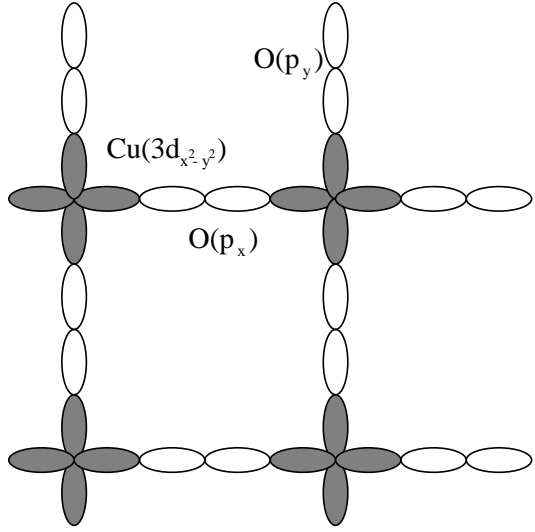,height=7cm,angle=0}
\fi
\caption{
Schematic drawing of the CuO$_2$ lattice consisting of the 
one-electron Cu($3d_{x^2-y^2}$) and the O($p_x,p_y$)
orbitals.
}
\label{1.1}
\end{figure}

The Hubbard model was introduced 
for describing the Coulomb correlation 
effects in the $d$-bands of the transition metals, 
which show both band-like and atomic-like behaviour
[Hubbard 1963].
The Hubbard Hamiltonian has a one-body kinetic term
and an interaction term which represents the onsite Coulomb
repulsion when two electrons occupy the same orbital.
In spite of studies covering four decades, 
questions remain about the Hubbard model.
In two dimensions, the Hubbard Hamiltonian 
with only nearest-neighbor hopping is given by
\begin{equation}
\label{Hubbard}
H = -t \sum_{\langle i,j\rangle, \sigma}
( c^{\dagger}_{i\sigma} c_{j\sigma} + 
c^{\dagger}_{j\sigma} c_{i\sigma} )
+ U \sum_i n_{i\sigma} n_{i-\sigma}
- \mu \sum_{i\sigma} n_{i\sigma},
\end{equation}
where the sum over $i$ and $j$ is done over the nearest-neighbour 
sites on a square lattice.
The one-electron hopping matrix element is $t$, 
the onsite Coulomb repulsion is $U$ 
and the chemical potential $\mu$ is used for 
controlling the electron occupation in the grand canonical 
ensemble. 
Here, $c^{\dagger}_{i\sigma}$ ($c_{i\sigma}$) 
creates (annihilates) an electron of spin $\sigma$ at site $i$
and $n_{i\sigma}=c^{\dagger}_{i\sigma} c_{i\sigma}$ 
is the occupation number of electrons with spin 
$\sigma$ at site $i$.
In addition to the 2D Hubbard model, the numerical results 
on the 2-leg Hubbard ladder with anisotropic hopping 
will be reviewed.
In this case, the Hamiltonian is 
\begin{eqnarray}
\label{Ladder}
H = -t \sum_{i,\lambda,\sigma}
( c^{\dagger}_{i,\lambda,\sigma} c_{i+1,\lambda,\sigma} + 
c^{\dagger}_{i+1,\lambda,\sigma} c_{i,\lambda,\sigma} )
-t_{\perp} \sum_{i,\sigma}
( c^{\dagger}_{i,1,\sigma} c_{i,2,\sigma} + 
c^{\dagger}_{i,2,\sigma} c_{i,1,\sigma} ) \nonumber \\ 
+ U \sum_{i,\lambda} 
n_{i,\lambda,\sigma} n_{i,\lambda,-\sigma}
- \mu \sum_{i,\lambda,\sigma} n_{i,\lambda,\sigma},
\end{eqnarray}
where $t$ is the hopping matrix element parallel to the 
chains and $t_{\perp}$ is the inter-chain hopping,
as illustrated in Fig.~1.2(b).
The operators $c^{\dagger}_{i,\lambda,\sigma}$ 
create an electron of spin $\sigma$ at site $i$ of the 
$\lambda$'th leg, and 
$n_{i,\lambda,\sigma}= c^{\dagger}_{i,\lambda,\sigma}
c_{i,\lambda,\sigma}$
is the electron number operator with spin $\sigma$
at site $(i,\lambda)$.

\begin{figure}
\centering
\iffigure
\mbox{
\subfigure[]{
\epsfysize=6cm
\epsffile[100 200 480 560]{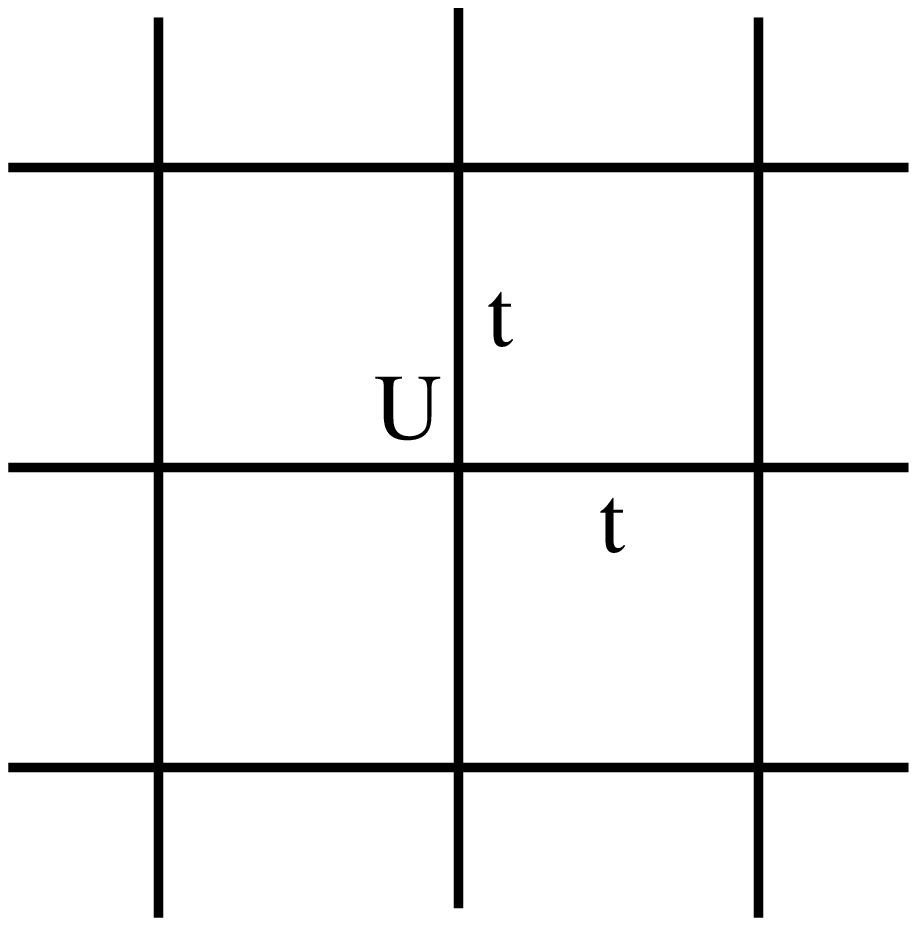}}
\quad
\subfigure[]{
\epsfysize=6cm
\epsffile[50 200 600 560]{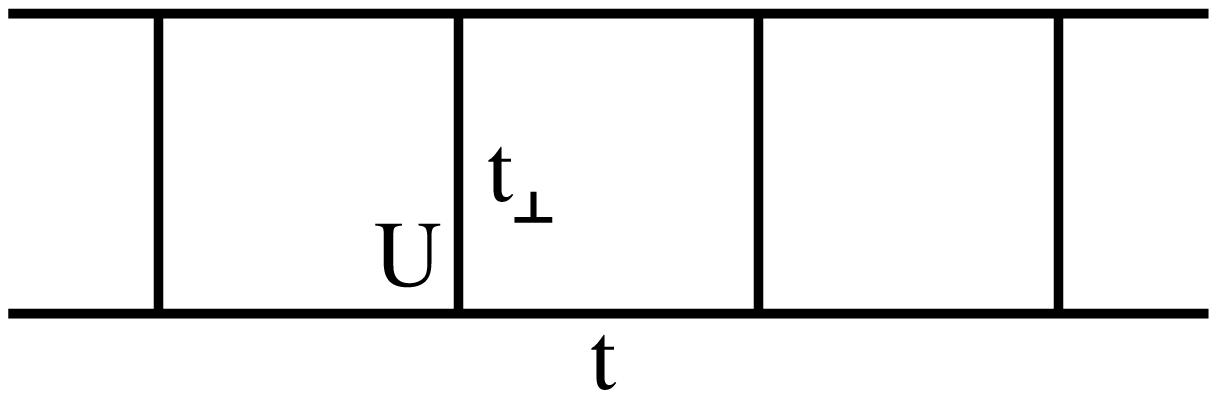}}}
\fi
\caption{
(a) Sketch of the 2D Hubbard lattice with 
isotropic nearest-neighbour hopping $t$ and the 
onsite Coluomb repulsion $U$.
(b) Sketch of the 2-leg Hubbard ladder with onsite Coulomb
repulsions $U$, and the intrachain and interchain
hopping matrix elements $t$ and $t_{\perp}$, respectively.
}
\label{1.2}
\end{figure}

The QMC and the DMRG are powerful numerical methods,
which have been developed for studying 
strongly-correlated systems such as the Hubbard model.
These techniques allow for the possibility to determine whether
superconductivity exists in the ground state of an interacting 
system without first having to construct an approximate 
theoretical framework for describing its elementary 
excitations.
The determinantal QMC method was introduced in 
Ref.~[Blankenbecler {\it et al.} 1981].
The QMC results reviewed here were obtained by using the 
algorithm described in Ref.~[White {\it et al.} 1989b].
Reviews of this method can be found in 
Refs.~[Scalapino 1993, Muramatsu 1999].
The DMRG method was developed by White 
[White 1992 and 1993],
and since then has been used to 
calculate the equal-time correlation functions in the 
ground state of interacting systems.
Here, 
in Sections 3 through 7,
numerical results on the 2D and 
the 2-leg Hubbard models, 
which were obtained by using these two algorithms, 
will be presented. 
In Appendix, 
the determinantal QMC and the DMRG techniques 
will be described briefly.

In Section 2, an introduction to 
the spin-fluctuation mediated $d_{x^2-y^2}$-wave 
superconductivity will be given.
In addition, here, the experimental evidence 
for $d_{x^2-y^2}$-wave superconductivity in the cuprates 
and the theoretical studies 
of $d_{x^2-y^2}$-wave pairing in the cuprates will be reviewed 
briefly.

In order to understand the microscopic 
many-body processes causing 
$d_{x^2-y^2}$-wave pairing correlations in the 
2D Hubbard model, it is useful to first discuss the magnetic, 
density and the single-particle excitations.
The review of the numerical studies on the 2D Hubbard model 
will begin in Section 3 by presenting QMC 
data on the magnetic properties.
Here, QMC results on the magnetic susceptibility will be shown
at and near half-filling.
In addition, an RPA-like model for the magnetic 
susceptibility will be introduced, and 
the effective particle-hole vertex will be discussed.
The NMR experiments which support the existence of short-range 
AF fluctuations in the high-$T_c$ cuprates will also be
reviewed briefly.

In Section 4, 
the QMC results on the charge susceptibility 
will be shown and its dependence on the strength of the 
Coulomb repulsion will be discussed.
Here, 
the possibility of any "$4{\bf k}_F$" 
charge-density-wave (CDW) fluctuations in the 
2D Hubbard model will be investigated.
In this section, 
numerical results on the optical conductivity will also be discussed.

The nature of the single-particle excitations in the 2D Hubbard 
model has received much attention within the context 
of the cuprates.
In Section 5, the single-particle 
density of states $N(\omega)$ and spectral weight 
$A({\bf p},\omega)$ obtained from the maximum-entropy 
analytic continuation of the QMC data on the single-particle
Green's function will be reviewed.
At half-filling, these calculations find 
a Mott-Hubbard gap, 
lower and upper Hubbard bands and quasiparticle bands.
These quasiparticle bands are similar to those found in a
spin-density-wave (SDW) insulator.
It will be seen that,
upon doping, a redistribution of the spectral weight takes
place forming a narrow metallic band at the Fermi level.
This band has unusually flat dispersion near the 
$(\pi,0)$ and $(0,\pi)$ points of the 
Brillouin zone leading to large amount of 
single-particle spectral weight available 
for scatterings in the particle-particle channel.
These results on $A({\bf p},\omega)$ will also be
compared with the results of the angular-resolved
photoemission spectroscopy (ARPES) 
measurements on the cuprates.

After reviewing the numerical results on the 
magnetic, density and the single-particle properties, 
in Section 6 the $d_{x^2-y^2}$-wave pairing correlations in the 
2D Hubbard model will be discussed.
The QMC simulations have found that 
when doped away from half-filling
there are short-range $d_{x^2-y^2}$-wave pairing correlations 
in the 2D Hubbard model, but no long-range superconducting order
has been observed in the parameter regime where the simulations 
are carried out.
The irreducible particle-particle
interaction $\Gamma_I$ has been also calculated 
with the QMC simulations and it gives
information on the microscopic many-body
processes causing the attraction in the 
$d_{x^2-y^2}$-wave pairing channel.
In Section 6, these QMC results will be reviewed.
It will be seen that $\Gamma_I$ is repulsive 
at ${\bf q}=(\pi,\pi)$ momentum transfers 
and that it is the short-range 
AF correlations which are responsible for this behavior.
Comparisons with the various 
diagrammatic approaches will also be carried out, 
which indicate that in the intermediate-coupling regime 
the momentum and the Matsubara-frequency structure in 
$\Gamma_I$ can be described by a properly-renormalized 
single spin-fluctuation exchange interaction. 
Using the QMC data on $\Gamma_I$ and the 
single-particle Green's functions, the Bethe-Salpeter
equation in the particle-particle channel has been solved. 
These calculations allow for a comparison of the strength of the 
pairing correlations in the various channels.
Here, it will be seen that as the temperature is lowered, 
the fastest growing pairing correlations occur in the 
singlet $d_{x^2-y^2}$-wave channel.

While these calculations are not carried out at sufficiently 
low temperatures to determine whether the
$d_{x^2-y^2}$-wave long-range superconducting order 
exists in the ground state of the doped 2D Hubbard model,
the results on $\Gamma_I$ are useful in the following sense.
Consider a phonon-mediated superconductor
such as Pb, where the effective interaction mediating 
the pairing forms already at temperatures of order the 
characteristic phonon frequency $\omega_D$.
Hence, in this case, 
at $T\sim \omega_D$ it would be possible to study 
the effective particle-particle interaction responsible 
for superconductivity, 
even though the superconducting long-range order 
takes place at much lower temperatures.
Similarly, 
the QMC simulations are carried out at temperatures 
of order or less than the characteristic 
energy scale $J\sim 4t^2/U$ of the AF correlations.
Hence, it is possible to probe $\Gamma_I$ in the 
temperature regime where short-range AF correlations have formed,
and see what type of many-body processes 
are important in mediating the pairing
and which pairing channels are favored.

The 2-leg Hubbard ladder is a model where the 
$d_{x^2-y^2}$-wave superconducting correlations can be 
studied in detail. 
In Section 7, the DMRG and the QMC results on the 
2-leg Hubbard ladder will be reviewed.
The DMRG calculations have found enhanced power-law decaying 
$d_{x^2-y^2}$-wave superconducting correlations in the 
ground state of the 2-leg Hubbard ladder. 
At the same time, with the QMC simulations,
it is possible to calculate the magnetic susceptibility, 
the single-particle spectral weight and the 
particle-particle vertex $\Gamma_I$ at sufficiently low
temperatures.
The comparisons of the DMRG and 
the QMC results show that it is
the strong short-range AF fluctuations which mediate 
the $d_{x^2-y^2}$-wave pairing correlations 
in this system.
Furthermore, in this case, it is possible to 
study the dependence of the strength of the 
superconducting correlations on the model parameters
such as $t_{\perp}/t$ and $U/t$.
It will be seen that the superconducting correlations are 
strongest in the intermediate-coupling regime and 
when $t_{\perp}/t$ is such that 
there is large amount of single-particle 
spectral weight pinned near the $(\pi,0)$ and 
$(0,\pi)$ points of the Brillouin zone.
In addition, the QMC calculations find that $\Gamma_I$
peaks at $(\pi,\pi)$ momentum transfer.
These features of $A({\bf p},\omega)$ and 
$\Gamma_I$ create optimum conditions for 
$d_{x^2-y^2}$-wave superconducting correlations.
In Section 7.3, these numerical results on the 2D 
and the 2-leg Hubbard models will be compared
with each other.
This comparison suggests that the $d_{x^2-y^2}$-wave 
pairing correlations seen in these models do not require
a particularly sharp peak in 
$\Gamma_I$ but rather simply weight at momentum transfers
near $(\pi,\pi)$.

In Section 8, 
the QMC and the DMRG results reviewed above
will be compared with the diagrammatic 
and the other numerical approaches to the Hubbard model
as well as with the numerical results on the $t$-$J$ model.
First, in Section 8.1, 
the QMC results on the 2D Hubbard model 
will be compared with the results 
of the fluctuation exchange (FLEX) approach.
Here, the purpose will be to make simple 
estimates for the maximum $T_c$ possible 
in the 2D Hubbard model, 
if a superconducting phase were to occur.
In particular, 
the effects of the system being near an AF Mott-Hubbard
insulator on the strength of the $d_{x^2-y^2}$-wave 
pairing will be explored.
In Section~8.2, the results from the variational and 
the projector Monte Carlo studies of the Hubbard model 
will be reviewed briefly.
Recently,
the dynamical cluster approximation (DCA) 
and the one-loop renormalization-group (RG)
technique employing a 2D Fermi surface were applied
to study $d_{x^2-y^2}$ pairing 
in the 2D Hubbard model.
The results of these calculations will be discussed 
in Section~8.3.
The origin of the normal state properties 
of the cuprates in the pseudogap regime 
remains an important problem in this field.
In Section~8.4, 
the findings of various calculations
in the low-doping regime of the Hubbard model 
will be compared with the pseudogap regime of the cuprates. 

The Hubbard model is closely related to the $t$-$J$ model,
for which various Monte Carlo, exact diagonalization and
DMRG calculations have been carried out.
There is much interest in the ground-state phase diagram
and the nature of the density and the pairing correlations 
in this model.
In Section~8.5, the numerical studies on phase separation, 
stripe correlations and the $d_{x^2-y^2}$-wave pairing 
correlations in the $t$-$J$ model will be briefly reviewed
and compared with the results on the Hubbard model.
Comparisons will also be made with the 2-leg $t$-$J$ model.

In Section~8.6, the implications of 
the numerical results on the Hubbard model for the 
nature of the 
$d_{x^2-y^2}$-wave pairing in the high-$T_c$ 
cuprates will be discussed.
In Section 9, 
the summary and the conclusions will be given.

\setcounter{equation}{0}\setcounter{figure}{0}
\section{{\boldmath$d_{x^2-y^2}$}-wave superconductivity 
in the cuprates}

\subsection{Spin-fluctuation mediated $d_{x^2-y^2}$-wave 
superconductivity} 

Since the development of the BCS theory,
it has been of interest to see whether the effective interaction 
which is responsible for pairing could be mediated 
by excitations other than phonons.
The superfluidity of $^3$He is an example where 
it is believed that the pairing is due to the exchange 
of ferromagnetic spin fluctuations resulting in triplet 
$p$-wave superconductivity [Leggett 1975].
Within the context of organic superconductors, 
the possibility of the pairing being mediated by spin fluctuations 
had been noted in Ref.~[Emery 1986].
For the heavy fermion materials, 
the possibility of $d$-wave superconductivity due to the exchange 
of AF spin fluctuations was proposed in 
Refs.~[Scalapino {\it et al.} 1986, 
Miyake {\it et al.} 1986, Cyrot 1986].
For the high-$T_c$ superconductors,
the possibility of $d_{x^2-y^2}$-wave superconductivity 
was first studied in
Ref.~[Bickers {\it et al.} 1987], 
where the framework of the two-dimensional Hubbard model 
was used near an SDW phase.

Here, a discussion of these ideas on how the exchange of the 
AF spin-fluctuations leads to a $d_{x^2-y^2}$-wave gap 
function will be given, 
since this picture will be useful for the 
remainder of the article.
In the paramagnetic phase
of the 2D Hubbard model and within RPA,
the single spin-fluctuation exchange interaction
between opposite-spin particles at zero frequency transfer is  
given by 
[Berk and Schrieffer 1966, Doniach and Engelsberg 1966]
\begin{equation} 
V({\bf p'}|{\bf p}) = U + 
{ U^3\chi_0^2({\bf p'}-{\bf p}) 
\over 1-U^2\chi_0^2({\bf p'}-{\bf p}) } +
{ U^2\chi_0({\bf p'}+{\bf p}) 
\over 1-U \chi_0({\bf p'}+{\bf p}) },
\end{equation}
where $\chi_0({\bf q})$ is the usual Lindhard function.
The longitudinal and the transverse spin fluctuations 
contributing to $V$ are illustrated in Fig.~2.1.
Here, $V({\bf p'}|{\bf p})$ is the interaction for the scattering of 
a pair of opposite-spin electrons at states $({\bf p},-{\bf p})$ 
to $({\bf p'},-{\bf p'})$.
Near half-filling and for a tight-binding band structure
$\varepsilon_{\bf p} = -2t (\cos{p_x} + \cos{p_y} ) - \mu $, 
where $\mu$ is the chemical potential, 
the Lindhard function peaks 
at ${\bf q}\sim (\pi,\pi)$.
Consequently,
for a system with Stoner-enhanced AF correlations,
the spin-fluctuation exchange interaction 
$ V({\bf p'}|{\bf p})$ 
is large and repulsive at 
${\bf p'}-{\bf p} \sim (\pi,\pi)$ momentum transfers.
Now, consider the BCS gap equation
\begin{equation}
\Delta_{\bf p} = - \sum_{\bf p'} 
{ V({\bf p}|{\bf p'})  \Delta_{\bf p'} \over 
2E_{\bf p'} },
\end{equation}
where ${\bf p}$ and ${\bf p'}$ are restricted to being 
on the Fermi surface of the doped system 
as illustrated in Fig.~2.2.
When this equation is solved,
it is found that 
the leading superconducting instability occurs for
a gap function with the $d_{x^2-y^2}$-wave symmetry,
\begin{equation}
\Delta_{\bf p} = {\Delta_0 \over 2}
( \cos{p_x} - \cos{p_y} ).
\end{equation}
In order to see why 
the $d_{x^2-y^2}$-wave form is a solution, consider 
a quasiparticle pair occupying $({\bf p},-{\bf p})$
with ${\bf p}$ near $(\pi,0)$,
as shown in Fig.~2.2.
This pair can scatter to 
$({\bf p'},-{\bf p'})$ where ${\bf p'}\sim(0,\pi)$
through the interaction $V({\bf p'}|{\bf p})$, 
which is enhanced and positive 
for momentum transfers near $(\pi,\pi)$,
because $\Delta_{\bf p}$ changes sign between 
${\bf p}$ and ${\bf p'}$.
This is basically the reason for why the $d_{x^2-y^2}$-wave gap 
symmetry is favored by the AF spin fluctuations. 
Since $V({\bf p'}|{\bf p})$ is always positive, 
there is no singlet solution with the 
usual $s$-wave symmetry.
For simplicity, 
these arguments were given at zero frequency,
where ${\bf p}$ and ${\bf p'}$ are restricted to being on the 
Fermi surface,
but the same arguments 
hold when the frequency dependencies of the 
gap equation and of the effective interaction are 
taken into account.
In Section~6, the QMC data on the irreducible particle-particle 
interaction $\Gamma_I$ 
and the solutions of the Bethe-Salpeter equation
will be reviewed for the 2D Hubbard model.
There, 
it will be seen that $\Gamma_I$ is similar to the 
single spin-fluctuation exchange interaction 
and 
the leading singlet pairing instability occurs
in the $d_{x^2-y^2}$-wave channel in the parameter regime where
the QMC simulations are carried out.

\begin{figure}
\centering
\iffigure
\epsfig{file=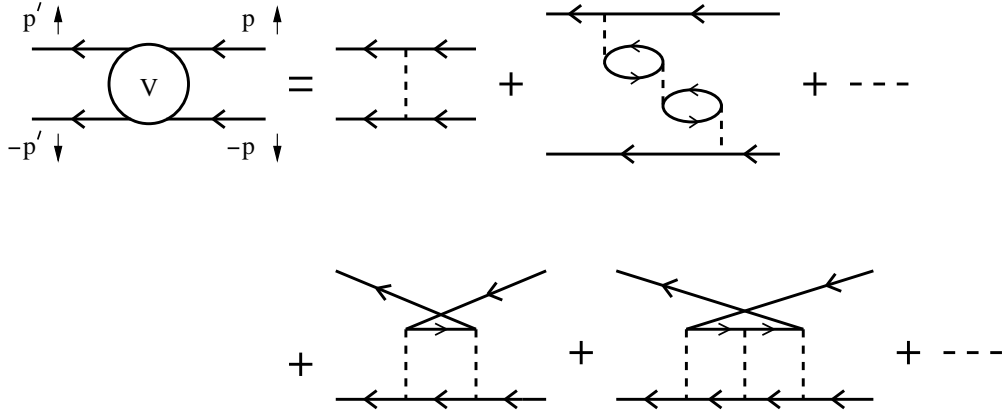,height=5.5cm,angle=0}
\fi
\caption{
Feynman diagrams illustrating the effective 
particle-particle interaction $V({\bf p'}|{\bf p})$ 
within the single spin-fluctuation exchange approximation. 
}
\label{2.1}
\end{figure}

\begin{figure}
\centering
\iffigure
\epsfysize=8cm
\epsffile[100 150 550 610]{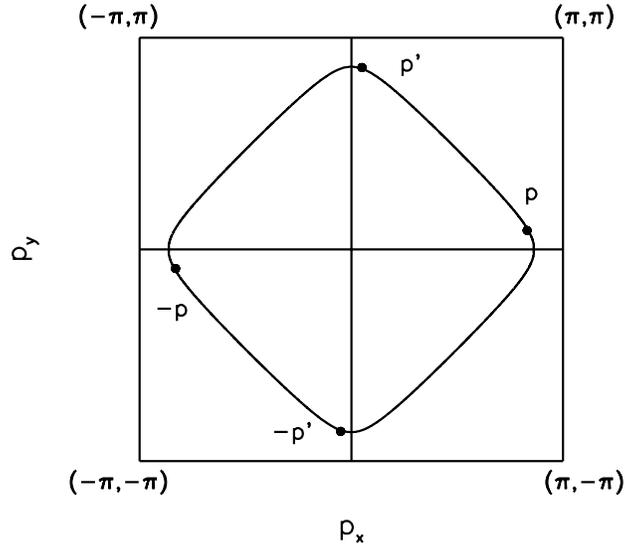}
\fi
\caption{
Sketch of the Fermi surface of the 2D tight-binding 
model doped near half-filling.
The two-particle momentum states $({\bf p},-{\bf p})$
and $({\bf p'},-{\bf p'})$ are indicated by the filled circles.
}
\label{2.2}
\end{figure}

\subsection{Experimental evidence for $d_{x^2-y^2}$-wave superconductivity
in the cuprates} 

Early experimental evidence for 
$d_{x^2-y^2}$-wave pairing in the cuprates arose from the 
measurement of the NMR longitudinal nuclear relaxation 
rate $T_1^{-1}$ for Cu(2) and O(2,3) nuclei in 
YBa$_2$Cu$_3$O$_7$.
In particular,
the anisotropy of $T_1^{-1}$ for Cu(2) 
below $T_c$ provided a signature for $d_{x^2-y^2}$-wave pairing
in this compound 
[Takigawa {\it et al.} 1991, 
Martindale {\it et al.} 1992, Bulut and Scalapino 1992].
These NMR measurements provided evidence for both the phase 
and the nodes of the $d_{x^2-y^2}$-wave gap function.
The temperature dependence of the Cu(2) transverse 
nuclear relaxation rate $T_2^{-1}$ was also used for the 
identification of the gap symmetry
[Bulut and Scalapino 1991, Itoh {\it et al.} 1992].
These NMR measurements were also supported by the 
ARPES experiments which extracted the 
magnitude of the gap function
$|\Delta_{\bf p}|$ on the Fermi surface
[Shen {\it et al.} 1993].
The microwave cavity measurements of the superconducting 
penetration depth $\lambda(T)$ in single crystals of 
YBa$_2$Cu$_3$O$_7$ found a linear temperature variation which also gave 
support to a $d_{x^2-y^2}$-wave gap [Hardy {\it et al.} 1993].
The phase-coherence measurements on 
YBCO-Pb dc SQUIDS [Wollman {\it et al.} 1993]
and the tricrystal experiments of Tsuei {\it et al.}
[Tsuei {\it et al.} 1994] established that in 
YBa$_2$Cu$_3$O$_7$ the order parameter has the 
$d_{x^2-y^2}$-wave symmetry.
Reviews of these results on the symmetry of the gap function 
are given in Refs.~[Schrieffer 1994, Scalapino 1995, 
van Harlingen 1995, Tsuei and Kirtley 2000].
Today, it is established that in a number of the hole 
and the electron doped cuprates the gap function has the 
$d_{x^2-y^2}$-wave symmetry [Tsuei and Kirtley 2000].

\subsection{Theoretical studies of $d_{x^2-y^2}$-wave pairing
in the cuprates} 

For the high-$T_c$ cuprates, 
the possibility of $d_{x^2-y^2}$-wave 
superconductivity was first studied by Bickers {\it et al.} 
[Bickers {\it et al.} 1987]
using the framework of the 2D Hubbard model.
In particular, the single spin-fluctuation exchange 
interaction was used for studying the $d_{x^2-y^2}$-wave pairing 
near an SDW instability within the random-phase approximation.
In these calculations, the effect of the single-particle self-energy 
corrections due to the spin fluctuations was taken into account.

The next level of calculations for 
the $d_{x^2-y^2}$-wave superconductivity 
were carried out using the fluctuation-exchange approximation (FLEX)
within the 2D Hubbard model 
[Bickers {\it et al.} 1989].
This method self-consistently treats the fluctuations in the 
magnetic, density and the pairing channels.
Within this approach it was found that the AF spin fluctuations
lead to a $d_{x^2-y^2}$-wave superconducting phase which
neighbors the SDW phase.
In these calculations, superconducting transition temperatures 
as high as $0.025t$ were found 
[Bickers {\it et al.} 1989, Bickers and White 1991].
Since the hopping matrix element within a one-band description 
of the cuprates is estimated to be about 0.45~eV
[Hybertson {\it et al.} 1990], this value of $T_c$ corresponds to 
about 130~K.
However, the need to carry out exact calculations 
was also noted.

Another approach to $d_{x^2-y^2}$-wave pairing in the cuprates 
was from a phenomenological point of view
[Moriya {\it et al.} 1990].
In this approach, a spin-fluctuation exchange 
interaction was used to calculate the 
superconducting transition temperatures. 
The spin susceptibility and the parameters 
used in the model were obtained from
fitting the NMR and the electrical resistivity 
data on the normal state of the cuprates
within the self-consistent renormalization theory.
In these calculations, 
the single-particle self-energy corrections due to the 
spin fluctuations were also included, and $T_c$'s as high as 
those in the cuprates were obtained.
A review of these calculations can be found in 
Ref.~[Moriya and Ueda 2000].

A similar phenomenological approach was taken 
in Ref.~[Monthoux {\it et al.} 1991].
In this approach, the spin susceptibility used in the 
effective interaction was also taken from fitting the NMR data, 
and $T_c$'s comparable to those in the cuprates were obtained. 
In these calculations, the $T_c$ was initially calculated without 
including the self-energy effects. 
Later, the self-energy corrections were included self-consistently
[Monthoux and Pines 1992].

These spin-fluctuation theories for superconductivity differ from 
phonon-mediated superconductivity in a fundamental way.
Here, the effective attractive interaction which 
is responsible for pairing the electrons is generated 
by the electronic correlations rather than by an external system
such as the lattice vibrations. 
Hence, it is an important question whether in a microscopic model
the electronic correlations can indeed produce 
an effective attractive interaction which leads to 
superconductivity, and, if so, what would be 
the nature and the strength 
of such pairing correlations.
The QMC and the DMRG methods were developed to address these
types of questions.
The numerical studies of the Hubbard model carried out 
with this purpose are the subject of this review article.

\setcounter{equation}{0}\setcounter{figure}{0}
\section{Magnetic correlations in the 2D Hubbard model}

The parent compounds of the cuprates are 
AF insulators, 
and the AF phase extends up to $\sim $5\% doping. 
For dopings beyond this, there is clear evidence from the 
NMR and the inelastic neutron scattering experiments that the 
system exhibits short-range AF spin fluctuations
[Pennington and Slichter 1990, Birgeneau 1990].
These are properties which support 
using the Hubbard framework for studying the 
electronic correlations of the cuprates.

The QMC calculations have found that 
the ground state of the Hubbard model at half-filling has 
AF long-range order 
[Hirsch and Tang 1989, White {\it et al.} 1989b].
Away from half-filling, the QMC calculations 
find short range AF correlations [Hirsch 1985].
In this section, 
the QMC results on the magnetic susceptibility 
$\chi$ of the 2D Hubbard model will be reviewed, and 
compared with an RPA-like approach for modelling it.
In addition, the QMC results on the irreducible 
particle-hole vertex will be discussed.
These will be useful for gaining a microscopic
understanding of the magnetic correlations in the Hubbard model.

\begin{figure}[ht]
\centering
\iffigure
\mbox{
\subfigure{
\epsfysize=8cm
\epsffile[100 150 480 610]{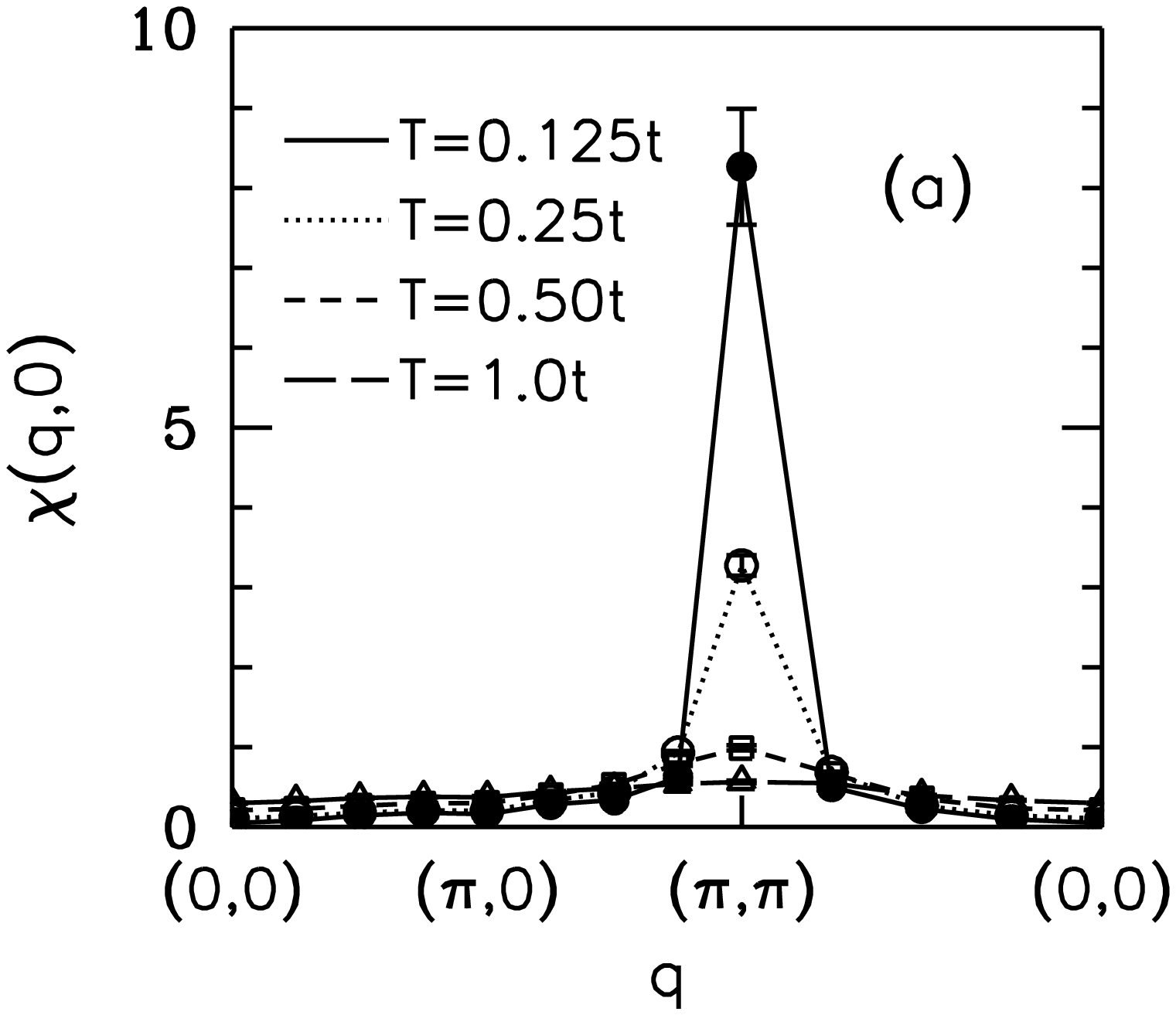}}
\quad
\subfigure{
\epsfysize=8cm
\epsffile[50 150 600 610]{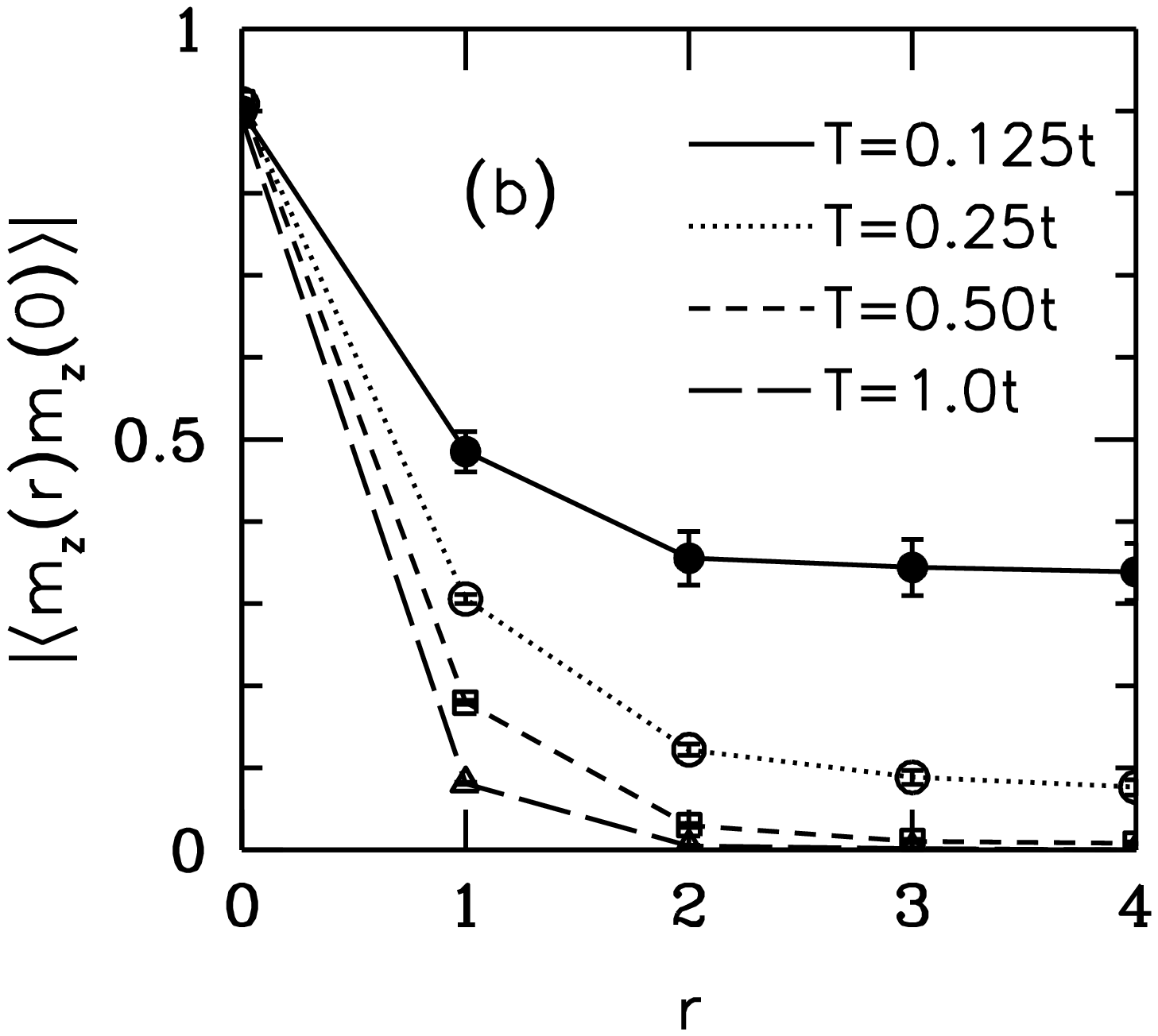}}}
\fi
\caption{
(a) Static magnetic susceptibility $\chi({\bf q},0)$ versus
${\bf q}$.
(b) Equal-time correlation function 
$|\langle m^z({\bf r})m^z(0)\rangle |$
versus ${\bf r}$ along $(1,0)$.
These results are for $U=8t$ 
and $\langle n\rangle=1.0$ on an $8\times 8$ lattice.
}
\label{3.1}
\end{figure}

\subsection{Antiferromagnetic long-range order at half-filling} 

The transverse magnetic susceptibility is defined by
\begin{equation}
\label{chiqw}
\chi({\bf q},\omega) = 
{1\over N} \sum_{\ell} \int_0^{\beta} \, d\tau \,
e^{i\omega_m\tau} \, e^{-i{\bf q}\cdot \ell} \,
\langle m^-(i+\ell,0) m^+(i,0) \rangle .
\end{equation}
Here, 
$m^+(i,0) = c^{\dagger}_{i\uparrow} c_{i\downarrow}$,
and $m^-(i,\tau) = e^{H\tau} m^-(i,0) e^{-H\tau}$,
where $m^-(i,0)$ is the hermitian conjugate of 
$m^+(i,0)$.
Ref.~[Mahan 1981] can be consulted for the finite-temperature 
Green's function formalism.
In the following, $\chi$ will be plotted in units of $t^{-1}$.
Figure 3.1(a) shows 
$\chi({\bf q},i\omega_m=0)$ versus ${\bf q}$ 
along various cuts in the Brillouin zone
as the temperature is lowered.
Here,
it is seen that as $T$ is lowered, 
$\chi({\bf q},0)$ develops a sharp peak at the antiferromagnetic 
wave vector ${\bf q}=(\pi,\pi)$.

Figure 3.1(b) shows the equal-time correlation function 
\begin{equation}
\label{mz}
|\langle m^z({\bf r}) m^z(0) \rangle |,
\end{equation}
where 
$m^z({\bf r}) = 
c^{\dagger}_{{\bf r}\uparrow} c_{{\bf r}\uparrow}
- c^{\dagger}_{{\bf r}\downarrow} c_{{\bf r}\downarrow}$,
versus ${\bf r}$ as the temperature is lowered on an $8\times 8$
lattice at half-filling. 
In this figure, it is seen that as $T$ is lowered below $0.5t$,
the AF correlation length $\xi_{AF}$ reaches 
the size of the system and long-range AF order is established 
on the $8\times 8$ lattice. 
For an infinite lattice, the long-range AF order 
would take place at $T=0$.

\begin{figure}[ht]
\centering
\iffigure
\mbox{
\subfigure{
\epsfysize=8cm
\epsffile[100 150 480 610]{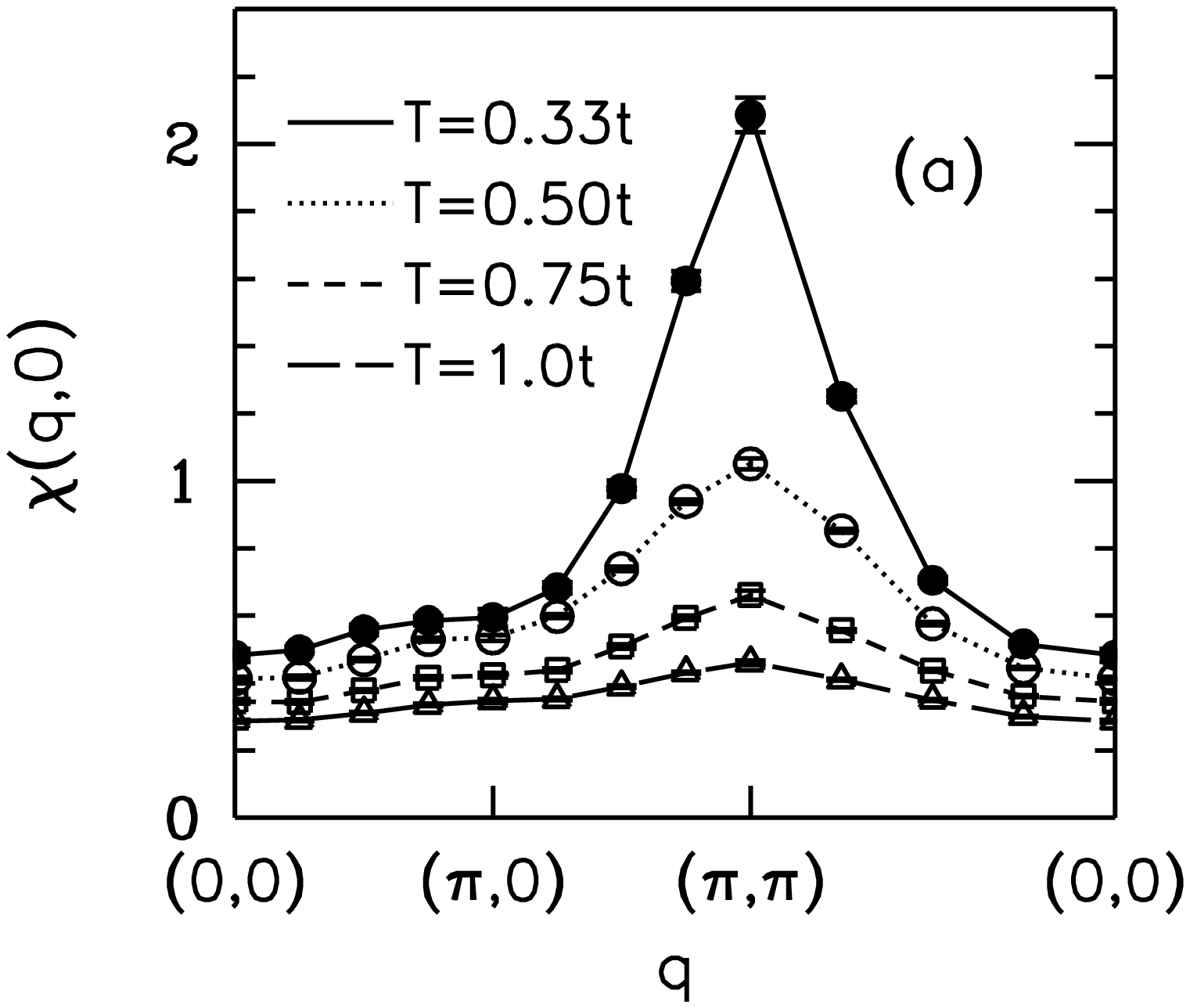}}
\quad
\subfigure{
\epsfysize=8cm
\epsffile[50 150 600 610]{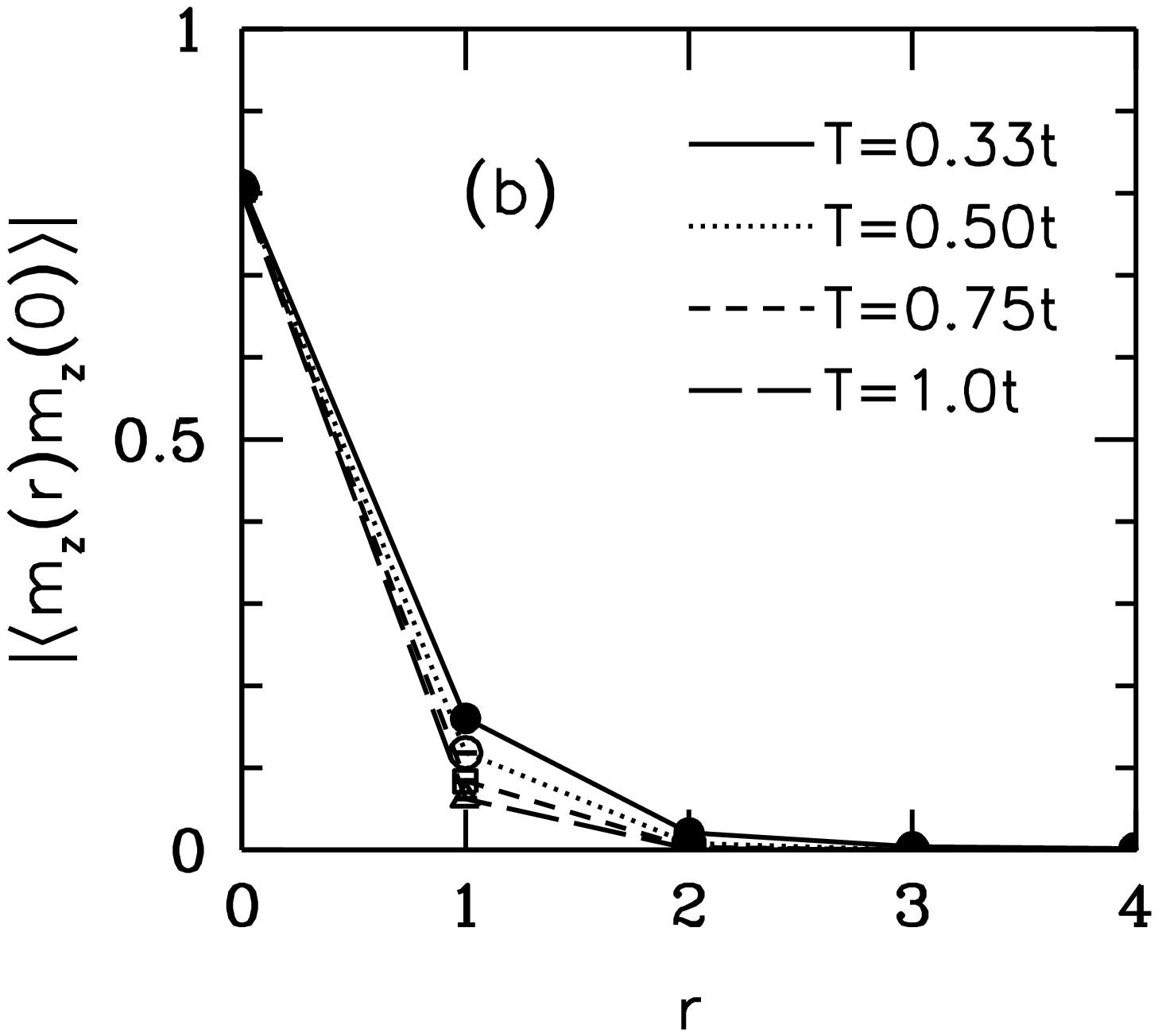}}}
\fi
\caption{
(a) Static magnetic susceptibility $\chi({\bf q},0)$ versus
${\bf q}$.
(b) Equal-time correlation function 
$|\langle m^z({\bf r})m^z(0)\rangle |$
versus ${\bf r}$ along $(1,0)$.
These results are for $U=8t$ and  
$\langle n\rangle=0.87$ on an $8\times 8$ lattice.
}
\label{3.2}
\end{figure}

\subsection{Short-range antiferromagnetic fluctuations 
away from half-filling}

Figure 3.2(a) shows $\chi({\bf q},0)$ 
versus ${\bf q}$ for $\langle n\rangle =0.87$ 
and $U=8t$ on the $8\times 8$ lattice.
Here, as $T$ decreases, 
the development of a broad peak centred at 
${\bf q}=(\pi,\pi)$ is seen. 
Figure 3.2(b) shows the corresponding 
$|\langle m^z({\bf r}) m^z(0) \rangle |$ 
versus ${\bf r}$. 
At the lowest temperature of 0.33t, 
the AF correlation length $\xi$ is about 0.6 
lattice spacing for this filling. 
Here, $\xi$ was obtained by fitting the decay 
of $|\langle m^z({\bf r})m^z(0)\rangle |$ 
to $e^{-|{\bf r}|/\xi}$.

The Monte Carlo results on $\chi({\bf q},i\omega_m)$ 
have been analytically continued to the real frequency axis 
using the Pade approximation.
The results on the spin-fluctuation spectral weight 
${\rm Im}\,\chi({\bf q},\omega)$ versus $\omega$
at ${\bf q}=(\pi,\pi)$ obtained in 
this way are shown in Fig.~3.3(a). 
While the Pade approximation can only resolve the crude features of the 
spectrum, the softening of the AF spin fluctuations at low temperatures
is clearly seen. 

\begin{figure}[ht]
\centering
\iffigure
\mbox{
\subfigure{
\epsfysize=8cm
\epsffile[100 150 480 610]{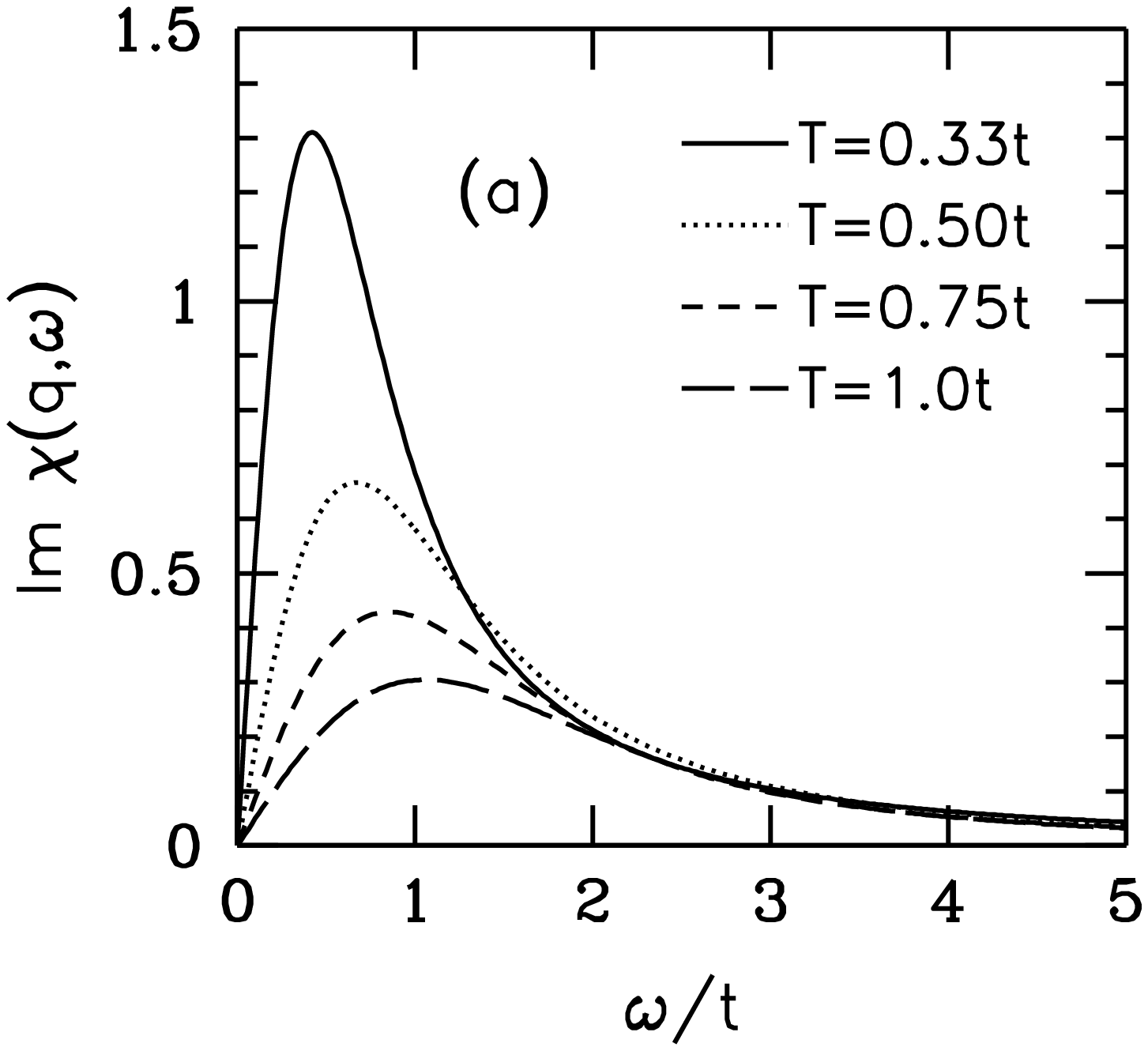}}
\quad
\subfigure{
\epsfysize=8cm
\epsffile[50 150 600 610]{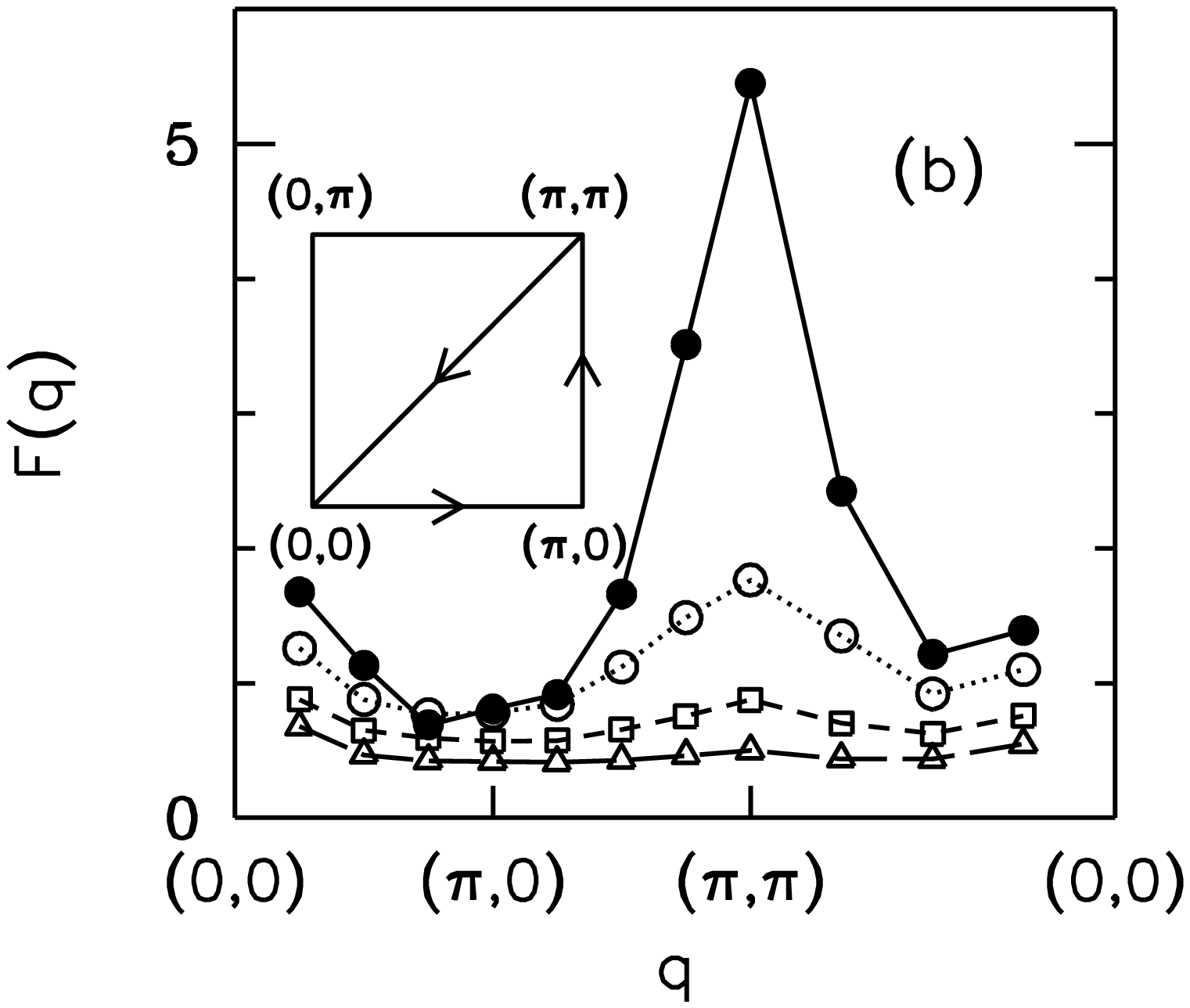}}}
\fi
\caption{
(a) Spin-fluctuation spectral weight 
${\rm Im}\,\chi({\bf q},\omega)$
versus $\omega$ at ${\bf q}=(\pi,\pi)$.
(b) $F({\bf q})$ versus ${\bf q}$
for the same temperatures as in (a).
These results are for $U=8t$ and  
$\langle n\rangle=0.87$ on an $8\times 8$ lattice.
}
\label{3.3}
\end{figure}

Another quantity which can be used for characterising 
the low-frequency magnetic correlations is 
\begin{equation}
\label{Fq}
F({\bf q}) = 
{ {\rm Im}\,\chi({\bf q},\omega) \over \omega}
\bigg|_{\omega\rightarrow 0}.
\end{equation}
This quantity is of interest because it is used to calculate 
the longitudinal nuclear relaxation rates of the 
planar Cu and O nuclear spins 
in the layered cuprates.
Figure 3.3(b) shows $F({\bf q})$ versus ${\bf q}$ for 
$\langle n\rangle =0.87$ and $U=8t$.
As $T$ is lowered, 
$F({\bf q})$ gets enhanced at $(\pi,\pi)$, and more weakly at 
${\bf q}\sim 0$.

\subsection{RPA-like model for the magnetic susceptibility}

An RPA-like approach has been used for modelling
$\chi({\bf q},\omega)$ 
in the intermediate coupling regime with the form
[Bulut {\it et al.} 1990, Bulut 1990, Chen {\it et al.} 1991, 
Bulut {\it et al.} 1993]
\begin{equation}
\label{chirpa}
\chi({\bf q},\omega) = 
{ \chi_0({\bf q},\omega) \over 
1 - \overline{U}\chi_0({\bf q},\omega) }
\end{equation}
where $\chi_0({\bf q},\omega)$ is the Lindhard function for the 
band electrons 
\begin{equation}
\label{lindhard}
\chi_0({\bf q},\omega) = 
{1 \over N} \sum_{\bf p}
{ 
f(\varepsilon_{{\bf p}+{\bf q}}) - f(\varepsilon_{\bf p}) 
\over 
\omega - ( \varepsilon_{{\bf p}+{\bf q}} - \varepsilon_{\bf p} ) + i\delta 
},
\end{equation}
$\varepsilon_{\bf p}=-2t(\cos{p_x} + \cos{p_y})-\mu$
and $f(\varepsilon_{\bf p})$ is the usual Fermi factor.
The parameter $\overline{U}$ entering Eq.~(\ref{chirpa})
represents the reduced Coulomb repulsion, which is assumed to 
be renormalized due to the electronic correlations.
In this simple approach, $\overline{U}$ is taken to be 
independent of momentum and temperature.
This turns out to be a reasonable approximation in the 
temperature range where the Monte Carlo simulations 
have been carried out.

Figure 3.4(a) compares this RPA-like approximation for 
$\chi({\bf q},0)$ with the Monte Carlo data for 
$U=4t$, $T=0.25t$ and $\langle n\rangle =0.87$. 
A similar comparison is given in Fig.~3.4(b) for the 
$T$ dependence of $\chi({\bf q}=(\pi,\pi),0)$.
In these figures $\overline{U}=2t$ has been used.
This choice will be discussed further in the following subsection,
where results on the irreducible particle-hole vertex 
will be reviewed.

\begin{figure}[ht]
\centering
\iffigure
\mbox{
\subfigure{
\epsfysize=8cm
\epsffile[100 150 480 610]{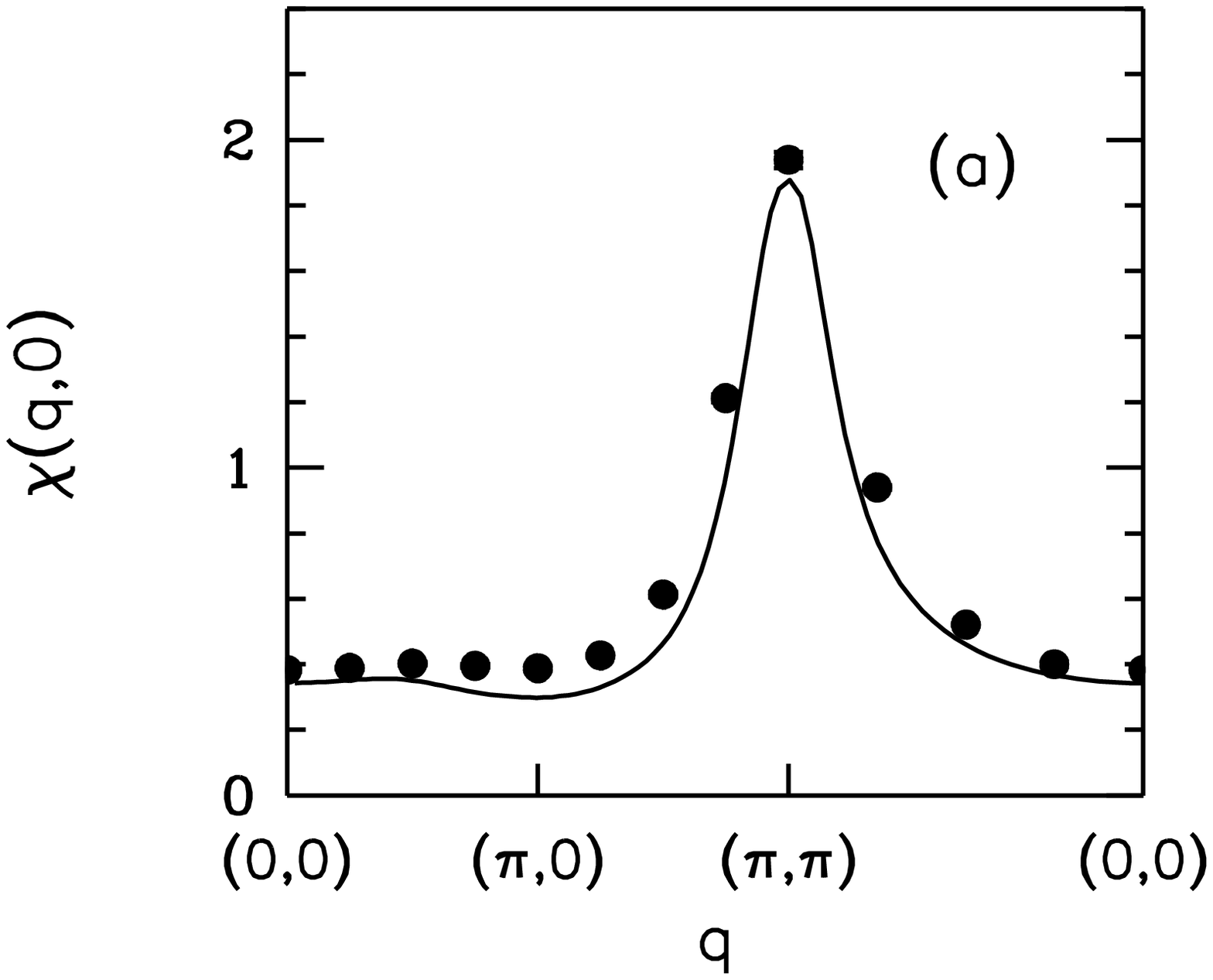}}
\quad
\subfigure{
\epsfysize=8cm
\epsffile[50 150 600 610]{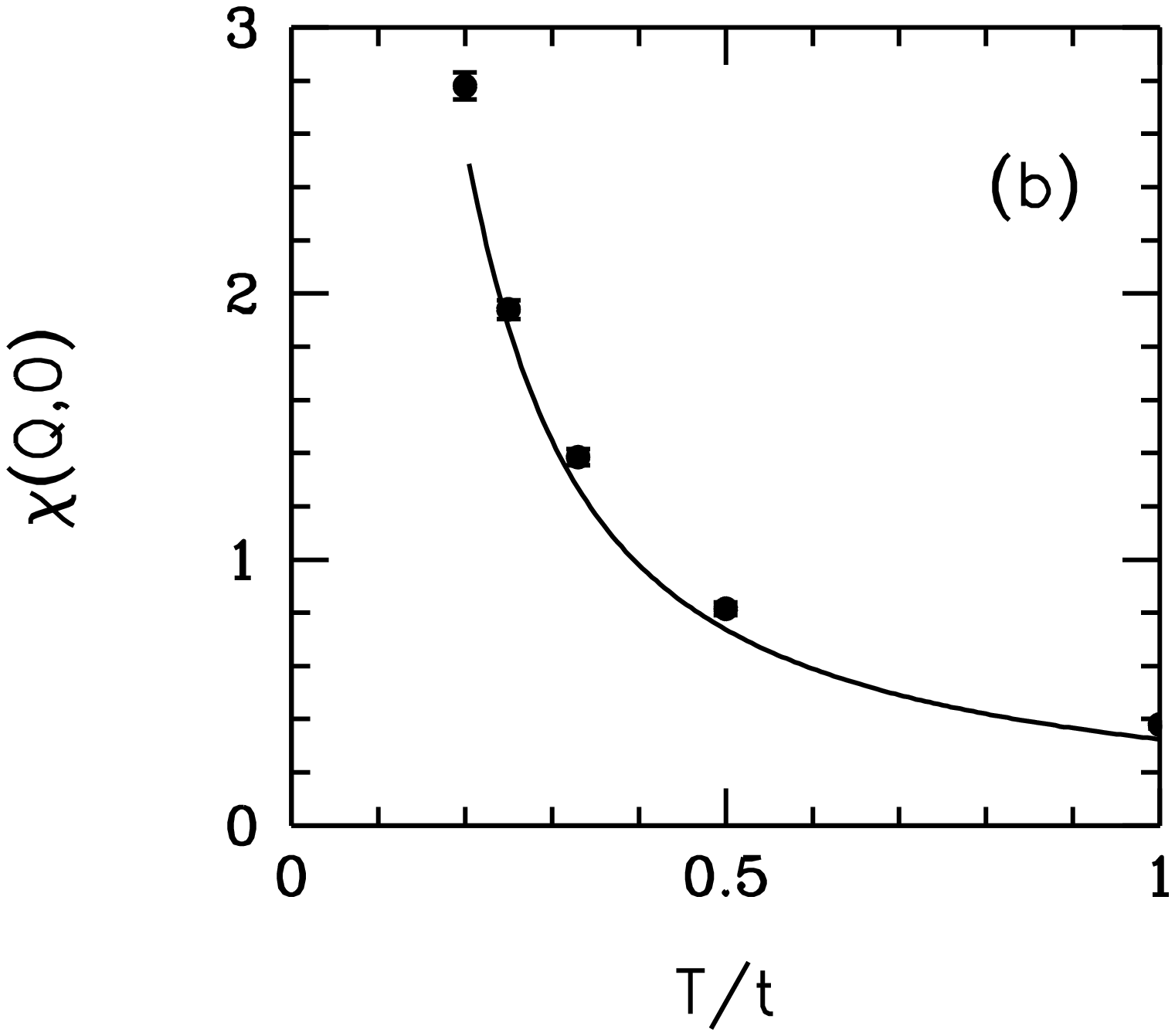}}}
\fi
\caption{
(a) Comparison of the QMC and the RPA 
results on $\chi({\bf q},0)$ versus ${\bf q}$ at $T=0.25t$.
(b) Comparison of the QMC and the RPA 
results on $\chi({\bf Q}=(\pi,\pi),0)$
versus $T/t$.
Here, the QMC data (filled circles) are for $U=4t$ and 
the RPA results (solid curves) were obtained 
using $\overline{U}=2t$.
In both figures, 
the results are shown for $\langle n\rangle=0.87$.
}
\label{3.4}
\end{figure}

\subsection{Irreducible particle-hole interaction}

In the RPA form discussed above, 
a reduced Coulomb repulsion $\overline{U}$ which is 
independent of ${\bf q}$ is used. 
In order to test this approximation, 
the effective particle-hole vertex defined by 
\begin{equation}
\overline{U}({\bf q},i\omega_m) = 
{1 \over \overline{\chi}({\bf q},i\omega_m)} - 
{1 \over \chi({\bf q},i\omega_m)},
\end{equation}
was calculated in Ref.~[Bulut {\it et al.} 1995].
Here, 
$\overline{\chi}({\bf q},i\omega_m)$ 
is the piece of the magnetic susceptibility 
which does not include the reducible particle-hole vertex,
and it is defined by 
\begin{equation}
\overline{\chi} ({\bf q},i\omega_m) = 
- {T\over N} \sum_{{\bf p},i\omega_m} 
G({\bf p}+{\bf q},i\omega_n + i\omega_m)
G({\bf p},i\omega_n),
\end{equation}
where $G({\bf p},i\omega_n)$ is the exact single-particle 
Green's function calculated with QMC.
In Fig.~3.5(a), 
the QMC data on $\overline{\chi}({\bf q},0)$ versus 
${\bf q}$ is compared with the Lindhard function
$\chi_0({\bf q},0)$ at $T=0.25t$.
Here, 
one sees that $\overline{\chi}({\bf q},0)$ 
is suppressed with respect to $\chi_0({\bf q},0)$,
and this is 
because of the single-particle self-energy effects.
The effective irreducible vertex 
$\overline{U}({\bf q},0)$ obtained 
from the QMC data on $\chi({\bf q},0)$ and 
$\overline{\chi}({\bf q},0)$ is plotted in Fig.~3.5(b) 
as a function of ${\bf q}$ at various temperatures.
In this figure it is seen that 
$\overline{U}({\bf q},0)$ shows little dependence on 
${\bf q}$ and $T$.
This is the reason for the good agreement between the QMC data 
and the RPA results obtained using a reduced $\overline{U}$
in Eq.~(3.4).
However, 
the values of $\overline{U}({\bf q},0)$ are larger than 
$\overline{U}=2t$ used in the RPA.
For instance, at $T=0.25t$ one has 
$\overline{U}({\bf q}=(\pi,\pi),0)=3.2t$.
This is because 
in the RPA form the Lindhard susceptibility $\chi_0$ is used rather 
than $\overline{\chi}$, which has the effects 
of the single-particle self-energy corrections. 
At this point, it should be noted that 
the momentum dependence of the effective particle-hole
vertex within the context of the cuprates was also
discussed in Ref.~[Anderson 1997].

\begin{figure}
\centering
\iffigure
\mbox{
\subfigure{
\epsfysize=8cm
\epsffile[100 150 480 610]{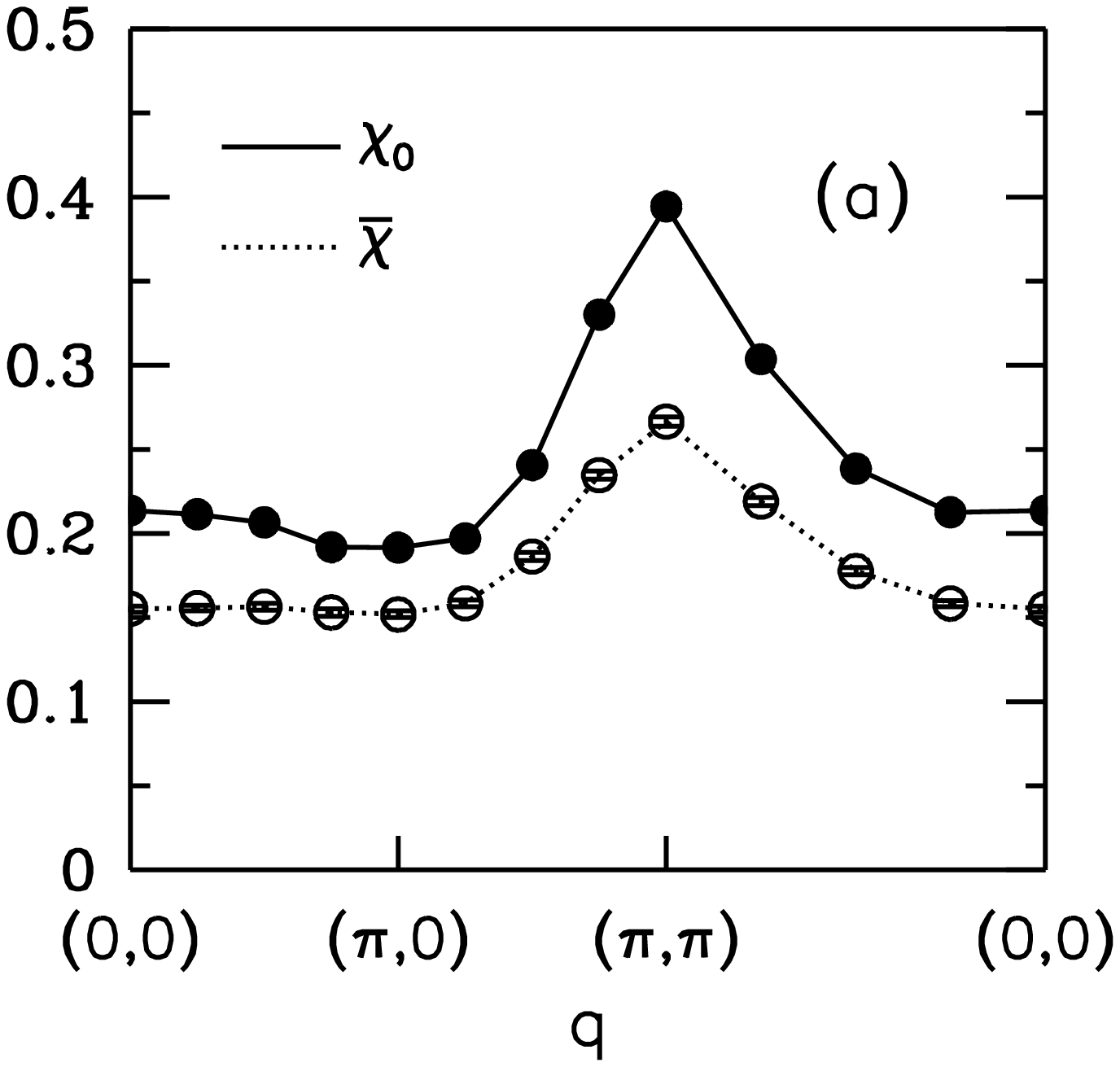}}
\quad
\subfigure{
\epsfysize=8cm
\epsffile[50 150 600 610]{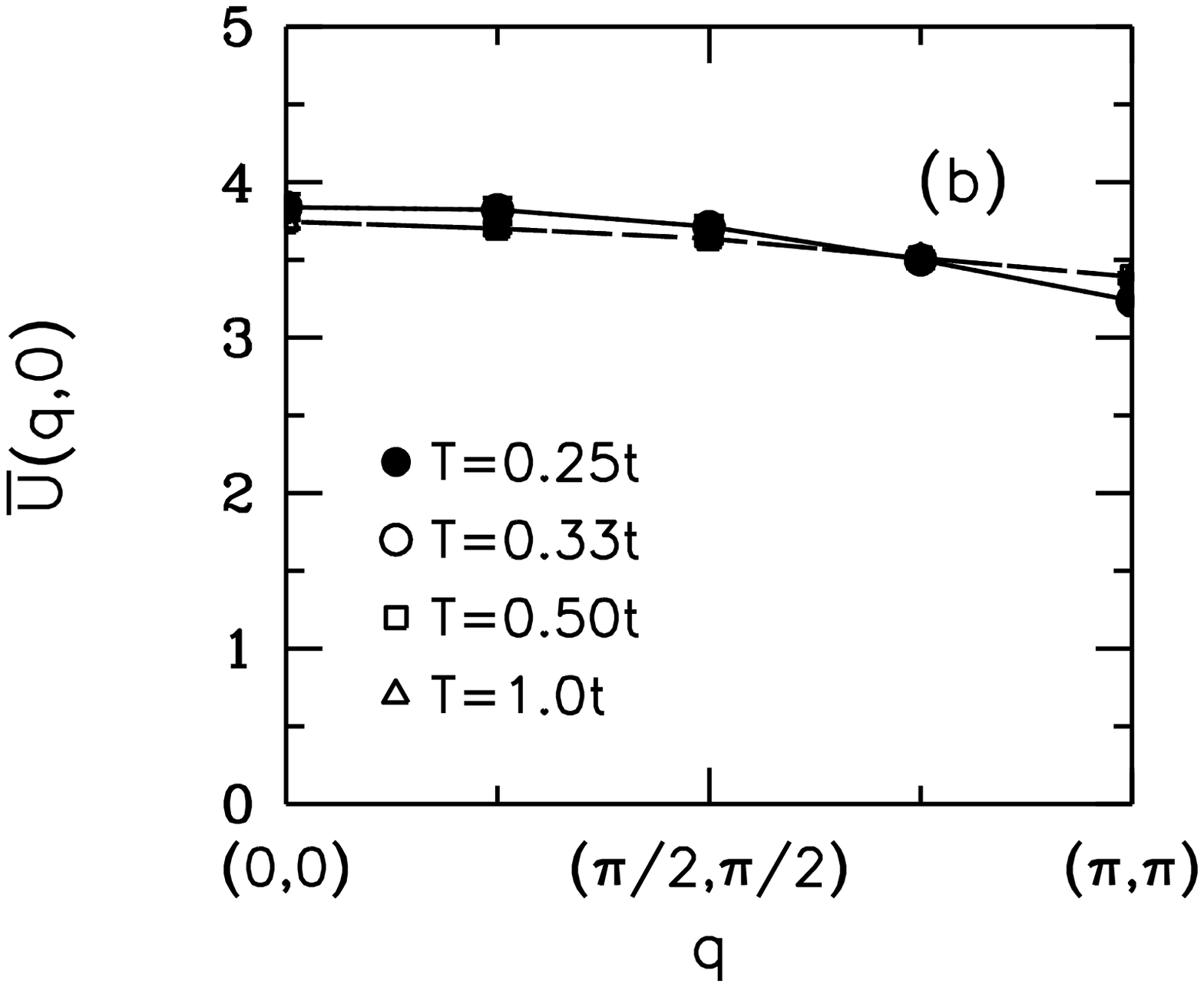}}}
\fi
\caption{
(a) Comparison of the QMC data on $\overline{\chi}({\bf q},0)$
versus ${\bf q}$ with the Lindhard function $\chi_0({\bf q},0)$ 
at $T=0.25t$.
(b) Effective particle-hole vertex
$\overline{U}({\bf q},0)$ versus ${\bf q}$ along $(1,1)$.
These results were obtained for 
$U=4t$ and $\langle n\rangle =0.87$ on an 
$8\times 8$ lattice.
}
\label{3.5}
\end{figure}

With the QMC simulations the irreducible 
particle-hole vertex $\overline{\Gamma}_I$ was also calculated
[Bulut {\it et al.} 1995].
The calculation of $\overline{\Gamma}_I$ is similar to 
that of the irreducible particle-particle vertex $\Gamma_I$,
which will be discussed in Section~6.
In this approach, $\overline{\Gamma}_I$ is 
obtained from the solution of the particle-hole $t$-matrix
equation illustrated in Fig.~3.6.
These calculations showed 
that the QMC data on $\overline{\Gamma}_I$
are in agreement with the results on the 
effective particle-hole vertex 
$\overline{U}({\bf q},0)$ discussed in this section.
In particular, $\overline{\Gamma}_I$ with the center-of-mass
momentum ${\bf q}=(\pi,\pi)$ and frequency $\omega_m=0$
was averaged over the incoming and outgoing 
momenta near the Fermi surface, 
and it was found that this average 
$\langle \overline{\Gamma}_I\rangle$ is $3.6t \pm 0.7t$
for $U=4t$, $T=0.25t$ and $\langle n\rangle=0.87$.
This agrees with 
$\overline{U}({\bf q}=(\pi,\pi),0)=3.2t$
found above.
Hence, the effective particle-hole vertex for 
center-of-mass momentum ${\bf q}=(\pi,\pi)$ and
frequency $\omega_m=0$ is about $3.2t$ for a bare Coulomb
repulsion of $U=4t$.
The difference reflects the renormalization of the bare
Coulomb repulsion in the particle-hole channel due to the 
higher-order many-body processes such as the Kanamori type 
of particle-particle scatterings
[Kanamori 1963].
However, in the RPA form of Eq.~(3.4), 
$\overline{U}=2t$ needs to be used in order to take 
into account the single-particle self-energy corrections 
which are neglected within RPA.

These results on the effective particle-hole vertex will 
also be useful when the irreducible particle-particle interaction
$\Gamma_I$ is compared with the single spin-fluctuation
exchange interaction $\Gamma^{SF}_{Is}$
in Section~6.4.
There,
it will be seen that when the effective coupling $gU$ 
between the electrons and the spin fluctuations is taken
to be about $3.2t$, the resulting $\Gamma^{SF}_{Is}$
agrees with the QMC data on $\Gamma_I$. 
One expects $gU$ to be closely related to 
$\overline{\Gamma}_I$ and, hence, 
the QMC data on the effective particle-particle 
and particle-hole vertices are consistent.

\begin{figure}
\centering
\iffigure
\epsfig{file=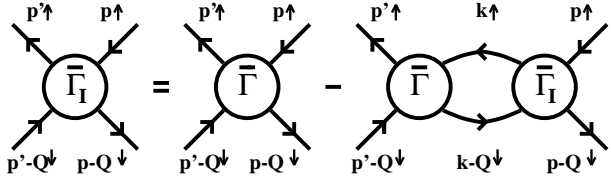,height=8cm,angle=90}
\fi
\caption{
Feynman diagrams for the particle-hole $t$-matrix equation.
}
\label{3.6}
\end{figure}

\subsection{Comparison with the NMR experiments on the cuprates}

The measurements of the longitudinal NMR rate 
$T_1^{-1}$ have provided evidence for the existence of 
short-range AF fluctuations in the normal state of the 
high-$T_c$ cuprates.
These experiments are reviewed in 
Refs.~[Pennington and Slichter 1990, Takigawa 1990].
The relaxation rate $T_1^{-1}$ is given by 
\begin{equation}
\label{T1}
T_1^{-1} = {T\over N} \sum_{\bf q} 
|A({\bf q})|^2 F({\bf q}),
\end{equation}
where $|A({\bf q})|^2$ is the hyperfine form factor of the 
particular nuclear spin, and $F({\bf q})$ has been defined in 
Section~3.2.
The hyperfine form factors 
used in calculating the various NMR rates for the cuprates 
were derived in Ref.~[Mila and Rice 1989].

An additional experimental quantity which supports the existence 
of the short-range AF correlations in the cuprates is the 
transverse relaxation rate $T_2^{-1}$, which is given by 
[Pennington and Slichter 1991]
\begin{equation}
\label{T2}
\bigg( 
{1\over T_2} \bigg)^2 =
{1\over N} \sum_{\bf q} 
|A({\bf q})|^4 \chi^2({\bf q},0)
- \bigg( {1\over N}\sum_{\bf q} 
|A({\bf q})|^2 \chi({\bf q},0) \bigg)^2.
\end{equation}
The measurements of $T^{-1}_2$ for $^{63}$Cu(2)
in the normal state of the YBa$_2$Cu$_3$O$_7$,
YBa$_2$Cu$_4$O$_8$ and the 
La$_{2-x}$Sr$_x$CuO$_4$ systems have 
provided evidence for the presence of the short-range 
AF correlations in these compounds.
[Pennington and Slichter 1991, Itoh {\it et al.} 1992, 
Imai {\it et al.} 1993].

Since the Monte Carlo calculations cannot be carried out at 
low temperatures, it is difficult to make direct comparisons with 
the $T_1^{-1}$ and the $T_2^{-1}$ measurements on the cuprates using 
the Monte Carlo data.
However, the RPA-like model described above in Section~3.3 
has been used to calculate the NMR rates in the normal state.
Even though it is a simple approximation,
this approach has been useful for analysing the 
$T_1^{-1}$ and the $T_2^{-1}$ data on 
YBa$_2$Cu$_3$O$_7$
[Bulut {\it et al.} 1990, 
Bulut and Scalapino 1991].
These calculations imply that the low-energy AF spin fluctuations 
estimated for the 2D Hubbard model have the right order of magnitude
for describing the general features of the AF correlations in 
optimally doped 
YBa$_2$Cu$_3$O$_7$.
The NMR data on YBa$_2$Cu$_3$O$_7$ were also fit 
phenomenologically
[Millis {\it et al.} 1990]
and by using self-consistent renormalization theory
[Moriya {\it et al.} 1990].

\setcounter{equation}{0}\setcounter{figure}{0}
\section{Charge fluctuations in the 2D Hubbard model}

In this section, the QMC data on the charge susceptibility 
and the optical conductivity of the 2D Hubbard model will be reviewed. 

\subsection{Charge susceptibility}

The momentum and frequency dependent charge susceptibility is defined by 
\begin{equation}
\label{Pi}
\Pi({\bf q},i\omega_m) = \int_0^{\beta} \, d\tau \,
e^{i\omega_m\tau} \, 
S({\bf q},\tau)
\end{equation}
where
\begin{equation}
S({\bf q},\tau) = -
\langle T \rho({\bf q},\tau) \rho^{\dagger} ({\bf q},0) \rangle ,
\end{equation}
with 
$\rho^{\dagger}({\bf q})= 
{1\over \sqrt{N}}  \sum_{{\bf p}\sigma} 
c^{\dagger}_{{\bf p}+{\bf q}\sigma} c_{{\bf p}\sigma}$
and $\rho({\bf q},\tau)= e^{H\tau} \rho({\bf q}) e^{-H\tau}$.
The equal-time density-density correlation function 
$S({\bf q})\equiv S({\bf q},\tau=0)$
and the charge compressibility 
$\kappa = \partial \langle n\rangle / \partial \mu$
given by $\Pi({\bf q}\rightarrow 0,0)$
are also of interest.
In the following $\Pi$ and $S$ 
will be plotted in units of $t^{-1}$.
The nature of the density correlations in the 2D Hubbard model was
studied with various numerical techniques.
The QMC 
[Moreo and Scalapino 1991, Moreo {\it et al.} 1991] 
and the exact diagonalization [Dagotto {\it et al.} 1992b] 
calculations did not find 
any indication of phase separation in the 2D Hubbard model.
The charge compressibility of the Hubbard model 
near half-filling was calculated 
by zero-temperature QMC simulations 
[Furukawa and Imada 1991 and 1992].
At fillings $\langle n\rangle=0.5$ and 0.2, 
$S({\bf q})$ was calculated with the determinantal QMC for 
$U=4t$ and $8t$ on up to $16\times 16$ lattices
[Chen {\it et al.} 1994].
The Green's function Monte Carlo (GFMC) method was also 
used for calculating $S({\bf q})$ for the 
2D Hubbard model on lattices with up to 162 sites in the ground state
[Becca {\it et al.} 2000].
These calculations were carried out for $U=4t$ and $10t$
at and near half-filling.
In this section, 
the QMC results on $\Pi({\bf q},\omega)$ and $S({\bf q})$
from Ref.~[Bulut 1996] will be shown
for various values of $U/t$ near half-filling.

Extensive numerical calculations have been 
also carried out for studying 
the density correlations in the ground state
of the $t$-$J$ model.
These calculations have found phase-separated, striped, and uniform
phases, and they will be briefly described in Section~8.
The results on $\Pi({\bf q},\omega)$ 
and $S({\bf q})$ presented here were obtained at relatively high
temperatures and it is difficult to reach conclusions about the 
ground state of the Hubbard model using them.
Nevertheless, here, the QMC data on 
the density correlations will be 
compared with the results on the $t$-$J$ model
obtained with the high-temperature series expansions
[Putikka {\it et al.} 1994].
This gives valuable information on the
dynamics of the density fluctuations in the 2D Hubbard model

The high-temperature series expansions have
found that the structure in $S({\bf q})$ 
can be described by the noninteracting spinless-fermion model.
This is due to the fact that the double-occupancy of the sites 
is not allowed in the $t$-$J$ model, and once this 
constraint is taken into account, the density 
correlations are similar to those of the noninteracting particles.
Consequently, for a filling $\langle n\rangle$, 
$S({\bf q})$ has structure at wave vectors which are 
associated with the Fermi surface corresponding to a filling 
of $2\langle n\rangle$.
In other words, 
the structure in $S({\bf q})$ follows the 
"$4{\bf k}_F$" wave vectors
rather than the "$2{\bf k}_F$" wave vectors
of the system.

First, results at half-filling will be presented.
Figure~4.1 shows QMC results on 
$\Pi({\bf q},0)$ versus ${\bf q}$ at various 
temperatures for $U=8t$.
Here, 
$\Pi({\bf q},0)$ is rather featureless in ${\bf q}$, and it gets 
further suppressed as $T$ is lowered.
This is due to the Mott-Hubbard gap which exists at 
half-filling.
Next, results for $\langle n\rangle=0.87$ are shown.
Figure~4.2(a) shows the temperature evolution of 
$\Pi({\bf q},0)$ versus ${\bf q}$ for $U=8t$.
In this case, the ${\bf q}\sim 0$ component of 
$\Pi({\bf q},0)$ gets enhanced as $T$ decreases.
The $T$ dependence of $\Pi({\bf q}\rightarrow 0,0)$ 
seen in Fig.~4.2(a) is consistent with the results of  
the zero-temperature QMC calculations
of the charge compressibility
[Furukawa and Imada 1991 and 1992].
Figure~4.2(b) shows ${\rm Im}\,\Pi({\bf q},\omega)$ versus
$\omega$ for ${\bf q}=(\pi/4,0)$,
where it is seen that the density fluctuations soften 
as $T$ is lowered.
The spectral weight 
${\rm Im}\,\Pi$ was obtained by the Pade analytic 
continuation of the QMC data.

The evolution of the ${\bf q}$ structure in $\Pi({\bf q},0)$ 
with $U/t$ is shown in Fig.~4.3(a) for $T=0.5t$
and $\langle n\rangle=0.87$.
For comparison, in Fig.~4.3(b) results  are shown for 
$\Pi_0({\bf q},i\omega_m=0)$ of the $U=0$ system
given by 
\begin{equation}
\label{P0}
\Pi_0({\bf q},i\omega_m) = 2
{1 \over N} \sum_{\bf p}
{ f(\varepsilon_{{\bf p}+{\bf q}}) - f(\varepsilon_{\bf p}) 
\over 
i\omega_m - ( \varepsilon_{{\bf p}+{\bf q}} 
- \varepsilon_{\bf p} ) + i\delta 
}.
\end{equation}
In this figure,
the solid curve represents $\Pi_0({\bf q},0)$ for 
$\langle n\rangle=0.87$ and the dotted curve is for 
$\langle n\rangle=1.74$, 
which corresponds to $\Pi({\bf q},0)$ 
of the noninteracting spinless-fermion system for
$\langle n\rangle=0.87$.
As $U/t$ increases, it is seen that 
$\Pi({\bf q},0)$ develops features which are more similar to 
those seen for the noninteracting spinless-fermion system.
In Fig.~4.3(a), 
it is seen that the peak in $\Pi({\bf q},0)$ shifts 
from ${\bf q}=(\pi,\pi)$ to ${\bf q}\sim 0$ as $U/t$ 
increases.
It is not possible to obtain this type of 
momentum dependence for $\Pi({\bf q},0)$ 
from the simple RPA form
\begin{equation}
\Pi_{\rm RPA}({\bf q},i\omega_m) =
{ \Pi_0({\bf q},i\omega_m) \over 
1 + {1\over 2} \overline{U} 
\Pi_0({\bf q},i\omega_m) },
\end{equation}
where $\overline{U}$ is a constant 
representing the effective coupling 
in the density channel.

\begin{figure}
\centering
\iffigure
\epsfysize=8cm
\epsffile[100 150 550 610]{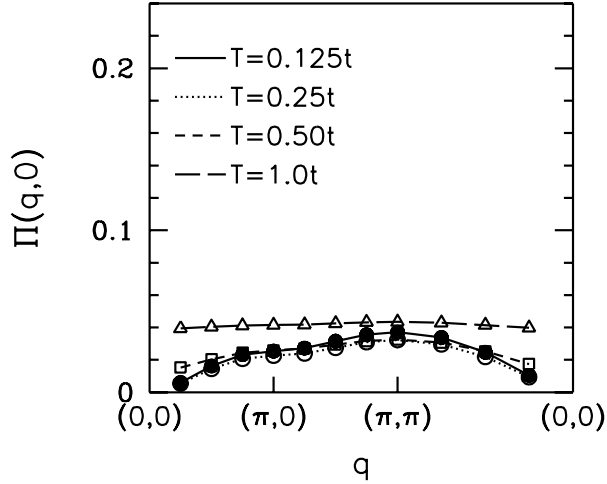}
\fi
\caption{
Momentum dependence of $\Pi({\bf q},0)$
at half-filling for $U=8t$ and various temperatures.
}
\label{4.1}
\end{figure}

\begin{figure}
\centering
\iffigure
\mbox{
\subfigure{
\epsfysize=8cm
\epsffile[100 150 480 610]{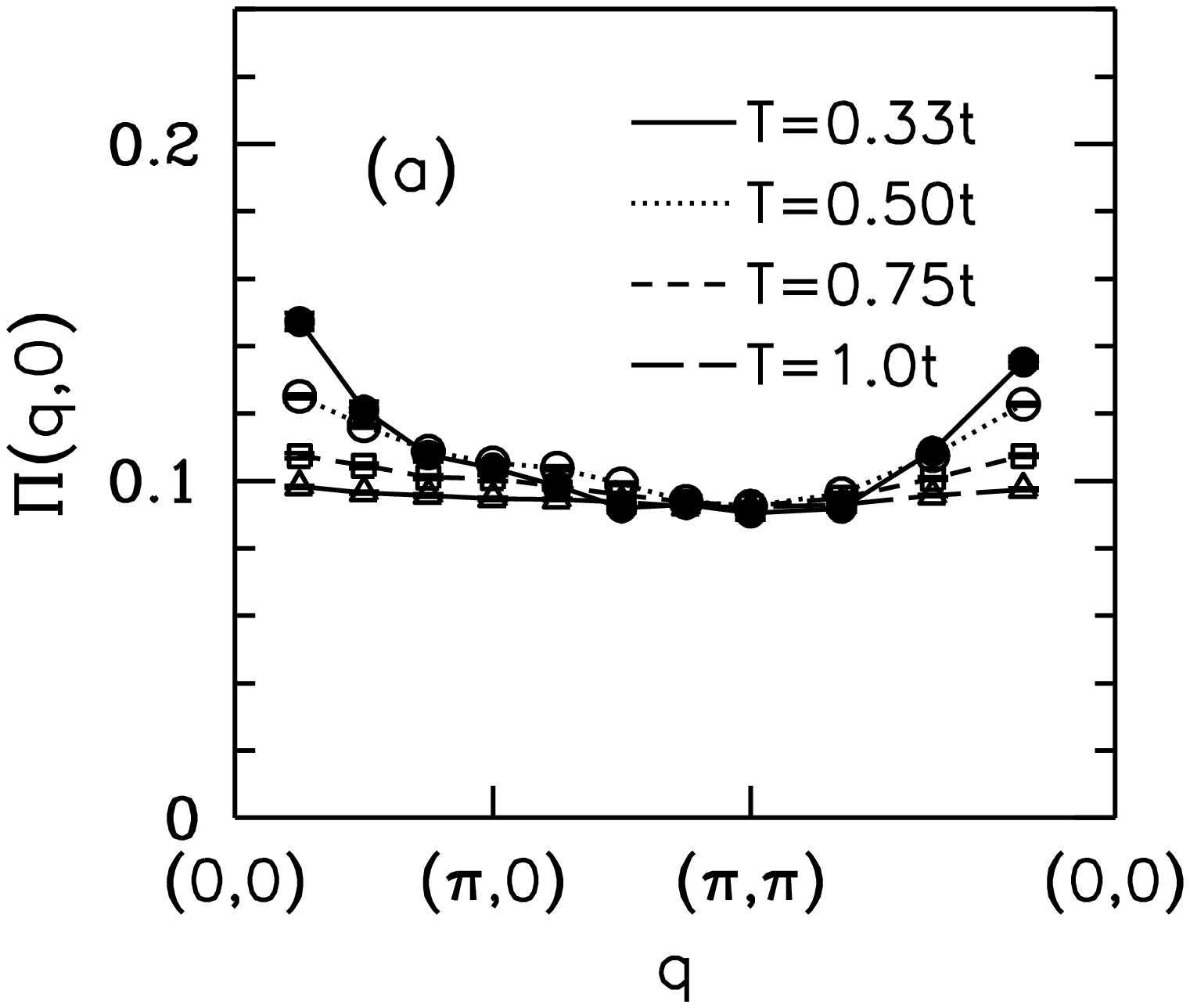}}
\quad
\subfigure{
\epsfysize=8cm
\epsffile[50 150 600 610]{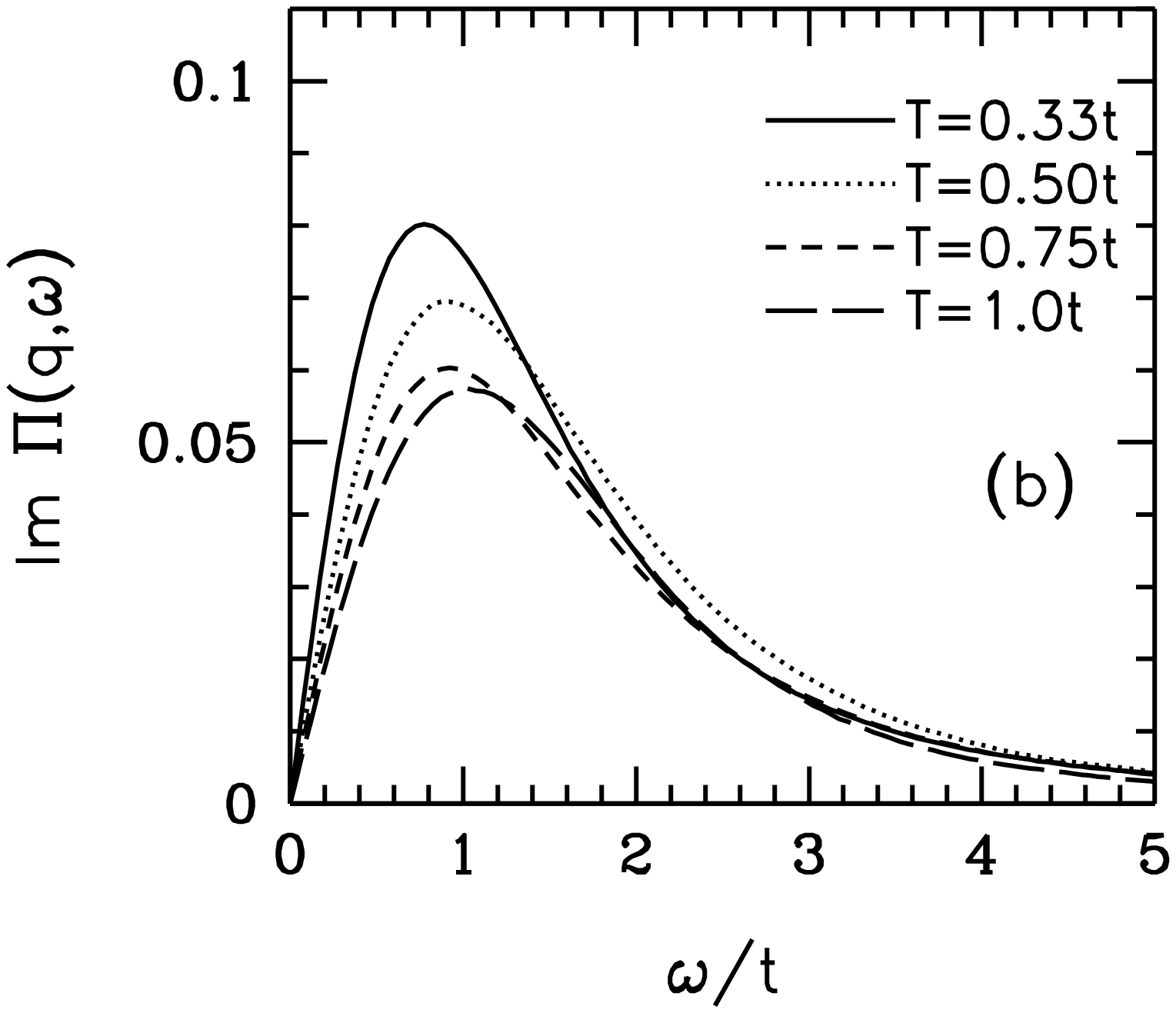}}}
\fi
\caption{
(a) Momentum dependence of $\Pi({\bf q},0)$.
(b) ${\rm Im}\,\Pi({\bf q},\omega)$ versus $\omega$
at ${\bf q}=(\pi/4,0)$.
These results are
for $\langle n\rangle=0.87$, $U=8t$ and various temperatures.
}
\label{4.2}
\end{figure}

\begin{figure}
\centering
\iffigure
\mbox{
\subfigure[]{
\epsfysize=8cm
\epsffile[100 150 480 610]{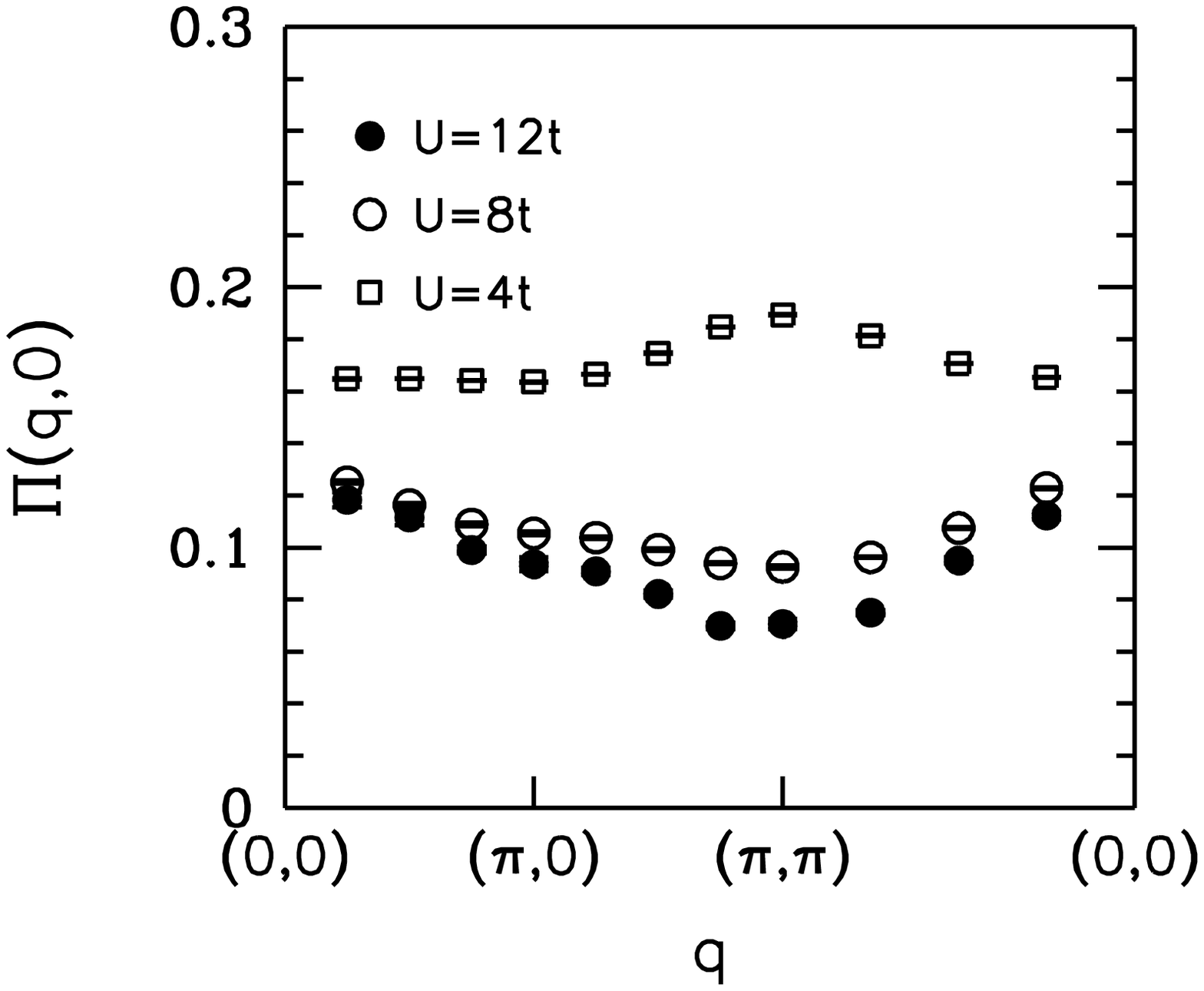}}
\quad
\subfigure[]{
\epsfysize=8cm
\epsffile[50 150 600 610]{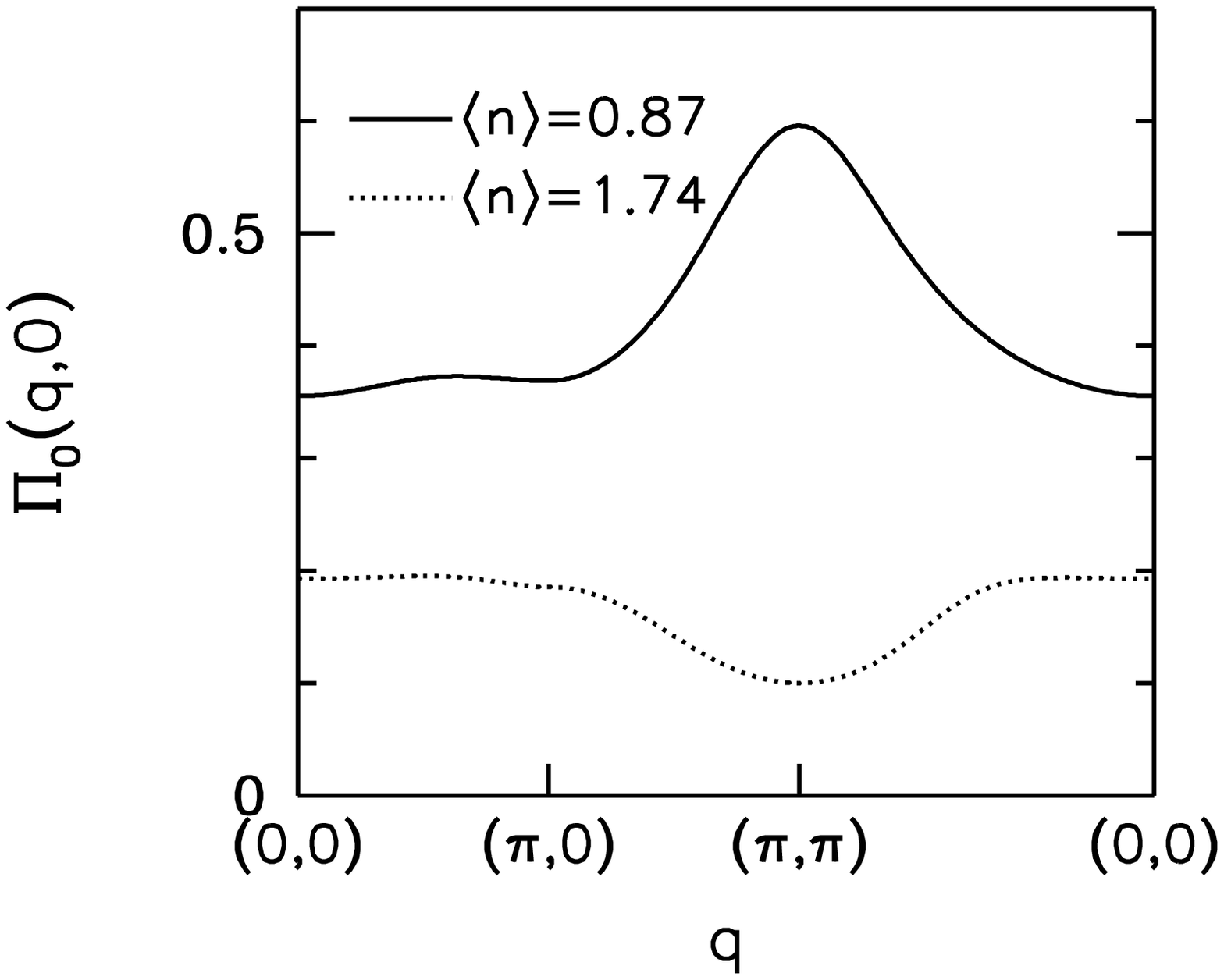}}}
\fi
\caption{
(a) $\Pi({\bf q},0)$ versus ${\bf q}$ 
at $\langle n\rangle=0.87$ 
for various values of $U$.
(b) $\Pi_0({\bf q},0)$ versus ${\bf q}$ 
for the $U=0$ system at
$\langle n\rangle=0.87$ (solid curve)
and $\langle n\rangle=1.74$ (dotted curve).
These results are for $T=0.5t$.
}
\label{4.3}
\end{figure}

Next, QMC results on $S({\bf q})$ will be discussed.
Figure~4.4 shows $S({\bf q})$ versus ${\bf q}$ at various 
temperatures for $U=8t$. 
Here, it is seen that 
$S({\bf q})$ exhibits small variation with $T/t$.
Figure~4.5(a) shows the evolution of $S({\bf q})$ with 
$U/t$ for $T=0.5t$ and $\langle n\rangle=0.87$.
Here, it is seen that with $U/t$ increasing, $S({\bf q})$ gets
suppressed.
However,
in contrast to the case of $\Pi({\bf q},0)$, 
there is no qualitative change in the ${\bf q}$ structure 
of $S({\bf q})$ as $U/t$ increases,
and the peak remains at $(\pi,\pi)$.
It is useful to compare these QMC results 
with $S_0({\bf q})$ of the noninteracting system.
Figure~4.5(b) shows $S_0({\bf q})$ versus ${\bf q}$ 
at $T=0.5t$ for $\langle n\rangle=0.87$ (solid curve)
and $\langle n\rangle=1.74$ (dotted curve).
Comparing Figs.~4.5(a) and (b), one observes that the QMC data on 
$S({\bf q})$ is more similar to $S_0({\bf q})$ for 
$\langle n\rangle=0.87$,
and the ${\bf q}$ structure in $S({\bf q})$ 
does not follow that of the 
noninteracting spinless-fermion system.
These results on $S({\bf q})$ are consistent with 
those found by the GFMC technique for the ground state
[Becca {\it et al.} 2000],
and by the QMC method at finite temperatures
[Chen {\it et al.} 1994].

\begin{figure}
\centering
\iffigure
\epsfysize=8cm
\epsffile[100 150 550 610]{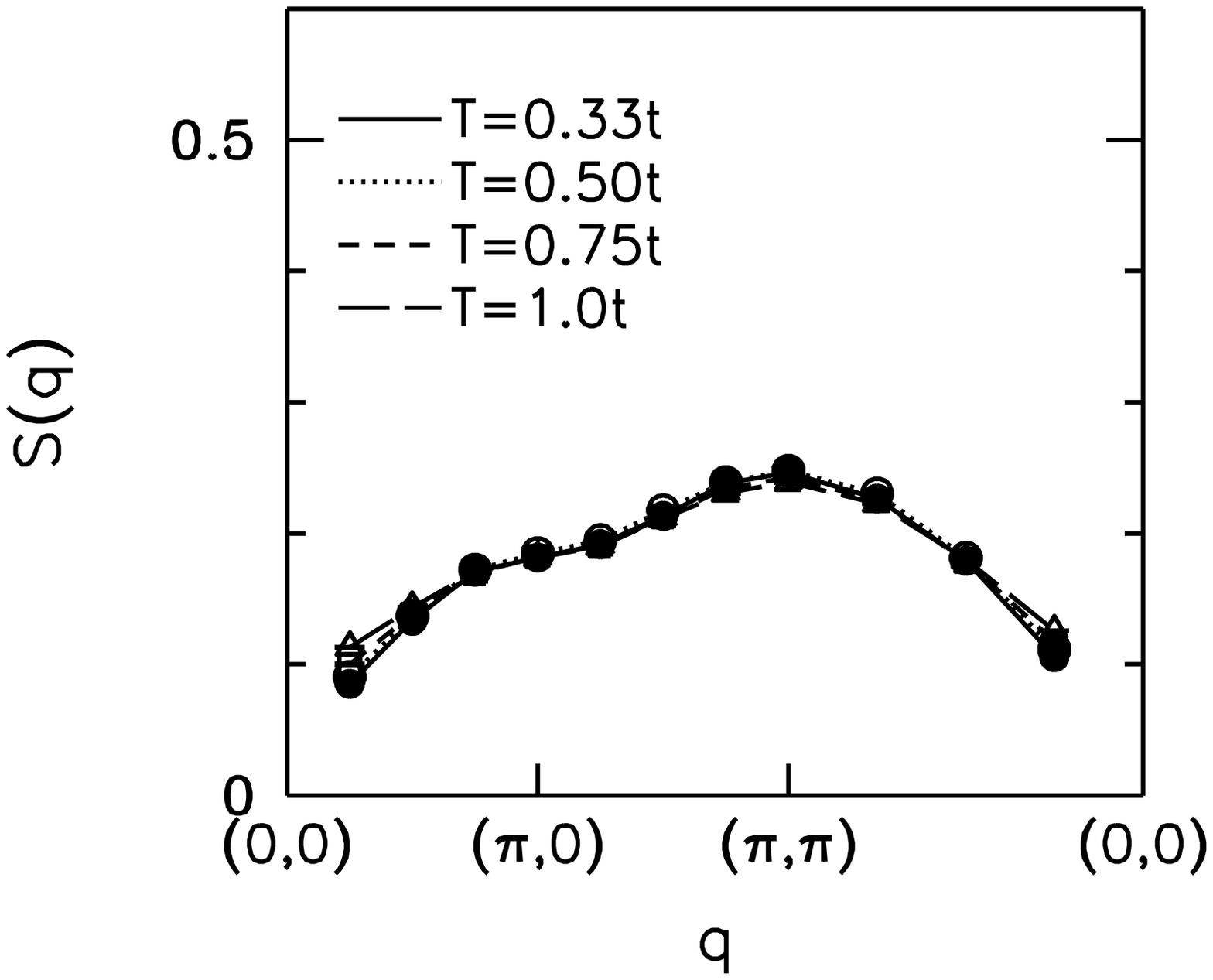}
\fi
\caption{
Equal-time density correlation function $S({\bf q})$
versus ${\bf q}$
at $\langle n\rangle=0.87$ for $U=8t$ and various temperatures.
}
\label{4.4}
\end{figure}

\begin{figure}
\centering
\iffigure
\mbox{
\subfigure[]{
\epsfysize=8cm
\epsffile[100 150 480 610]{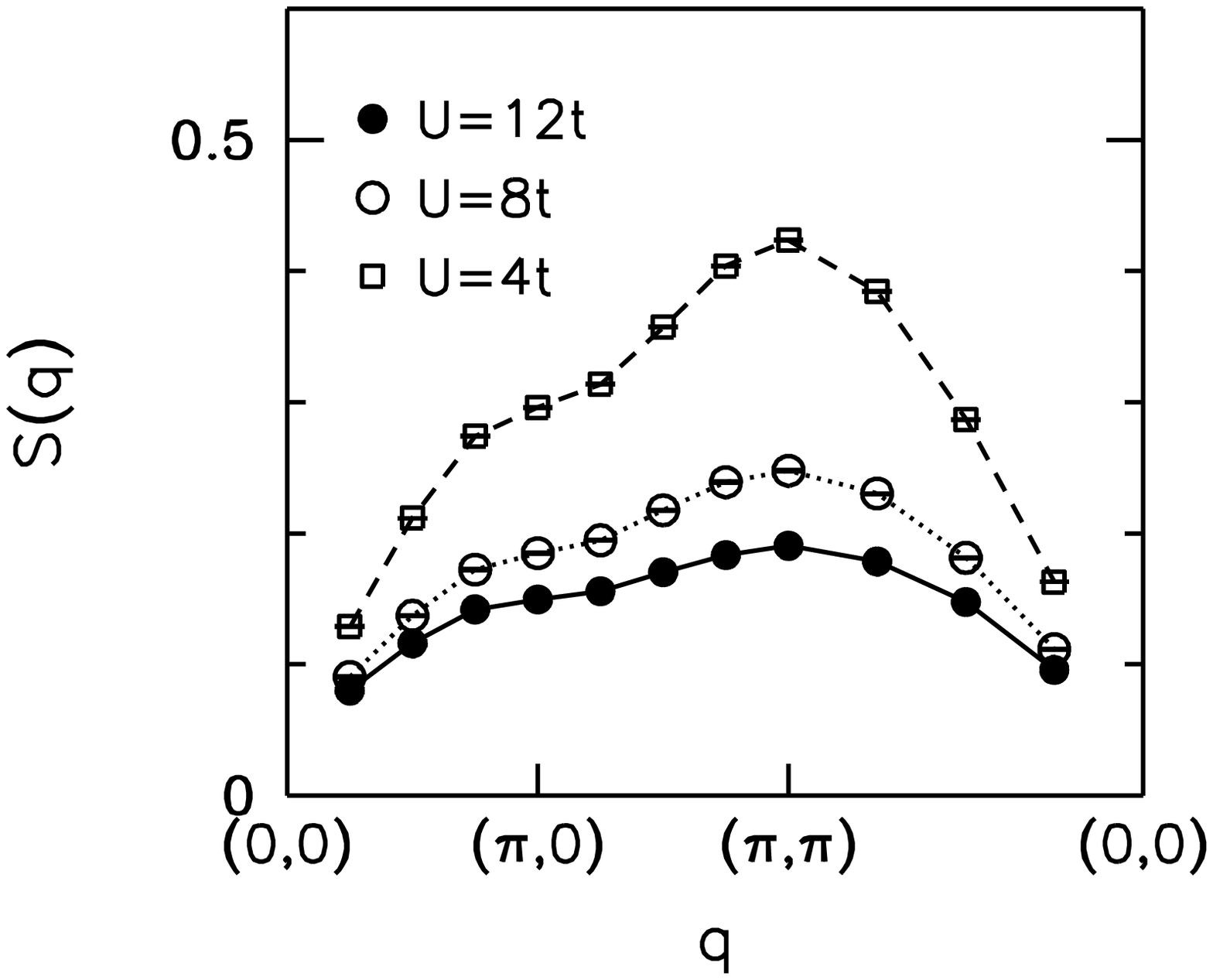}}
\quad
\subfigure[]{
\epsfysize=8cm
\epsffile[50 150 600 610]{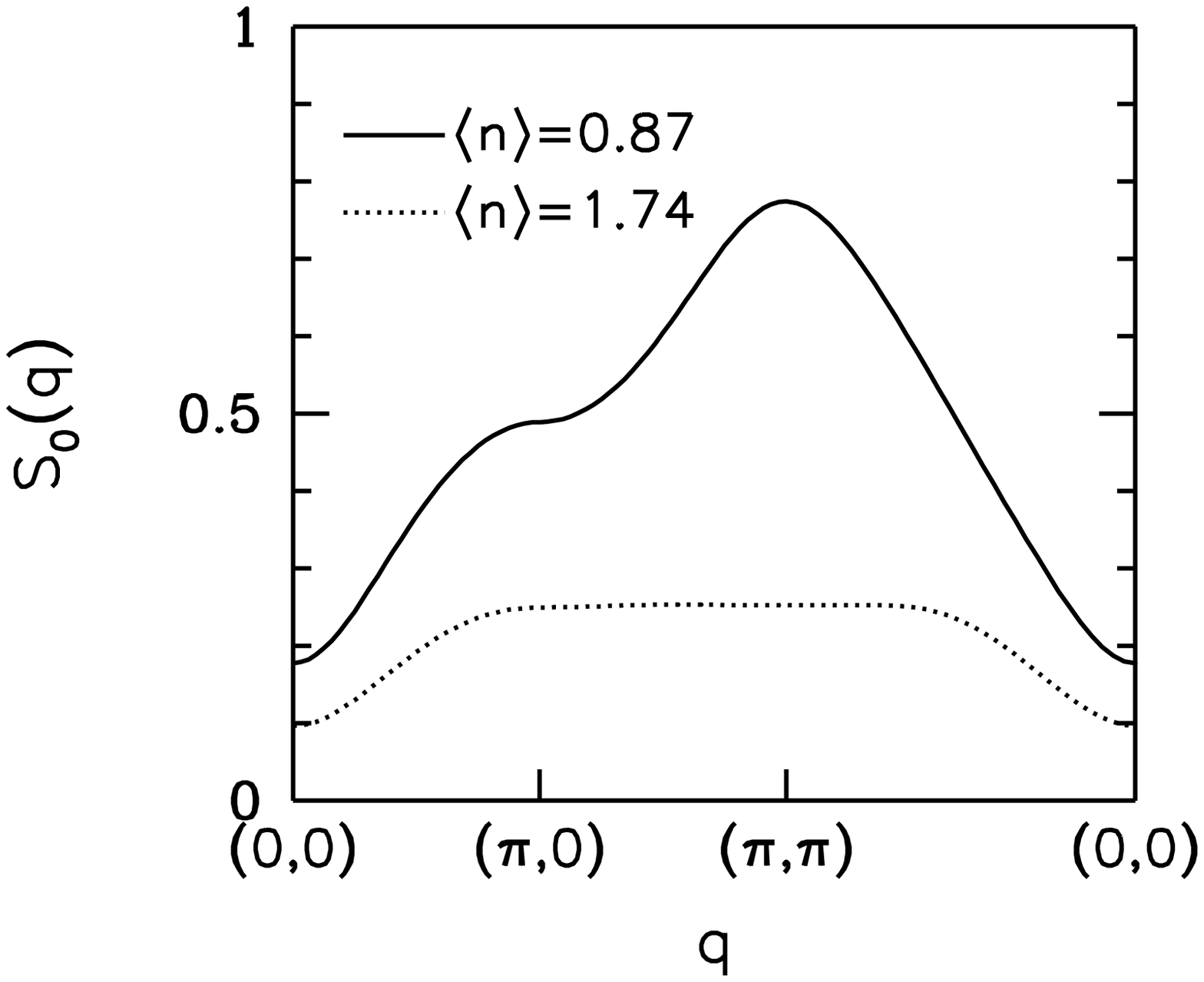}}}
\fi
\caption{
(a) $S({\bf q})$ versus ${\bf q}$ 
at $\langle n\rangle=0.87$ 
for various values values of $U$.
(b) $S_0({\bf q})$ versus ${\bf q}$ 
of the $U=0$ system for 
$\langle n\rangle=0.87$ (solid curve)
and $\langle n\rangle=1.74$ (dotted curve).
These results are for $T=0.5t$.
}
\label{4.5}
\end{figure}

Here, it has been seen that
the results on $S({\bf q})$ for the 2D Hubbard model  
are quite different from those calculated
for the 2D $t$-$J$ model for $J\sim 0.5t$ with the 
high-temperature series expansion
[Putikka {\it et al.} 1994]
or by the GFMC technique [Calandra {\it et al.} 1998].
In the 2D Hubbard model near half-filling,
$S({\bf q})$ does not exhibit 
obvious features which might be associated
with the "$4{\bf k}_F$" wave vectors of the Fermi surface.
However, $\Pi({\bf q},0)$ has structure, 
which might be associated with the "$4{\bf k}_F$" wave vectors, 
when $U/t$ is large.
Hence, 
the dynamical density fluctuations in 2D Hubbard model
with large $U/t$ have features 
which might be related to the "$4{\bf k}_F$" wave vectors,
while the static density correlations 
appear to follow the 
"$2{\bf k}_F$" wave vectors.
However, results on $\Pi({\bf q},\omega)$ 
at lower temperatures and on bigger lattices
are necessary before reaching conclusions.

The nature of the density correlations 
in the Hubbard model will be discussed further
in the remainder of the article.
In Section~7.3, comparisons will be made with the 2-leg
Hubbard ladder, in which case power-law decaying 
"$4{\bf k}_F$" CDW correlations have been found
[Noack {\it et al.} 1996].
In addition, in Section~8
there will be further comparisons with the 
density correlations in the 2D $t$-$J$ model.

\subsection{Optical conductivity}

In this section, 
the results of the numerical calculations on
the optical conductivity 
$\sigma_1(\omega)$ of the 2D Hubbard model will be discussed.
The optical conductivity for the 2D Hubbard model 
was calculated with the exact diagonalization technique 
for a $4\times 4$ lattice with $U=10t$
at half-filling and in the doped case
[Dagotto {\it et al.} 1992c].
In these calculations,
the insulating gap in $\sigma_1(\omega)$
at half-filling is clearly seen. 
Upon doping to $\langle n\rangle = 0.875$,
the amount of the spectral weight 
above the insulating gap is reduced. 
In this case, the additional features found in $\sigma_1(\omega)$
are a Drude peak at $\omega=0$ and spectral weight induced 
at intermediate energies.
The Drude weight of the Hubbard model was also calculated 
with the QMC simulations
[Scalapino {\it et al.} 1992 and 1993].
A review of the exact diagonalization calculations 
for $\sigma_1(\omega)$ and of comparisons with the 
experimental data on the cuprates can be found 
in Ref.~[Dagotto 1994].

The real part of the frequency-dependent 
${\bf q}=0$ conductivity is given by 
\begin{equation} 
\sigma_1(\omega) = 
{\rm Re}\,{\Lambda_{xx}(i\omega_m) \over i\omega_m}
\bigg|_{i\omega_m \rightarrow \omega+i\delta}
\end{equation}
where the current-current correlation function 
$\Lambda_{xx}$ is defined as
\begin{equation}
\Lambda_{xx}(i\omega_m) = 
\int_0^{\beta} d\tau\, 
e^{-i\omega_m\tau} 
\langle j_x(\tau) j_x(0) \rangle
\end{equation}
with
\begin{equation}
j_x = -itea\sum_{is} 
( c^{\dagger}_{is} c_{i+xs} - 
c^{\dagger}_{i+xs}c_{is} ).
\end{equation}
The analytic continuation in Eq.~(4.5) is carried out using 
the maximum entropy procedure. 
In the following, the hopping $t$, the electron charge $e$ and 
the lattice constant $a$ are set to unity.

The solid curve in Fig.~4.6 shows 
$\sigma_1(\omega)$ versus $\omega$ 
at half-filling for $U=8t$, $T=0.125t$ 
and an $8\times 8$ lattice. 
These are data from Ref.~[Bulut {\it et al.} 1994a].
Here, 
in spite of the limited resolution 
of the analytic continuation procedure, 
the insulating gap in $\sigma_1(\omega)$ is clearly seen. 
The dotted curve in this graph 
is the mean-field SDW result given by 
\begin{eqnarray} 
\sigma_1(\omega) = {2\pi \over N}
\sum_{\bf p} \, 
{1\over 2}
\left( 1 - {\varepsilon^2_{\bf p} - \Delta^2 \over E^2_{\bf p}} \right) 
\left( 1 - 2f(E_{\bf p}) \over E_{\bf p} \right)
\sin^2{p_x}
\end{eqnarray}
which was calculated using $\Delta=2.4t$.  
Near the threshold $\omega \simeq 2\Delta$, 
the SDW coherence factor 
${1\over 2}[ 1 - (\varepsilon^2_{\bf p}-\Delta^2)/E^2_{\bf p} ]$ 
in Eq.~4.8 goes to 1 for $\varepsilon_{\bf p}$ near zero, and 
hence the peak at the threshold does not vanish within SDW. 
A check of the maximum entropy analytic continuation 
is the $f$-sum rule which has the form
\begin{equation}
\int_0^{\infty} d\omega\, \sigma_1(\omega) = 
{\pi \over 2} \langle -k_x \rangle, 
\end{equation}
where $\langle k_x \rangle$ is the average kinetic 
energy per site associated 
with hopping in the $x$ direction. 
The area under the solid curve in Fig.~4.6 is 0.74, 
where the separate QMC measurement 
of ${\pi \over 2} \langle -k_x \rangle$ gives 0.77.
The difference reflects the difficulty in analytically continuing 
the numerical data. 
When the system is doped, 
it is difficult to estimate the accuracy 
of the analytic continuation procedure for $\sigma_1(\omega)$,
especially in the limit $\omega\rightarrow 0$.
Hence, 
away from half-filling, 
the maximum entropy or the Pade techniques were not used to 
extract $\sigma_1(\omega)$ or the dc resistivity $\rho(T)$
from the QMC data on 
the current-current correlation function $\Lambda_{xx}$.

\begin{figure}
\centering
\iffigure
\epsfysize=8cm
\epsffile[100 150 550 610]{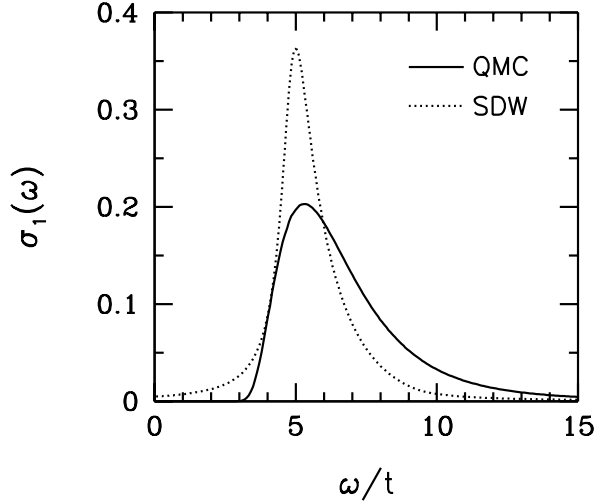}
\fi
\caption{
Real part of the frequency dependent conductivity of the half-filled
Hubbard model with $U=8t$ (solid curve).
These results were obtained for an $8\times 8$ lattice
at $T=0.125t$.
The dotted curve is the mean-field SDW result with $\Delta=2.4t$.
A finite broadening of $\Gamma=0.5t$ has been used 
in plotting the SDW result.
}
\label{4.6}
\end{figure}

Even though it has not been possible to reliably extract 
$\rho(T)$ for the 2D Hubbard model,
for 2D $t$-$J$ clusters
$\sigma_1(\omega)$ and $\rho(T)$ 
were calculated at various fillings
by using the finite-temperature Lanczos technique
[Jaklic and Prelovsek 2000].
For $J/t=0.3$, the results for $\sigma_1(\omega)$ and $\rho(T)$ 
were compared with the experimental data 
on cuprates in various doping regimes. 
In some cases, 
the agreement obtained with the experimental data
was at a quantitative level. 
In the intermediate doping regime, 
where $0.75 < \langle n\rangle < 0.85$, 
it was shown that $\sigma_1(\omega)$ 
is not consistent with the usual Drude
form but rather with the marginal Fermi 
liquid concept where the only $\omega$ 
scale is given by $T$ 
[Varma {\it et al.} 1989].
The finite-temperature Lanczos calculations 
find also that $\rho(T)$ is proportional 
to $T$ in this regime but the slope of 
$\rho(T)$ changes at $T \sim J$.
The underdoped and the overdoped regimes 
were also studied with this technique.
Jaklic and Prelovsek concluded that 
the main features of the unusual normal state properties 
of the cuprates could be attributed 
to the large degeneracy of states 
and the frustration induced by doping the antiferromagnet. 

\setcounter{equation}{0}\setcounter{figure}{0}
\section{Single-particle properties of the 2D Hubbard model}

The angular-resolved photoemission spectroscopy (ARPES) measurements
on the high-$T_c$ cuprates probe the single-particle properties 
of these compounds and have found a number of 
interesting features.
In the normal state of the optimally doped cuprates,
the ARPES experiments 
have found that there are quasiparticle-like bands which cross
the Fermi level leading to a large Fermi surface.
An unusual feature is that these bands have
extended flat dispersion near the $(\pi,0)$ and 
$(0,\pi)$ points in the Brillouin zone
[Dessau {\it et al.} 1993, Gofron {\it et al.}
1993].
The ARPES experiments have also provided valuable 
information about the evolution of the single-particle
spectral weight with doping in the underdoped regime as the 
insulating state is approached.
Reviews of these experiments can be found in 
Refs.~[Shen and Dessau 1995, Damascelli {\it et al.} 2001].
Below, the numerical studies of the single-particle 
spectral weight in the 2D Hubbard model will be reviewed.
In particular, 
the origin of the correlated metallic band 
which develops at the Fermi level 
upon doping the AF Mott-Hubbard insulator 
will be investigated.
It will be seen that the 2D Hubbard model provides an explanation 
for some of the features found in the ARPES data.

The single-particle properties of the 2D Hubbard model 
were studied with various many-body techniques.
The single-particle spectral weight 
was calculated phenomenologically by taking into account the 
scattering of the quasiparticles by the AF spin fluctuations
[Kampf and Schrieffer 1990a and 1990b].
The evolution of the single-particle spectral weight 
with doping was 
also studied with the exact diagonalization technique 
[Dagotto {\it et al.} 1991].
There have been extensive numerical calculations 
of $A({\bf p},\omega)$ by analytically continuing the QMC data 
on the imaginary-time single-particle Green's function
[White {\it et al.} 1989c, White 1991, Vekic and White 1993,
Bulut {\it et al.} 1994a, Bulut and Scalapino 1995, 
Moreo {\it et al.} 1995, Haas {\it et al.} 1995, 
Duffy and Moreo 1995, Preuss {\it et al.} 1995 and 1997, 
Gr\"ober {\it et al.} 2000].
In these studies, 
the maximum-entropy technique has been the main
algorithm for the analytic continuation procedure.
This technique
was also used for calculating $A({\bf p},\omega)$ 
for the three-band CuO$_2$ model [Dopf {\it et al.} 1992b] and 
the one-dimensional Hubbard model 
[Preuss {\it et al.} 1994, Zacher {\it et al.} 1998].
These studies have provided valuable 
information about the single-particle spectral weight in the 
Hubbard model.
In the following, the maximum-entropy results on the 
2D Hubbard model will be reviewed.
Here, the emphasis will be on the 
narrow correlated band which forms at the Fermi level 
upon doping the insulator, 
and on the flat bands which are
observed near the $(\pi,0)$ and $(0,\pi)$ points 
in the Brillouin zone.

The single-particle spectral weight 
$A({\bf p},\omega)$ is given by
\begin{equation}
\label{Apw}
A({\bf p},\omega) = - {1\over \pi}
{\rm Im}\, G({\bf p},i\omega_n \rightarrow \omega+i\delta)
\end{equation}
where
\begin{equation}
G({\bf p},i\omega_n) = \int_0^{\beta} \, d\tau \, 
e^{i\omega_n \tau} \,
G({\bf p},\tau)
\end{equation}
and
\begin{equation}
\label{Gpt}
G({\bf p},\tau) = - \langle T \,
c_{{\bf p}\sigma}(\tau) 
c^{\dagger}_{{\bf p}\sigma}(0) \rangle.
\end{equation}
With the maximum-entropy technique, one uses the QMC data on 
$G({\bf p},\tau)$ to invert the integral equation
\begin{equation}
G({\bf p},\tau) = - \int_{-\infty}^{\infty} \, 
d\omega  \,
{ e^{-\omega\tau} \over 1 + e^{-\beta\omega} } \,
A({\bf p},\omega),
\end{equation}
in order to obtain $A({\bf p},\omega)$. 
By applying the same procedure to 
\begin{equation}
G_{ii}(\tau) = - \langle T\, 
c_{i\sigma}(\tau) c^{\dagger}_{i\sigma}(0) \rangle,
\end{equation}
the single-particle density of states 
$N(\omega)$ is obtained. 
The maximum-entropy technique for analytically continuing
QMC data is described in 
Refs.~[Silver {\it et al.} 1990, Jarrell and Gubernatis 1996].
The statistical errors for $G_{ii}(\tau)$ are smaller than for 
$G({\bf p},\tau)$,
hence the results for $N(\omega)$ shown below have higher resolution. 
Here, 
first the results on the evolution of $N(\omega)$ with doping 
will be reviewed, and later $A({\bf p},\omega)$ will be discussed.
In the following, 
$N(\omega)$ and $A({\bf p},\omega)$
will be plotted in units of $t^{-1}$.
The data which will be shown in this section are from 
Refs.~[Bulut {\it et al.} 1994a, Bulut and Scalapino 1995].

\begin{figure}[ht]
\centering
\iffigure
\epsfysize=8cm
\epsffile[100 150 550 610]{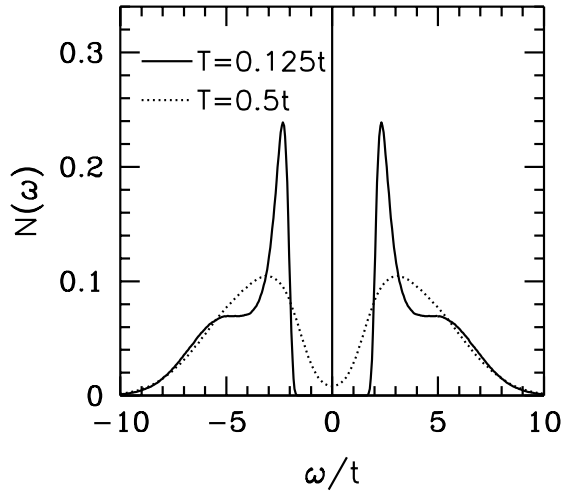}
\fi
\caption{
Single-particle density of states $N(\omega)$ versus
$\omega$.
These results were obtained for 
$U=8t$ and $\langle n\rangle =1.0$ on a
$12\times 12$ lattice.
}
\label{5.1}
\end{figure}

\subsection{Single-particle density of states}

Figure 5.1 shows the single-particle 
density of states $N(\omega)$ versus $\omega$
at half-filling for $U=8t$ on a $12\times 12$ lattice
at $T=0.5$ and $0.125t$.
Here, the chemical potential $\mu$ 
is located at $\omega=0$.
In this figure, 
it is seen that at $T=0.5t$, 
$N(\omega)$
consists of two peaks corresponding to the lower 
and the upper Hubbard bands, 
which are separated by the Mott-Hubbard pseudogap.
At this temperature, the AF correlation length 
is still less than the system size, 
as it was seen in Section 3.1.
As $T$ is lowered down to $0.125t$, 
the AF correlation length reaches the system size, and 
in this case additional sharp peaks appear at the upper edge of the 
lower Hubbard band and at the lower edge of 
the upper Hubbard band.
At $T=0.125t$,
the pseudogap has become a full gap 
in the single-particle spectrum 
with a magnitude of $2\Delta \approx 4.5t$.
The size of the gap $2\Delta$ was also calculated within 
an SDW approach where the single-particle self-energy 
corrections were included [Schrieffer {\it et al.} 1989].
Within this approach, for $U=8t$, 
$2\Delta$ is found to be about $4.8t$, while for
$U=4t$, $2\Delta$ is about $2t$.

The sharp peaks which are located below and above the 
Mott-Hubbard gap appear when 
the system has long-range AF order.
These narrow bands have a bandwidth of about $1t$, 
corresponding to $\approx 2J$ where 
$J=4t^2/U$, and they exhibit SDW-like dispersion, 
which will be discussed in the next section.
It is known that in the half-filled 2D $t$-$J$ model 
the quasiparticle bandwidth is about $2J$
[Liu and Manousakis 1992].
The exact diagonalization calculations for the 2D $t$-$J$ 
model also find SDW-like quasiparticle
excitations with a bandwidth of about $2.2J$ at half-filling
[Dagotto 1994].
In the $t$-$J$ model at half-filling, 
one doped hole propagates by flipping the spins in the AF background 
around it.
For this reason, 
the bandwidth for the hole motion is determined by 
the magnetic exchange $J$ rather 
than the hopping matrix element $t$.
The QMC data implies that 
the hole propagation is accompanied by 
similar many-body processes in the 
half-filled 2D Hubbard model.

\begin{figure}[ht]
\centerline{
\epsfysize=6cm \epsffile[-207 164 367 598]{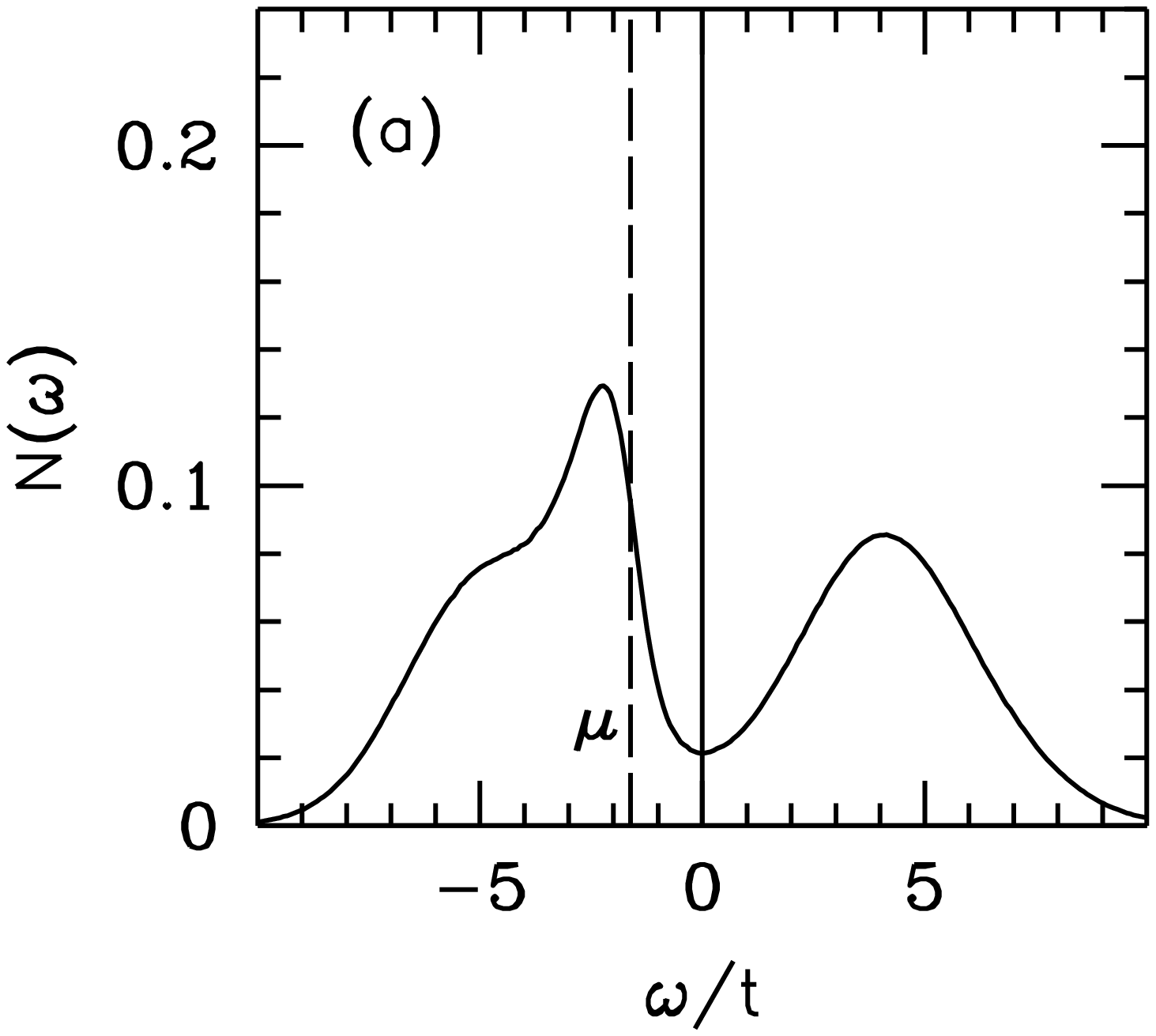}
\epsfysize=6cm \epsffile[18 164 592 598]{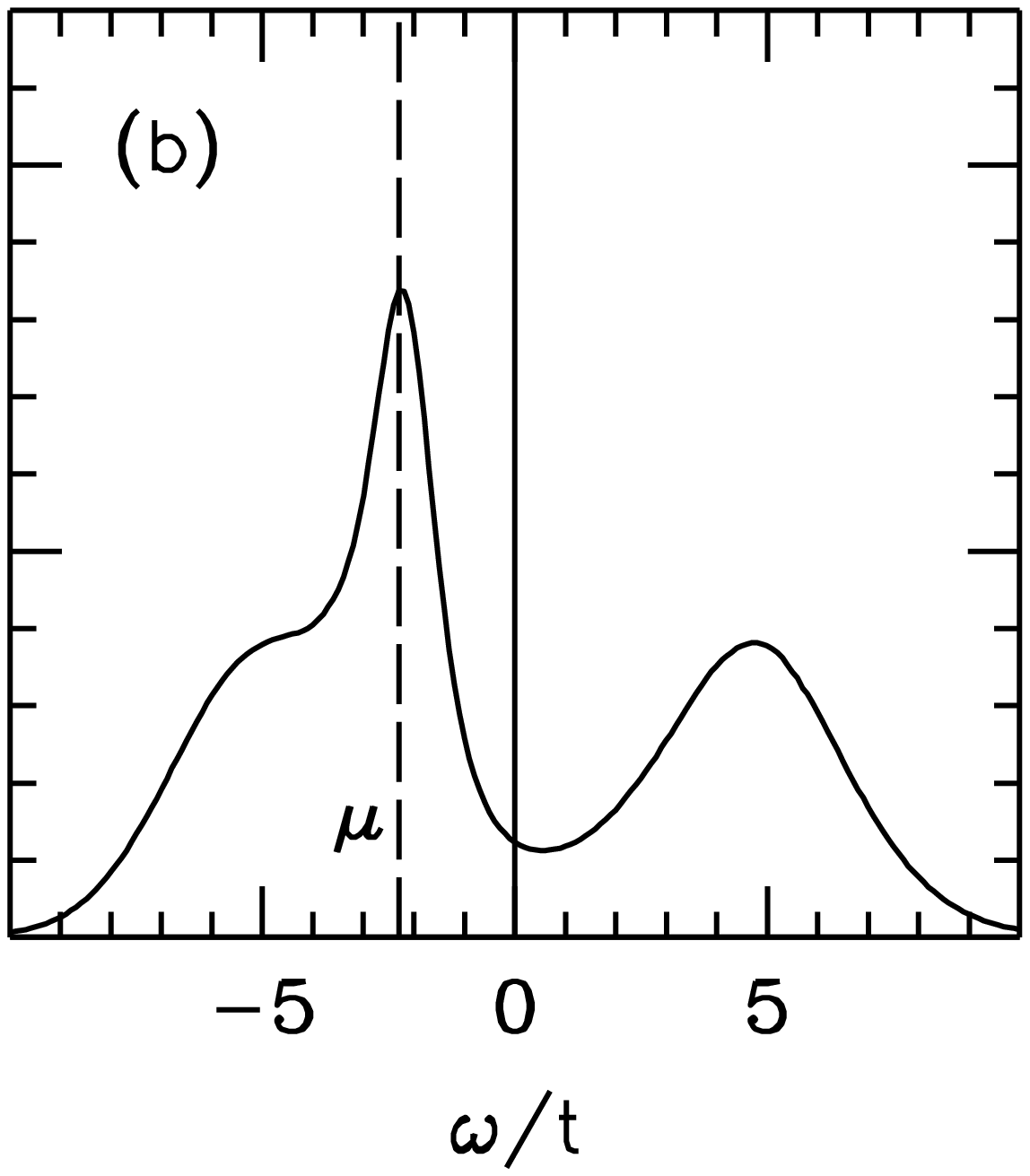}
\epsfysize=6cm \epsffile[243 164 817 598]{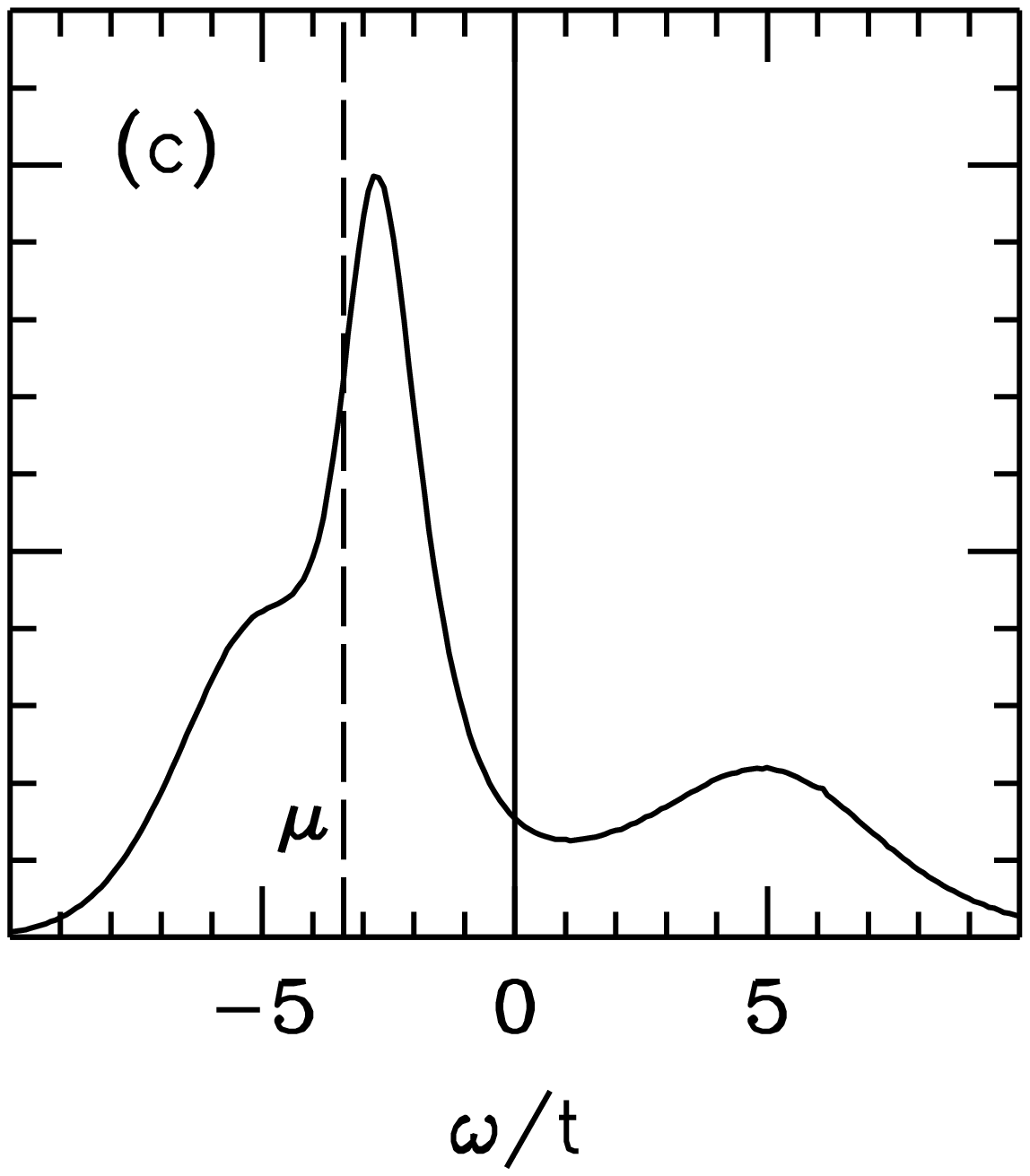}}
\caption{
Single-particle density of states $N(\omega)$ versus
$\omega$ at fillings (a) $\langle n\rangle=0.94$, 
(b) 0.87 and (c) 0.70.
These results are for $T=0.5t$ and $U=8t$.
}
\label{5.2}
\end{figure}

Next, results on the evolution of $N(\omega)$ 
with doping are shown.
Figure~5.2 shows $N(\omega)$ versus $\omega$ at fillings 
$\langle n\rangle =0.94$, 0.87 and 0.70
for $U=8t$ and $T=0.5t$.
Upon hole doping, the chemical potential moves
rapidly from $\omega=0$ to the top of the lower Hubbard band.
This is accompanied with a gradual transfer of spectral
weight from above the Mott-Hubbard gap to the Fermi level.
The transferred spectral weight goes into forming 
a narrow metallic band at the Fermi level.
These data are in agreement 
with the exact diagonalization calculations
[Dagotto {\it et al.} 1991].
In the overdoped regime at $\langle n\rangle=0.70$, 
the Fermi level lies below the metallic band.
The temperature variation of $N(\omega)$ is shown in 
Fig.~5.3 for $\langle n\rangle=0.87$ and $U=8t$.
Here, the build up of a narrow band with a width 
between $4J$ and $5J$ is clearly seen as $T$ decreases.
The comparison of Figs.~5.3 and 5.1 indicates that 
this correlated metallic band forms out
of the SDW-like bands of the half-filled case.

At this point,
it should be noted that the general features of 
$N(\omega)$ seen in Fig.~5.3 are similar to what is found for the 
infinite-dimensional Hubbard model away from half-filling
[Jarrell 1992].
$N(\omega)$ for the infinite-dimensional Hubbard model 
is calculated by first mapping the model to an 
Anderson impurity problem, which is then solved exactly.
In this case,
a narrow peak at the Fermi level is also found in addition to 
the lower and upper Hubbard bands.

\begin{figure}[ht]
\centering
\iffigure
\epsfysize=8cm
\epsffile[100 150 550 610]{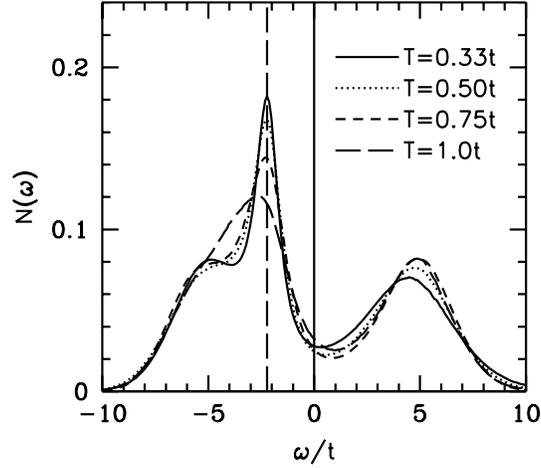}
\fi
\caption{
Development of the 
single-particle density of states $N(\omega)$ as the 
temperature is lowered for 
$\langle n\rangle=0.87$ and $U=8t$.
}
\label{5.3}
\end{figure}

\begin{figure}
\centering
\iffigure
\mbox{
\subfigure{
\epsfysize=8cm
\epsffile[100 150 480 610]{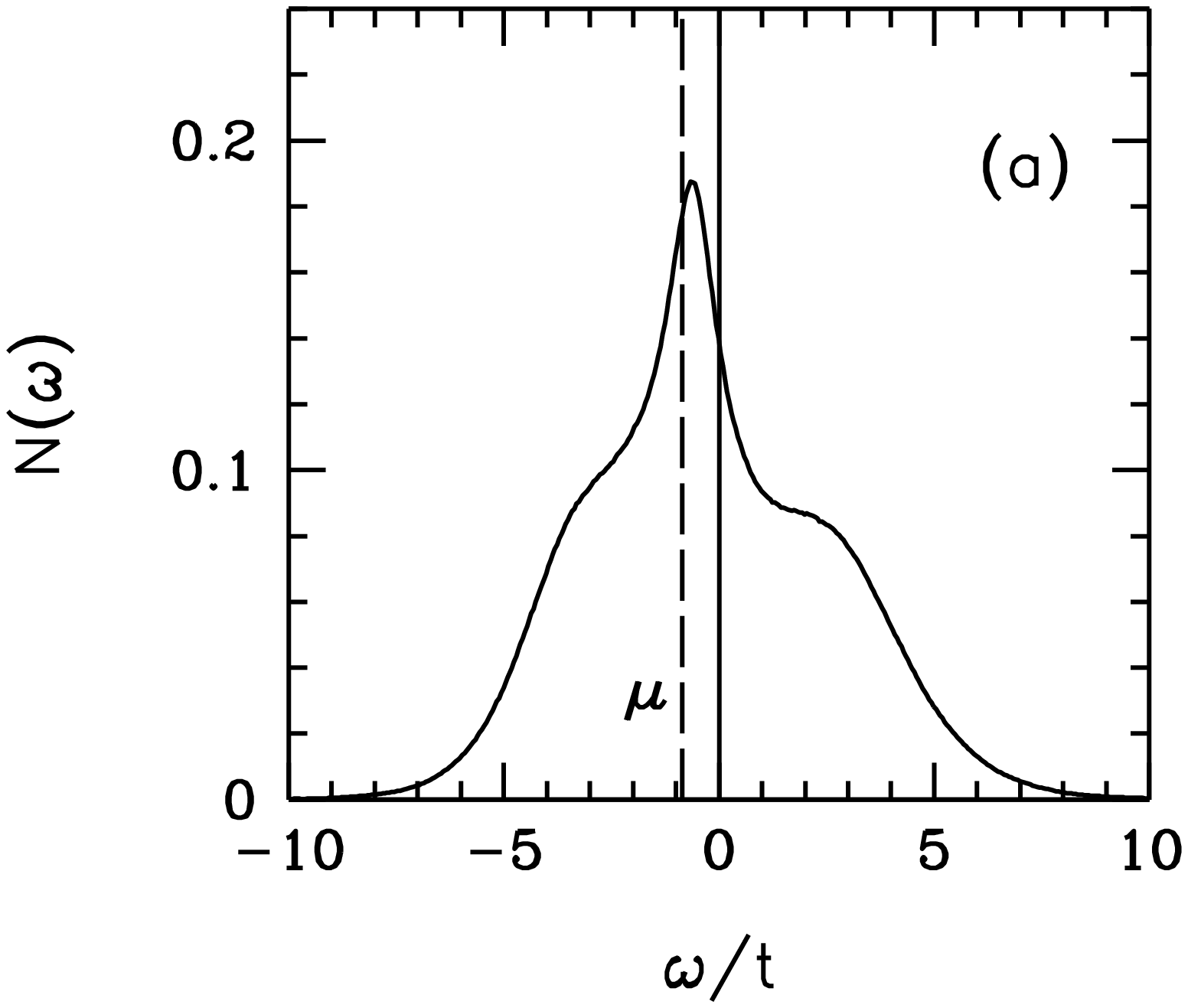}}
\quad
\subfigure{
\epsfysize=8cm
\epsffile[50 150 600 610]{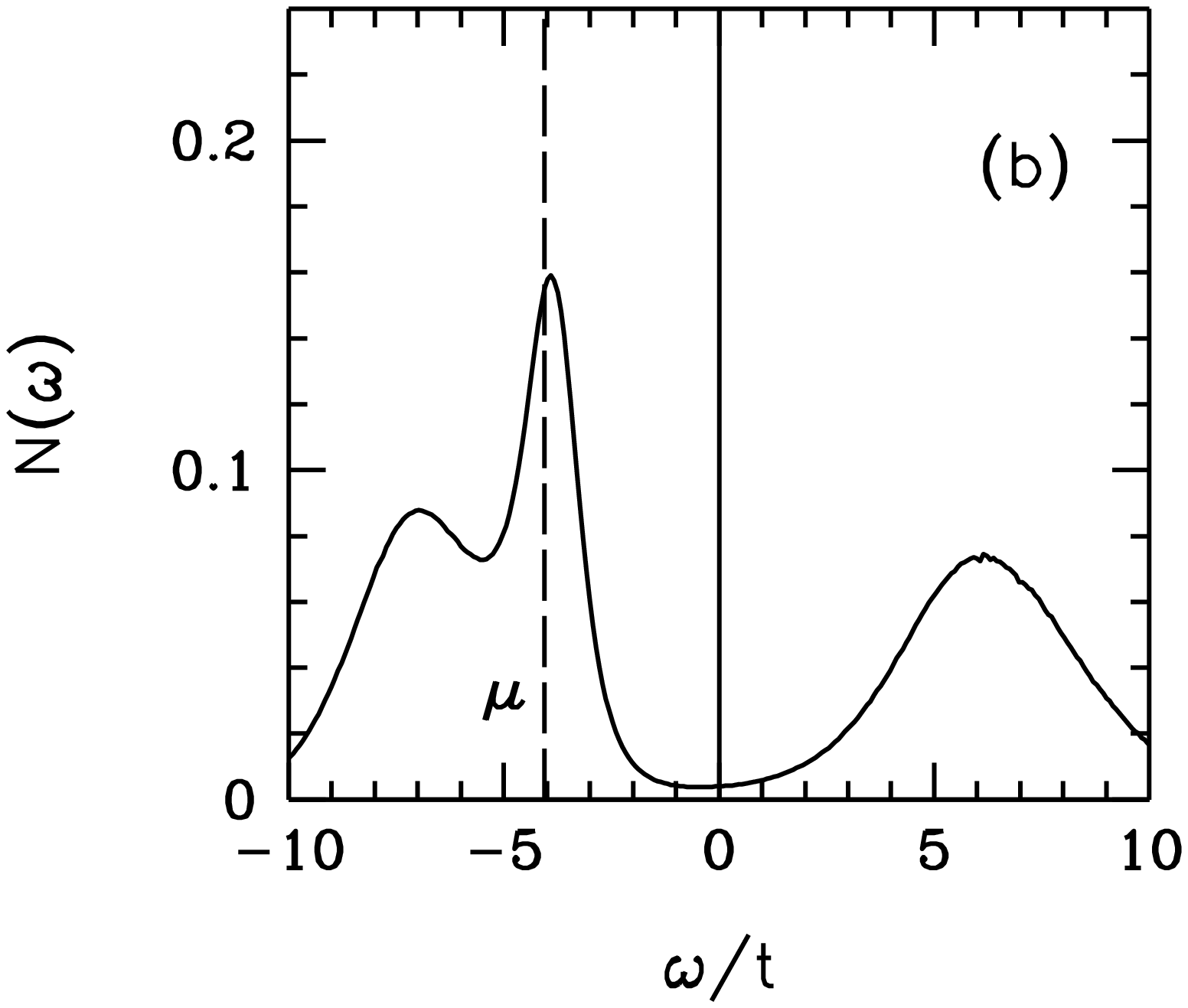}}}
\fi
\caption{
$N(\omega)$ versus $\omega$ for (a) $U=4t$ and 
(b) $U=12t$ at $\langle n\rangle=0.87$ and $T=0.5t$.
}
\label{5.4}
\end{figure}

It is useful to compare these results on $N(\omega)$ for 
$U=8t$ with the results for $U=4t$ and $12t$, 
which are shown in Fig.~5.4.
For $U=12t$, the narrow metallic band is further separated 
from the lower and the upper Hubbard bands,
and the pseudogap is bigger.
On the other hand,
for $U=4t$, the pseudogap is not observed at these 
temperatures, even though there are hump-like structures which 
might be attributed to the lower and the upper Hubbard bands.
In addition, the peak at the Fermi level is broader.
Hence, as $U$ increases from $4t$ to $12t$, 
the distribution of the weight in $N(\omega)$ changes
considerably.
However, the density of states at the Fermi level 
$N(\mu)$ varies by a small amount.

It is also useful to discuss how $N(\mu)$ varies with doping 
at fixed $U$ and $T$.
Figure~5.5 shows $N(\mu)$ versus $\langle n\rangle$ for 
$U=8t$ at $T=0.5t$ (filled circles) and $1.0t$ (open circles).
It is clearly seen that at these temperatures $N(\mu)$ 
has a peak at finite doping, and 
a depression in $N(\mu)$ exists near half-filling.
However, it is not possible to extract 
the $T\rightarrow 0$ behavior.

\begin{figure}[ht]
\centering
\iffigure
\epsfysize=8cm
\epsffile[100 150 550 610]{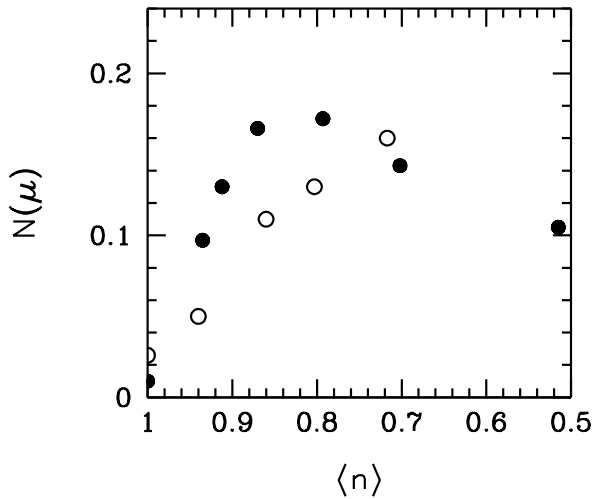}
\fi
\caption{
Filling dependence of the density of states at the Fermi level
$N(\mu)$ for $U=8t$ at temperatures $T=0.5t$ (filled circles)
and $1.0t$ (open circles).
}
\label{5.5}
\end{figure}

\subsection{Single-particle spectral weight}

In this section, results on the single-particle spectral weight 
$A({\bf p},\omega)$ will be discussed.
It is useful to first study the effects of varying $U$ on 
$A({\bf p},\omega)$.
Figure~5.6 shows $A({\bf p},\omega)$ versus $\omega$ at various
temperatures for $U=8t$ and $4t$.
These results were obtained for the ${\bf p}=(\pi/2,\pi/2)$ 
point on an $8\times 8$ lattice.
For both values of $U$, a heavily-damped 
quasiparticle-like peak develops near the
Fermi level as the temperature decreases.
However, for $U=8t$ this peak is broader and it has reduced weight.
In addition, in this case, the peak location varies with the
temperature.
Even though the maximum-entropy technique has an intrinsic broadening 
which is difficult to estimate, 
these results suggest that the damping of the quasiparticles
is stronger for larger $U/t$.

\begin{figure}
\centering
\iffigure
\mbox{
\subfigure{
\epsfysize=8cm
\epsffile[100 150 480 610]{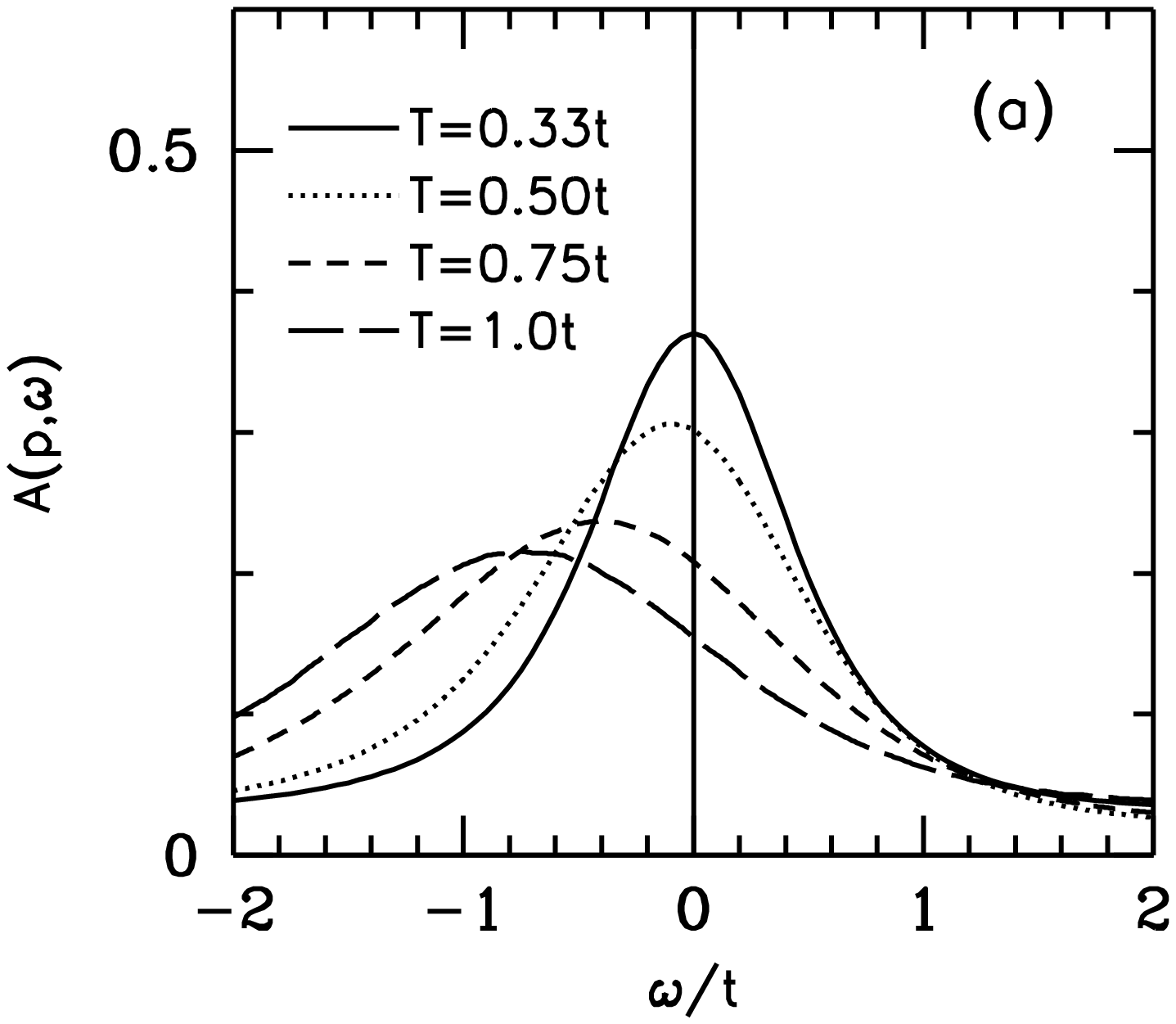}}
\quad
\subfigure{
\epsfysize=8cm
\epsffile[50 150 600 610]{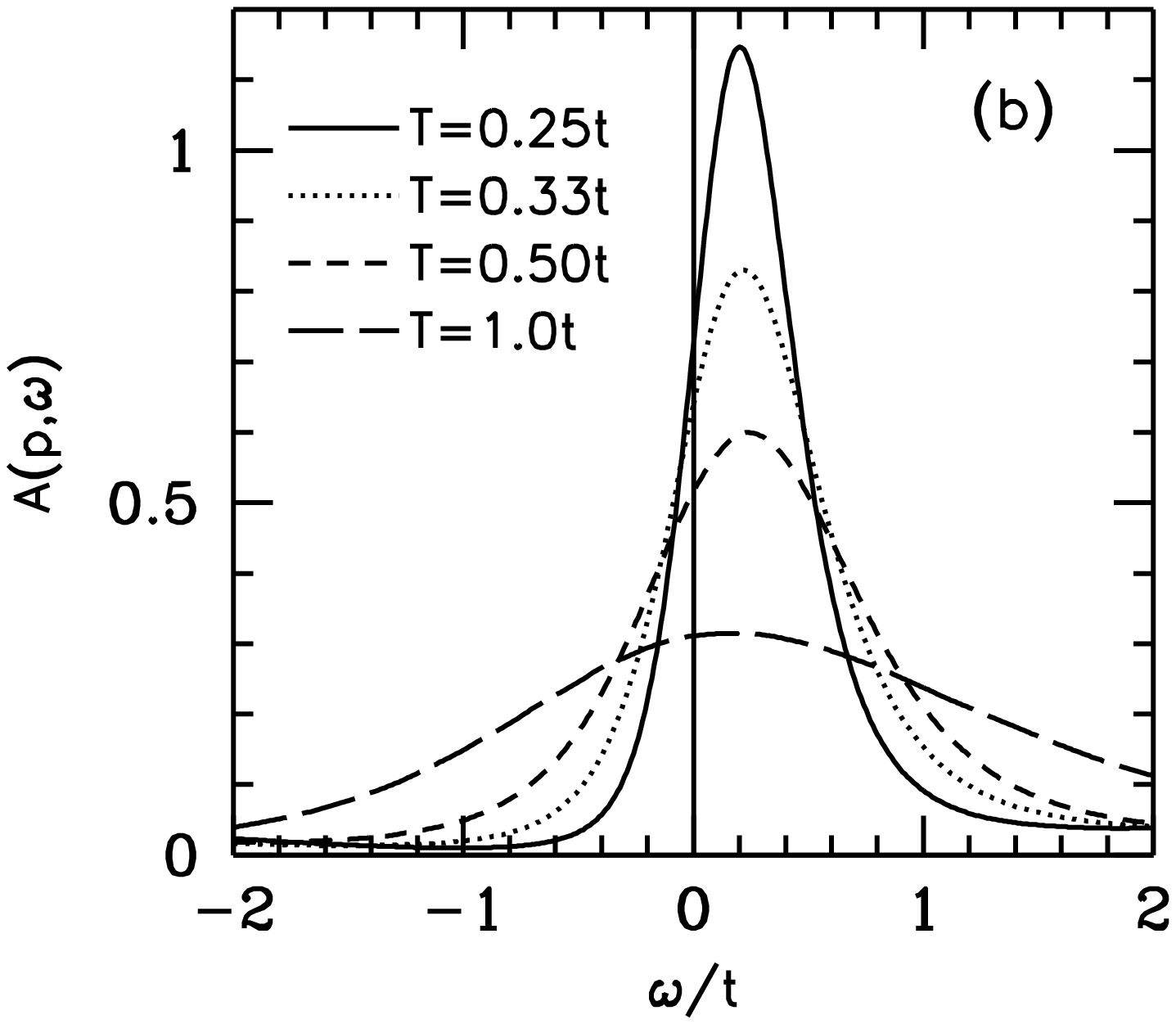}}}
\fi
\caption{
Comparison of
$A({\bf p}=(\pi/2,\pi/2),\omega)$ versus $\omega$ 
for (a) $U=8t$ and (b) $U=4t$.
Here results are shown at various temperatures 
for $\langle n\rangle=0.87$, 
and the frequency axis is shifted such that the Fermi level 
occurs at $\omega=0$.
}
\label{5.6}
\end{figure}

In the previous section, it was seen that for 
$U=8t$ a narrow metallic band develops at the Fermi level 
upon doping.
In order to gain insight into the origin of this band, 
in Fig.~5.7, $A({\bf p},\omega)$ versus $\omega$ is plotted for 
${\bf p}$ taken along various cuts in the Brillouin zone 
as indicated in the insets.
These data were obtained on a $12 \times 12$ lattice for
$U=8t$, $\langle n\rangle=0.87$ and T$=0.5t$.
In this figure, spectral weight representing the upper 
and the lower Hubbard bands are seen in addition to 
a quasiparticle band which crosses the Fermi level.
The quasiparticle peak is especially of interest.
When ${\bf p}$ is on the Fermi surface, this peak is 
centred at $\omega=0$.
As ${\bf p}$ moves outside of the Fermi surface, 
the quasiparticle peak is clearly resolved.
However, when ${\bf p}$ is below the Fermi surface, 
for instance at ${\bf p}=(0,0)$,
the quasiparticle peak is obscured by the lower Hubbard band, 
even though the exact diagonalization calculations
[Dagotto {\it et al.} 1992a]
find a quasiparticle peak at this momentum in addition to the 
lower Hubbard band.
This is because the maximum-entropy algorithm used here has 
poor resolution when $\omega$ is away from the Fermi level.
Indeed, 
similar calculations with higher resolution were able to observe
the quasiparticle peak below the Fermi surface
[Preuss {\it et al.} 1995].
An unresolved issue is whether the dispersing band near the Fermi level 
corresponds to a true quasiparticle band.
However, the exact determination of the energy and the temperature
scaling of the spectral weight cannot be carried out 
because the maximum-entropy technique has finite resolution and, 
in addition, bigger lattices and lower temperatures
are required.

\begin{figure}
\centerline{
\epsfysize=9cm \epsffile[-207 124 367 778]{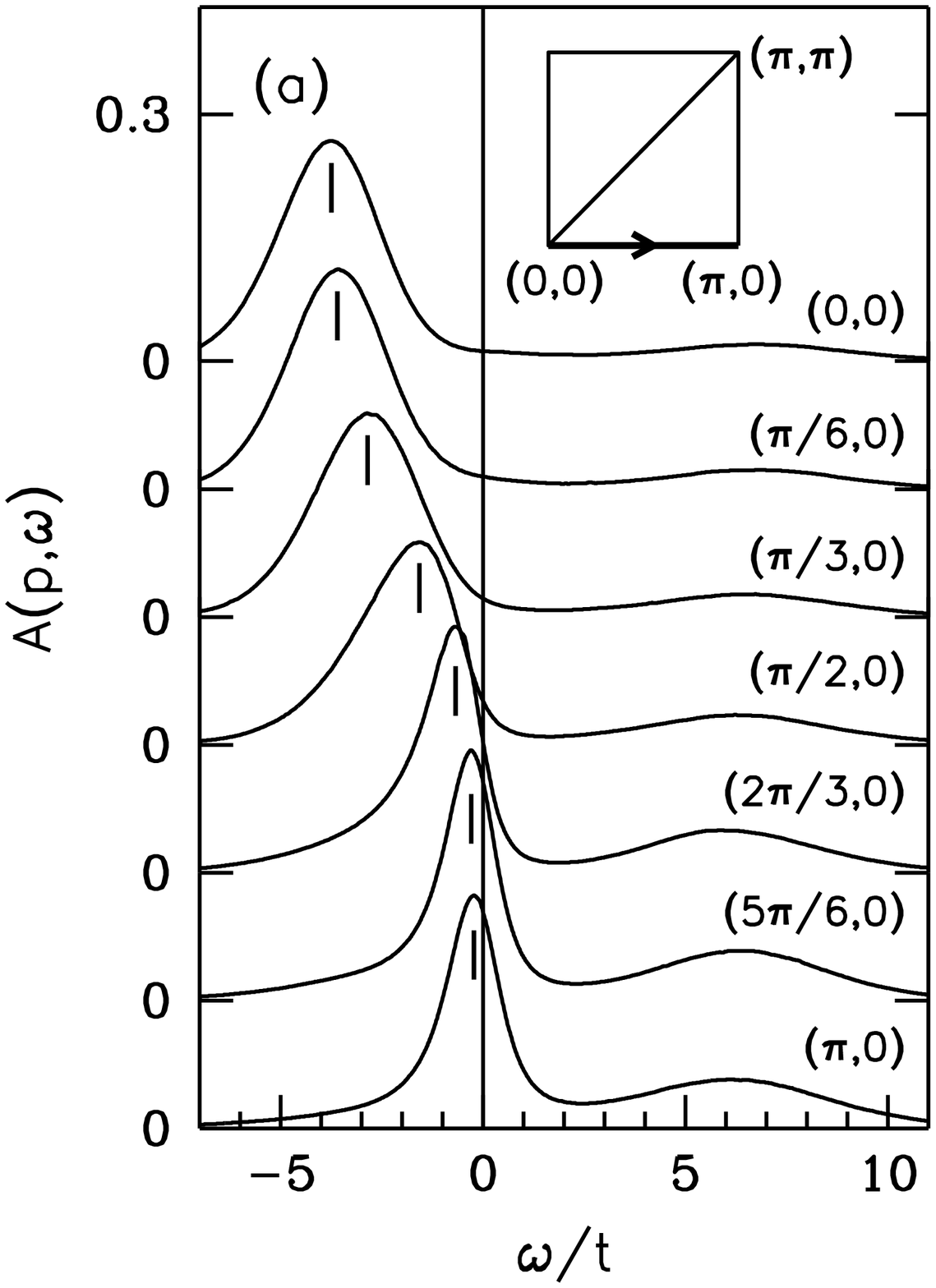} 
\epsfysize=9cm \epsffile[18 124 592 778]{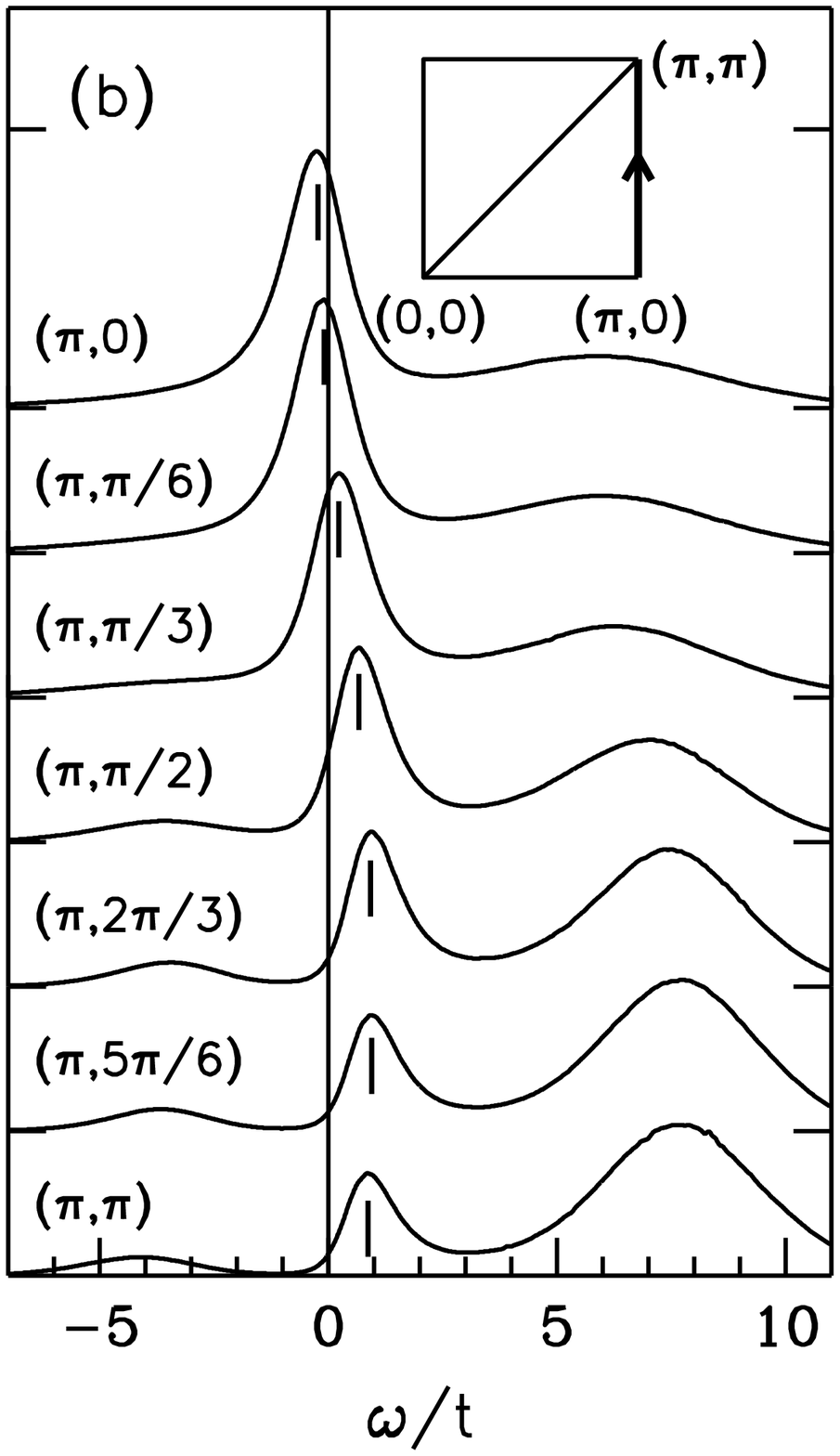} 
\epsfysize=9cm \epsffile[243 124 817 778]{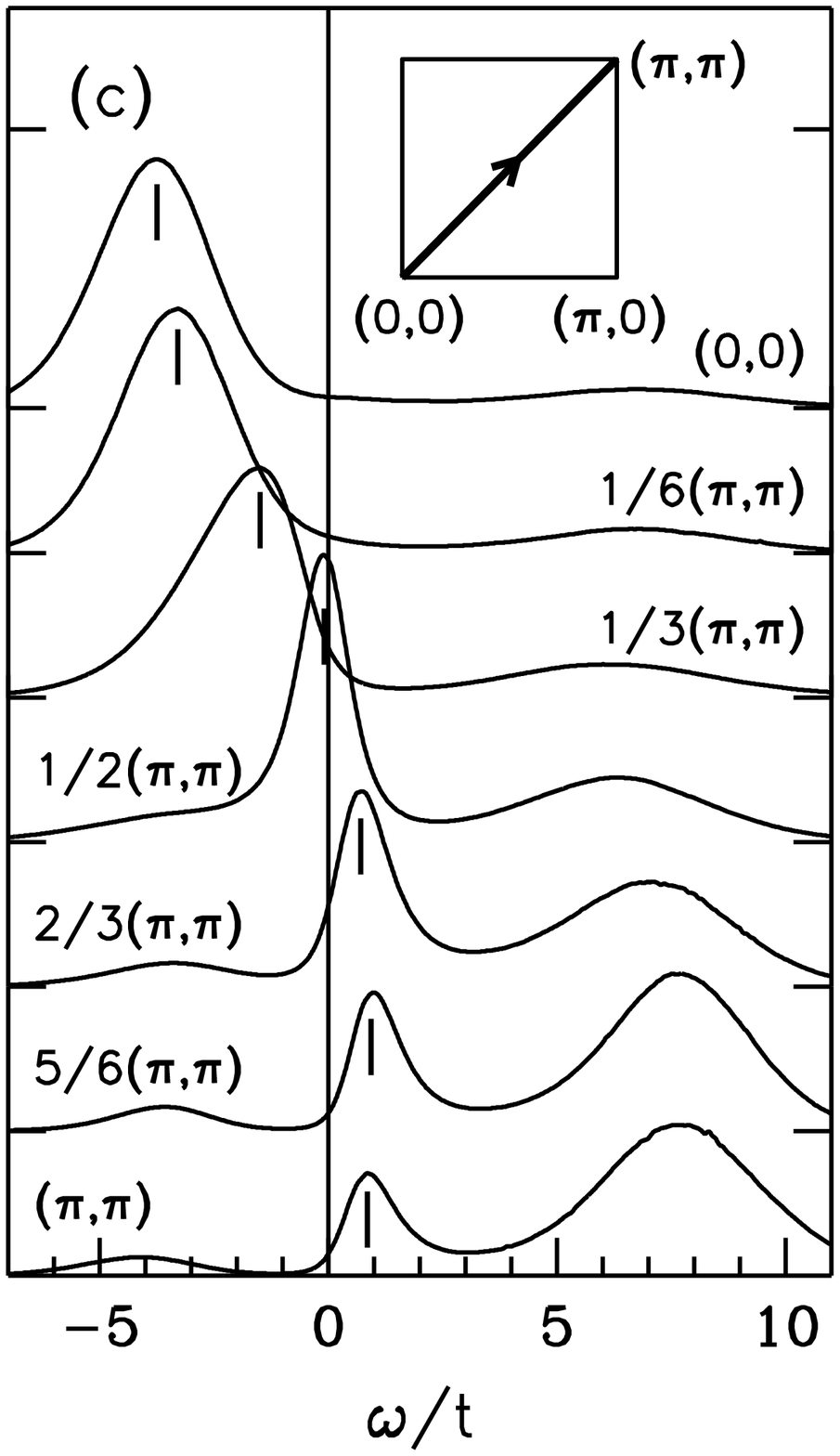}}
\caption{
Single-particle spectral weight $A({\bf p},\omega)$ versus
$\omega$ for various momentum cuts through the 
Brillouin zone for $U=8t$ and $\langle n\rangle = 0.87$.
Here, the frequency axis is shifted such that the Fermi level 
occurs at $\omega=0$.
}
\label{5.7}
\end{figure}

An important feature of the quasiparticle band 
is that it is unusually flat near the $(\pi,0)$ 
and $(0,\pi)$ points of the Brillouin zone. 
This causes large amount of spectral weight 
to be near the Fermi level.  
Even though the weight in the quasiparticle peak near 
the Fermi level decreases
as $U$ increases from $4t$ to $8t$, 
the density of states at
the Fermi level stays nearly the same. 
This is because of the flat bands which are pinned near 
the Fermi level. 
The presence of the flat bands is especially important 
because they generate phase space 
for the scattering of the quasiparticles 
in the $d_{x^2-y^2}$-wave pairing channel.
These features found in the 2D Hubbard model 
are similar to the ARPES data on the cuprates 
[Dessau {\it et al.} 1993, Gofron {\it et al.} 1993].

In addition to the QMC data,
in Ref.~[Dagotto {\it et al.} 1994], 
the relation with the flat bands 
seen in the ARPES data were noted by using 
the dispersion of one hole in the half-filled $t$-$J$
model calculated on a $16\times 16$ lattice.
It should also be noted that Beenen and Edwards
have found good agreement with the main features of the 
QMC data using a two-pole approximation in 
calculating the single-particle Green's function
[Beenen and Edwards 1995].
Dorneich {\it et al.} have extended this type of calculations 
for the Hubbard model for the case of $\langle n\rangle=1$
[Dorneich {\it et al.} 2000].
Finally, it should be noted that 
a number of transport measurements
on the cuprates have been interpreted in terms of the Fermi
level being close to a van Hove singularity
[Tsuei {\it et al.} 1992, Newns {\it et al.} 1994].
Here, 
it is seen that in the 2D Hubbard model 
there can be extended flat bands near the Fermi level and 
they are produced by the many-body effects 
rather than being due to the one-electron band structure.

These calculations for $A({\bf p},\omega)$ were also 
carried out at half-filling, and the results were fitted 
with the SDW form [Bulut {\it et al.} 1994a],
\begin{equation}
A({\bf p},\omega) = u^2_{\bf p} \delta(\omega-E_{\bf p}) +
v^2_{\bf p}\delta(\omega+E_{\bf p})
\end{equation}
where $u^2_{\bf p}={1\over 2}( 1 + \gamma_{\bf p}/E_{\bf p} )$,
$v^2_{\bf p} = {1\over 2}(1 - \gamma_{\bf p}/E_{\bf p})$,
$\gamma_{\bf p} = -2t(\cos{p_x} + \cos(p_y))$, 
and $E_{\bf p}=\sqrt{\gamma^2_{\bf p} + \Delta^2}$,
For large $\Delta/t$,
\begin{equation}
E_{\bf p} = \sqrt{ \gamma^2_{\bf p} + \Delta^2 }
\simeq \Delta + J(\cos{p_x} + \cos{p_y})^2. 
\end{equation}
Similar calculations with higher resolution 
[Preuss {\it et al.} 1995] were carried out, which
showed that, in fact, the spectrum is better described by 
$E_{\bf p}\simeq \Delta + {J \over 2}( \cos{p_x} + \cos{p_y} )^2$,
hence, 
the quasiparticle band at half-filling has a width of $2J$ 
in agreement with the results on 
$N(\omega)$ at half-filling and 
with the calculations on the $t$-$J$ model with one hole.

Based on the results discussed above, 
schematic dispersion 
relations were constructed for the 2D Hubbard model, 
which are shown in Fig.~5.8
for $\langle n\rangle=0.87$ and $\langle n\rangle=1.0$.
In Fig.~5.8(a),
the dashed horizontal line represents the chemical potential,
and the solid curve denotes the narrow metallic band located
at the Fermi level.
The shaded regions represent the lower and the upper 
Hubbard bands.
The width of the quasiparticle band is of order $4J$.
Figure~5.8(b) illustrates the distribution of the spectral 
weight at half-filling.
Here, the solid and the dotted lines denote 
the SDW-like quasiparticle bands.
The picture which emerges from these calculations is that the 
narrow metallic band which lies at the Fermi level
for $\langle n\rangle=0.87$ forms out of the SDW-like bands
which exist in the half-filled case.

The QMC results seen in Fig.~5.7 were obtained at $T=0.5t$ for 
$\langle n\rangle =0.87$, where the AF correlation length 
$\xi$ is less than one lattice spacing. 
The QMC simulations were also carried out at $T=0.25t$
for $\langle n \rangle =0.93$, where $\xi$ is about 1.2
lattice spacing 
[Preuss {\it et al.} 1997].
In this case, 
spectral weight has been observed at wave vectors 
which are displaced by ${\bf Q}=(\pi,\pi)$ with respect to the 
narrow metallic band at the Fermi level.
These results are consistent with the calculations 
where "shadow bands" due to the scattering 
of the quasiparticles by the AF spin fluctuations were found
[Kampf and Schrieffer 1990b].
This type of shadow structures were also observed in the 
photoemission experiments 
[Aebi {\it et al.} 1995].
Detailed single-particle spectra 
in the low-doping regime of the 2D Hubbard 
model were obtained recently
[Gr\"ober {\it et al.} 2000].

\begin{figure}[ht]
\centering
\iffigure
\mbox{
\subfigure[]{
\epsfysize=8cm
\epsffile[100 150 480 610]{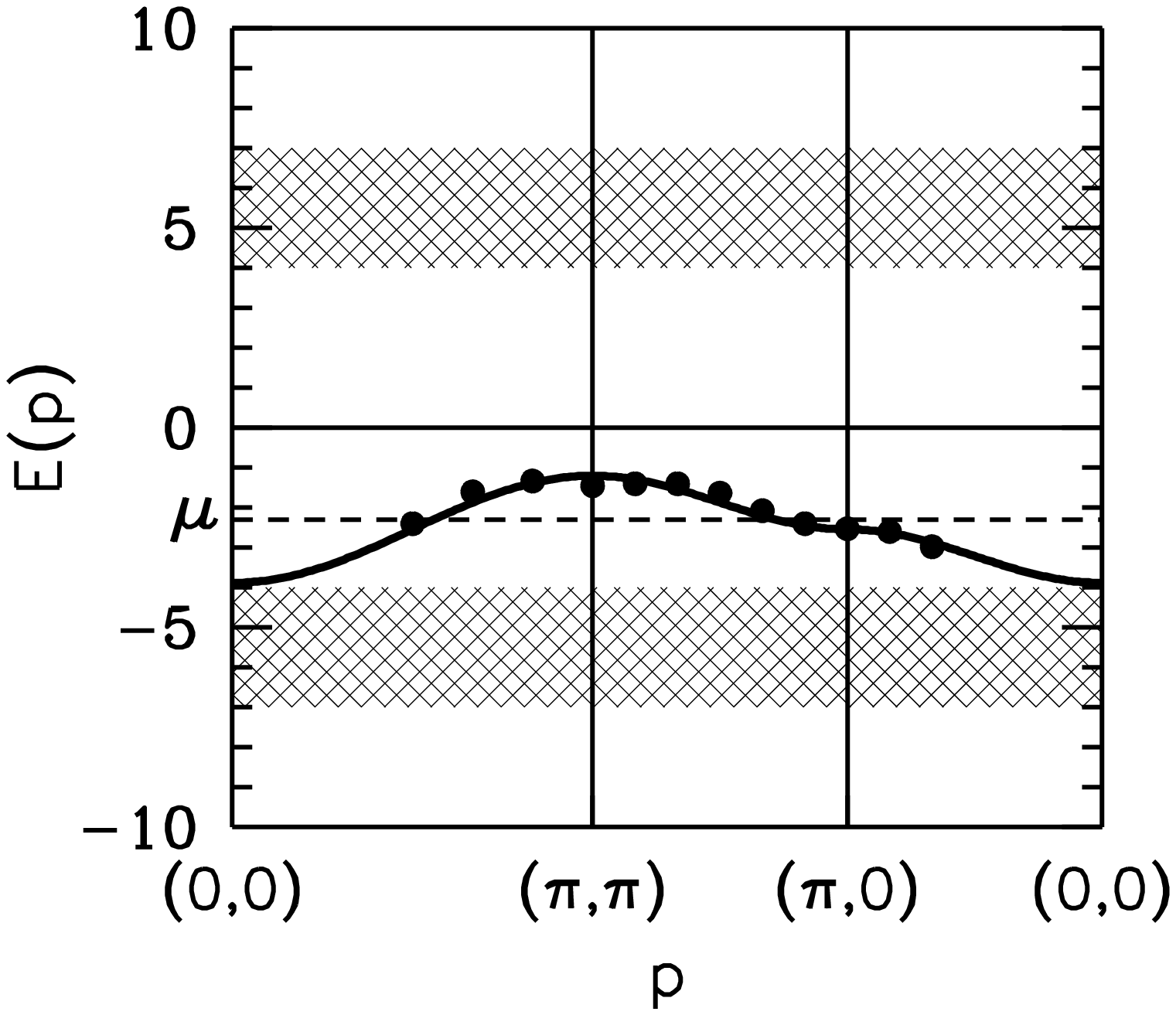}}
\quad
\subfigure[]{
\epsfysize=8cm
\epsffile[50 150 600 610]{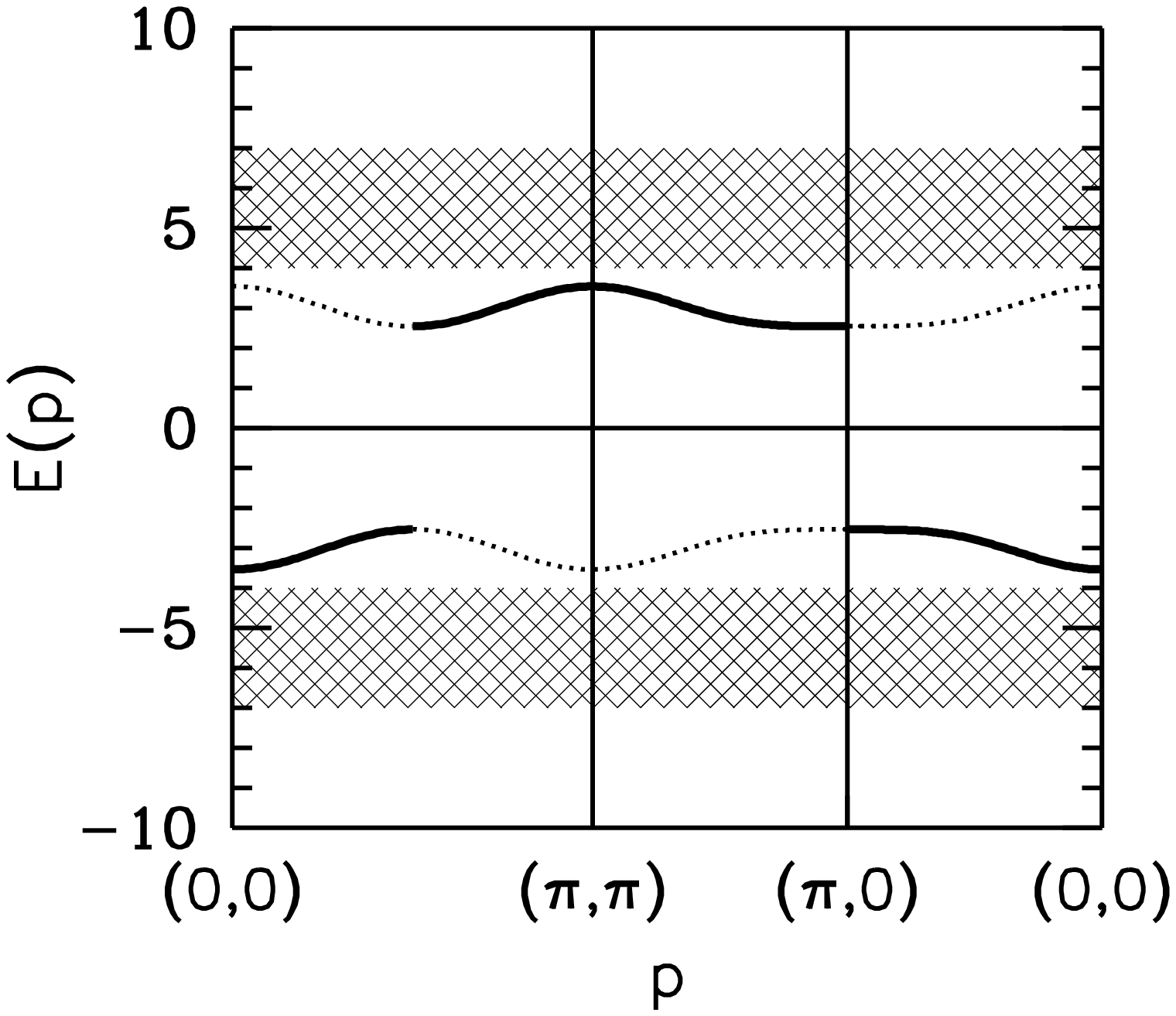}}}
\fi
\caption{
Schematic drawing of 
$E_{\bf p}$ versus ${\bf p}$ at $U=8t$ 
for (a) $\langle n\rangle =0.87$ 
and (b) $\langle n\rangle=1.0$.
Here, the shaded areas represent the lower and the upper Hubbard
bands, and the solid and the dotted curves indicate the 
quasiparticle bands.
In (a), the solid points indicate the position 
of the quasiparticle peak obtained from Fig.~5.7
when the quasiparticle peak is not obscured by the lower Hubbard band,
and the horizontal dashed line denotes the chemical potential
$\mu$.
}
\label{5.8}
\end{figure}

In this section, 
it was seen that the single-particle 
properties in the 2D Hubbard model 
are determined by the AF spin fluctuations and the 
Coulomb correlations.
An especially important feature of the single-particle 
spectral weight is the flat bands near 
${\bf p}=(\pi,0)$ and $(0,\pi)$.
They create large amount of phase space for the scattering of the 
quasiparticles in the $d_{x^2-y^2}$-wave pairing channel.
In Section 7, it will be seen that similar flat bands exist in the 
2-leg Hubbard ladder, and they play a key role 
in determining the strength of the $d_{x^2-y^2}$-like 
superconducting correlations in this system.
In Section~8.1, these QMC results on the single-particle 
spectral weight will be compared with those obtained from the 
FLEX approximation, and the implications for the 
$d_{x^2-y^2}$-wave pairing in the 2D Hubbard model will be 
discussed.

\setcounter{equation}{0}\setcounter{figure}{0}
\section{Pairing correlations in the 2D Hubbard model}

An important question for the high-$T_c$ cuprates is 
whether the 2D Hubbard model exhibits superconductivity, 
and if does, what is the nature of the pairing. 
Various views were expressed on this subject, and 
some of them are as follows.
In one view, 
the effective interaction between the particles is dominated 
by the exchange of an AF spin fluctuation, and 
this leads to $d_{x^2-y^2}$-wave pairing
[Bickers {\it et al.} 1987].
In an alternative view, 
the magnetic correlations are responsible for the pairing,
but it is because of the "bag effect" which leads to pairing
with the extended $s$ or the $d_{x^2-y^2}$-wave symmetry
[Schrieffer {\it et al.} 1988 and 1989].
In this approach, the effective pairing interaction is
qualitatively different than the single spin-fluctuation
exchange interaction.
Another view is that in the ground state of the 
doped 2D Hubbard model there are no single-particle 
excitations carrying both charge and spin and 
the one-layer Hubbard model does not exhibit superconductivity
[Anderson 1987, Anderson {\it et al.} 1987, Anderson and Zou 1988].
In this view, 
the pairing is mediated by interlayer hopping.
In order to differentiate among these theories, 
QMC simulations were carried out for the 2D Hubbard model.
In this section, these QMC calculations will be reviewed.

In Section~6.1, the QMC results on the pair-field susceptibilities
will be reviewed. 
These calculations have found that there is an effective 
attractive interaction in the singlet $d_{x^2-y^2}$-wave channel
[White {\it et al.} 1989a].
In addition, it is found that there is an 
attractive interaction in the extended $s$-wave channel.
However, at the temperatures where the simulations are carried out,
the pairing correlations are short range and do not show scaling with 
the lattice size
[Moreo and Scalapino 1991, Moreo 1992].
QMC simulations were also carried out for the pair-field 
susceptibilities in the three-band CuO$_2$ model
and similar results were found
[Scalettar {\it et al.} 1991, Dopf {\it et al.} 1992a].

Using QMC simulations, the irreducible particle-particle 
interaction $\Gamma_I$ was calculated
[Bulut {\it et al.} 1993, 1994b, 1995] 
and these results will be reviewed in Section~6.2.
In these calculations, it is found that the momentum and the 
Matsubara-frequency structure in $\Gamma_I$ follows closely 
that of the magnetic susceptibility, 
which means that the AF spin fluctuations dominate the 
effective interaction in the parameter regime 
where the simulations are carried out.
In Section~6.3, the Bethe-Salpeter equation in the 
particle-particle channel will be solved using the 
QMC data on $\Gamma_I$ and the single-particle Green's function $G$.
The solution of the Bethe-Salpeter equation makes 
it possible to determine the strength of the various pairing channels
quantitatively.
Finite-size scaling of these results shows that as $T$ is lowered, 
the fastest growing pairing instability 
is in the singlet $d_{x^2-y^2}$-wave channel.
Here, it is also found that as $U/t$ increases from 4 to 8, 
the strength of the $d_{x^2-y^2}$ pairing correlations grows.
Later in Section~8.1, these calculations will be compared with the 
results of the FLEX approximation.
In Section~6.4, the results on $\Gamma_I$ will be compared with the
single spin-fluctuation exchange approximation.
In Section~6.5, comparisons will be made with the perturbation theory
results for $\Gamma_I$, which are third order in $U$.
These comparisons are useful for gaining insight into the 
effects of the various subgroups of many-body scattering processes 
contributing to $\Gamma_I$.

\subsection{Pair-field susceptibilities}

In order to study the pairing correlations in the 2D Hubbard model,
the pair-field susceptibilities defined by 
\begin{equation}
\label{P}
P_{\alpha} = \int_0^{\beta} \, d\tau \,
{1\over N} \sum_{\ell} \,
\langle \Delta_{\alpha}(\ell,\tau) 
\Delta^{\dagger}_{\alpha}(0,0) \rangle,
\end{equation}
where
$\Delta^{\dagger}_{\alpha}$ is a pair creation 
operator with $\alpha$ symmetry,
were calculated.
The QMC calculations of the pair-field susceptibilities 
for the Hubbard model were first carried out 
in Ref.~[Hirsch 1985].
In these calculations, it was found that there is an 
attractive effective interaction between two electrons 
with antiparallel spins when they are separated by one lattice 
spacing.
In this reference, 
the pair-field susceptibilities in the singlet $s$ and 
extended $s$-wave ($s^*$) channels and the triplet channels were 
calculated, however the singlet $d_{x^2-y^2}$-wave 
channel was not considered.
Later in Ref.~[White {\it et al.} 1989a], 
various pair-field susceptibilities including the 
one with the singlet $d_{x^2-y^2}$-wave symmetry were 
calculated with QMC.
In this section, results from Ref.~[White {\it et al.} 1989a]
will be shown for $P_{\alpha}$ which were calculated using the 
following pair-field operators in the 
$d_{x^2-y^2}$, $s$, extended $s$ ($s^*$) and 
$p$-wave channels,
\begin{eqnarray}
\Delta^{\dagger}_d && = {1\over 4} \sum_{\ell=1}^4 \,
(-1)^{\ell} 
c^{\dagger}_{i+\ell \uparrow}
c^{\dagger}_{i \downarrow}, \\
\Delta^{\dagger}_s && = 
c^{\dagger}_{i \uparrow}
c^{\dagger}_{i \downarrow}, \\
\Delta^{\dagger}_{s^*} && = {1\over 4} \sum_{\ell=1}^4 \,
c^{\dagger}_{i+\ell \uparrow}
c^{\dagger}_{i \downarrow}, \\
\Delta^{\dagger}_{p_x} && = {1\over 2} 
(c^{\dagger}_{i+{\bf x} \uparrow}
c^{\dagger}_{i \downarrow} 
- c^{\dagger}_{i-{\bf x} \uparrow}
c^{\dagger}_{i \downarrow}).
\end{eqnarray}
Here $\ell$ sums over the four neighbours
of site $i$. 

\begin{figure}
\centering
\iffigure
\epsfig{file=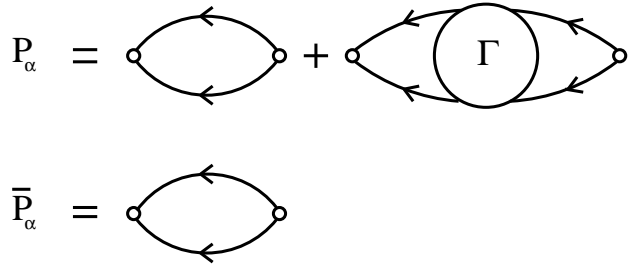,height=3.5cm}
\fi
\caption{
Feynman diagrams illustrating the pair-field
susceptibilities 
(a) $P_{\alpha}$ and
(b) $\overline{P}_{\alpha}$.
Here, 
$\Gamma$ represents the reducible particle-particle interaction.
}
\label{6.1}
\end{figure}

In order to see whether a given pairing channel 
with the symmetry $\alpha$ is attractive, 
$P_{\alpha}$ was compared with 
$\overline{P}_{\alpha}$, 
where $\overline{P}_{\alpha}$ is the 
component of $P_{\alpha}$ which does not include
the reducible particle-particle interaction $\Gamma$
as illustrated in Fig.~6.1.
Figure 6.2 compares $P_{\alpha}$ and $\overline{P}_{\alpha}$ 
as a function of $T/t$ for $U=4t$ and $\langle n\rangle=0.87$
on an $8\times 8$ lattice.
Here, it is seen that $P_{\alpha}$ is enhanced 
with respect to $\overline{P}_{\alpha}$ by the particle-particle
reducible vertex for the $d_{x^2-y^2}$-wave and weakly for the extended
$s$-wave symmetries.
On the other hand, in the $s$- and $p$-wave 
channels the pair-field susceptibility is suppressed
with respect to $\overline{P}_{\alpha}$.
This means that the effective particle-particle interaction 
is attractive in the $d_{x^2-y^2}$ and $s^*$  channels,
and it is repulsive in the $s$ and $p$-wave channels.
However, at the lowest temperatures where 
the QMC simulations can be carried out, 
the $d_{x^2-y^2}$-wave pairing correlations are only short range:
$P_d$ does not grow as the system size increases.  
Furthermore, 
while $P_d$ is enhanced with respect to $\overline{P}_d$,
it is suppressed with respect to $P_d^0$, 
the $d_{x^2-y^2}$-wave pair-field susceptibility of the 
$U=0$ system, as seen in Fig.~6.2(a).
So, at these temperatures, the $d_{x^2-y^2}$-wave  
pair-field susceptibility gets suppressed when the Coulomb 
repulsion is turned on,
which is not encouraging for $d_{x^2-y^2}$-wave 
superconductivity in the 2D Hubbard model.

\begin{figure}
\centering
\iffigure
\mbox{
\subfigure{
\epsfysize=8cm
\epsffile[100 150 480 610]{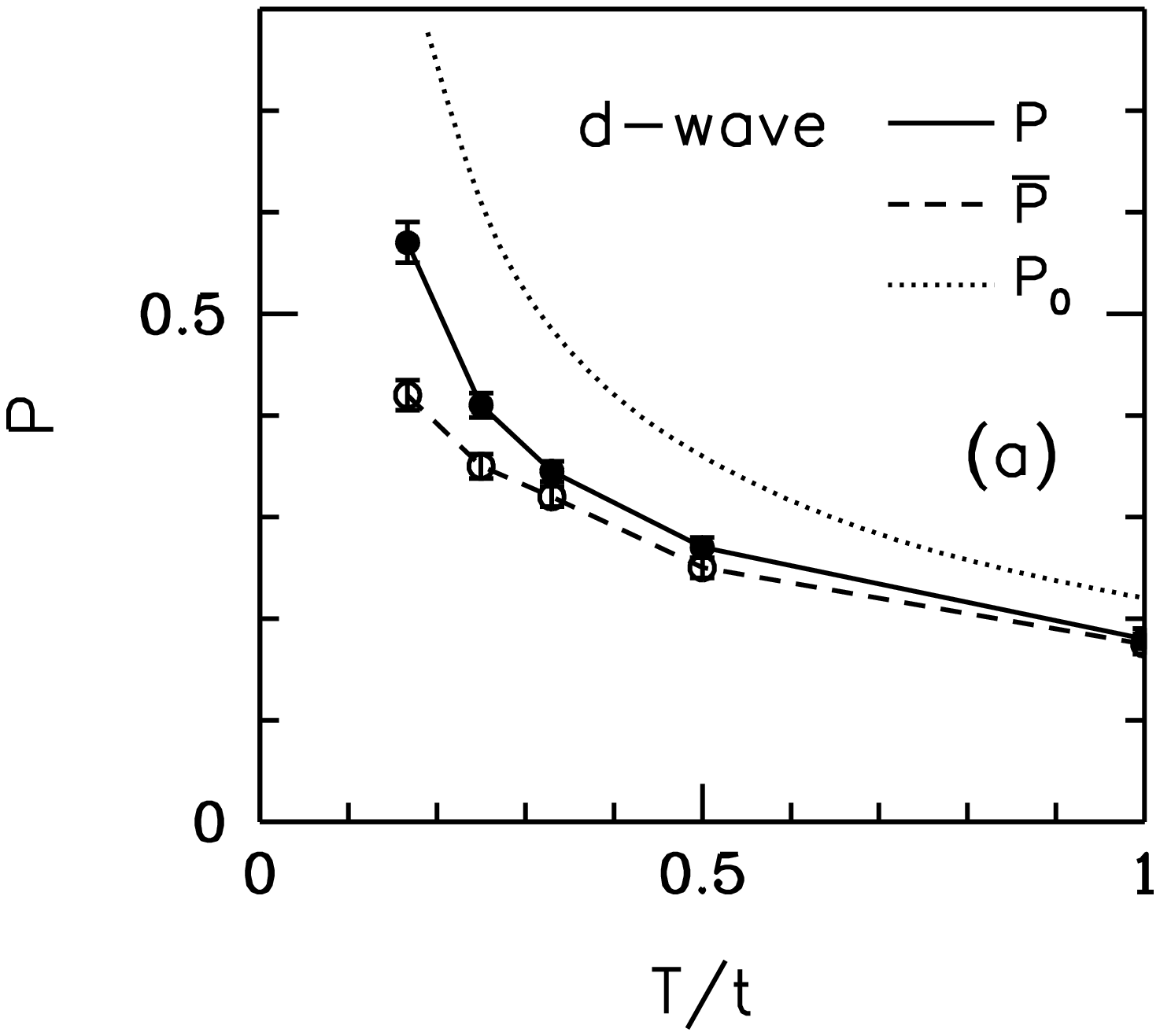}}
\quad
\subfigure{
\epsfysize=8cm
\epsffile[50 150 600 610]{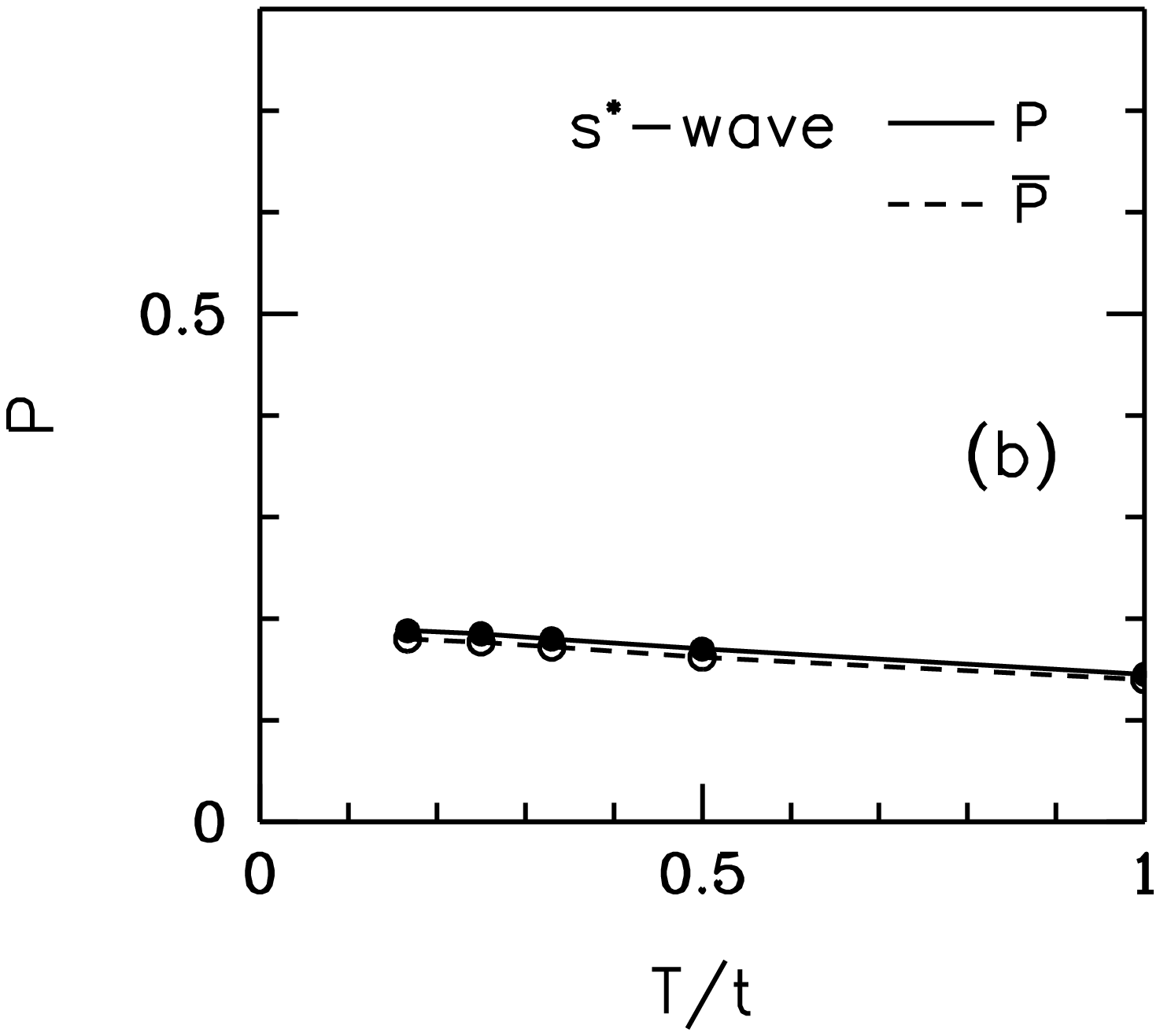}}}
\fi
\centering
\iffigure
\mbox{
\subfigure{
\epsfysize=8cm
\epsffile[100 150 480 610]{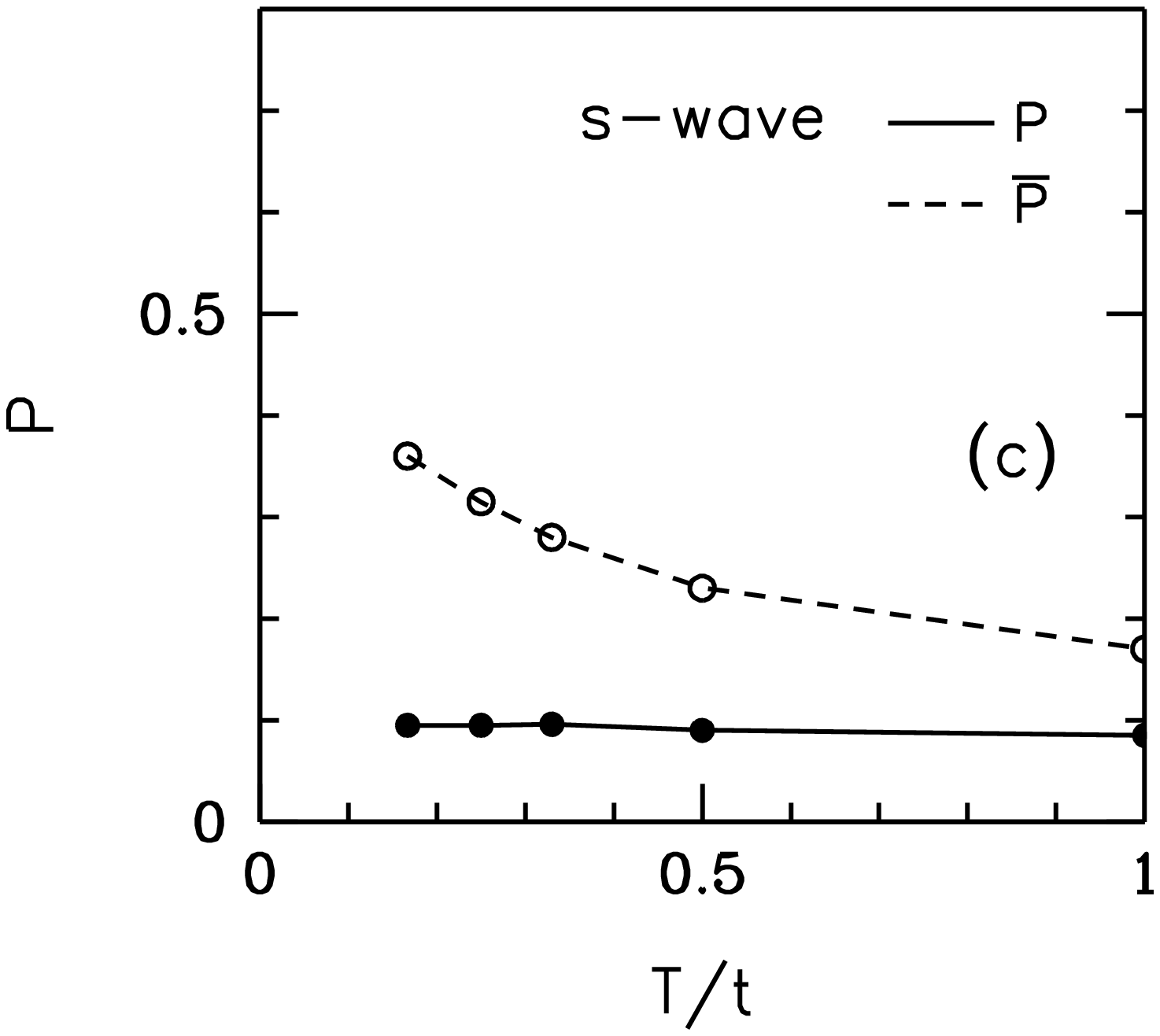}}
\quad
\subfigure{
\epsfysize=8cm
\epsffile[50 150 600 610]{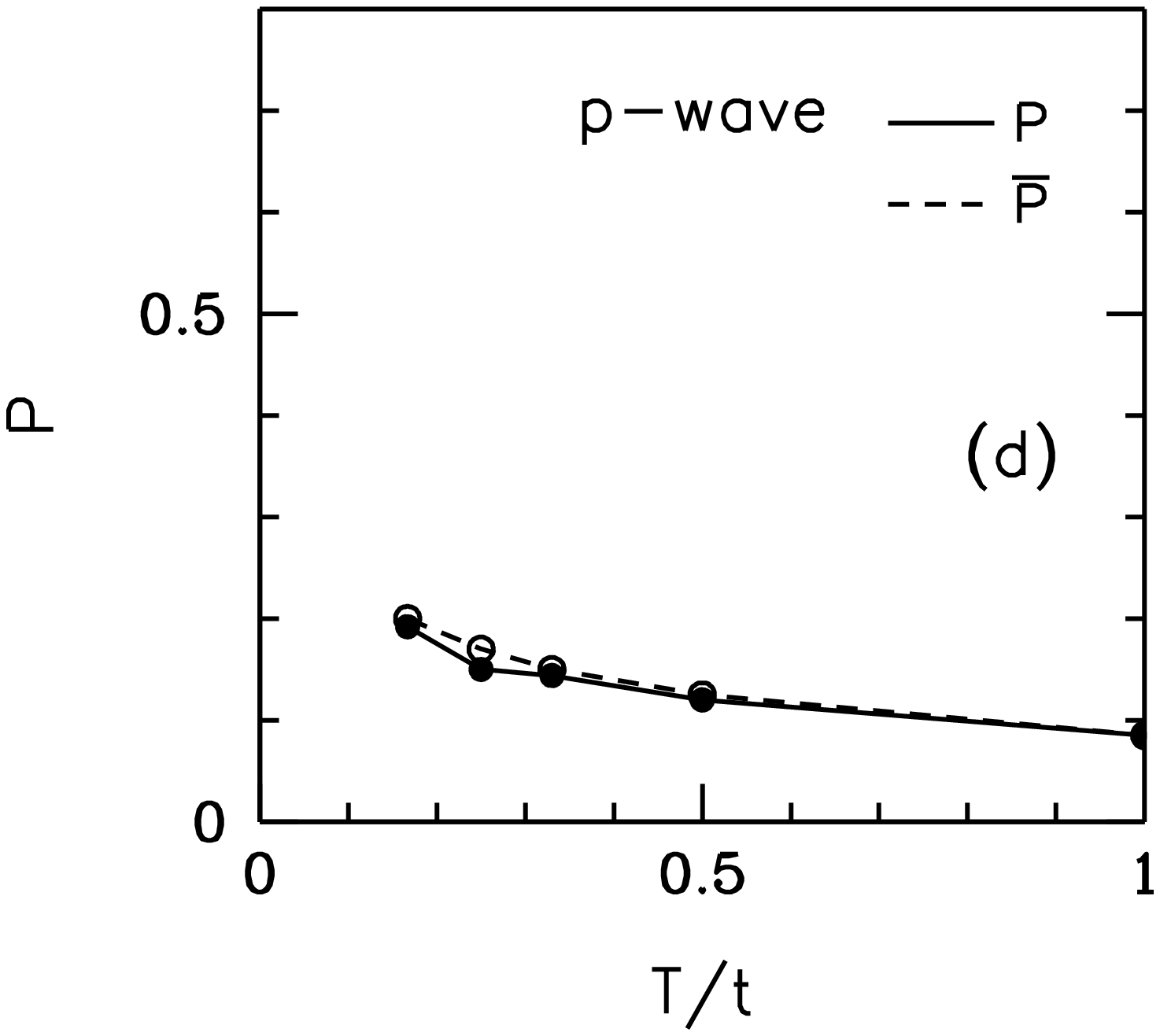}}}
\fi
\caption{
Pair-field susceptibilities $P_{\alpha}$ and
$\overline{P}_{\alpha}$ versus $T/t$ for 
the $d_{x^2-y^2}$, $s^*$, $s$ and $p$-wave 
pairing channels.
These results are for 
$U=4t$ and $\langle n\rangle=0.87$ on an $8\times 8$ lattice.
In (a), $P_0$ denotes the $d_{x^2-y^2}$-wave 
pair-field susceptibility of the $U=0$ system.
From [White {\it et al.} 1989a].
}
\label{6.2}
\end{figure}

\subsection{Irreducible particle-particle interaction}

In this section, the QMC data on the 
irreducible particle-particle vertex $\Gamma_I$
from Ref.~[Bulut {\it et al.} 1993, 1994b, 1995] 
will be reviewed.
Using QMC simulations it is possible 
to calculate the two-particle Green's function 
\begin{equation}
\Lambda(x_4,x_3|x_2,x_1) = - \langle T \,
c_{\uparrow}(x_4) c_{\downarrow}(x_3)
c^{\dagger}_{\downarrow}(x_2) c^{\dagger}_{\uparrow}(x_1)
\rangle,
\end{equation}
where $c^{\dagger}_{\sigma}(x_i)$ with $x_i=({\bf x}_i,\tau_i)$ 
creates an electron with spin $\sigma$ at site ${\bf x}_i$
and imaginary time $\tau_i$.
By Fourier transforming on both the space and the 
imaginary-time variables one obtains
\begin{equation}
\Lambda(p',k'|p,k) = -\delta_{pp'} \delta_{kk'} 
G_{\uparrow}(p) G_{\downarrow}(k) + 
{T\over N} \delta_{k',p+k-p'}
G_{\uparrow}(p') G_{\downarrow}(k') 
\Gamma(p',k'|p,k) 
G_{\uparrow}(p) G_{\downarrow}(k),
\end{equation}
where $p=({\bf p},i\omega_n)$,
$G_{\sigma}(p)$ is the single-particle Green's function, and 
$\Gamma(p',k'|p,k)$ is the reducible particle-particle
vertex.
Hence, using the QMC data on $\Lambda$ and $G$,
$\Gamma$ can be calculated.
This equation is illustrated in Fig.~6.3 in terms of 
the Feynman diagrams.
Here, the particle-particle interaction will be studied in the 
zero center-of-mass momentum and energy channel, 
and hence $k$ and $k'$ will be set to $-p$ and $-p'$, respectively.
At the next stage, 
the irreducible particle-particle vertex 
$\Gamma_I$ is obtained from 
the Monte Carlo results on $\Gamma$ and $G$ 
by solving the particle-particle $t$-matrix equation,
\begin{equation}
\Gamma_I(p'|p) = \Gamma(p'|p) 
+ {T\over N} \sum_k \,
\Gamma_I(p'|k) G_{\uparrow}(k) G_{\downarrow}(-k)
\Gamma(k|p), 
\end{equation}
which is illustrated in Fig.~6.4.
In Eq.~(6.8), $\Gamma(p'|p)$ is used as a short notation for 
$\Gamma(p',-p'|p,-p)$.
This procedure for calculating $\Gamma_I$ is essentially the 
opposite of the usual diagrammatic approach in which $\Gamma_I$ 
is used to solve for $\Gamma$.
In solving the $t$-matrix equation, 
an upper frequency cut-off of order the bandwidth 
is used.
Finally, 
the singlet component of the irreducible vertex 
is obtained from 
\begin{equation} 
\Gamma_{Is}(p'|p) = {1\over 2} 
\bigg[ \Gamma_I(p'|p) + \Gamma_I(-p'|p) \bigg].
\end{equation}
In the following, the QMC data on $\Gamma_{Is}$ and 
$\Gamma_s$ will be plotted 
in units of $t$ as a function of the momentum 
transfer ${\bf q}={\bf p'}-{\bf p}$
and of the Matsubara-frequency transfer
$\omega_m=\omega_{n'}-\omega_n$.
Here, ${\bf p}$ will be kept fixed at $(\pi,0)$ 
and $\omega_n$ at $\pi T$.

\begin{figure}
\centering
\iffigure
\epsfig{file=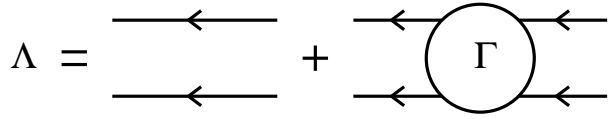,height=1.5cm}
\fi
\caption{
Feynman diagrams illustrating the 
correlation function 
$\Lambda(p',k'|p,k)$
in terms of the single-particle Green's functions $G$ and 
the reducible particle-particle interaction $\Gamma$.
}
\label{6.3}
\end{figure}

\begin{figure}
\centering
\iffigure
\epsfig{file=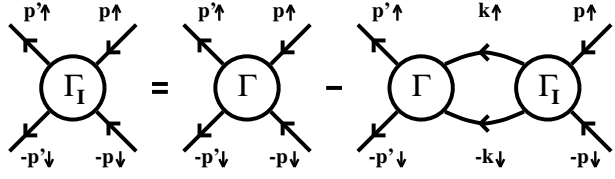,height=8cm,angle=+90}
\fi
\caption{
Feynman diagrams illustrating the $t$-matrix equation 
which relates the irreducible particle-particle interaction
$\Gamma_I$ to the reducible particle-particle interaction
$\Gamma$. 
}
\label{6.4}
\end{figure}

It is useful to start the discussion by first showing 
results on the singlet reducible vertex $\Gamma_s$.
In Fig.~6.5, 
$\Gamma_s({\bf q},i\omega_m=0)$ 
versus ${\bf q}$ is shown for $U=4t$ and $8t$.
As the temperature is lowered, 
the ${\bf q}\sim (\pi,\pi)$ component of 
$\Gamma_s$ increases,
and the ${\bf q}\sim 0$ component is suppressed.
Here, it is seen that $\Gamma_s$ becomes quite large especially 
for $U=8t$.
In Fig.~6.5(b), the ${\bf q}=(0,0)$ point is not shown 
because of large error bars
for this point due to the way $\Gamma$ is calculated 
from $\Lambda$.
It is also not possible to show results at lower temperatures
since the error bars grow rapidly.
For instance, at $T=0.33t$ and $U=8t$, 
$\Gamma_s({\bf q}=(\pi,\pi),0)$ was calculated to be 
$60t\pm 20t$ after long simulation times,
while $\Gamma_s({\bf q}=(\pi/4,\pi/4),0)$
was $10t\pm 10t$. 

\begin{figure}
\centering
\iffigure
\mbox{
\subfigure[]{
\epsfysize=8cm
\epsffile[100 150 480 610]{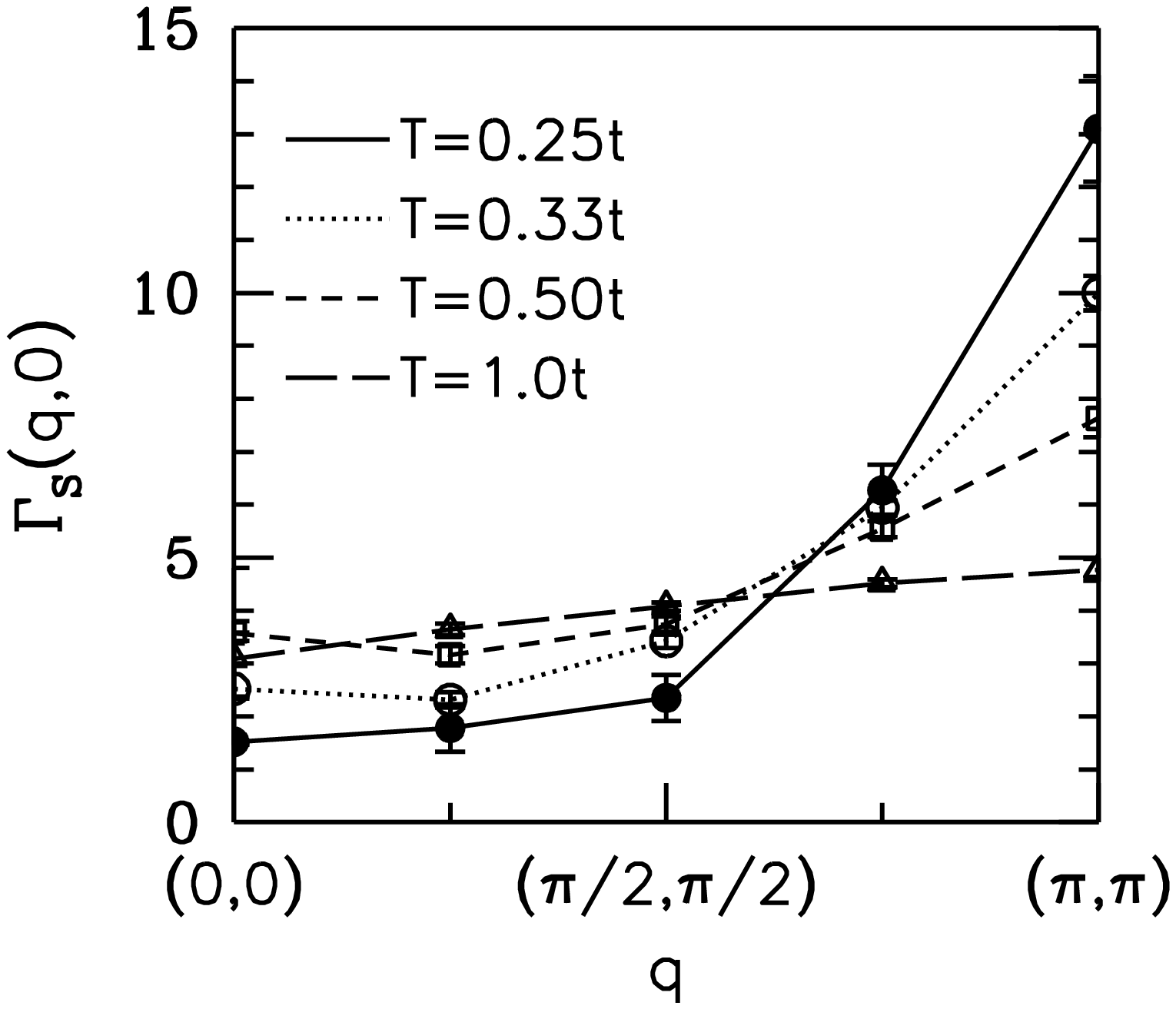}}
\quad
\subfigure[]{
\epsfysize=8cm
\epsffile[50 150 600 610]{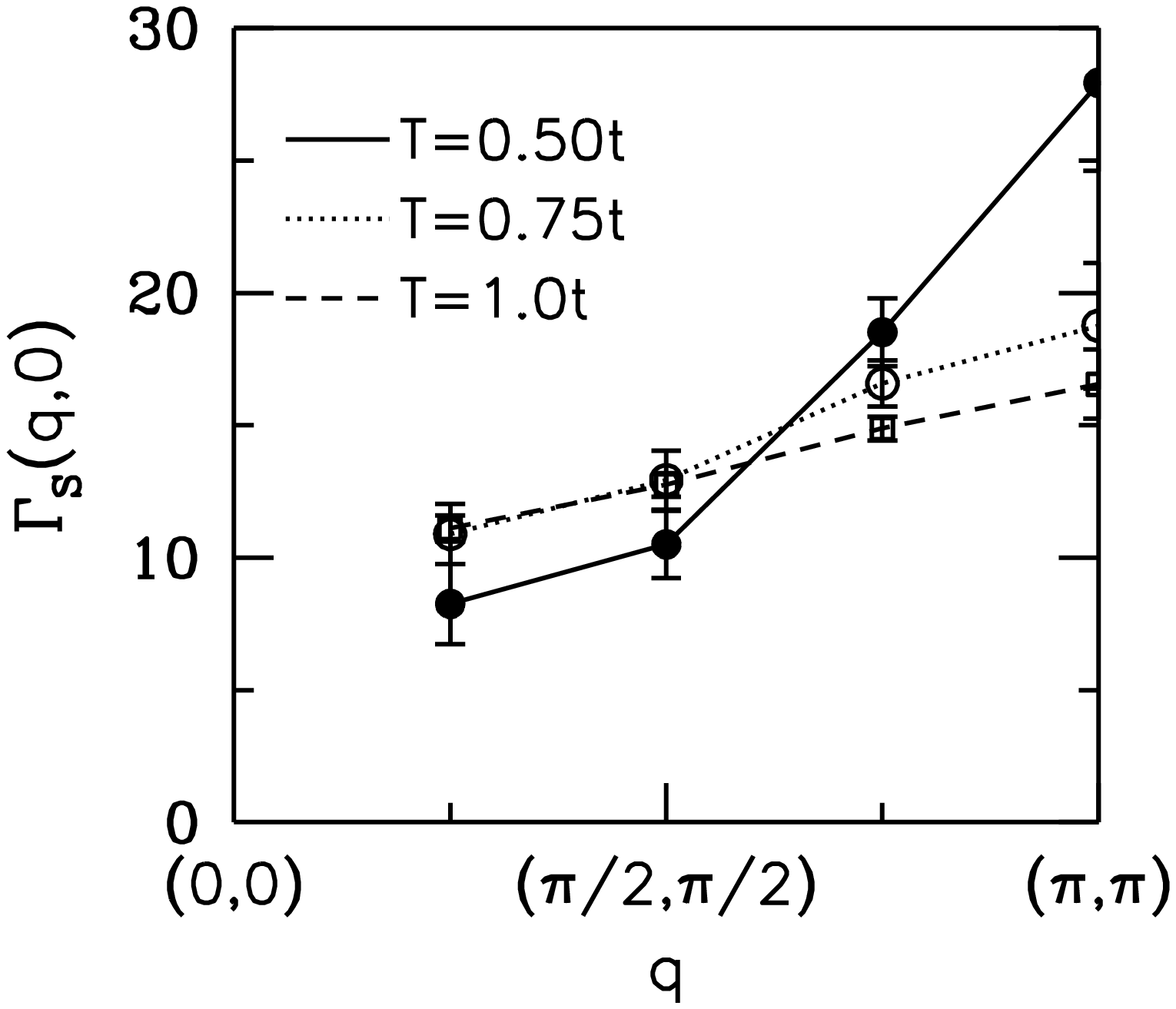}}}
\fi
\caption{
Reducible particle-particle scattering vertex in the singlet channel
$\Gamma_s({\bf q},i\omega_m=0)$ versus ${\bf q}$
for (a) $U=4t$ and (b) $U=8t$.
Here, ${\bf q}={\bf p'}-{\bf p}$ and ${\bf p}$ is kept fixed at 
$(\pi,0)$.
These results were obtained 
at $\langle n\rangle=0.87$ on an $8\times 8$ lattice.
For $U=8t$, 
$\Gamma_s({\bf q},0)$ at ${\bf q}=(0,0)$ is not shown 
because of large error bars in the data in this case.
}
\label{6.5}
\end{figure}

Next, in Fig.~6.6(a) and (b), the momentum 
and the Matsubara-frequency 
dependence of the irreducible particle-particle vertex
$\Gamma_{Is}({\bf q},i\omega_m)$ is shown for $U=4t$.
At ${\bf q}=(\pi,\pi)$ momentum transfer, 
$\Gamma_{Is}({\bf q},0)$ reaches values larger than twice the 
bandwidth. 
It is useful to compare Figs.~6.6(a) and 6.5(a)
in order to understand the effect of the repeated 
particle-particle scatterings on $\Gamma_{Is}$.
At these temperatures, the effect of these scatterings is 
basically to suppress the momentum and frequency-independent 
background in $\Gamma_{Is}$.
For instance, at $T=0.25t$ the difference in the magnitude of 
$\Gamma_{Is}({\bf q},0)$ between 
${\bf q}=(\pi,\pi)$ and $(\pi/4,\pi/4)$ is about $10t$,
which is the same as in $\Gamma_s({\bf q},0)$.
This is similar to the suppression of the screened Coulomb
repulsion, which varies more slowly in frequency 
compared to the phonon propagator, 
in the usual phonon-mediated pairing.
As a $d_{x^2-y^2}$-wave superconducting instability 
is approached, 
the expected behaviour for an infinite system is that 
$\Gamma_s({\bf q}=(\pi,\pi),0) \rightarrow +\infty$
while $\Gamma_s({\bf q}=(0,0),0) \rightarrow -\infty$.
These QMC data show that such resonant scattering in the 
$d_{x^2-y^2}$-wave channel is not taking place 
at these temperatures.

\begin{figure}
\centering
\iffigure
\mbox{
\subfigure{
\epsfysize=8cm
\epsffile[100 150 480 610]{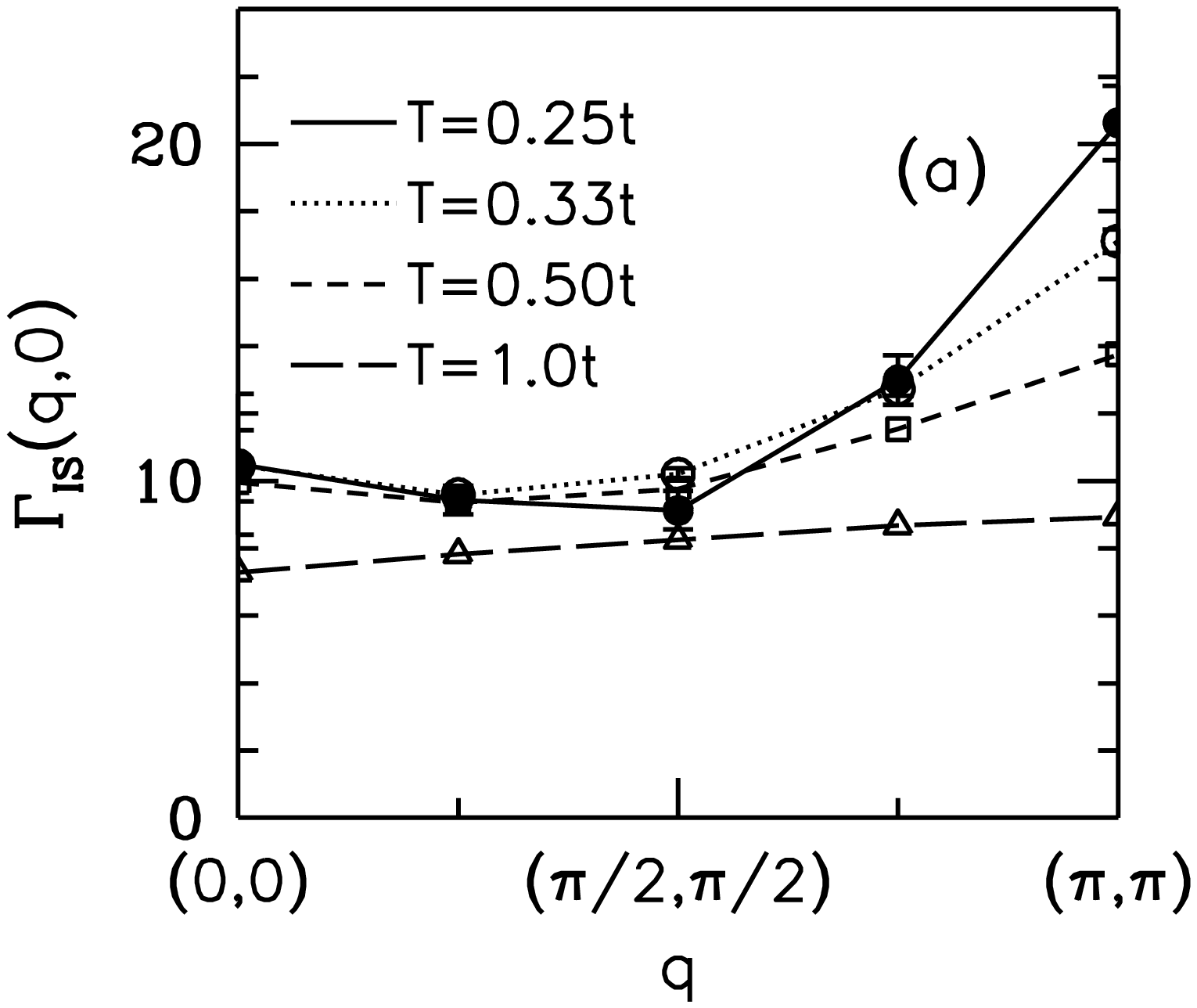}}
\quad
\subfigure{
\epsfysize=8cm
\epsffile[50 150 600 610]{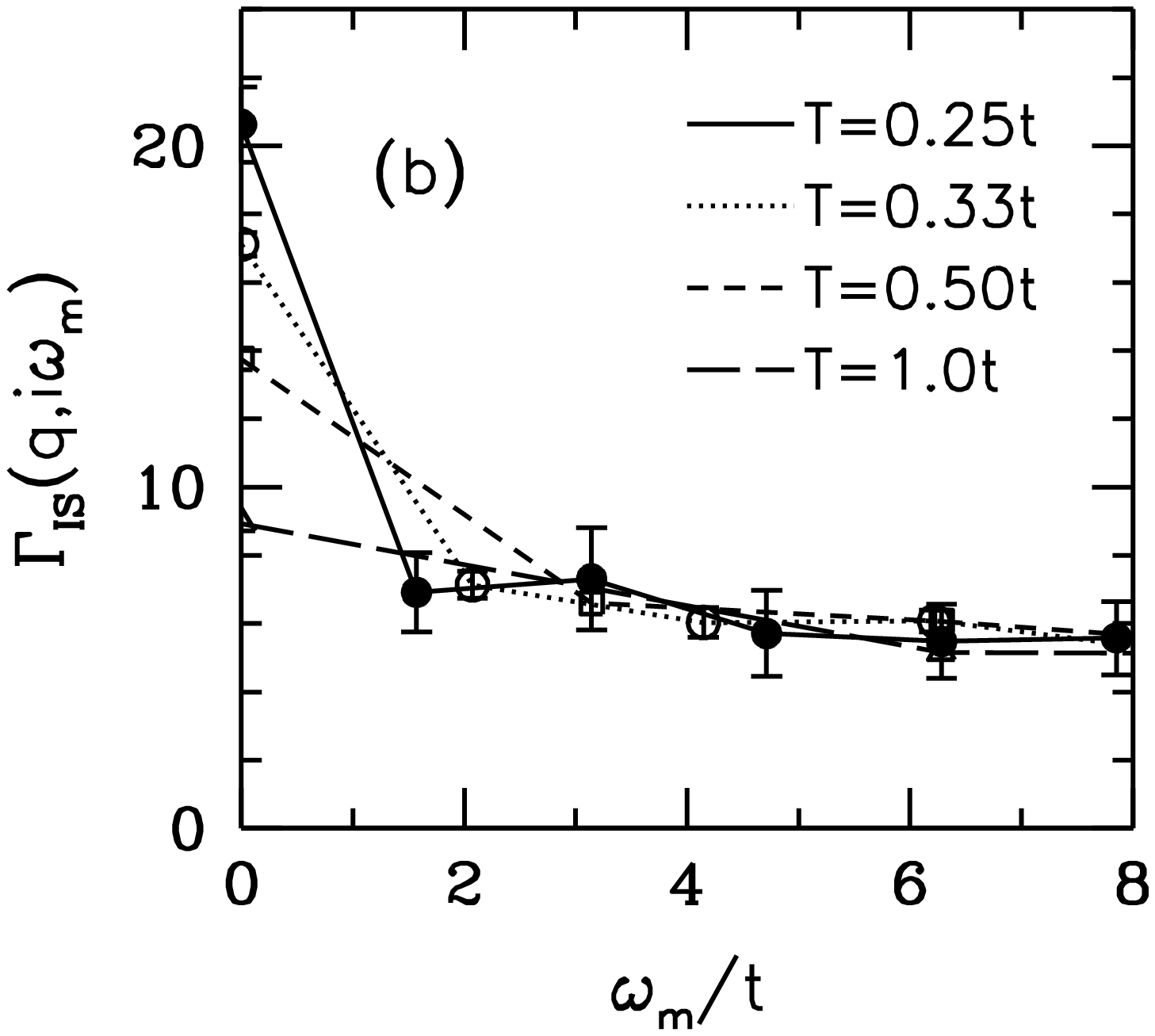}}}
\fi
\caption{
(a) Momentum and (b) the Matsubara-frequency dependence of the
irreducible particle-particle scattering vertex in the singlet channel
$\Gamma_{Is}({\bf q},i\omega_m)$. 
In (a), $\Gamma_{Is}$ versus ${\bf q}={\bf p'}-{\bf p}$ 
is plotted where ${\bf p}$ is kept fixed at $(\pi,0)$.
In (b), $\Gamma_{Is}$ versus $\omega_m$ 
is plotted for ${\bf q}=(\pi,\pi)$.
Here, $\omega_m=\omega_{n'}-\omega_n$ and 
$\omega_n$ is kept fixed at $\pi T$.
These results are for $U=4t$ and $\langle n\rangle=0.87$
on an $8\times 8$ lattice.
}
\label{6.6}
\end{figure}

It is desirable to know how $\Gamma_{Is}$ 
varies as $U/t$ increases.
However, it has not been possible to obtain $\Gamma_{Is}$ for $U=8t$
from the $t$-matrix equation because 
of the larger error bars for this case.
Nevertheless, in Fig.~6.5(b) it is seen that  
$\Gamma_s({\bf q},0)$ for $U=8t$ and $T=0.5t$ exhibits 
large variation of order $20t$ 
between points $(\pi/4,\pi/4)$ and $(\pi,\pi)$.
If at these temperatures the effect of the $t$-matrix 
scattering is only to suppress the 
${\bf q}$ and $\omega_m$ independent background in 
$\Gamma_{Is}$, then this means that $\Gamma_{Is}$ grows
considerably as $U/t$ increases from 4 to 8.

\begin{figure}
\centering
\iffigure
\mbox{
\subfigure[]{
\epsfysize=8cm
\epsffile[100 150 480 610]{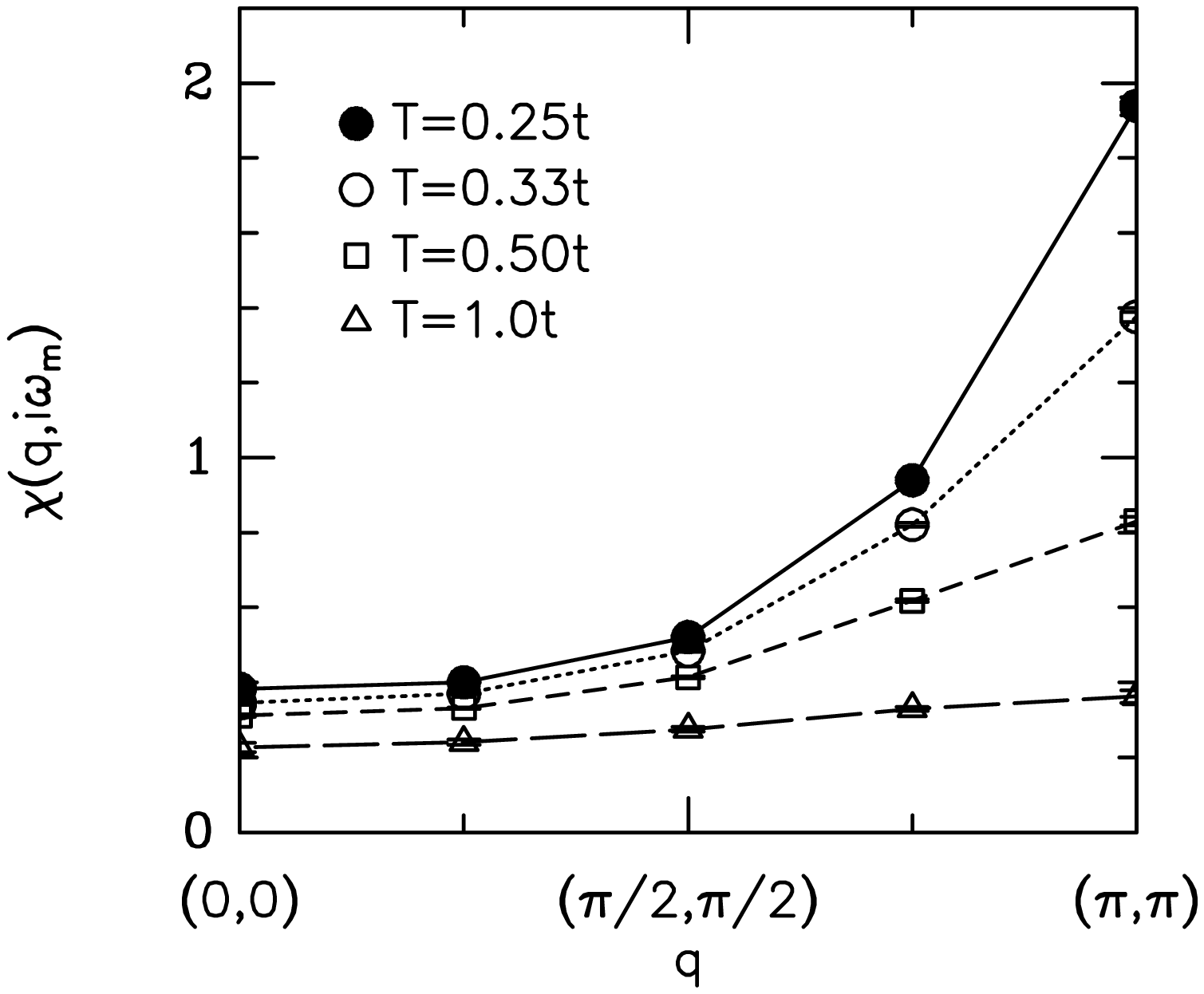}}
\quad
\subfigure[]{
\epsfysize=8cm
\epsffile[50 150 600 610]{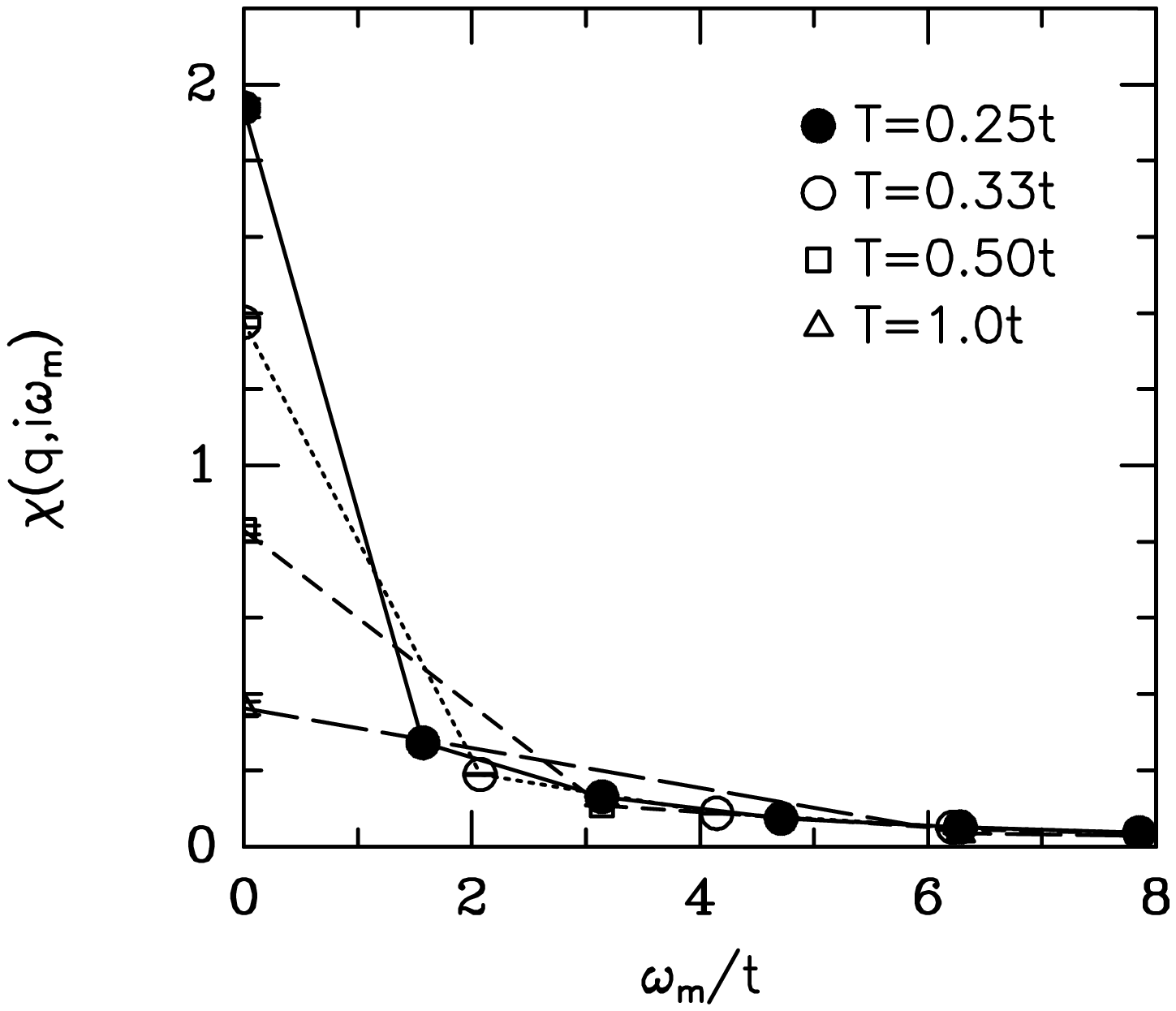}}}
\fi
\caption{
(a) Momentum and (b) the Matsubara-frequency dependence of 
the magnetic susceptibility $\chi({\bf q},i\omega_m)$
at various temperatures.
In (a), $\chi({\bf q},i\omega_m)$ versus ${\bf q}$ is plotted
for $\omega_m=0$.
In (b), $\chi({\bf q}=(\pi,\pi),i\omega_m)$ versus $\omega_m$ 
is plotted.
These results are for $U=4t$ and $\langle n\rangle=0.87$
on an $8\times 8$ lattice.
}
\label{6.7}
\end{figure}

Next,
the temperature evolution of 
$\Gamma_{Is}({\bf q},i\omega_m)$ 
for $U=4t$ is compared with 
that of the magnetic susceptibility 
$\chi({\bf q},i\omega_m)$.
Figure 6.7(a) shows Monte Carlo data on $\chi({\bf q},0)$ 
versus ${\bf q}$ at the same temperatures as in Fig.~6.6.
The Matsubara frequency dependence of 
$\chi({\bf q}=(\pi,\pi),i\omega_m)$ is shown in Fig.~6.7(b).
Comparing Figs.~6.6 and 6.7, one sees that the temperature 
evolution of $\Gamma_{Is}({\bf q},i\omega_m)$ 
closely follows that of $\chi({\bf q},i\omega_m)$.
Both of these quantities peak at ${\bf q}=(\pi,\pi)$, 
and as $\omega_m$ increases $\Gamma_{Is}$ goes to 
the bare $U$ value while $\chi$ decays to zero.
The relation between $\Gamma_{Is}$ and the spin fluctuations
will be studied in more detail in Section~6.4, 
where the QMC results on $\Gamma_{Is}$ will be compared 
with the single spin-fluctuation exchange interaction.

\begin{figure}
\centering
\iffigure
\epsfysize=8cm
\epsffile[100 200 550 560]{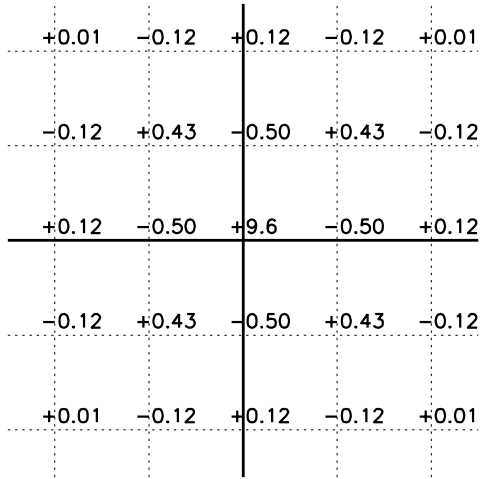}
\fi
\caption{
Real-space pattern of $\Gamma_{Is}({\bf R})$ for 
$U=4t$, $T=0.25t$ and $\langle n\rangle=0.87$.
Here, it is seen that for ${\bf R}=0$, 
$\Gamma_{Is}({\bf R})$ is strongly repulsive while 
being attractive when the singlet electron pair is 
separated by one lattice spacing.
}
\label{6.8}
\end{figure}

At this point, it should be noted that 
the ${\bf q}$ dependence of the effective pairing 
interaction for the cuprates was also studied 
within the spin-bag approach where a dip at ${\bf q}\sim 0$
is found in addition to the peak at ${\bf q}\sim (\pi,\pi)$
[Kampf and Schrieffer 1990a].
The dip is due to the cross-exchange of two AF spin 
fluctuations and it is responsible for the effective attraction 
between two spin bags.
At $T=0.25t$, 
which is the lowest temperature where the 
QMC calculation of $\Gamma_{Is}$ can be carried out, 
a dip in $\Gamma_{Is}$ at ${\bf q}\sim 0$ is not found.

In order to gain insight into the 
real-space structure of the effective
particle-particle interaction, 
it is useful to consider the Fourier transform
\begin{equation}
\Gamma_{Is}({\bf R})=
{1\over N^2} \, \sum_{{\bf p},{\bf p'}} \,
e^{ i({\bf p}-{\bf p'})\cdot {\bf R}} \,
\Gamma_{Is}({\bf p'},i\omega_{n'} | {\bf p},i\omega_n)
\end{equation}
for the lowest Matsubara frequencies 
$\omega_n=\omega_{n'}=\pi T$.
Figure~6.8 shows the Monte Carlo data on 
$\Gamma_{Is}({\bf R})$ as a function of ${\bf R}$ 
for $T=0.25t$.
At ${\bf R}=0$,  
$\Gamma_{Is}$ is strongly repulsive, as expected, 
but for a singlet electron pair separated by one lattice
spacing, $\Gamma_{Is}$ is attractive. 
As the pair separation increases further,
$\Gamma_{Is}$ oscillates in sign and its magnitude 
decreases rapidly, reflecting the short-range nature of the interaction.
Pairing correlations with the proper space-time structure 
can avoid the large onsite repulsion while taking 
advantage of the near-neighbour attraction. 
Thus, the interaction $\Gamma_{Is}$ is attractive 
in the $d_{x^2-y^2}$-wave channel.

In Fig.~6.8, it is seen that the effective 
attractive interaction is $-0.5t$ at the nearest-neighbour site. 
Here, one might argue that the long-range Coulomb 
repulsion between the electrons, 
which is not taken into account in the Hubbard model, 
could overcome this attraction. 
However, 
one would expect
the long-range Coulomb repulsion to have 
weak $\omega_m$ dependence
compared to that seen in Fig.~6.6(b), 
and after the $t$-matrix scatterings 
it should have weak influence on the strength of the 
pairing in the $d_{x^2-y^2}$-wave channel.

\subsection{Bethe-Salpeter equation in the particle-particle channel}

In this section,  
the particle-particle Bethe-Salpeter equation
will be solved
using the QMC data on the irreducible interaction $\Gamma_I$ 
and the one-electron Green's function.
This way one can determine the magnitude of the eigenvalues 
and the $({\bf p},i\omega_n)$ structure of the leading pair-field
eigenfunctions. 
This approach is useful for studying the leading
scattering channels in the $t$-matrix quantitatively.
For instance, one 
can determine how close 
the system is to a 
Kosterlitz-Thouless superconducting transition,
and compare the strength of the pairing  
in various channels.
The particle-particle Bethe-Salpeter equation is 
\begin{equation}
\label{BS}
\lambda_{\alpha} \phi_{\alpha}(p) = - {T\over N} 
\sum_{p'} \,
\Gamma_I(p|p') G_{\uparrow}(p') G_{\downarrow}(-p')
\phi_{\alpha}(p'),
\end{equation}
where $\phi_{\alpha}(p)$ is the pair-field eigenfunction and 
$\lambda_{\alpha}$ is the corresponding eigenvalue.
The Feynman diagram representing Eq.~(\ref{BS}) is 
shown in Fig.~6.9.
When the leading eigenvalue reaches one,
the superconducting transition takes place,
and the Bethe-Salpeter equation for the corresponding 
eigenfunction becomes equivalent to the superconducting gap 
equation.

\begin{figure}[ht]
\centering
\iffigure
\epsfig{file=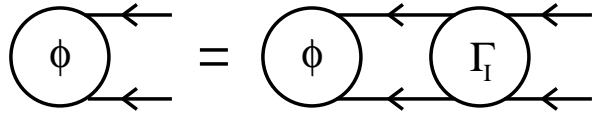,height=1.5cm}
\fi
\caption{
Feynman diagram representing the Bethe-Salpeter equation.
}
\label{6.9}
\end{figure}

Here, data from Refs.~[Bulut {\it et al.} 1993]
will be shown.
Both the triplet and the singlet solutions of the 
Bethe-Salpeter equation will be considered.
The momentum and the Matsubara-frequency structures of the 
leading eigenfunctions will be presented. 
In addition to the singlet $d_{x^2-y^2}$-wave eigenfunction 
$\phi_d({\bf p},i\omega)$, 
solutions which are odd in $\omega_n$ and have $s$ and $p$-wave
symmetries are found. 
The odd-frequency eigenfunctions with $s$-wave and 
$p$-wave symmetries correspond to 
triplet and singlet solutions, respectively.
The finite-size scaling of the eigenvalues indicates
that as $T$ is lowered the singlet $d_{x^2-y^2}$-wave 
eigenvalue grows fastest,
and at low $T$ the dominant singlet pairing channel has 
the $d_{x^2-y^2}$-wave symmetry.
In these calculations, the even-frequency extended $s$-wave channel 
was not found to be one of the leading pairing channels.

The case of the odd-frequency pairing channel is interesting.
The possibility of pairing in the triplet odd-frequency $s$-wave 
channel was first studied within the context of $^3$He in 
Ref.~[Berezinskii 1974].
The singlet odd-frequency $p$-wave channel was studied 
initially by using an effective interaction which is
mediated by phonons [Balatsky and Abrahams 1992].
The possibility that an effective attraction in 
the singlet odd-frequency $p$-wave channel 
could be generated by the spin fluctuations 
was noted in Ref.~[Abrahams {\it et al.} 1993] 
at about the same time the QMC calculations reviewed here 
were carried out.
The general properties of the odd-frequency superconductors 
were discussed in Ref.~[Abrahams {\it et al.} 1995].

In addition, here 
the effects of increasing $U/t$ 
on the leading pairing channels will be discussed.
As seen in the previous section, it was not possible to calculate
$\Gamma_{Is}$ for $U=8t$.
However, if $\Gamma$ is used rather than $\Gamma_I$ 
in Eq.~(\ref{BS}), then the corresponding eigenvalues are given 
by $\lambda_{\alpha}/(1-\lambda_{\alpha})$,
where $\lambda_{\alpha}$ are the eigenvalues for the
irreducible vertex.
Hence, through this indirect way it is possible to study
the leading Bethe-Salpeter eigenvalues for $U=8t$.
It will be shown that as $U/t$ increases from 4 to 8,
the leading eigenvalues grow, 
since the effective particle-particle interaction gets stronger,
but the momentum and the frequency structure of their
eigenfunctions does not exhibit qualitative changes.

In general, the Bethe-Salpeter equation can have both singlet and
triplet solutions corresponding to a pair-wave function that has 
overall even or odd parity when 
$p=({\bf p},i\omega_n)$ goes to $(-{\bf p},-i\omega_n)$.
Here, the pair wave functions are characterized by 
its symmetry in momentum and spin space. 
The usual singlet $s$ and $d_{x^2-y^2}$-wave states are even in 
frequency and even in momentum,
$\phi({\bf p},-i\omega_n)=\phi({\bf p},i\omega_n)$ and 
$\phi(-{\bf p},i\omega_n)=\phi({\bf p},i\omega_n)$,
while the usual triplet $p_x$ or ($p_y$) state is even in frequency 
and odd when $p_x$ goes to $-p_x$.
There are also odd-frequency pair-wave functions. 
In this case, one can have an odd-frequency $s$-wave 
triplet for which 
$\phi({\bf p},-i\omega_n)=-\phi({\bf p},i\omega_n)$ and 
$\phi(-{\bf p},i\omega_n)=\phi({\bf p},i\omega_n)$, or an 
odd-frequency $p_x$ (or $p_y$)-wave singlet with 
$\phi({\bf p},-i\omega_n)=-\phi({\bf p},i\omega_n)$ and 
$\phi(-{\bf p},i\omega_n)=-\phi({\bf p},i\omega_n)$. 
In the regime of the Hubbard model that is being studied,
the $s$-wave triplet, and the $p$ and $d_{x^2-y^2}$-wave 
singlet solutions are dominant.

The temperature evolution of the 
four largest eigenvalues for $U=4t$ and $\langle n\rangle=0.87$
are given in Table~1.
The momentum and the frequency dependence of the corresponding 
pair-wave functions are shown in Figs.~6.10 and 6.11
for $T=0.5t$.
At this temperature, 
an $s$-wave triplet state has the largest eigenvalue
$\lambda_s \approx 0.23$.
As seen in Figs.~6.10(a) and 6.11(a) (solid circles), 
the pair-wave function $\phi_s({\bf p},i\omega_n)$ of the
$s$-wave triplet state is even in ${\bf p}$ and odd in $\omega_n$. 
The open circles in Figs.~6.10(a) and 6.11(a) represent the pair-wave 
function $\phi_{s'}({\bf p},i\omega_n)$ which has the second 
largest eigenvalue. 
This is also an odd-frequency $s$-wave triplet state.
The $d_{x^2-y^2}$-wave singlet state shown as the solid 
circles in Figs.~6.10(b) and 6.11(b) has the third largest eigenvalue.
The fourth largest eigenvalue corresponds to a state which is 
odd in both $\omega_n$ and ${\bf p}$, having $p_y$ (or $p_x$)
symmetry [open circles in Figs.~6.10(b) and 6.11(b)], 
hence it is also a singlet. 

\begin{table}[ht]
\centering
\begin{tabular} {| c | c | c | c | c |} \hline
$T/t$  & $\lambda_s$ & $\lambda_{s'}$ & $\lambda_p$ & $\lambda_d$ \\ \hline 
$1.0$  & $0.261$     &  0.037        & $0.023$     & $0.022$ \\ 
$0.50$ & $0.228$     &  0.090         & $0.066$     & $0.076$ \\ 
$0.33$ & $0.251$     &  0.104         & $0.095$     & $0.130$ \\ 
$0.25$ & $0.264\pm 0.007$ & $0.124\pm 0.006$ & $0.129\pm 0.007$ & 
$0.182\pm 0.006$ \\ 
\hline
\end{tabular}
\caption{Temperature dependence of the Bethe-Salpeter eigenvalues
for $U=4t$ and $\langle n\rangle=0.87$ on an $8\times 8$ lattice.
The error bars represent the uncertainty due to the 
Monte Carlo sampling. 
The error bars were not calculated at all temperatures, 
since it requires considerably more computer time.
When the error bars are not indicated, 
they are estimated to be less than 10\%.
}
\end{table}

\begin{figure}
\centering
\iffigure
\mbox{
\subfigure{
\epsfysize=8cm
\epsffile[100 150 480 610]{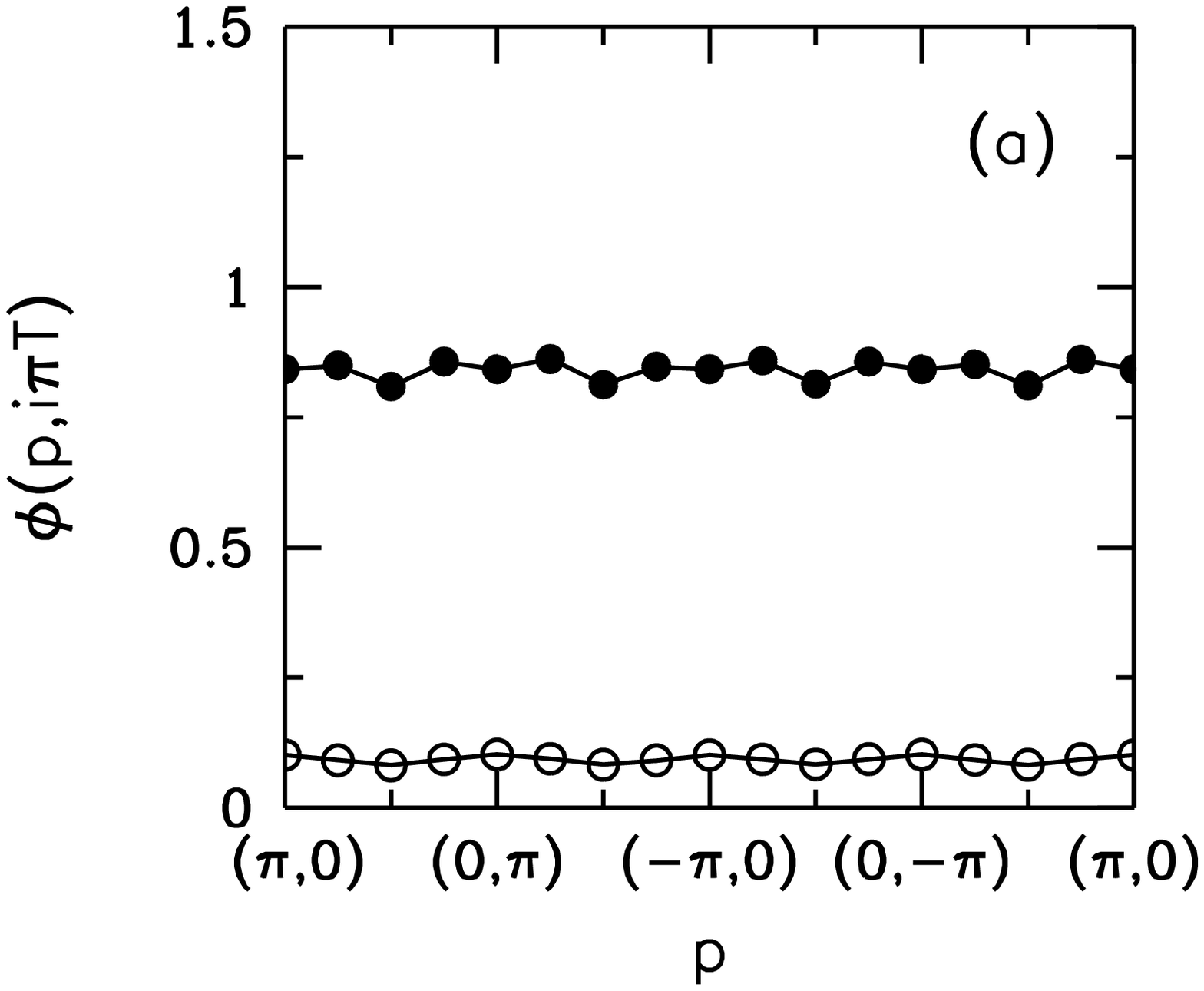}}
\quad
\subfigure{
\epsfysize=8cm
\epsffile[50 150 600 610]{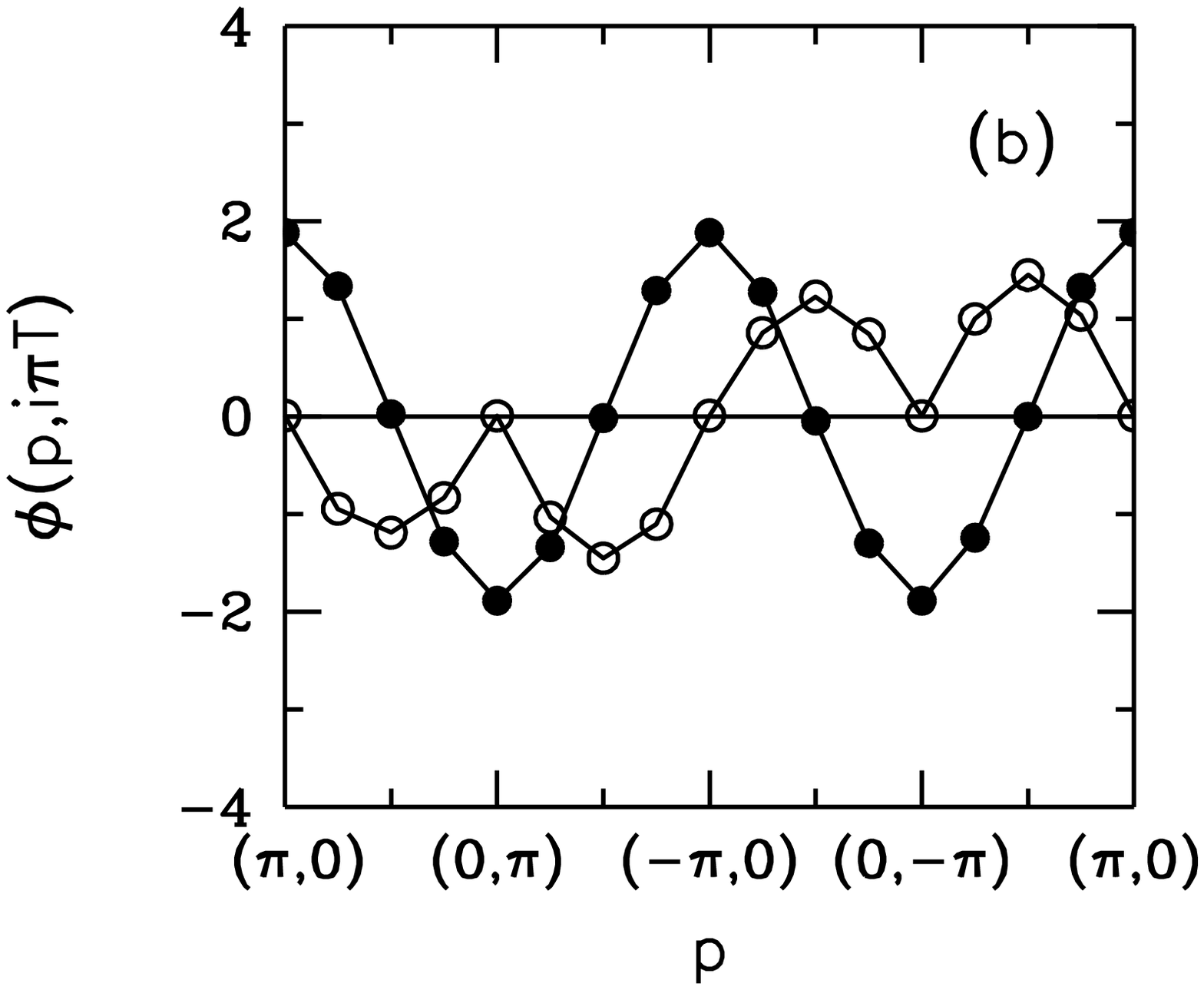}}}
\fi
\caption{
Momentum dependence of the leading 
Bethe-Salpeter eigenfunctions 
$\phi_{\alpha}({\bf p},i\omega_n)$.
These results are for $\omega_n=\pi T$,
$U=4t$, $T=0.5t$ and $\langle n\rangle=0.87$
on an $8\times 8$ lattice.
In (a), two odd-frequency $s$-wave eigenfunctions are plotted, 
and in (b) the $d_{x^2-y^2}$-wave (filled circles) and 
$p_y$-wave (open circles) are plotted.
}
\label{6.10}
\end{figure}

\begin{figure}
\centering
\iffigure
\mbox{
\subfigure{
\epsfysize=8cm
\epsffile[100 150 480 610]{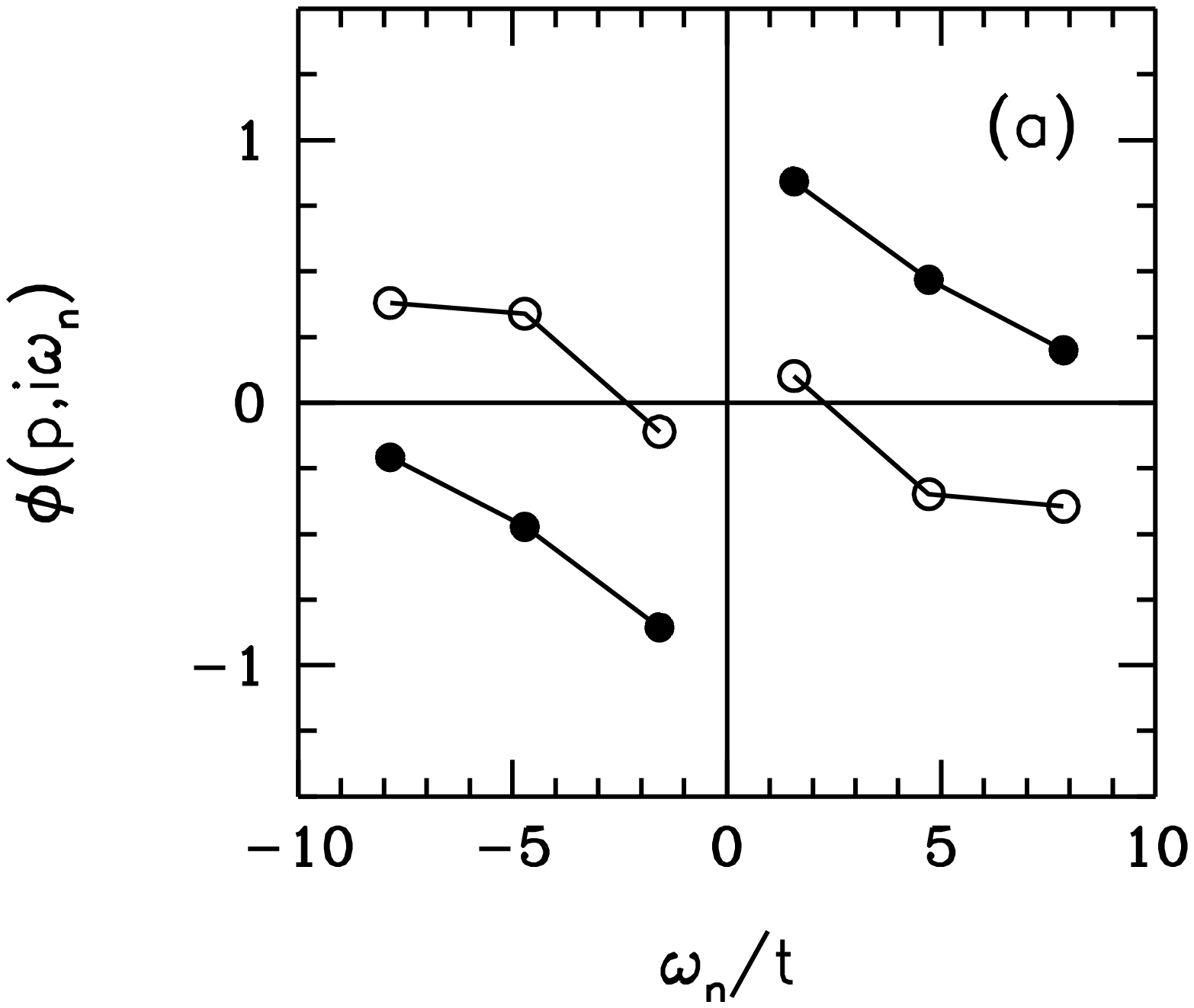}}
\quad
\subfigure{
\epsfysize=8cm
\epsffile[50 150 600 610]{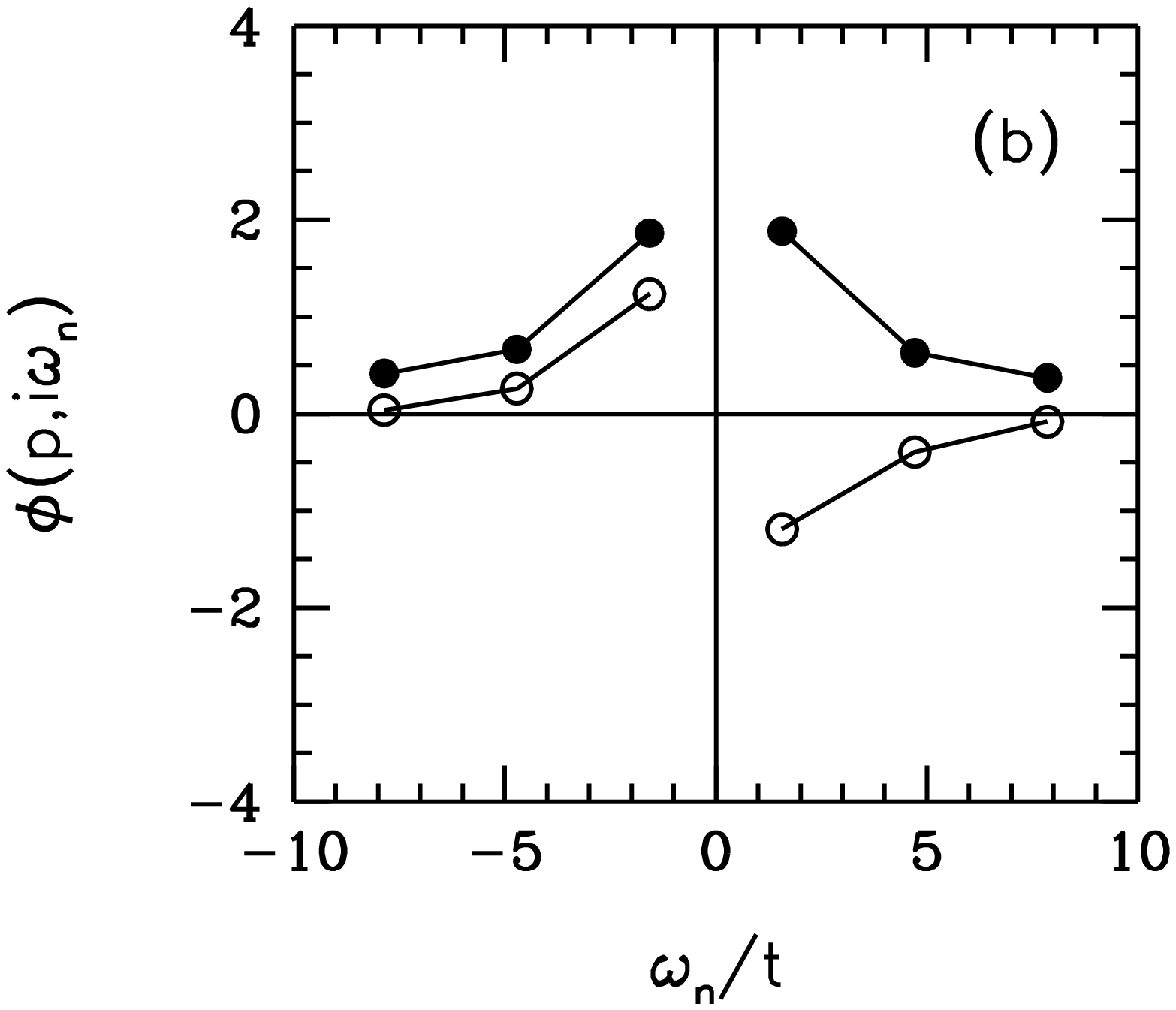}}}
\fi
\caption{
Matsubara frequency dependence of the leading 
Bethe-Salpeter eigenfunctions 
$\phi_{\alpha}({\bf p},i\omega_n)$.
These results are for
$U=4t$, $T=0.5t$ and $\langle n\rangle=0.87$
on an $8\times 8$ lattice.
In (a), 
$\phi({\bf p}=(\pi,0),i\omega_n)$ 
is shown for the odd-$\omega_n$ $s$ and $s'$ channels.
In (b), 
$\phi({\bf p},i\omega_n)$
is shown for the singlet $d_{x^2-y^2}$-wave channel
at ${\bf p}=(\pi,0)$ and for the 
singlet $p_y$-wave channel at ${\bf p}=(\pi/2,\pi/2)$.
}
\label{6.11}
\end{figure}

\begin{table}
\centering
\begin{tabular} {| c | c | c | c |} \hline
$L\times L$ & $\lambda_s$      & $\lambda_p$      & $\lambda_d$ \\ \hline
$4\times 4$ & $0.296\pm 0.010$ & $0.204\pm 0.014$ & $0.184\pm 0.016$ \\ 
$8\times 8$ & $0.264\pm 0.007$ & $0.129\pm 0.007$ & $0.182\pm 0.006$ \\ 
\hline
\end{tabular}
\caption{Finite-size dependence of the Bethe-Salpeter eigenvalues
for  $T=0.25t$, $U=4t$ and $\langle n\rangle=0.87$.}
\end{table}

\begin{figure}
\centering
\iffigure
\epsfysize=8cm
\epsffile[100 150 550 610]{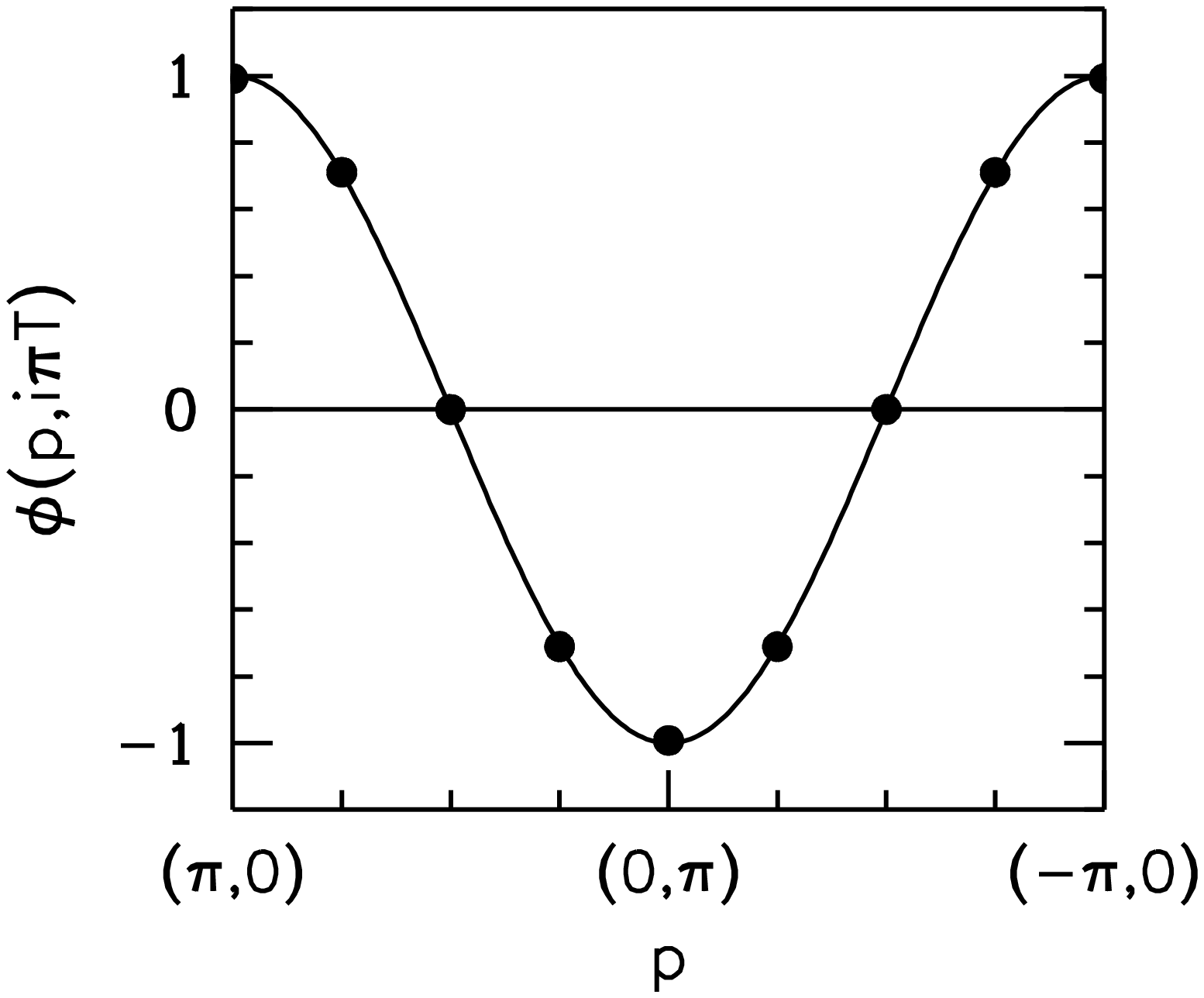}
\fi
\caption{
Comparison of the $d_{x^2-y^2}$-wave eigenfunction 
$\phi_d({\bf p},i\pi T)$ 
with the usual $d_{x^2-y^2}$-wave form  
$\Delta_{\bf p} = (\Delta_0/2)(\cos{p_x} + \cos{p_y})$
for $U=4t$, $T=0.25t$ and $\langle n\rangle=0.87$.
Here, both 
$\phi_d({\bf p},i\pi T)$ and 
$\Delta_{\bf p}$ have been normalised to 1 at
${\bf p}=(\pi,0)$.
}
\label{6.12}
\end{figure}

As $T$ is lowered from $1.0t$ down to $0.25t$, 
the largest eigenvalue $\lambda_s$ stays nearly the same, 
while $\lambda_d$ increases 
by about a factor of seven. 
This can be understood in terms of the temperature 
dependence of $\Gamma_I(p'|p)$ which enters the Bethe-Salpeter 
equation. 
The pair-wave functions which are smooth in ${\bf p}$
but odd in $\omega_n$ make optimum use of the 
$(\omega_n,\omega_{n'})$ frequency structure of the 
repulsive $\Gamma_I(p'|p)$ for pairing.
However, as the temperature is lowered 
and $\Gamma_I(p'|p)$ for ${\bf p'}-{\bf p}=(\pi,\pi)$ grows,
the $d_{x^2-y^2}$ and $p$-wave solutions can make better 
use of the momentum structure in $\Gamma_I$, 
and their eigenvalues get enhanced.

Table~2 shows the finite-size effects on the leading 
eigenvalues at $T=0.25t$.
Here, it is seen that the finite-size effects 
are especially large for the $p$-wave channel, 
and $\lambda_p$ decreases as the system size grows
from $4\times 4$ to $8\times 8$.
The finite size effects for the $d_{x^2-y^2}$-wave case
are small.
Hence, these results show that as $T$ is lowered, 
$\lambda_d$ grows fastest and at low $T$ 
the dominant singlet pairing channel has the 
$d_{x^2-y^2}$-wave symmetry.

It is useful to compare the momentum dependence of the 
$d_{x^2-y^2}$-wave eigenvalue 
$\phi_d({\bf p},i\omega_n)$ at $\omega_n=\pi T$
with the usual $d_{x^2-y^2}$-wave gap form
$\Delta_{\bf p}=(\Delta_0/2)(\cos{p_x} - \cos{p_y})$,
since this is often used in modelling the 
superconducting state of the cuprates.
Figure 6.12
shows that the QMC data on $\phi_d({\bf p},i\pi T)$
taken at $T=0.25t$
follow the usual $d_{x^2-y^2}$-wave form closely. 

Finally, 
the filled circles in Fig.~6.13 show the growth of 
$\lambda_d$ for $U=4t$ as the temperature is lowered,
while the open circles represent $\lambda_d$ for 
$U=8t$.
The results on $\lambda_d$ for $U=8t$ were obtained 
as described above.
Here, it is seen that $\lambda_d$ increases with $U/t$.

\begin{figure}
\centering
\iffigure
\epsfysize=8cm
\epsffile[100 150 550 610]{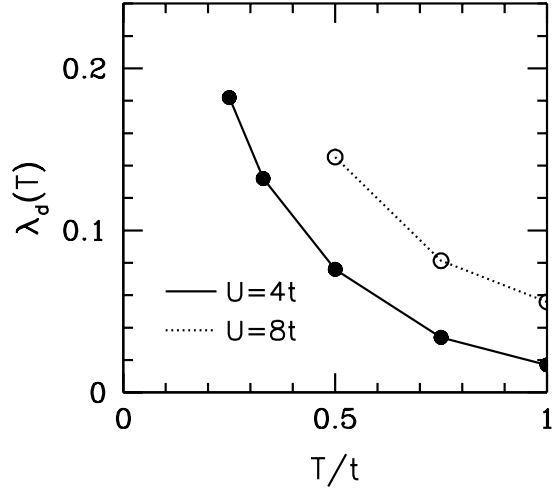}
\fi
\caption{
Temperature dependence of the $d_{x^2-y^2}$-wave eigenvalue
$\lambda_d$ for $U=4t$ and $8t$
at $\langle n\rangle=0.87$.
}
\label{6.13}
\end{figure}

At the lowest temperature where $\lambda_d$ can be calculated,
the system is far from a 
Kosterlitz-Thouless superconducting transition
which would be signalled by $\lambda_d\rightarrow 1$.
Hence, while these data imply that at the temperatures
where the QMC simulations are carried out, 
the singlet $d_{x^2-y^2}$-wave pairing correlations are becoming dominant, 
it is not known whether $\lambda_d\rightarrow 1$ 
at lower temperatures.
Below in Section~8.1, these results on $\lambda_d$ will be 
compared with the results of the FLEX calculations.

\begin{figure}
\centering
\iffigure
\mbox{
\subfigure{
\epsfysize=8cm
\epsffile[100 150 480 610]{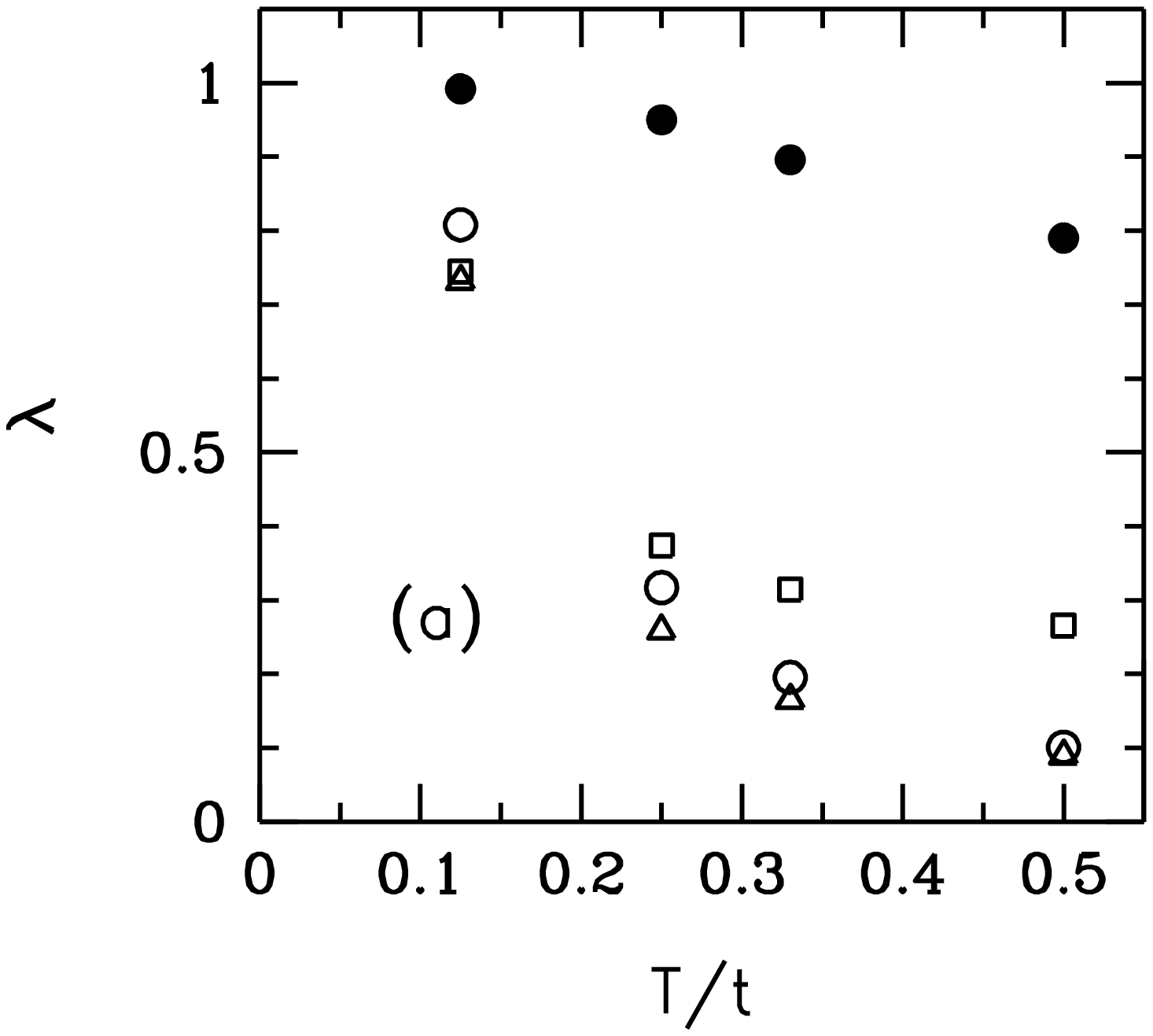}}
\quad
\subfigure{
\epsfysize=8cm
\epsffile[50 150 600 610]{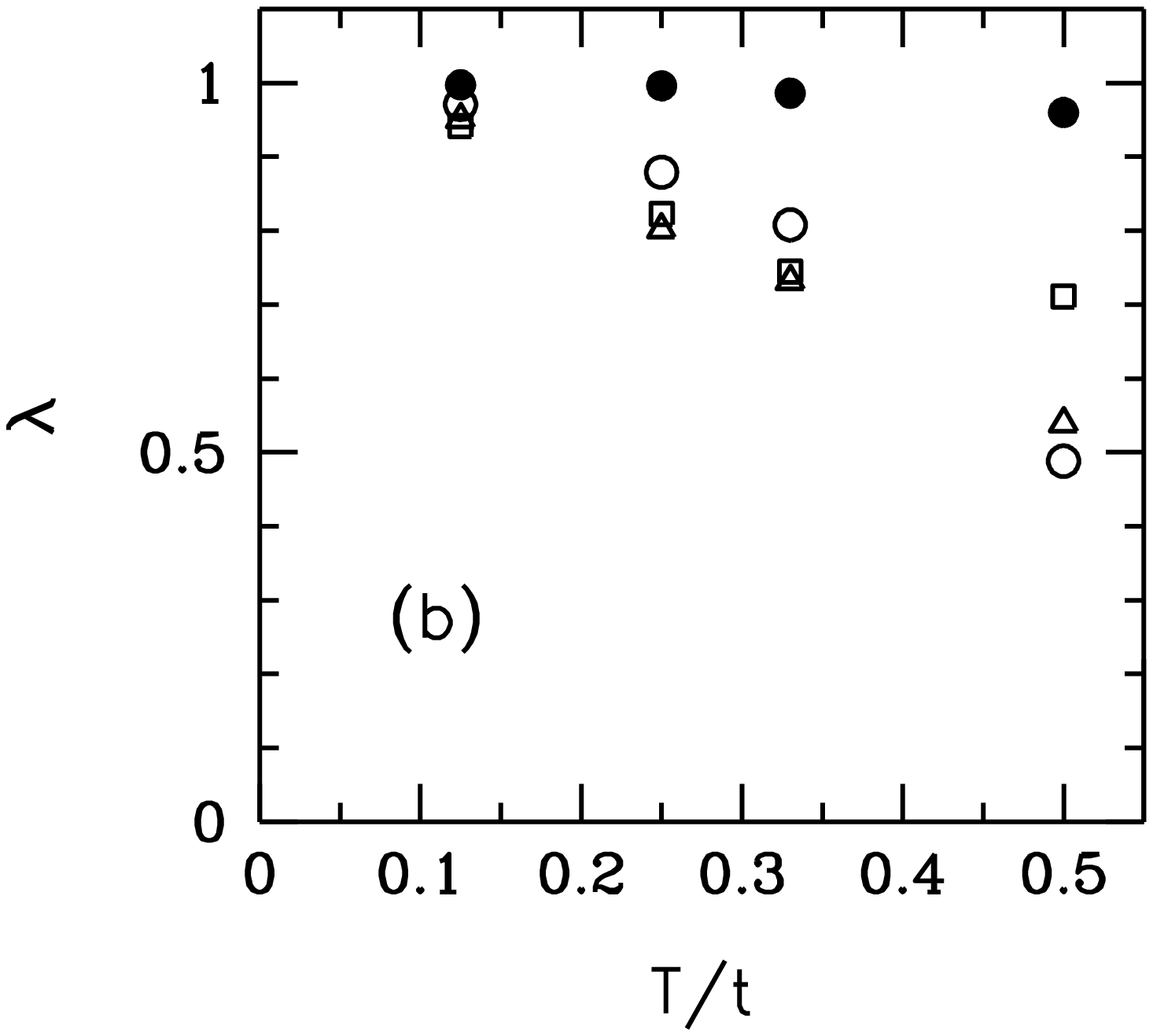}}}
\fi
\caption{
Eigenvalues of the 
particle-hole and the particle-particle 
Bethe-Salpeter equations versus $T$
for (a) $U=4t$ and (b) $U=8t$ for an $8\times 8$ 
half-filled lattice.
The solid points ($\bullet$) are for the leading eigenvalue 
$\overline{\lambda}_1$ of the Bethe-Salpeter 
equation in the AF particle-hole channel
with the center-of-mass momentum ${\bf Q}=(\pi,\pi)$.
The open symbols denote the 
even-frequency $d_{x^2-y^2}$-wave ($\circ$), and 
the odd-frequency $p$ ($\triangle$) and $s$-wave ($\Box$)
eigenvalues of the 
particle-particle Bethe-Salpeter equation.
}
\label{6.14}
\end{figure}

The QMC calculation of $\lambda_d$ for the doped case cannot be 
carried out at any lower temperatures.
However, for half-filling it is possible to calculate the various 
pairing eigenvalues at low $T$
and compare them with the eigenvalues of 
the Bethe-Salpeter equation in the 
AF particle-hole channel,
$\overline{\lambda}_{\alpha}$.
The leading magnetic eigenvalue $\overline{\lambda}_1$
occurs in an even-frequency $s$-wave channel
with center-of-mass momentum ${\bf Q}=(\pi,\pi)$.
In Fig.~6.14, $\overline{\lambda}_1$ is
compared with the various pairing eigenvalues
for $U=4t$ and $8t$ at half-filling.
As expected, at low temperatures $\overline{\lambda}_1$ 
reaches 1 asymptotically, signalling the phase transition to the 
AF order state on the $8\times 8$ lattice.
Here, it is also seen that as the AF correlations develop
at half-filling,
$\lambda_d$ becomes the leading pairing eigenvalue
while always staying smaller than the magnetic eigenvalue 
$\overline{\lambda}_1$.

\subsection{Comparison with the spin-fluctuation exchange approximation}

Various spin-fluctuation exchange theories have been used 
for studying $d_{x^2-y^2}$-wave pairing for the cuprates
[Bickers {\it et al.} 1987 and 1989,
Moriya {\it et al.} 1990, 
Monthoux {\it et al.} 1991].
In this context, it is of interest to see to 
what extent the Monte Carlo results for the irreducible 
vertex can be modelled by a single 
spin-fluctuation exchange interaction. 
For this purpose, here we compare the Monte Carlo data 
with the approximate form
\begin{equation}
\label{GSF}
\Gamma_I^{SF}({\bf q},i\omega_m) = 
U + {3\over 2} g^2U^2 \, 
\chi({\bf q},i\omega_{m}).
\end{equation}
This form is motivated by the 
single spin-fluctuation exchange interaction
[Berk and Schrieffer 1966, Doniach and Engelsberg 1966], 
which basically has this form with $g=1$ 
near the antiferromagnetic instability. 
The Feynman diagrams illustrating the 
single spin-fluctuation exchange 
interaction were shown in Fig.~2.1.
The factor of $3/2$ arises from the two transverse 
and one longitudinal spin fluctuations. 
In calculating $\Gamma_{Is}^{SF}$ with 
Eq.~(\ref{GSF}), we will use Monte Carlo results for 
$\chi({\bf q},i\omega_m)$ and also set $g=0.8$.
The corresponding value of $3.2t$ for the effective coupling 
$gU$ is consistent with the results of the Monte Carlo
calculations of the effective irreducible 
vertex in the particle-hole channel,
$\overline{U}({\bf q},0)$, 
which were discussed in Section~3.4.
Formally, Eq.~(\ref{GSF}) is analogous to the effective 
interaction in the electron-phonon superconductor
\begin{equation}
V({\bf q},i\omega_m) = U + 
\sum_{\lambda} \, |g_{{\bf q}\lambda}|^2 \, 
D_{\lambda}({\bf q},i\omega_m),
\end{equation}
where $D_{\lambda}({\bf q},i\omega_m)$ is the dressed 
phonon propagator and $|g_{{\bf q}\lambda}|^2$ 
is the renormalised electron-phonon coupling.

\begin{figure}
\centering
\iffigure
\mbox{
\subfigure{
\epsfysize=8cm
\epsffile[100 150 480 610]{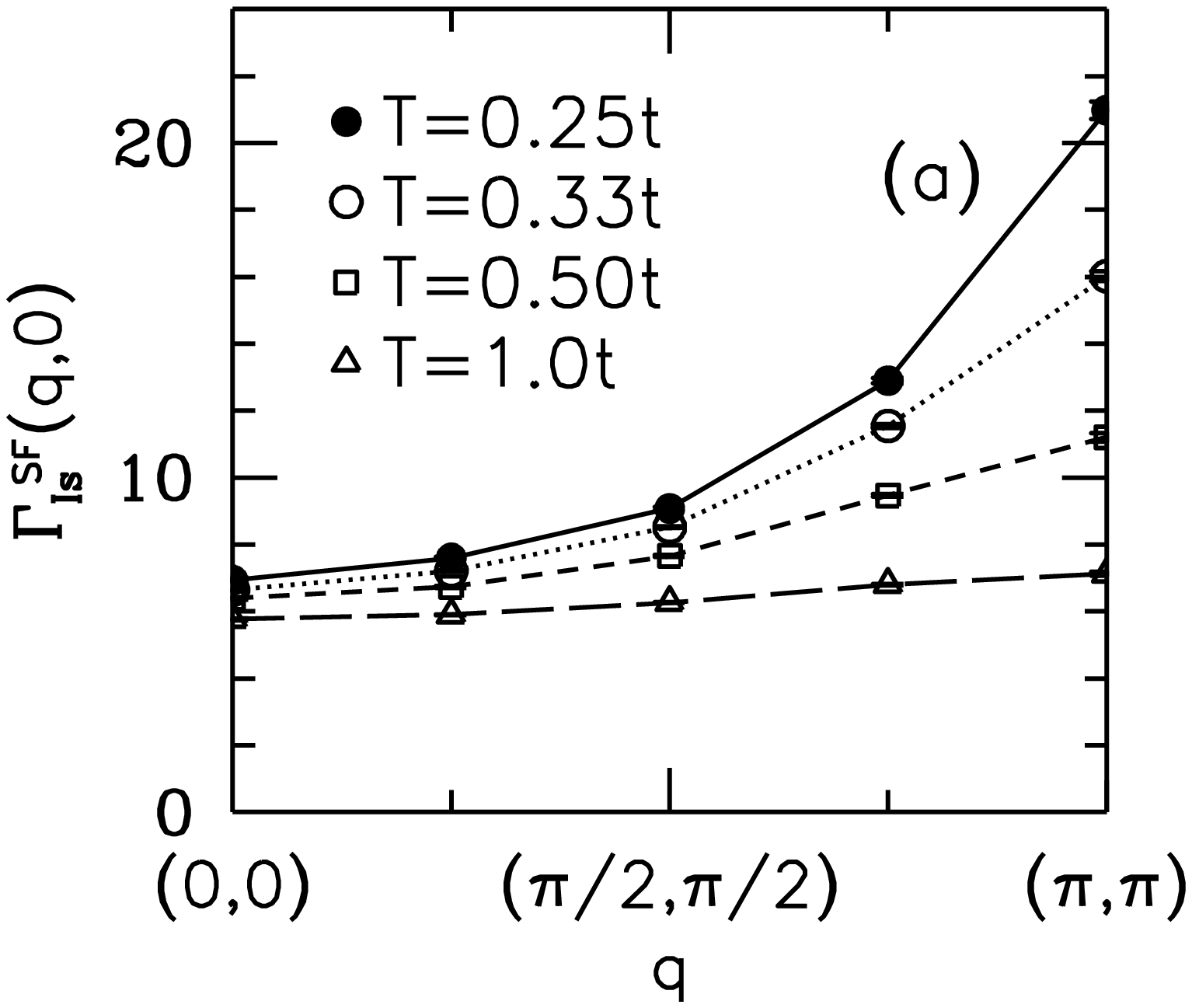}}
\quad
\subfigure{
\epsfysize=8cm
\epsffile[50 150 600 610]{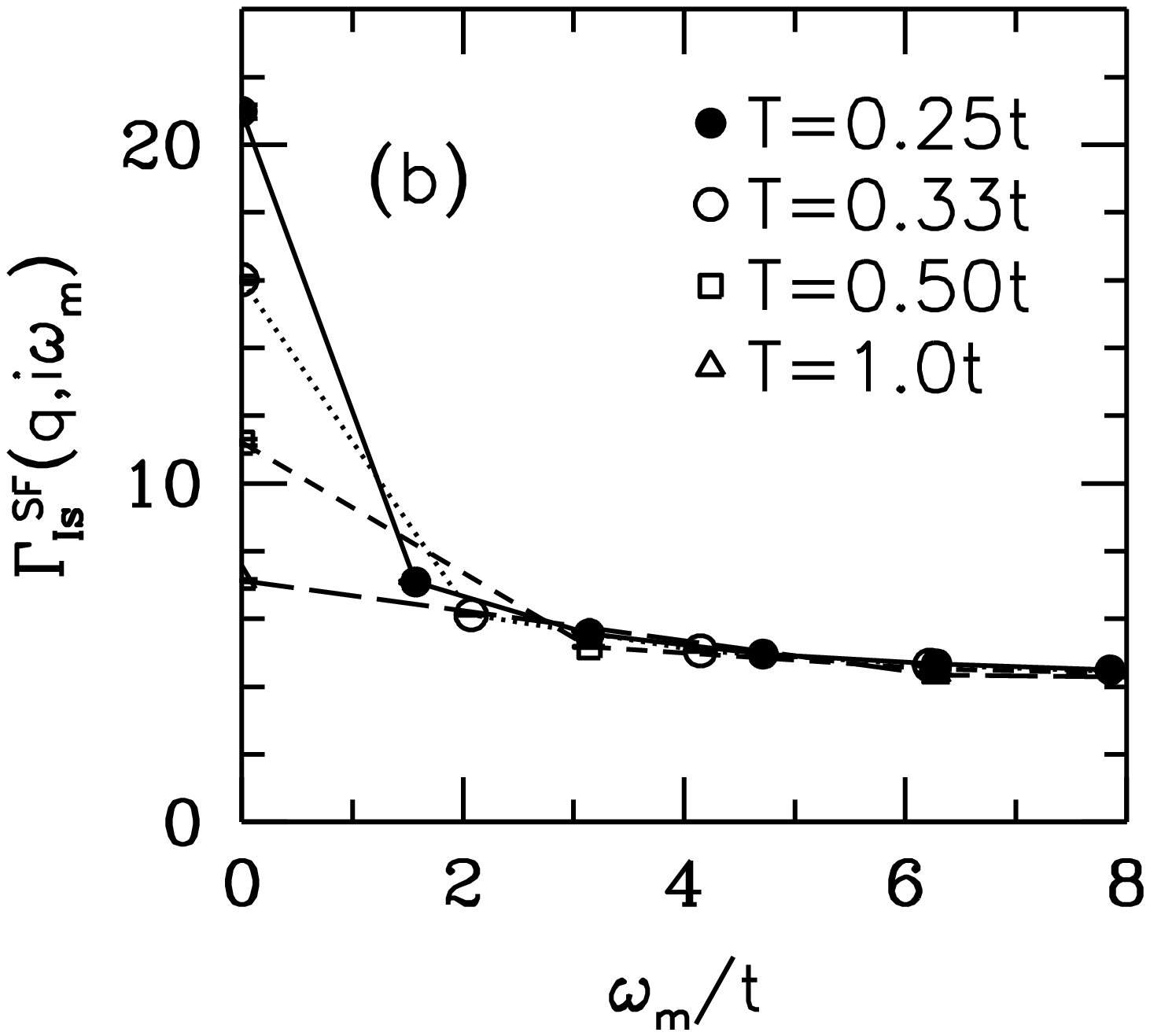}}}
\fi
\caption{
Single-spin-fluctuation exchange approximation for 
the irreducible particle-particle scattering vertex 
in the singlet channel
$\Gamma_{Is}({\bf q},i\omega_m)$. 
In (a), $\Gamma_{Is}$ versus ${\bf q}$ is plotted.
In (b), $\Gamma_{Is}$ versus $\omega_m$ 
is plotted for ${\bf q}=(\pi,\pi)$.
These results have been obtained 
for $U=4t$ and $\langle n\rangle=0.87$
using the Monte Carlo data on $\chi({\bf q},i\omega_m)$.
}
\label{6.15}
\end{figure}

Figure~6.15(a) shows the single spin-fluctuation interaction 
in the singlet channel 
$\Gamma_{Is}^{SF}({\bf q},i\omega_m)$ versus ${\bf q}$ 
at various temperatures.
These results compare well with 
$\Gamma_{Is}({\bf q},i\omega_m=0)$ seen in Fig.~6.6(a).
Likewise, 
the comparison of Fig.~6.15(b) with Fig.~6.6(b) 
shows that the frequency dependence of 
$\Gamma_{Is}^{SF}$ is in agreement with the Monte Carlo data.
Considering the simplicity of Eq.~(\ref{GSF}), this agreement 
with the Monte Carlo data is quite good.
These comparisons suggest that a properly renormalized 
single-spin-fluctuation exchange interaction 
is capable of reproducing the basic features 
of the effective particle-particle interaction in the 
weak-to-intermediate coupling Hubbard model
at temperatures greater or of order $J/2$.

\subsection{Comparison with the perturbation theory}

\begin{figure}
\centering
\iffigure
\epsfig{file=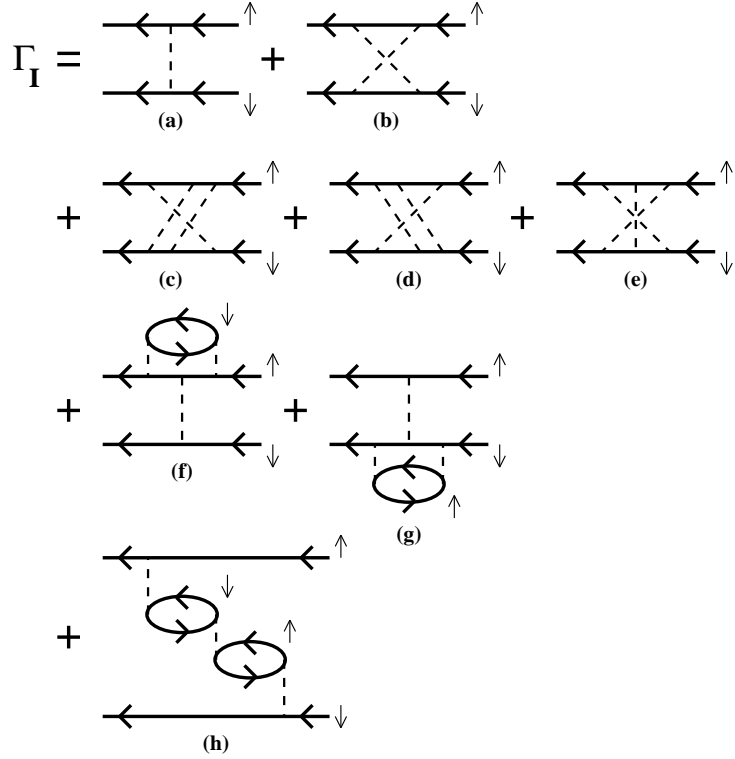,height=10cm}
\fi
\caption{
Feynman diagrams contributing to the irreducible particle-particle
interaction through third order in $U$.
}
\label{6.16}
\end{figure}

In this section, the perturbation theory results for 
$\Gamma_{Is}$ through third order in $U$ 
from Ref.~[Bulut {\it et al.} 1995] will be shown.
These provide insight into the 
various subprocesses contributing to $\Gamma_{Is}$.
Figure 6.16 shows the diagrams contributing to $\Gamma_{Is}$ 
up to third order in $U$.
The dashed line in this figure represents the bare Coulomb repulsion. 
The diagrams (a) and (h) represent the first two terms in an
RPA series corresponding to the exchange of a longitudinal
spin fluctuation.
Similarly, 
the low-order contributions arising from the exchange 
of a transverse spin fluctuation are represented by the
diagrams (b) and (e).
The diagrams (c) and (d) can be considered 
as corrections to diagram
(b), where the bare particle-hole irreducible vertex is 
renormalized through Kanamori type of particle-particle
scatterings [Kanamori 1963].
The diagrams (f) and (g) represent vertex corrections 
to the bare interaction (a).

\begin{figure}
\centering
\iffigure
\mbox{
\subfigure[]{
\epsfysize=8cm
\epsffile[100 150 480 610]{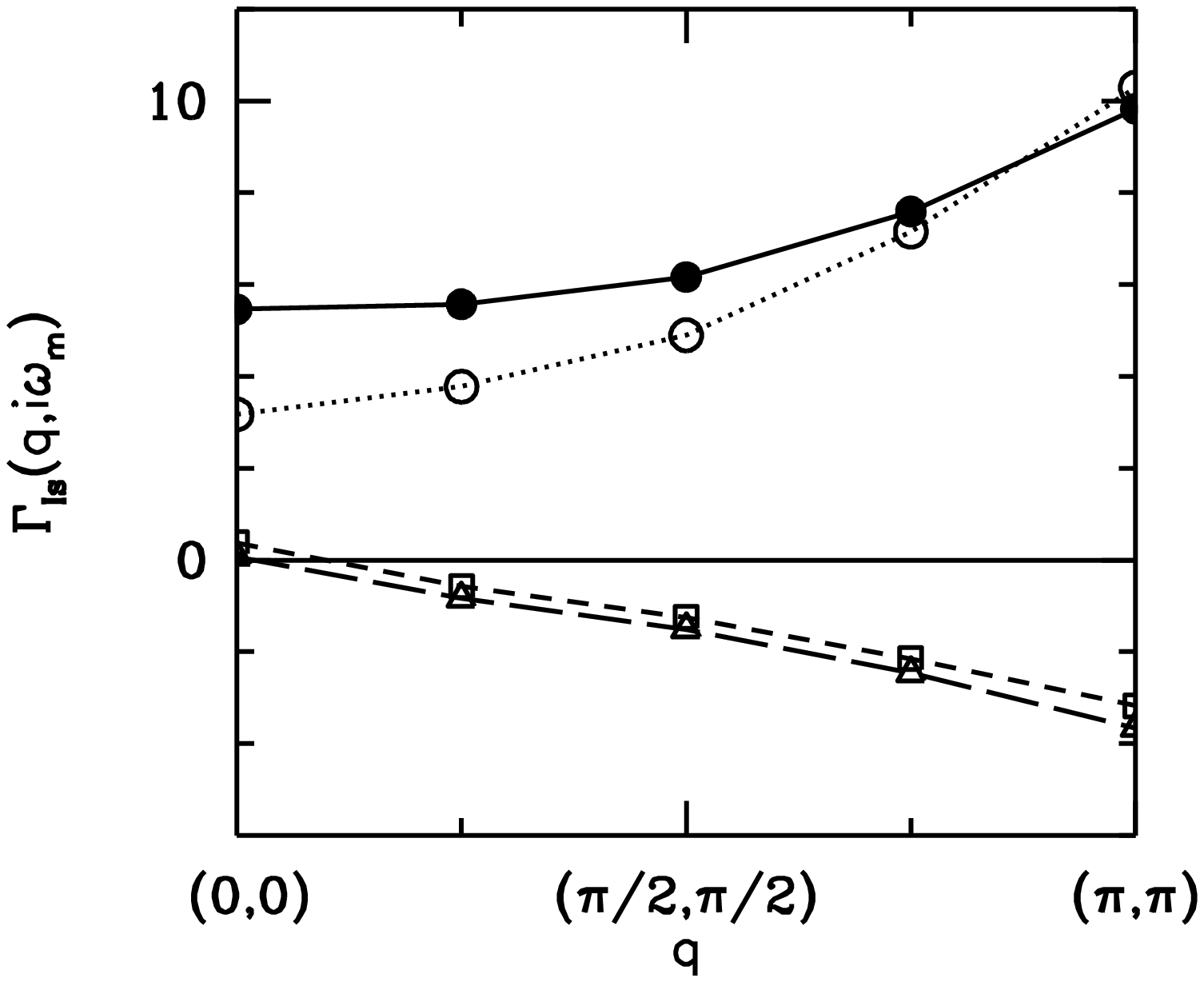}}
\quad
\subfigure[]{
\epsfysize=8cm
\epsffile[50 150 600 610]{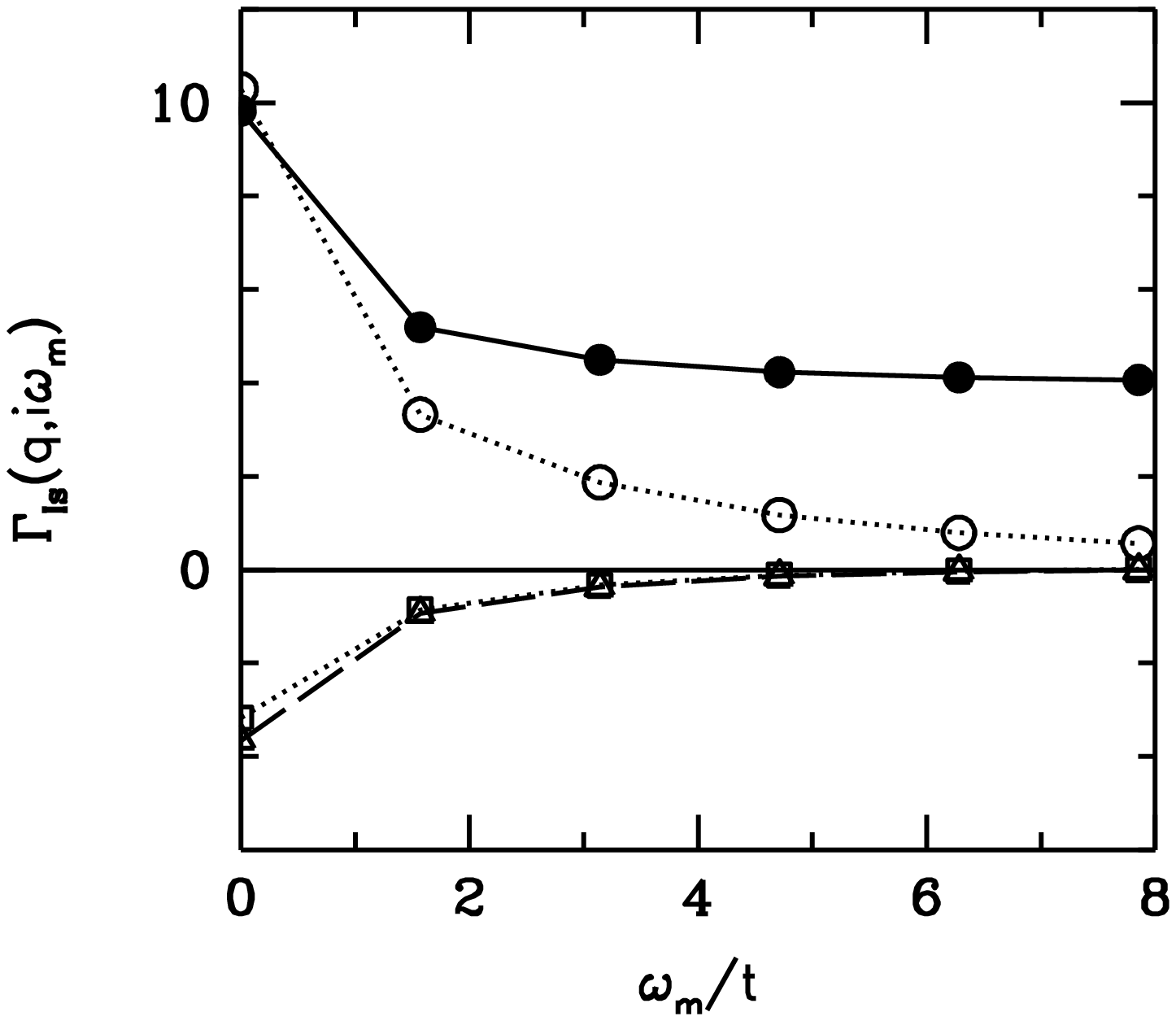}}}
\fi
\caption{
Various diagrams contributing to the 
irreducible particle-particle interaction 
$\Gamma_{Is}({\bf q},i\omega_m)$ through third order in $U$.
In (a) the momentum dependence is shown for 
$\omega_m=0$, and in (b) the frequency dependence is
shown for ${\bf q}=(\pi,\pi)$.
Here, the filled circles represent the contribution of the bare $U$ 
and the longitudinal spin-fluctuation exchange 
(diagrams of type (a) and (h) in Fig.~6.16),
the open circles represent the contribution of 
the transverse spin fluctuations
(Fig.~6.16(b) and (e)), 
the open squares show the ordinary vertex corrections 
(Fig.~6.16(f) and (g)), and the open triangles show the 
Kanamori type of vertex corrections 
(Fig.~6.16(c) and (d)).
}
\label{6.17}
\end{figure}

\begin{figure}
\centering
\iffigure
\mbox{
\subfigure[]{
\epsfysize=8cm
\epsffile[100 150 480 610]{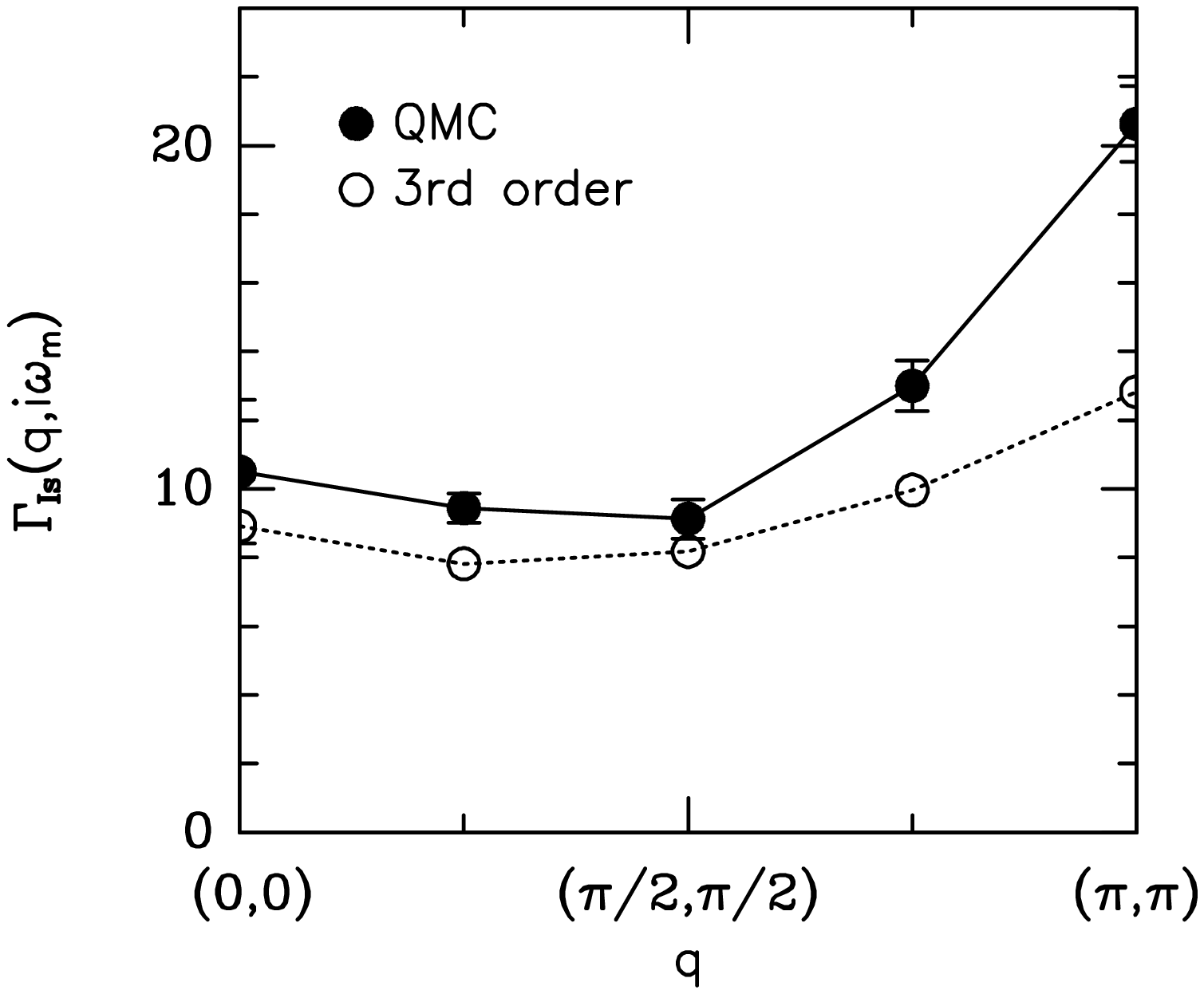}}
\quad
\subfigure[]{
\epsfysize=8cm
\epsffile[50 150 600 610]{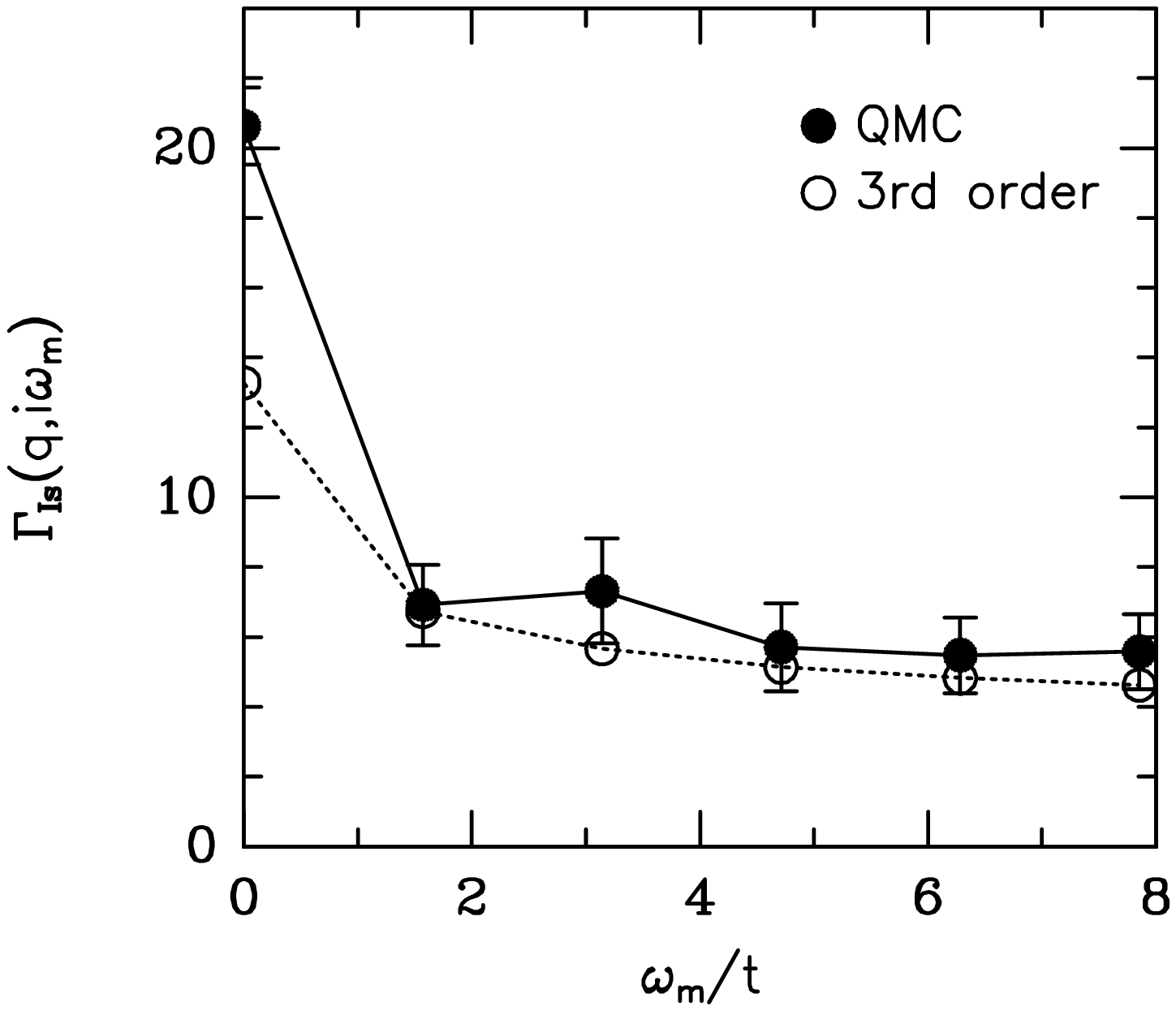}}}
\fi
\caption{
Comparison of the perturbation theory results on 
$\Gamma_{Is}({\bf q},i\omega_m)$ through third order in $U$
with the QMC data.
In (a) the momentum dependence is shown for 
$\omega_m=0$, and in (b) the frequency dependence is
shown for ${\bf q}=(\pi,\pi)$.
}
\label{6.18}
\end{figure}

These diagrams have been evaluated on an $8\times 8$ 
lattice with $U=4t$, 
and the results of the various contributions are shown as a function 
of ${\bf q}$ for $\omega_m=0$ in Fig.~6.17(a).
The frequency dependence for ${\bf q}=(\pi,\pi)$ 
is shown similarly in Fig.~6.17(b).
Here, the results on $\Gamma_{Is}$ are plotted 
in the same way as in the previous section.
In these figures, the filled circles represent the 
diagrams (a) and (h) in Fig.~6.16, and the open circles 
are for the diagrams (b) and (e).
The triangles and the squares in Figs.~6.17(a) and (b) 
represent the contributions of the (c)-(d) and 
(f)-(g), respectively.
The dominant contributions arise from the leading contributions 
to the spin-fluctuation exchange processes (a), (b), (e) and (h),
and they give contributions which peak at ${\bf q}=(\pi,\pi)$.
For $\omega_m=0$, both the Kanamori renormalization graphs 
and the vertex corrections act to reduce the strength of the 
${\bf q}=(\pi,\pi)$ interaction by about one-third.
In Fig.~6.18, the results obtained by summing the diagrams shown 
in Fig.~6.16 are compared with the Monte Carlo results for 
$T=0.25t$.
At this temperature, 
the low-order graphs in Fig.~6.16 are not adequate to represent 
the large momentum behaviour of the effective interaction. 

The importance of the vertex corrections to the single 
spin-fluctuation exchange interaction was pointed out in 
Ref.~[Schrieffer 1995].
In particular, it was noted that for large values of the ratio 
$\chi({\bf q}=(\pi,\pi),0)/\chi({\bf q}\rightarrow 0,0)$,
the vertex corrections should suppress the spin-fluctuation 
exchange interaction at ${\bf q}=(\pi,\pi)$ momentum transfer.
The diagrams (f) and (g) in Fig.~6.16 are the lowest order vertex
corrections, and, indeed, at this level they act to suppress 
the ${\bf q}=(\pi,\pi)$ component of $\Gamma_{Is}$ 
as seen in Fig.~6.17.
However, the QMC data indicates that 
the suppression of the peak at ${\bf q}=(\pi,\pi)$
in $\Gamma_{Is}$ is partial.
This is because the ratio
$\chi({\bf q}=(\pi,\pi),0)/\chi({\bf q}\rightarrow 0,0)$
is about 5 rather than being of order 100
at the lowest temperature $\Gamma_{Is}$ 
was calculated with QMC.
Hence, a large AF correlation length 
is not necessary, and simply weight in $\Gamma_{Is}$ at large momentum 
transfers is sufficient to yield a sizeable attractive interaction in 
the $d_{x^2-y^2}$-wave channel.

The results reviewed in this section constitute 
what has been learned 
about $d_{x^2-y^2}$-wave pairing in the 2D Hubbard model
from the determinantal QMC simulations.
These simulations are not carried out at lower temperatures
because of the sign problem, and, hence, 
it is not possible to know whether 
superconducting long-range order develops at lower $T$
[Loh {\it et al.} 1990].
In Fig.~6.19(a), 
the average sign of the fermion determinants,
$\langle {\rm sign}\rangle$,
which is defined in the Appendix, 
is plotted as a function of $T$ 
at $\langle n\rangle=0.87$ for $U=8t$ and $4t$.
As the value of $\langle {\rm sign}\rangle$ decreases below 1,
the statistical error in the QMC data grows rapidly requiring 
exponentially long simulation times.
In Fig.~6.19(b), the filling dependence of 
$\langle {\rm sign}\rangle$ is shown for $U=8t$ at
$T=0.5t$ and $0.33t$.
These figures show the boundary of the parameter regime 
of the Hubbard model which cannot be probed because of the 
sign problem.
However, the DMRG calculations 
[White 1992] are carried out at zero temperature,
and they have provided valuable information 
about this regime in the 2-leg Hubbard ladder.
The DMRG studies of the 2-leg Hubbard ladder 
will be reviewed in the next section. 

\begin{figure}
\centering
\iffigure
\mbox{
\subfigure{
\epsfysize=8cm
\epsffile[100 150 480 610]{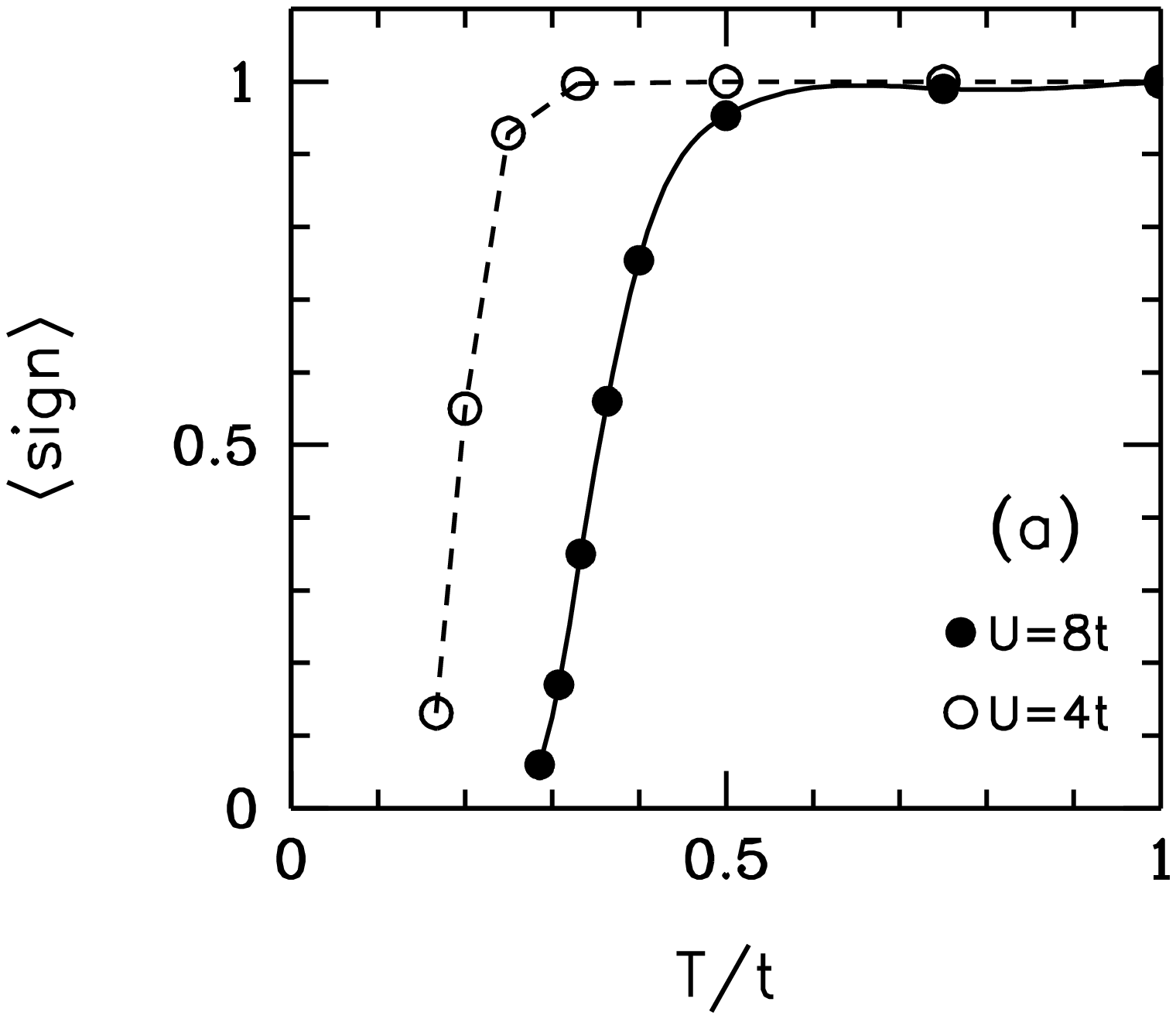}}
\quad
\subfigure{
\epsfysize=8cm
\epsffile[50 150 600 610]{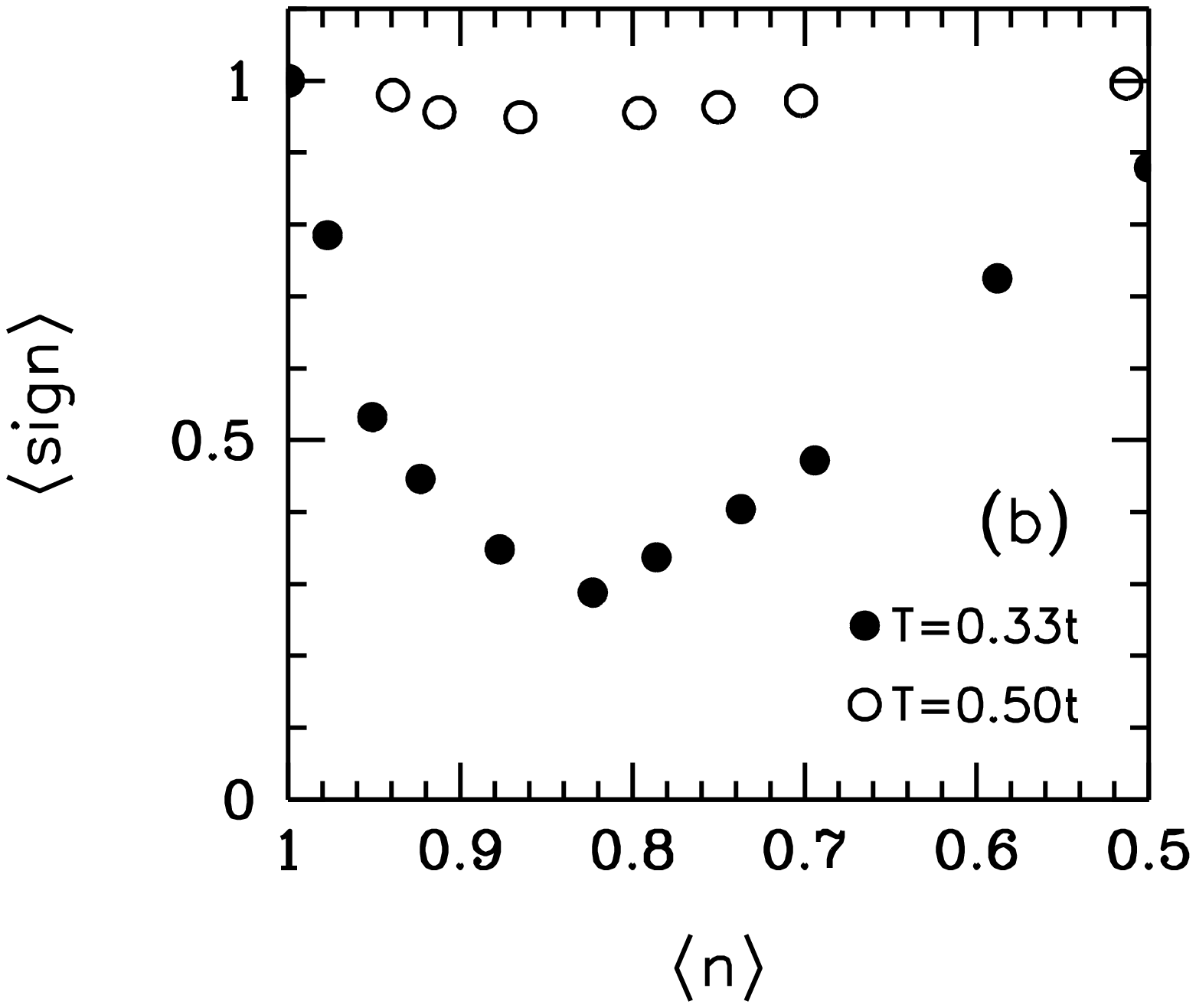}}}
\fi
\caption{
(a) Temperature and (b) the filling dependence of 
$\langle {\rm sign}\rangle$.
In (a), the results are shown for 
$U=8t$ and $4t$ at $\langle n\rangle=0.87$,
and in (b) for $U=8t$.
}
\label{6.19}
\end{figure}

\setcounter{equation}{0}\setcounter{figure}{0}
\section{2-leg Hubbard ladder}

The 2-leg Hubbard ladder 
has been studied using various many-body techniques 
such as the DMRG 
[Noack {\it et al.} 1994, 1995, 1996 and 1997], 
the QMC [Dahm and Scalapino 1997], 
the exact diagonalization [Yamaji and Shimoi 1994], 
and the weak-coupling renormalization group
[Balents and Fisher 1996].
The results of the DMRG and 
the QMC calculations will be reviewed here.

At half-filling, the ground state of this system does not have AF
long-range order, rather it has a spin gap.
Near half-filling, this model exhibits short-range AF correlations and 
power-law decaying $d_{x^2-y^2}$-like superconducting and 
"$4{\bf k}_F$" CDW correlations
[Noack {\it et al.} 1994 and 1996].
The DMRG calculations have found that in the ground state of 
this system the pair-field correlations for the interacting 
system can get enhanced with respect to those of the noninteracting 
($U=0$) system for a range of the model parameters.
This is the first time ever in an exact ground state calculation 
for a bulk system that 
by turning on an onsite Coulomb repulsion the superconducting 
correlations get enhanced. 
For this reason, the 2-leg Hubbard model is quite important.

Another reason for studying this model is that here it is possible to 
understand the mechanism mediating the $d_{x^2-y^2}$-like
superconducting correlations by comparing the DMRG results with the 
QMC data obtained at relatively low temperatures.
These comparisons indicate that it is the short-range AF fluctuations 
which mediate the pairing.    
Furthermore, 
the pairing is strongest when the model parameters 
are such that there is enhanced single-particle spectral weight 
near the $(\pi,0)$ and the $(0,\pi)$ points of the Brillouin zone.
For $U=8t$ and $\langle n\rangle =0.875$, 
this occurs when $t_{\perp}/t \sim 1.5$.
In this case, the irreducible particle-particle vertex peaks at 
momentum transfers near $(\pi,\pi)$ creating optimum conditions for 
$d_{x^2-y^2}$ pairing.

It is also interesting to study the 2-leg Hubbard ladder 
because the half-filled insulating state is spin gapped while 
in the 2D case there is long-range AF order. 
In the cuprates, on the other hand, the undoped system has 
long-range AF order, and a spin gapped phase lies between the 
superconducting and the insulating phases. 
The 2-leg Hubbard model is a system where the relation between the 
spin gap and the superconducting correlations as well as the 
density correlations can be studied exactly.

In section 7.1 below, the DMRG results on the pair-field correlation 
function from Ref.~[Noack {\it et al.} 1997] will be shown for various 
values of the model parameters.
In order to understand these results better, 
in Section~7.2 the QMC results 
on the single-particle spectral weight 
from Ref.~[Noack {\it et al.} 1997], 
the irreducible particle-particle interaction 
and the solution of the Bethe-Salpeter equation 
from Ref.~[Dahm and Scalapino 1997] will be presented. 
In Section 7.3, the results on the 2-leg Hubbard ladder will be 
compared with those on the 2D case.
Later, in Section 8.5.3, comparisons will be made with the
superconducting correlations found in the 2-leg $t$-$J$ ladder.

\subsection{DMRG results}

\begin{figure}
\centering
\iffigure
\epsfig{file=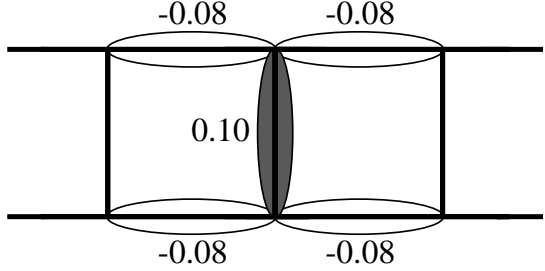,height=6cm}
\fi
\caption{
Schematic drawing of the pair-wave function
showing the values of the off-diagonal matrix element 
$\langle N_2 | ( 
c^{\dagger}_{{\bf r}\uparrow} c^{\dagger}_{{\bf r'}\downarrow} 
-
c^{\dagger}_{{\bf r}\downarrow} c^{\dagger}_{{\bf r'}\uparrow} )
| N_1\rangle$
for creating a singlet pair between near-neighbor sites.
Here, it is seen that 
the matrix element for this process is negative
when the singlet-pair is created along a chain,
while it is positive across a rung. 
This shows the $d_{x^2-y^2}$-like nature of the pairing correlations 
in the 2-leg Hubbard ladder.
}
\label{7.1}
\end{figure}

The DMRG calculations found that there are power-law decaying 
$d_{x^2-y^2}$-wave pair-field correlations 
in the ground state of the 2-leg
Hubbard ladder.
The $d_{x^2-y^2}$-like internal structure of the pairs
can be seen by considering the pair-creation amplitude
\begin{equation}
\label{amplitude}
\langle N_2 | ( 
c^{\dagger}_{{\bf r}\uparrow} c^{\dagger}_{{\bf r'}\downarrow} 
-
c^{\dagger}_{{\bf r}\downarrow} c^{\dagger}_{{\bf r'}\uparrow} )
| N_1\rangle
\end{equation}
for adding a singlet pair on near-neighbor sites along 
and across the legs.
Here, $|N_1\rangle$ is the ground state with four holes relative 
to the half-filled band and $|N_2\rangle$ is the ground state 
with two holes on a $2\times 16$  ladder.
The results on the pair amplitude 
are shown in Fig.~7.1
for $U/t=8$ and $t_{\perp}/t=1.5$.
Note the $d_{x^2-y^2}$-like change 
in the sign of this matrix element. 

Using the DMRG method it is possible to calculate 
the rung-rung correlation function
\begin{equation}
\label{D}
D(i,j) = \langle \Delta(i) \Delta^{\dagger}(j) \rangle
\end{equation}
in the ground state of the 2-leg Hubbard ladder with
the open boundary conditions.
Here, 
\begin{equation}
\Delta^{\dagger}(i) = 
c^{\dagger}_{i1\uparrow} c^{\dagger}_{i2\downarrow} -
c^{\dagger}_{i1\downarrow} c^{\dagger}_{i2\uparrow} 
\end{equation}
creates a singlet pair across the $i$'th rung,
and $c^{\dagger}_{ik\sigma}$ is the electron creation operator 
with spin $\sigma$
at the $i$'th site of the $k$'th leg of the 2-leg ladder.
The fact that the matrix elements shown in Fig.~7.1 
are finite means that this bare pair-creation operator,
which is composed of the bare electron-creation 
operators $c^{\dagger}_{ik\sigma}$,
has finite overlap with the true pair-creation
eigen-operator for this system.
In order to minimize the effects of the boundaries, 
here $D(i,j)$ is averaged over six $(i,j)$ pairs with 
$\ell=|i-j|$ fixed. 
This averaging starts with symmetrically placed $(i,j)$ 
values and then proceeds to shift 
these to the left and right of the center. 
By comparing results obtained on different size lattices, 
it is possible to control the finite size effects. 
In the following, 
$D(\ell)$ calculated on the $2\times 32$ lattice for 
$\ell<20$ will be shown. 
In this case, the finite size effects are negligible.

\begin{figure}
\centering
\iffigure
\epsfig{file=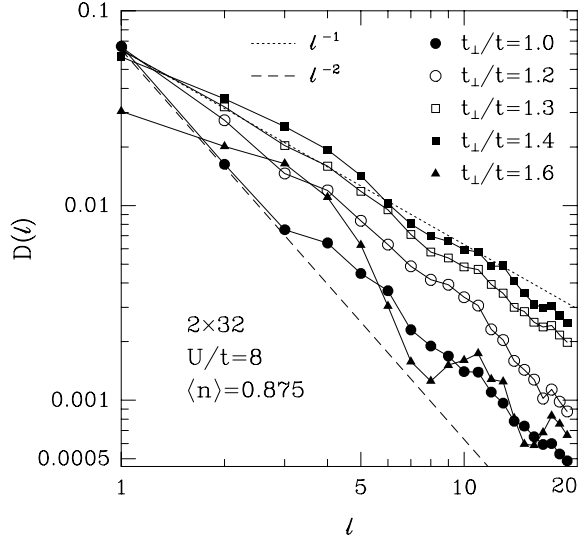,height=8cm}
\fi
\caption{
Pair-field correlation function 
$D(\ell)$ versus $\ell$ for various values of 
$t_{\perp}/t$ with $U=8t$ and $\langle n\rangle =0.875$.
}
\label{7.2}
\end{figure}

Figure ~7.2 shows $D(\ell)$ versus $\ell$ for various values of 
$t_{\perp}/t$ with $U=8t$ and $\langle n\rangle =0.875$.
The dashed and the dotted lines represent power-law decays of 
$\ell^{-2}$ and $\ell^{-1}$, respectively. 
In this figure, it is seen that 
$D(\ell)$ exhibits a power-law decay for 
$t_{\perp}/t< 1.6$.
For $t_{\perp}/t=1.0$, $D(\ell)$ decays as $\ell^{-2}$.
When $t_{\perp}/t$ is increased from
1.0 to 1.4, the strength of $D(\ell)$ 
gets enhanced and it decays more slowly.
For $t_{\perp}/t=1.6$, $D(\ell)$ is reduced and it decays faster.
Below in Section 7.1, it will be seen that when $t_{\perp}/t > 1.4$,
the antibonding single-particle band becomes unoccupied,
and the decrease in the strength of the pairing correlations
is due to this effect. 
Using the data in Fig.~7.2 for 
$1 < \ell <18$, $D(\ell)$ has been fitted to a form
\begin{equation}
D(\ell) = {1 \over {\ell^{\theta}} }
\end{equation}
with a linear least-squares approximation.
The resulting $\theta$ values are plotted as a function of 
$t_{\perp}/t$ for various fillings in Fig.~7.3.
Here, it is seen that the minimum in $\theta$
versus $t_{\perp}/t$ depends on the filling, but it occurs
for $t_{\perp}/t\sim 1.4$ near half-filling. 
For $t_{\perp}/t > 1.4$, the antibonding band is no longer 
occupied and the pairing correlations decrease rapidly.
Thus the pairing correlations are enhanced near the point 
at which the antibonding band moves through the Fermi level
[Noack {\it et al.} 1995 and 1997, Yamaji and Shimoi 1994].

\begin{figure}[ht]
\centering
\iffigure
\epsfig{file=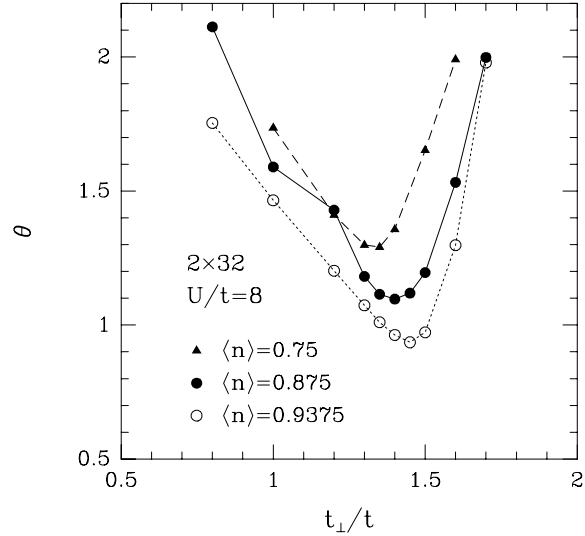,height=8cm}
\fi
\caption{
Exponent $\theta$ versus $t_{\perp}/t$ for 
$U/t=8$ and various values of $\langle n\rangle$.
}
\label{7.3}
\end{figure}

Another measure of the strength of the pair-field 
correlations which can be used is the average of 
$D(\ell)/D(1)$ for rung separations $\ell=8$ to 12:
\begin{equation}
\label{Dbar}
\overline{D} = {1\over 5}
\sum_{\ell=8}^{12} \,
{ D(\ell) \over D(1) }.
\end{equation}
Figure~7.4 shows $\overline{D}$ versus 
$t_{\perp}/t$ for $U=8t$ at various fillings.
This clearly shows how sensitively the pairing correlations 
depend on the value of $t_{\perp}/t$.
Next, 
the variation of $\overline{D}$ with $U/t$ is shown 
in Fig.~7.5.
In this figure, the crosses represent the results for $U=0$.
Hence, the onsite Coulomb repulsion can significantly
enhance the pairing correlations for a range of $t_{\perp}/t$
values.
Here, one also observes that the enhancement of 
$\overline{D}$ is strongest 
for $U/t$ between 3 and 8, which is in
the intermediate coupling regime.
In addition, as $U/t$ increases the value of $t_{\perp}/t$ 
at which the peak in $\overline{D}$ occurs shifts towards
smaller values.
Hence, the strength of the pairing correlations is a 
sensitive function of $t_{\perp}/t$, $U/t$ and
$\langle n\rangle$. 

\begin{figure}[ht]
\centering
\iffigure
\epsfig{file=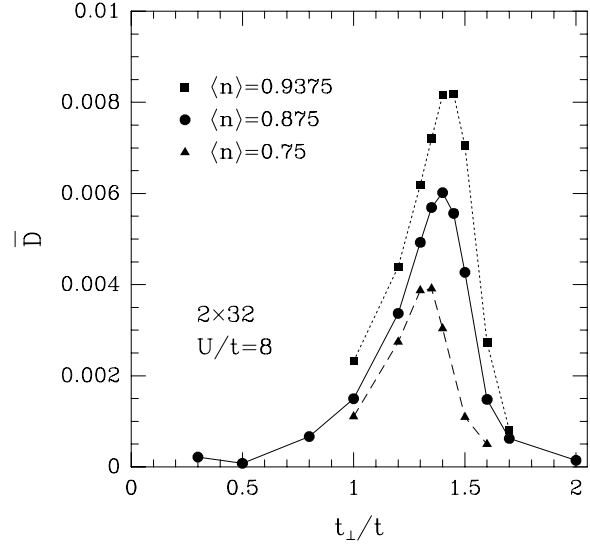,height=8cm}
\fi
\caption{
Averaged pair-field correlation function 
$\overline{D}$ versus $t_{\perp}/t$
for $U=8t$ and various values of 
$\langle n\rangle$.
}
\label{7.4}
\end{figure}

\begin{figure}
\centering
\iffigure
\epsfig{file=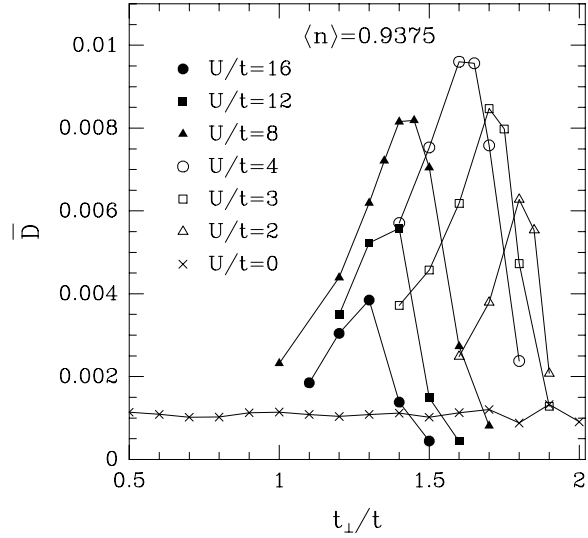,height=8cm}
\fi
\caption{
Averaged pair-field correlation function 
$\overline{D}$ versus $t_{\perp}/t$
for $\langle n\rangle=0.9375$ and various values of $U/t$.
}
\label{7.5}
\end{figure}

These DMRG results have a special place in many-body physics.
They represent the first case in an exact ground-state 
calculation for a bulk system 
where the pair-field correlation function gets enhanced 
by turning on an onsite Coulomb repulsion.
These calculations were carried out on ladders with up to 32 rungs 
resulting in negligible finite-size effects.
As seen in Section 6.1, 
in the 2D Hubbard model, 
it was found that by turning 
on the Coulomb repulsion the $d_{x^2-y^2}$-wave pair-field 
susceptibility $P_d$ gets enhanced with respect to 
$\overline{P}_d$.
However, $P_d$ was always found to be suppressed with respect 
to the pair-field susceptibility $P_d^0$ of the $U=0$ system.
This meant that, at these temperatures, the effective attractive 
interaction in the $d_{x^2-y^2}$-wave channel is not sufficiently 
strong to overcome the suppression 
of $P_d$
induced by the single-particle self-energy effects.
Here, it is seen that for the 2-leg Hubbard ladder 
in the ground state, 
$\overline{D}$ can get enhanced over the $U=0$ result for 
a set of the model parameters.

\subsection{QMC results}

In order to understand these DMRG results better, in this section 
QMC data on the 2-leg Hubbard ladder will be shown.

\subsubsection{Single-particle spectral weight}

The single-particle properties of the 2-leg Hubbard 
ladder at half-filling were studied in 
Ref.~[Endres {\it et al.} 1996].
Here,
the evolution of the 
single-particle spectral weight $A({\bf k},\omega)$ 
with $t_{\perp}/t$ will be shown for the doped case. 
These data, 
which were obtained for a $2\times 16$ ladder 
with periodic boundary conditions along the chains, 
are from Ref.~[Noack {\it et al.} 1997]. 
The comparison of the results 
on $A({\bf k},\omega)$ and $D(\ell)$ will show 
that the pairing correlations are enhanced 
for $t_{\perp}/t$ such that the bottom 
of the antibonding band moves through the Fermi level.

\begin{figure}
\centering
\iffigure
\mbox{
\subfigure{
\epsfig{file=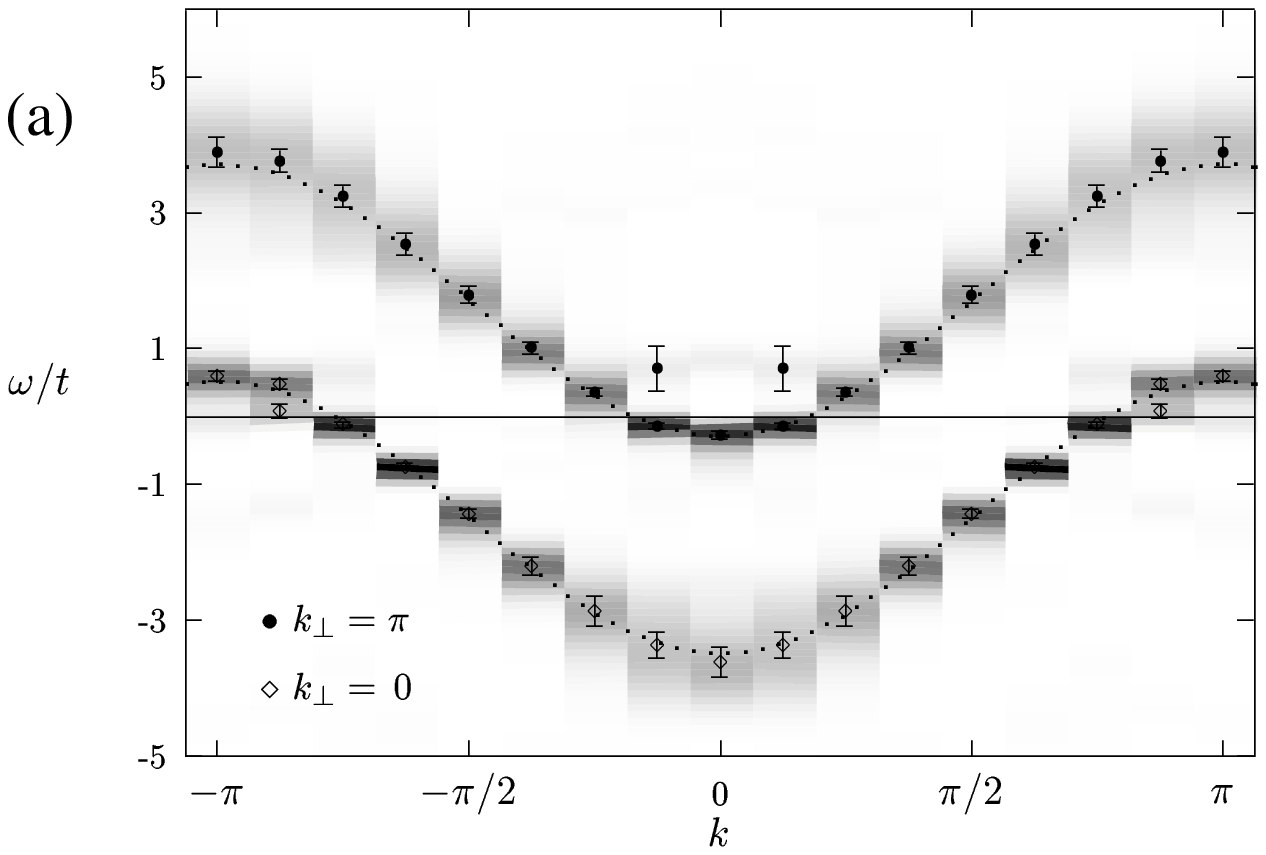,height=5cm}}
\quad
\subfigure{
\epsfig{file=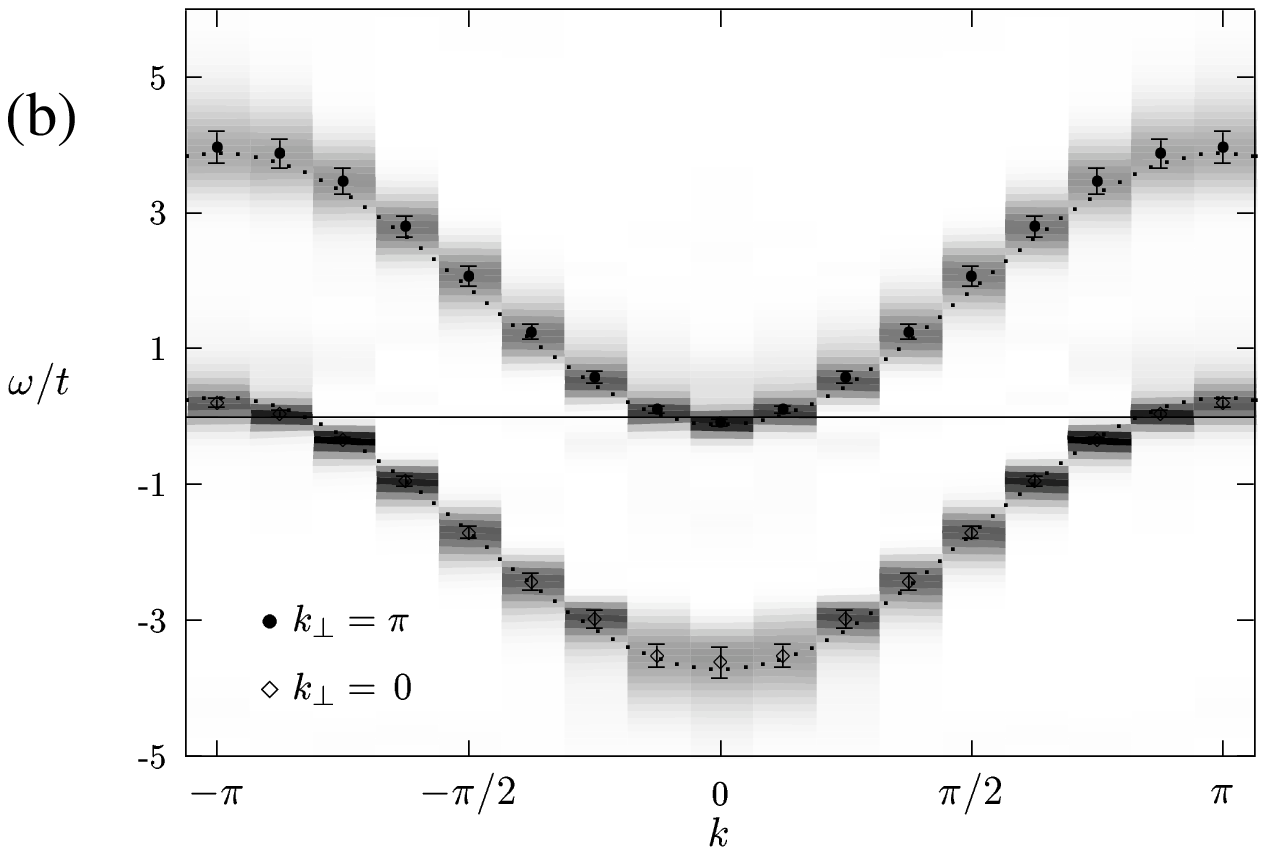,height=5cm}}}
\fi
\iffigure
\epsfig{file=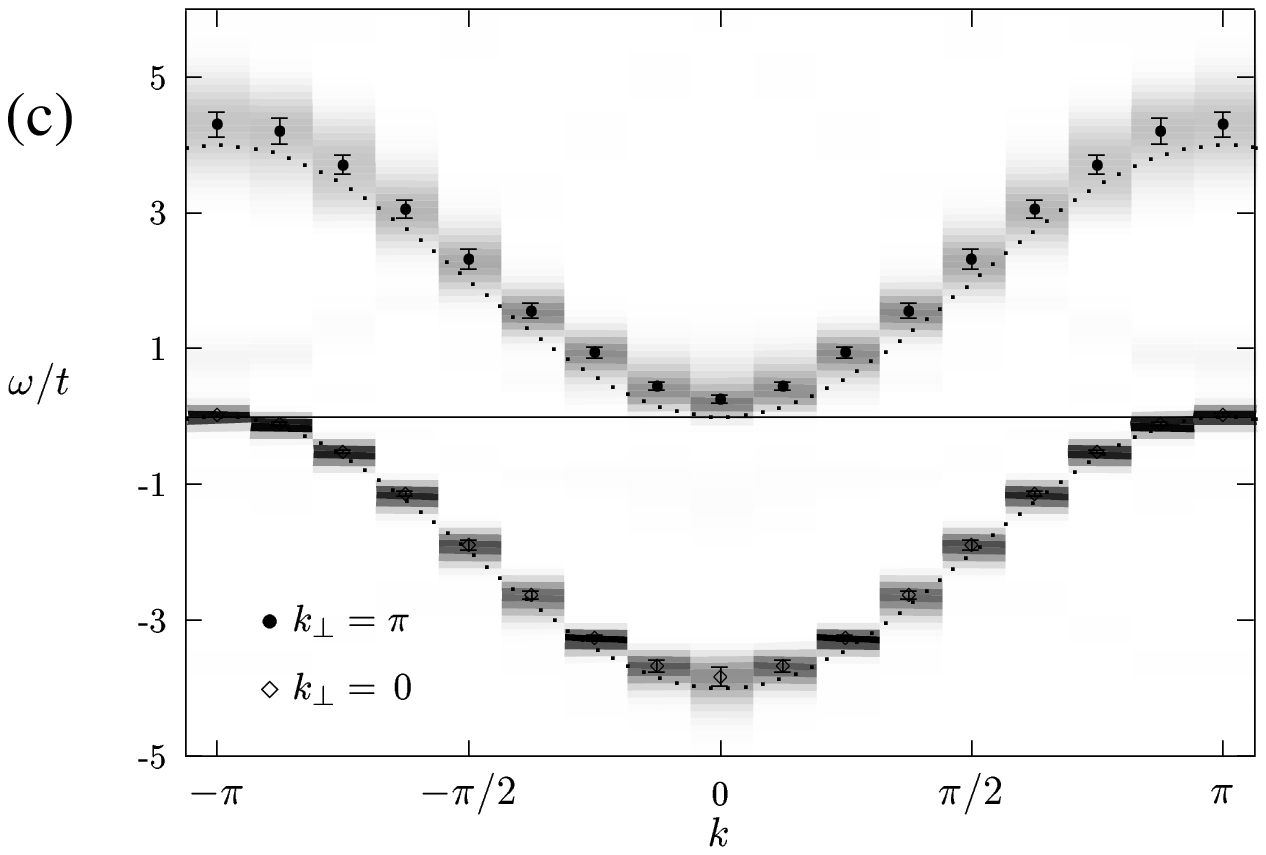,height=5cm}
\fi
\caption{
Distribution of the 
single-particle spectral weight $A({\bf k},\omega)$ 
in the ${\bf k}$ and $\omega$ plane.
The intensity of the shading indicates the amount
of the spectral weight.
These results are for $U=2t$, 
$\langle n\rangle=0.94$, $T=0.125t$ and 
(a) $t_{\perp}/t=1.6$, (b) 1.8 and (c) 2.0. 
}
\label{7.6}
\end{figure}

The results on $A({\bf k},\omega)$ were obtained by the maximum-entropy 
analytic continuation of the Monte Carlo data.
Figures 7.6 and 7.7 show data for $U/t=2$ and 4, respectively.
In these figures $\langle n\rangle=0.94$ and 
$T=0.125t$, and the results are given for various values of 
$t_{\perp}/t$. 
Here, $A({\bf k},\omega)$ is shown for both 
$k_{\perp}=0$ (bonding) and $k_{\perp}=\pi$ (antibonding)
bands as a contour plot in the 
$\omega$-$k$ plane where the intensity of the 
shading represents the magnitude of $A({\bf k},\omega)$ and
$k$ is the momentum along the chains. 
In the $U=0$ system, 
the quasiparticle dispersion consists of the bonding and the 
antibonding bands given by 
\begin{equation}
\varepsilon_{\bf k} = - 2t \cos{k} \pm 2 t_{\perp},
\end{equation}
where ${\bf k}=(k,k_{\perp})$.
The dotted curves in Fig.~7.6 represent 
$\varepsilon_{\bf k}$ for $U=0$.
Here, 
it is seen that the dispersion obtained for
$U/t=2$ closely follows that of the $U=0$ system.
Note also that for $t_{\perp}/t=1.8$ the bottom 
of the antibonding band is located right at the Fermi level while 
for $t_{\perp}/t=2.0$ the antibonding band becomes unoccupied.
In Fig.~7.5, it was seen that the pairing correlations for $U/t=2$ 
are strongest when $t_{\perp}/t=1.8$.
This comparison suggests that it is the variation 
in the single-particle spectral weight with $t_{\perp}/t$ 
which controls the dependence 
of the pairing correlations on $t_{\perp}/t$.

\begin{figure}
\centering
\centering
\iffigure
\mbox{
\subfigure{
\epsfig{file=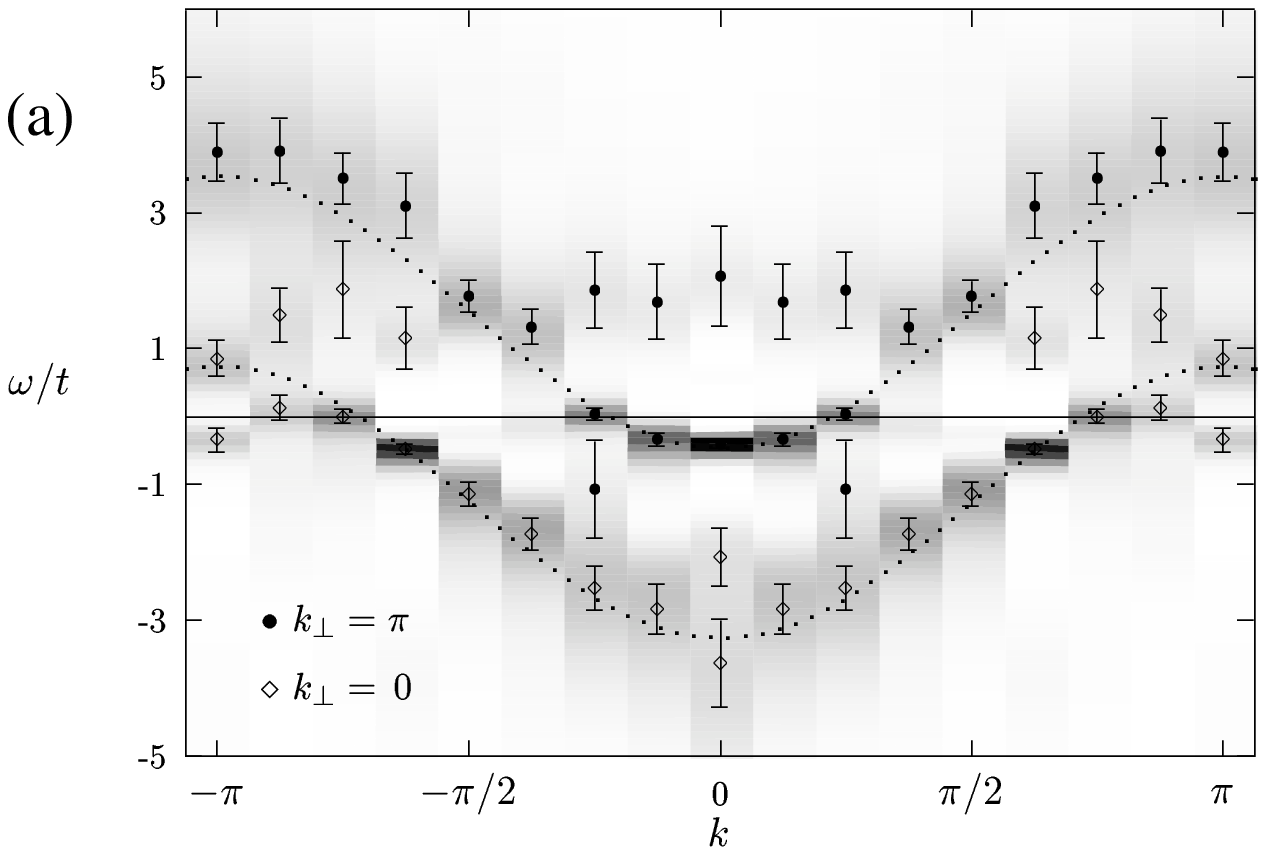,height=5cm}}
\quad
\subfigure{
\epsfig{file=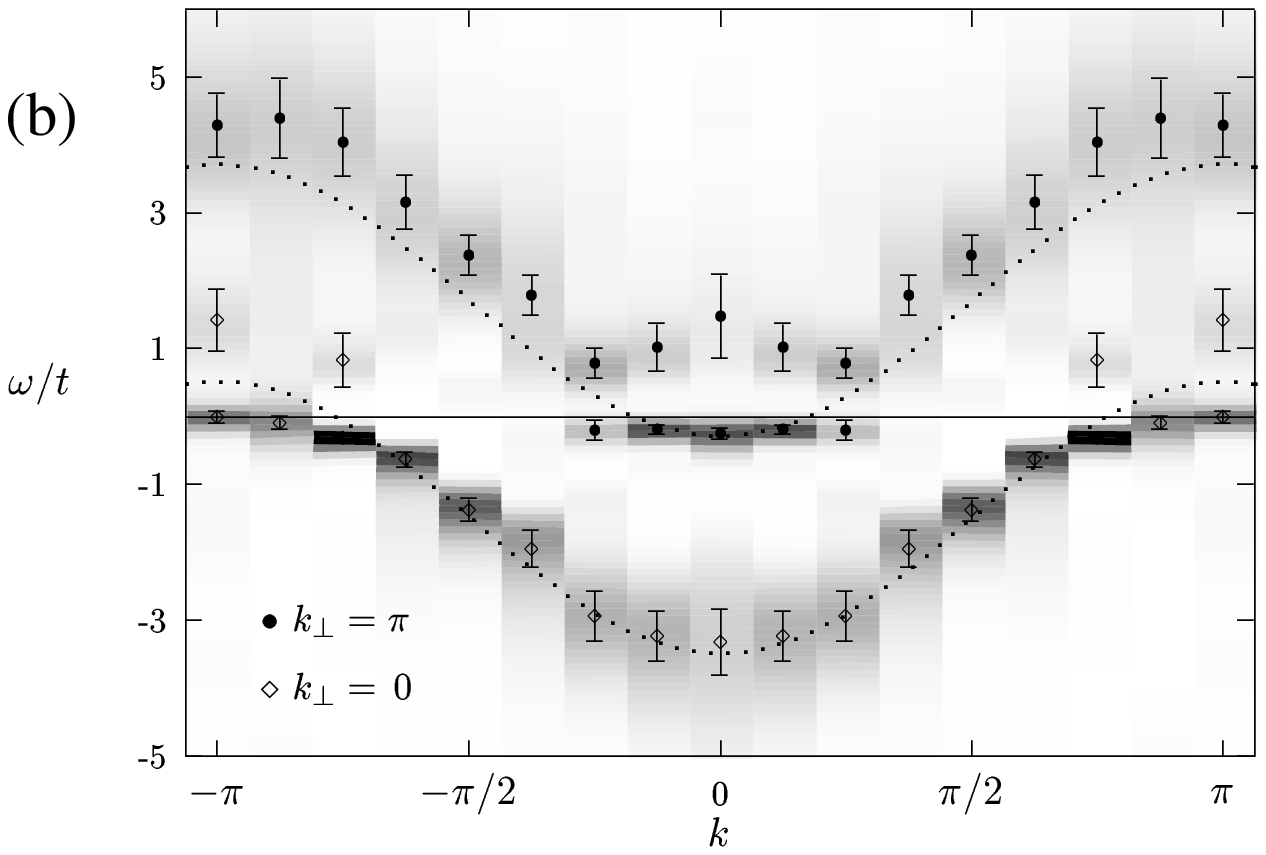,height=5cm}}}
\fi
\iffigure
\epsfig{file=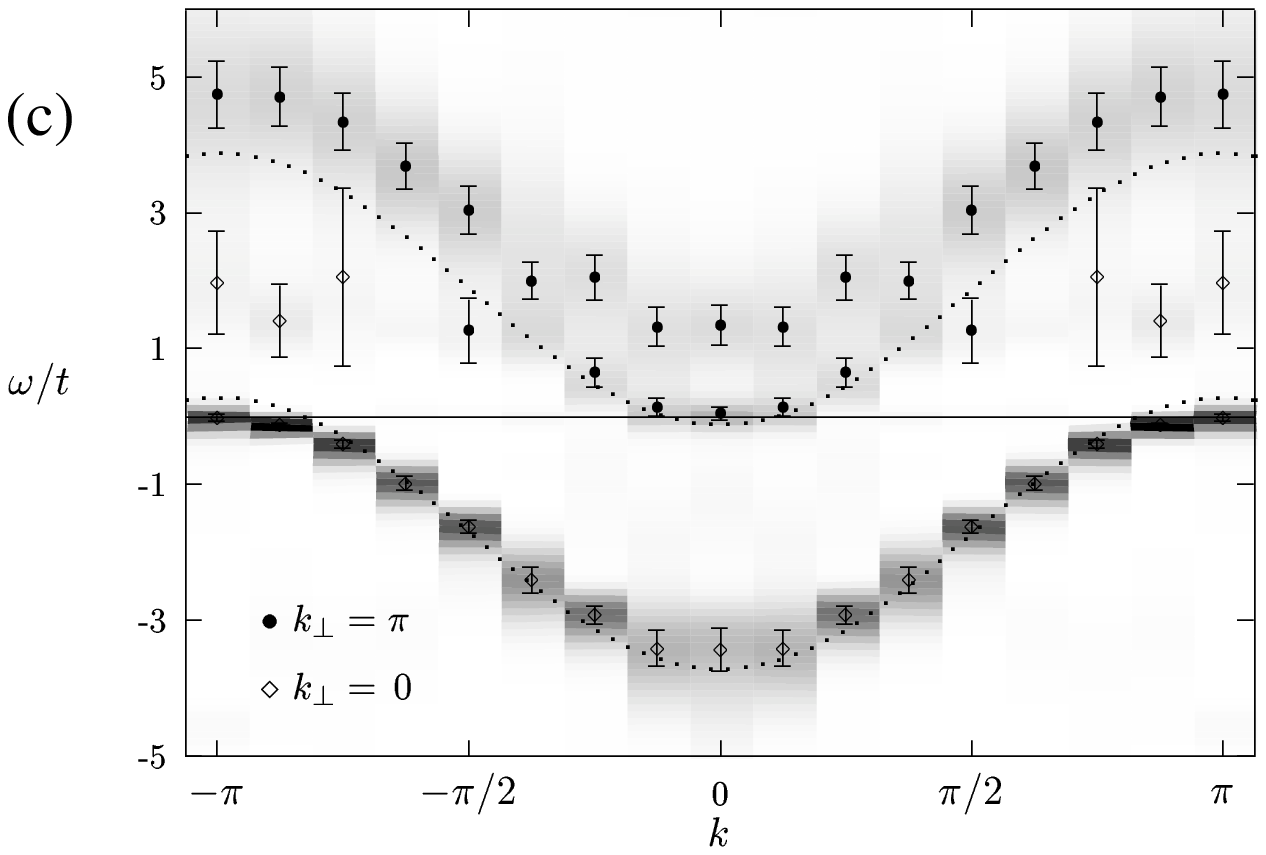,height=5cm}
\fi
\caption{
Distribution of the 
single-particle spectral weight $A({\bf k},\omega)$ 
in the ${\bf k}$ and $\omega$ plane.
The intensity of the shading indicates the amount
of the spectral weight.
These results are for $U=4t$, 
$\langle n\rangle=0.94$, $T=0.125t$ and 
(a) $t_{\perp}/t=1.4$, (b) 1.6 and (c) 1.8. 
}
\label{7.7}
\end{figure}

Figure~7.7 shows similar data on $A({\bf k},\omega)$ 
for $U/t=4$ and $t_{\perp}/t=1.4$, 1.6 and 1.8.
In this case, 
there are more differences 
between the QMC data and the $U=0$ results
denoted by the dotted curves.
Here,
the antibonding band becomes unoccupied for a smaller value of 
$t_{\perp}/t=1.8$.
One also notices that the top of the bonding and the bottom of the 
antibonding bands are flattened, 
especially for $t_{\perp}/t=1.6$, 
increasing the amount of 
the single-particle spectral weight near the Fermi level.
This behaviour is similar to the build up of spectral weight near
$(\pi,0)$ in the 2D Hubbard model as discussed in 
Section~5, and in the ARPES experiments on the 
high-$T_c$ cuprates which are reviewed by [Shen and Dessau 1995].

In the next section, 
it will be seen that for the 2-leg 
Hubbard ladder the irreducible particle-particle 
scattering vertex $\Gamma_I$ peaks at $(\pi,\pi)$ 
momentum transfer as for the 2D Hubbard model.
These results suggest that the peaking of $\Gamma_I$ 
near $(\pi,\pi)$ momentum transfer along with the enhanced 
single-particle spectral weight enhances the 
pairing correlations, and this is the reason for the strong
dependence of $\overline{D}$ on $t_{\perp}/t$ seen in 
Fig~7.4.

\subsubsection{Irreducible particle-particle interaction}

In the previous sections, 
it has been seen that the 2-leg 
Hubbard ladder has power-law $d_{x^2-y^2}$-wave pairing correlations 
in its ground state, and for a range of the model parameters 
the pairing correlations can get strong. 
It is useful to gain insight into the mechanism which leads
to the $d_{x^2-y^2}$-wave pairing correlations in this model.
For this reason, here the Monte Carlo results on the 
irreducible particle-particle interaction $\Gamma_I$ of the
2-leg Hubbard ladder will be reviewed.
These are results from [Dahm and Scalapino 1997].
Comparisons with the magnetic susceptibility $\chi({\bf q},0)$ 
will show that the short-range AF spin-fluctuations are responsible 
for the momentum structure in $\Gamma_I$
for ${\bf q}$ near $(\pi,\pi)$.
Using the data on $\Gamma_I$ and the single-particle 
Green's function $G$, the Bethe-Salpeter equation will be solved 
in the particle-particle channel
and the leading singlet pairing channel will be shown to have 
$d_{x^2-y^2}$-like symmetry.

\begin{figure}
\centering
\iffigure
\mbox{
\subfigure[]{
\epsfysize=8cm
\epsffile[100 150 480 610]{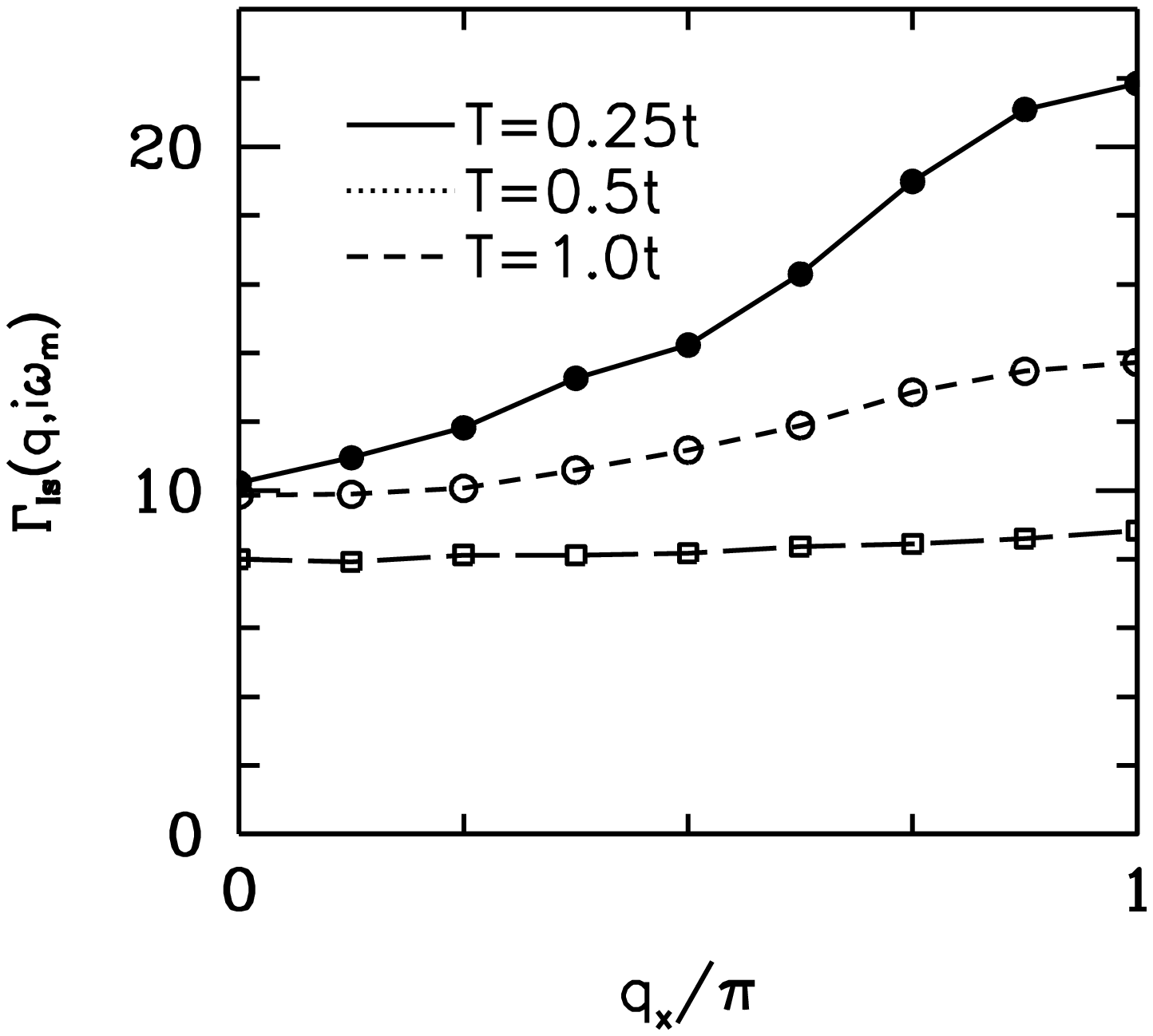}}
\quad
\subfigure[]{
\epsfysize=8cm
\epsffile[50 150 600 610]{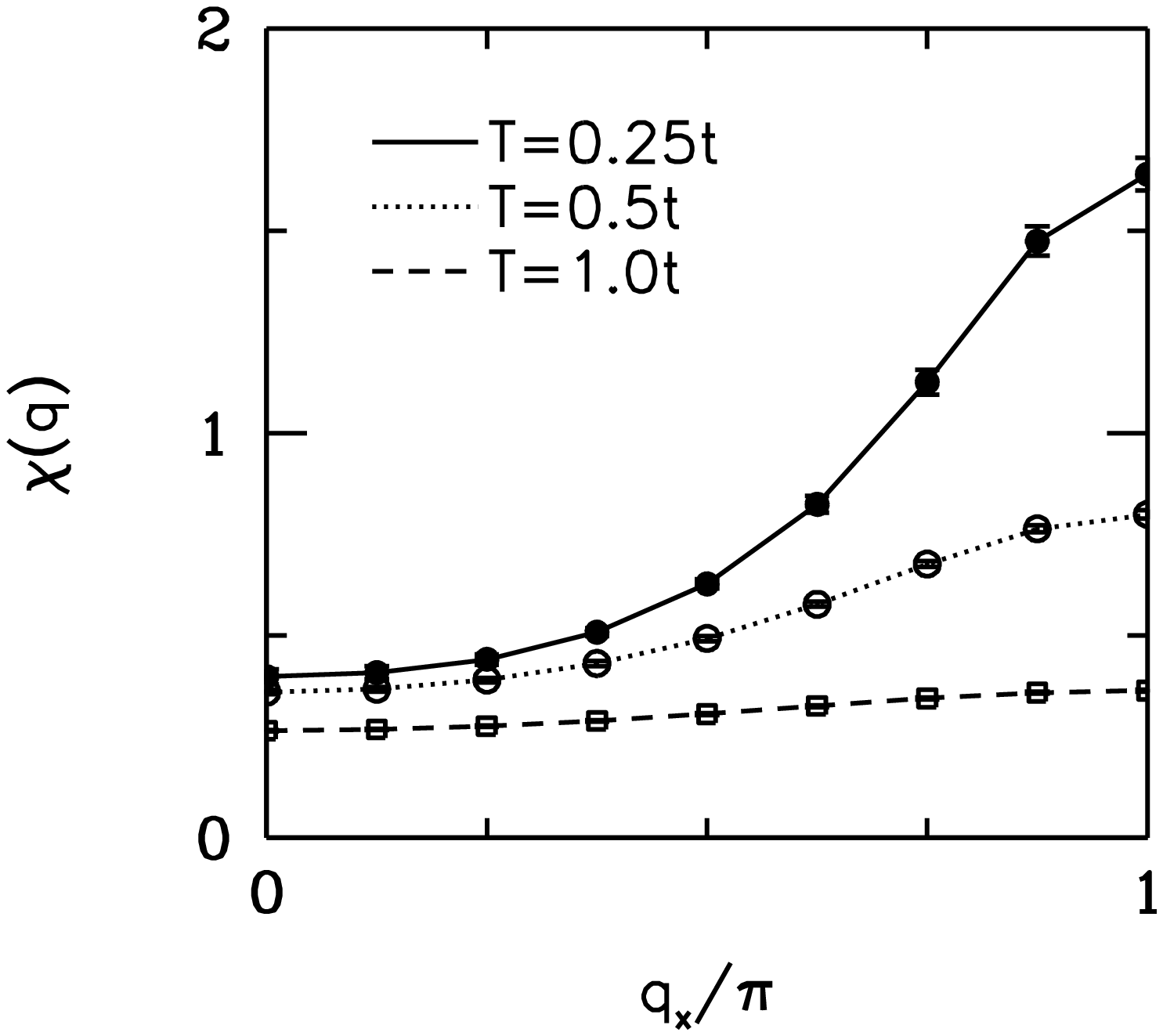}}}
\fi
\caption{
Monte Carlo results on the momentum dependence of 
(a) the irreducible particle-particle interaction 
in the singlet channel
$\Gamma_{Is}({\bf q},i\omega_m=0)$ and 
(b) the magnetic susceptibility $\chi({\bf q},i\omega_m=0)$. 
These results were obtained on a $2\times 16$ lattice  
for $U=4t$, $T=0.25t$, $\langle n\rangle=0.87$
and $t_{\perp}=1.5t$.
}
\label{7.8}
\end{figure}

Using Monte Carlo simulations,
the singlet irreducible vertex 
$\Gamma_{Is}({\bf q},i\omega_m)$ has been calculated on a 
$2\times 16$ lattice for $U/t=4$ and 
$\langle n\rangle = 0.875$
in the same way as discussed in Section 6.2.
In addition, here $t_{\perp}/t=1.5$ was chosen so that the 
system has strong pairing correlations in the ground state. 
In Fig.~7.8(a), 
$\Gamma_{Is}({\bf q},i\omega_m=0)$ is plotted 
as a function of $q$ where ${\bf q}=(q,\pi)$
at various temperatures.
Here, as in Section~6, ${\bf q}={\bf p}-{\bf p'}$ is the momentum 
transfer and ${\bf p'}$ is kept fixed at $(\pi,0)$ while ${\bf p}$ 
is scanned.
At high temperatures, 
$\Gamma_{Is}({\bf q})$ is flat in momentum space with 
a magnitude varying between $8t$ and $10t$,
and as $T$ is lowered to $0.25t$, 
$\Gamma_{Is}({\bf q},0)$ develops significant amount of 
weight at ${\bf q}=(\pi,\pi)$ momentum transfer 
becoming of order $20t$.
This behaviour is similar to what has been seen in Section~6 for the 
2D Hubbard model.

Figure~7.8(b) shows the momentum dependence of the magnetic susceptibility 
$\chi({\bf q},0)$ for the 2-leg Hubbard ladder 
for the same model parameters.
Here, 
$\chi({\bf q},0)$ is also plotted as a function of $q$
where ${\bf q}=(q,\pi)$.
Comparing with Fig.~7.8(a), 
one observes that the evolution of 
$\Gamma_{Is}$ with temperature is 
closely related to that of $\chi({\bf q},0)$,
which implies that the short-range antiferromagnetic correlations 
are responsible for the momentum structure in 
$\Gamma_{Is}({\bf q},0)$
for ${\bf q}$ near $(\pi,\pi)$.

\begin{figure}
\centering
\iffigure
\epsfysize=8cm
\epsffile[100 150 550 610]{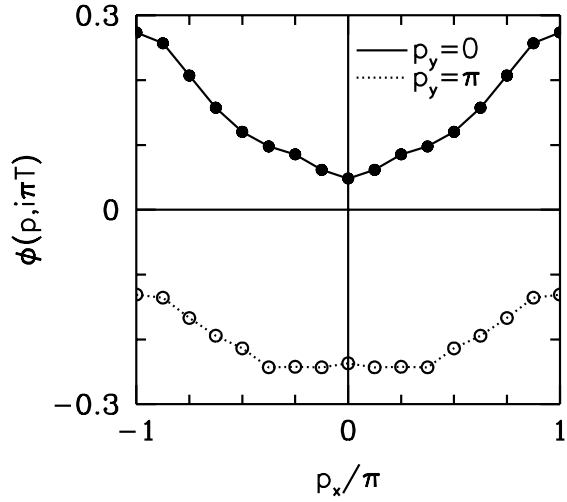}
\fi
\caption{
The leading singlet eigenfunction 
$\phi({\bf p},i\omega_n)$
versus $p_x$ where ${\bf p}=(p_x,p_y)$ and $\omega_n=\pi T$.
These results were obtained on a $2\times 16$ lattice 
for $U=4t$, $T=0.25t$, $\langle n\rangle=0.87$ and
$t_{\perp}=1.5t$.
}
\label{7.9}
\end{figure}

\subsubsection{Bethe-Salpeter equation}

In order to see what type of pairing state is favored by the
irreducible particle-particle vertex 
seen in the previous section, 
here results from the solution of the Bethe-Salpeter equation
\begin{equation}
\lambda_{\alpha}\phi_{\alpha}({\bf p},i\omega_n) = 
- {T\over N} \, \sum_{\bf p'} \,
\Gamma_I({\bf p}-{\bf p'},
i\omega_n-i\omega_{n'})
|G({\bf p'},i\omega_n{n'})|^2
\phi_{\alpha}({\bf p'},i\omega_{n'})
\end{equation}
from Ref.~[Dahm and Scalapino 1997] will be shown.
Figure~7.9 shows the momentum dependence of the leading
eigenfunction $\phi({\bf p},i\omega_n)$ 
at $\omega_n=\pi T$ for $T=0.25t$.
Figures~7.10(a) and (b) show the dependence of the leading 
singlet eigenvalue $\lambda_1$ on $T/t$ and $t_{\perp}/t$,
respectively.
Note that $\phi({\bf p},i\pi T)$ peaks 
near $(\pi,0)$ and $(0,\pi)$, and it changes sign 
between these two points. 
Hence, in this sense it is $d_{x^2-y^2}$-like, 
but it does not have a node since 
$\phi({\bf p},i\pi T)$ does not vanish 
near any of the four Fermi surface points.
In the previous section, we have seen that for 
$t_{\perp}/t\sim 1.5$, the bonding band has spectral weight 
near the Fermi level for ${\bf p}\sim(\pi,0)$, and 
the antibonding band has spectral weight near the Fermi level
for ${\bf p}\sim (0,\pi)$.
Hence, these Fermi points can be connected by scatterings 
involving ${\bf q}=(\pi,\pi)$ momentum transfer.
Since $\Gamma_{Is}$ is large and repulsive for 
${\bf q}\sim (\pi,\pi)$, the leading singlet eigenfunction 
$\phi$ of the Bethe-Salpeter equation has opposite signs for 
${\bf p}$ near $(\pi,0)$ and 
$(0,\pi)$.

\begin{figure}
\centering
\iffigure
\mbox{
\subfigure{
\epsfysize=8cm
\epsffile[100 150 480 610]{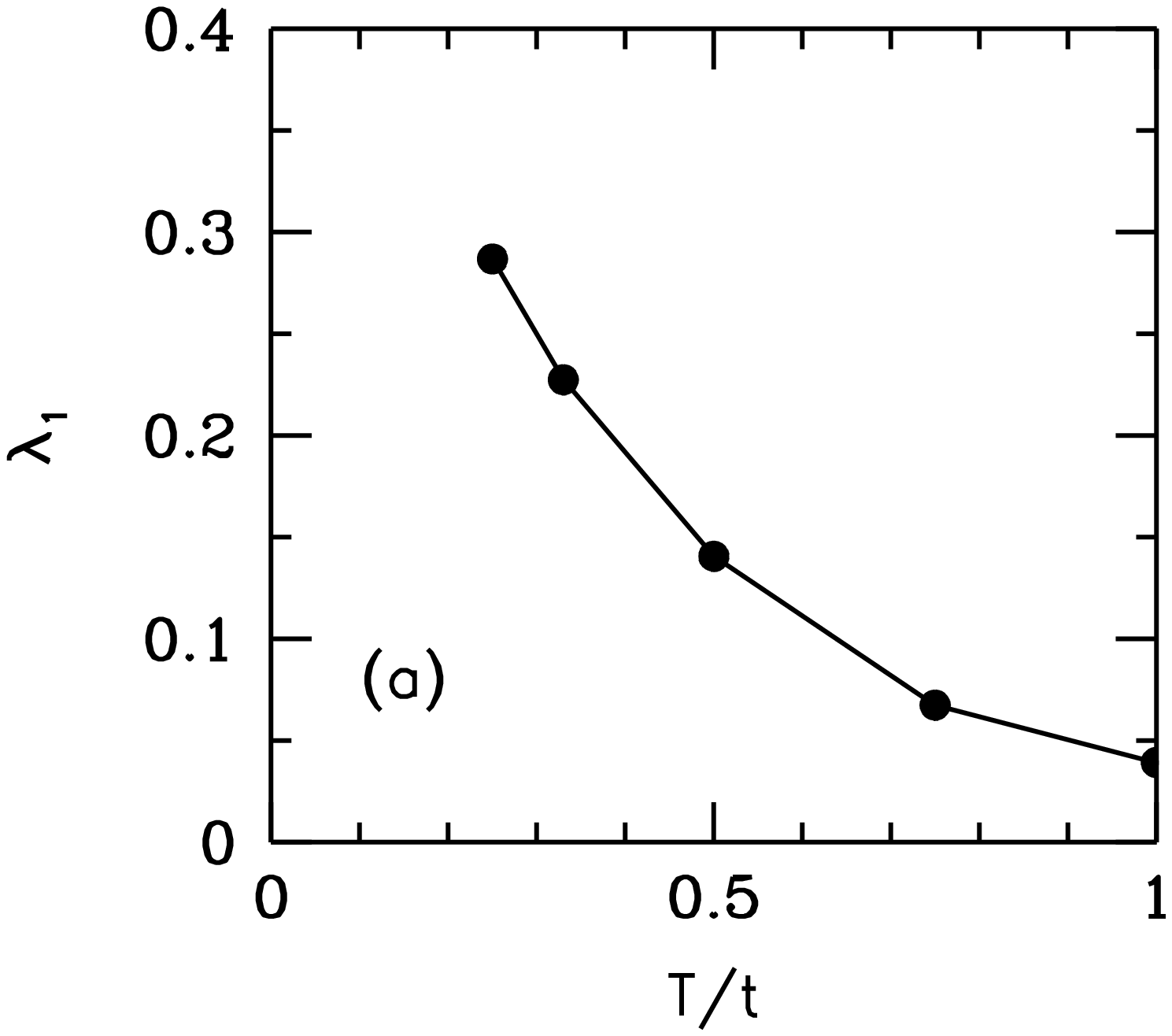}}
\quad
\subfigure{
\epsfysize=8cm
\epsffile[50 150 600 610]{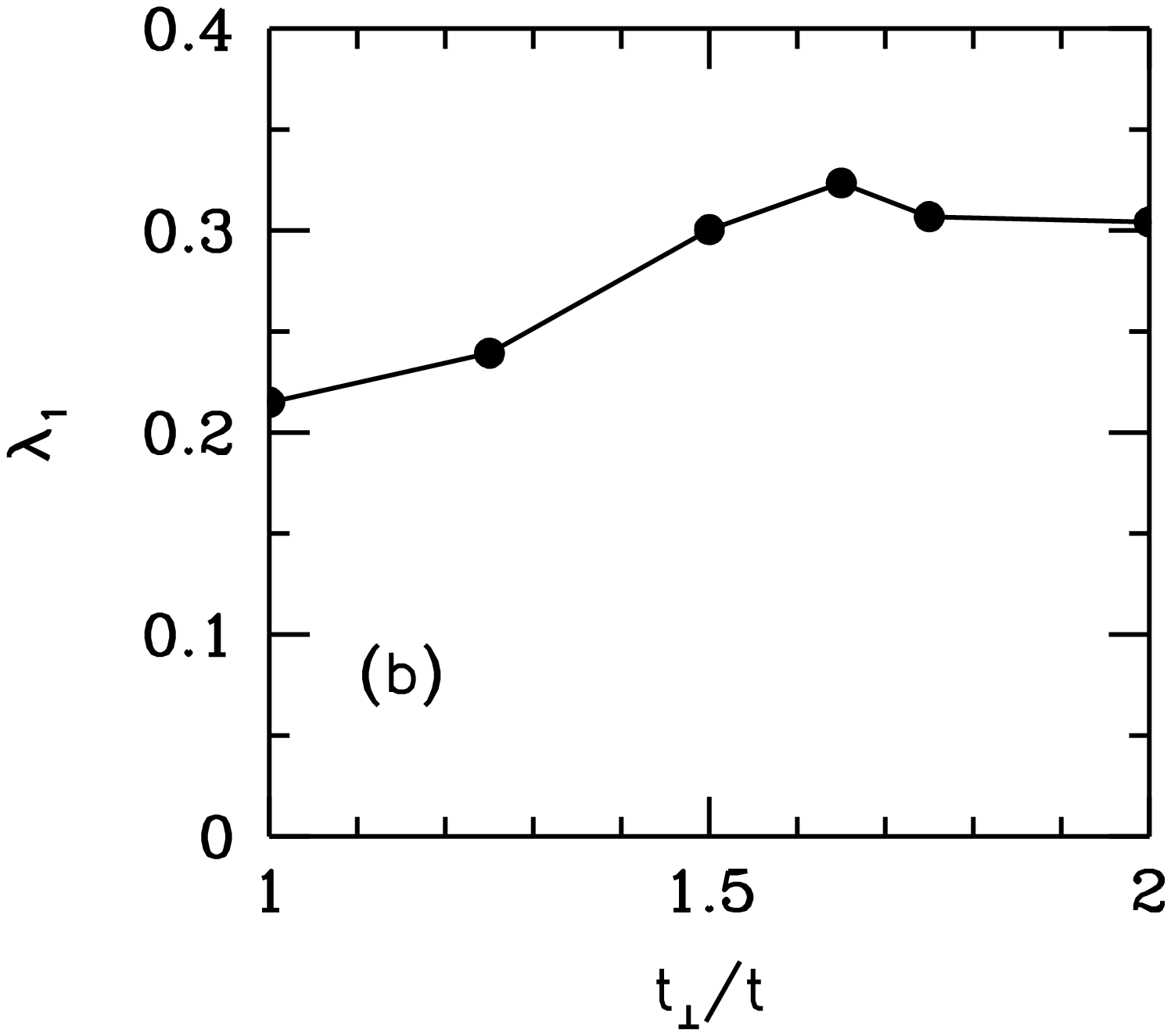}}}
\fi
\caption{
(a) Temperature and (b) $t_{\perp}/t$ 
dependence of the leading singlet
eigenvalue $\lambda_1$.
These results were obtained on a $2\times 16$ lattice 
for $U=4t$ and $\langle n\rangle=0.87$.
In (a) $t_{\perp}/t=1.5$ and in (b) $T=0.25t$ were used.
}
\label{7.10}
\end{figure}

In Fig.~7.10(a), it is seen that the eigenvalue of the 
$d_{x^2-y^2}$-like eigenfunction increases by about an order of magnitude 
as $T$ is lowered from $1t$ to $0.25t$.
However, even at $T=0.25t$, which is the lowest temperature 
where the Monte Carlo calculation of the Bethe-Salpeter 
eigenvalues can be carried out, the leading singlet eigenvalue 
is only about 0.3.

Figure~7.10(b) shows that at $T=0.25t$ the leading singlet 
eigenvalue has in fact a weak dependence on $t_{\perp}/t$, 
and only a broad peak at $t_{\perp}/t=1.6$ is seen.
This is unexpected considering the strong dependence of 
$\overline{D}$ on $t_{\perp}/t$
seen in the previous section.
It means that at $T=0.25t$ the pairing interaction has not reached
its full strength and, in addition, the thermal smearing effects 
significantly weaken the pairing correlations
which are observed in the ground state by the DMRG technique.

\subsection{Comparison of the 2-leg and the 2D Hubbard models}

It is interesting to compare these results    
on the 2-leg ladder with those on the
2D case seen in Section~6.
In Fig.~7.10(b), one sees that at $T=0.25t$, $U/t=4$ 
and $t_{\perp}/t=1.0$ 
the leading singlet eigenvalue is 0.22, 
which is slightly larger than
$\lambda_d=0.18$ at the same $U/t$, $\langle n\rangle$ and 
$T/t$ values for the 2D case.
For $t_{\perp}/t=1.0$, the DMRG calculations find that 
$D(\ell)$ varies approximately as $\ell^{-2}$ 
in the ground state and, hence, 
the pairing correlations are only as strong as in the $U=0$ case.
If one just uses this comparison of the eigenvalues for the 
2-leg and the 2D cases, then one would expect weak
pairing correlations in the ground state of the 2D Hubbard model.
However, in 2D,
$\lambda_d$ for $U=8t$ is found to be higher 
than for $U=4t$ at $T=0.5t$.
Furthermore, 
from the DMRG calculations one knows that when 
$t_{\perp}/t$ is tuned to the right value, in the 2-leg ladder
strong pairing correlations can develop.
A similar dependence on the model parameters,
such as the second-nearest-neighbour hopping $t'$,
could exist in the 2D Hubbard model. 

It has been seen above that the QMC results 
on the irreducible particle-particle vertex 
of the 2-leg ladder are similar to those on the 
2D Hubbard model.
In the ground state of the 2-leg ladder, the DMRG calculations 
find enhanced $d_{x^2-y^2}$-wave pairing correlations 
in a certain range of the parameters.
Both in the 2D and the 2-leg models, we have seen that, when doped
with holes there are short-range AF correlations and
that they strongly influence the low-energy 
single-particle properties.
In both cases, 
short-range AF correlations cause 
$\Gamma_I({\bf q},0)$ to peak at $(\pi,\pi)$ momentum transfer.
It needs to be noted that the pairing correlations observed 
in these two models do not require a particularly sharp peak in 
$\Gamma_{I}({\bf q},0)$ at ${\bf q}=(\pi,\pi)$, 
but rather simply weight at large momentum transfers.

Currently, it is not possible to determine whether
the doped 2D Hubbard model has long-range 
$d_{x^2-y^2}$-wave superconducting order in its ground state,
or, if it did, 
whether it would be sufficient to explain superconducting
transition temperatures as high as those found in the cuprates.
In spite of this, 
the results discussed above show that effects which 
increase the single-particle spectral weight 
near the $(\pi,0)$ and $(0,\pi)$ points of the 
Brillouin zone as well as effects which increase 
the strength of the particle-particle interaction 
at $(\pi,\pi)$ momentum transfer will act to enhance 
$d_{x^2-y^2}$-wave pairing.

It is possible that the strength of the $d_{x^2-y^2}$
pairing correlations in the ground state of the 2D Hubbard
model depends sensitively on the model parameters such as
the second-near-neighbour hopping $t'$ in a way similar
to what is seen for the 2-leg ladder in Section 7.
In the 2-leg case, the pairing correlations are 
as weak as those of the $U=0$ system when $t_{\perp}=t$.
However, 
when $t_{\perp}/t$ is tuned so that the flat bands 
are located near the Fermi level, the system exhibits 
enhanced pairing correlations.
It is possible that the 2D Hubbard model 
with only near-neighbour hopping similarly exhibits 
weak pairing correlations in the ground state, 
but there can be enhanced pairing when an additional 
parameter such as $t'$ is tuned.
In the 2-leg ladder case,
the value of $t_{\perp}/t$ for which 
the flat bands are near the Fermi level is renormalized
by the Coulomb repulsion. 
Similarly,  
in the 2D case, 
the optimum value of $t'$ could be renormalized,
and it is difficult to estimate it in advance.
In a bilayer Hubbard model, 
the bilayer coupling could also play a role.
These are issues which need to be resolved 
by exact techniques in the future.

Beyond these,
one would expect that any additional contribution 
to the irreducible electron-electron interaction 
which is repulsive for ${\bf q}\sim (\pi,\pi)$
momentum transfers or attractive for ${\bf q}\sim 0$ 
would act to enhance the $d_{x^2-y^2}$ pairing, 
when added to the 2D or the 2-leg Hubbard system. 
For instance, 
when a phonon mediated interaction which is most attractive for 
${\bf q}\sim 0$ momentum transfers 
is added to an AF spin-fluctuation exchange interaction, 
it is found within the $t$-matrix approximation 
that the $d_{x^2-y^2}$ eigenvalue gets enhanced
[Bulut and Scalapino 1996].
However, 
it is necessary to study such effects using exact techniques.

It could also be possible to design flat band dispersion
near the Fermi level by controlling the lattice geometry
and parameters [Imada and Kohno 2000].
Imada and Kohno have carried out exact diagonalization calculations
for a 1D 16-site $t$-$J$ model with additional
three-site terms and longer range hoppings.
By tuning the longer-range hopping parameters,
they have created flat band dispersion near the Fermi level and,
in this case, 
they find an enhanced spin gap and an enhanced tendency 
for pairing.
They have also proposed various multiband models 
which could exhibit flat bands near the Fermi level 
and enhanced pairing.

Even though for the 2D case the low-doping 
and the low-temperature regime where a spin gap could 
exist is beyond the reach of the exact techniques,
in the 2-leg case the spin gap 
$\Delta_s$ can attain large values. 
In the doped 2-leg ladder, 
$\Delta_s$ is maximum when the bottom of the 
antibonding band is near the Fermi level
[Noack {\it et al.} 1996].
For instance, 
for $U=8t$, $\langle n\rangle=0.875$ and $t_{\perp}=1.5t$,
the spin gap has the value of $0.06t$,
which corresponds to $\approx 300K$
for a $t$ of order 0.45~eV.

It is also useful to compare the density correlations 
seen in the 2-leg and the 2D Hubbard models.
With the DMRG method [Noack {\it et al.} 1996], 
the following density-density correlation 
function has been calculated for the $2\times 32$ Hubbard ladder,
\begin{equation}
S(i,j,\lambda) = \langle  n_{i\lambda} n_{j1} \rangle 
-\langle n_{i\lambda} \rangle   \langle n_{j1}\rangle. 
\end{equation}
Here, $n_{i\lambda}$ is the electron occupation number at the 
$i$'th site of chain $\lambda$.
By Fourier transforming,
$S({\bf q})$ has been obtained for 
$U=8t$, $\langle n\rangle =0.875$ and various values of 
$t_{\perp}/t$.
No obvious feature is found in $S({\bf q})$ 
at the "$2{\bf k}_F$" wave vector
of the 2-leg Hubbard ladder, which is ${\bf q^*}=(q^*,\pi)$ 
with $q^*=\pi \langle n\rangle$. 
On the other hand, 
a feature is observed at the "$4{\bf k}_F$" wave vector, 
which corresponds to $(\pi/4,0)$ for 
$\langle n\rangle=0.875$.
Especially for $t_{\perp}/t = 1.5$, 
this feature becomes more obvious. 
In order to isolate the "$4{\bf k}_F$" component of the 
density correlations, a correlation function $N({\bf q})$ 
involving four density operators has been calculated.
This correlation function exhibits a clear peak at $(\pi/4,0)$ 
for both $t_{\perp}/t=1.0$ and 1.5, 
and in real space it decays as power law,
while $S(i,j,\lambda)$ decays exponentially.

The results on the 2-leg Hubbard ladder show that, when doped, 
this model exhibits simultaneously short-range AF correlations 
and power-law decaying $d_{x^2-y^2}$-like superconducting and 
"$4{\bf k}_F$" density correlations.
For $t_{\perp}/t \sim 1.5$, 
the superconducting correlations decay more slowly than the 
"$4{\bf k}_F$" density correlations.
The 2D Hubbard model also exhibits short-range AF and 
$d_{x^2-y^2}$ superconducting correlations. 
In addition, the QMC data on the 2D case seen in Section~4 imply 
that the features found in the density susceptibility 
$\Pi({\bf q},i\omega_m=0)$ might be related to the "$4{\bf k}_F$" 
wave vectors rather than "$2{\bf k}_F$" for large 
$U/t$.
Hence, the 2-leg and the 2D Hubbard models, 
when doped, appear to 
have similar magnetic, superconducting and density properties. 
Furthermore, in both cases, the single-particle dispersion 
near the $(\pi,0)$ and $(0,\pi)$ points get flattened 
by the many-body effects.
These flat bands also seem to play a key role 
in determining the strength of the pairing correlations 
in both models.
At this point, it is necessary to note that in order to compare 
the 2-leg Hubbard model with the 2D case, one should not use 
isotropic hopping, $t_{\perp}/t=1.0$, since in this case 
the Fermi surface points are not connected by scatterings involving 
${\bf Q}\sim (\pi,\pi)$ momentum transfers.
Instead, anisotropic hopping, where the bonding and the antibonding 
Fermi surface points can be connected by $(\pi,\pi)$ scatterings,
need to be used.

Finally, 
an important question is whether the 2D Hubbard model 
exhibits spin-charge separation as in the 1D case 
[Kivelson {\it et al.} 1987, Anderson {\it et al.} 1987].
While the 2D case cannot be resolved currently, 
in the 2-leg Hubbard ladder 
no indications of spin-charge separation 
as in the 1D Hubbard model have been found
[Noack {\it et al.} 1996].

\setcounter{equation}{0}\setcounter{figure}{0}
\section{Discussion}

The numerical results discussed above point out that 
the short-range AF spin fluctuations are responsible for the 
$d_{x^2-y^2}$-wave superconducting correlations.
The relation of the AF spin fluctuations to the
$d_{x^2-y^2}$-wave superconductivity has been studied 
using various diagrammatic approaches.
Perhaps, 
the most commonly used of these approaches
is the fluctuation exchange (FLEX) approximation 
[Bickers {\it et al.} 1989], 
which self-consistently treats the fluctuations in the 
magnetic, density and the pairing channels.
The FLEX technique
has been used for obtaining possible phase diagrams and the 
estimates of the superconducting transition temperatures.
In Section~8.1 below, 
a comparison of the QMC data with the results of the FLEX 
approach to the 2D Hubbard model will be carried out for the 
single-particle and the pairing properties.
The purpose of this comparison is to have an idea for the range 
of applicability of the FLEX approximation, and see how it 
should be extended further.
Another reason for carrying out such a comparison is because 
the Eliashberg type of calculations of the $T_c$'s using the 
spin fluctuations for the cuprates [Monthoux and Pines 1992]
or the similar self-consistent spin-fluctuation exchange calculations 
[Moriya {\it et al.} 1990] are basically 
at the same level with the FLEX approximation.
Hence, 
it is of interest to compare with the exact QMC calculations.

It will be seen that for $U=4t$, 
the FLEX provides results in excellent 
agreement with the QMC data on the density of states $N(\omega)$ 
and the Bethe-Salpeter eigenvalues in the pairing channel
at the temperatures where the QMC calculations are carried out, 
as found earlier [Bickers {\it et al.} 1989].
For stronger coupling $U=8t$, there are differences between the 
FLEX and the QMC results.
The correlated metallic band which develops by doping the 
AF Mott-Hubbard insulator is not obtained within FLEX 
and this affects the strength of the $d_{x^2-y^2}$-wave 
pairing correlations.

Following this, 
in Section~8.2, 
other types of Monte Carlo approaches to the Hubbard model,
in particular the variational and the projector 
Monte Carlo algorithms, will be briefly discussed.
In Section 8.3, 
the results of the recent dynamical-cluster approximation 
and the one-loop RG calculations for $d_{x^2-y^2}$ pairing 
in the 2D Hubbard model will be discussed.
There is also much interest in understanding 
the unusual normal state properties of the cuprates 
in the pseudogap regime. 
The origin and the implications of the pseudogap
are important issues in this field.
In Section 8.4, 
various calculations on the low-doping regime
of the 2D Hubbard model will be discussed briefly
and their results will be compared with the pseudogap
seen in the cuprates.

The $t$-$J$ model which is closely related to the Hubbard model 
has also drawn broad attention.  
It is useful to compare the QMC and the DMRG results 
on the Hubbard model
with the numerical studies of the $t$-$J$ model.
Below, 
in Section~8.5, such a comparison will be given.  
Here, the attention will be concentrated on the density 
and the pairing properties since that is where there are 
unresolved issues.
A question of interest is whether in the ground state of the doped
2D Hubbard model there are special density correlations
as seen in the $t$-$J$ model,
for instance, phase separation or stripes,  
and, if so, whether they would suppress the $d_{x^2-y^2}$-wave 
pairing correlations observed at higher temperatures with the 
QMC simulations.

In Section~8.6, what these numerical studies of the Hubbard model 
imply for the mechanism of the $d_{x^2-y^2}$-wave superconductivity 
seen in the high-$T_c$ cuprates will be discussed.
The issue which will be addressed here is whether 
the 2D Hubbard model, or some variation of it, is sufficient 
for explaining why the values of the $T_c$'s found
in the cuprates are so high.

\subsection{Comparisons with the fluctuation-exchange approach to 
the 2D Hubbard model}

The FLEX approximation was used first by Bickers {\it et al.} 
to study $d_{x^2-y^2}$-wave superconductivity in the 
2D Hubbard model 
[Bickers {\it et al.} 1989, Bickers and White 1991].
This approach self-consistently incorporates the fluctuations 
in the magnetic, density and the pairing channels.
It is an approximation around the band limit, and it is conserving 
in the sense that the microscopic conservation laws for the particle 
number, energy, and momentum are obeyed. 
Within FLEX, it is found that the 
$d_{x^2-y^2}$ pairing correlations are mediated  
by the AF spin fluctuations.
The phase diagram of the $U=4t$ Hubbard model within the 
FLEX approximation is shown in Fig.~8.1.
The FLEX calculations find a finite mean-field Neel temperature 
at half-filling and at small dopings up to 6\%.
Neighboring the SDW phase is a $d_{x^2-y^2}$-wave 
superconducting phase which is stabilized for dopings 
between 6\% and 20\%.
Note that in the 2D Hubbard model, a finite Neel 
temperature is not possible and only a Kosterlitz-Thouless type
of superconducting transition can occur. 
Hence, the finite transition temperatures seen in Fig.~8.1 
represent transitions to regimes 
where the $d_{x^2-y^2}$ correlations have a 
power law decay at finite $T$.
In order to induce true long-range order
at finite temperature, 
three dimensional couplings would be required.

\begin{figure}[ht]
\centering
\iffigure
\epsfysize=8cm
\epsffile[100 150 550 610]{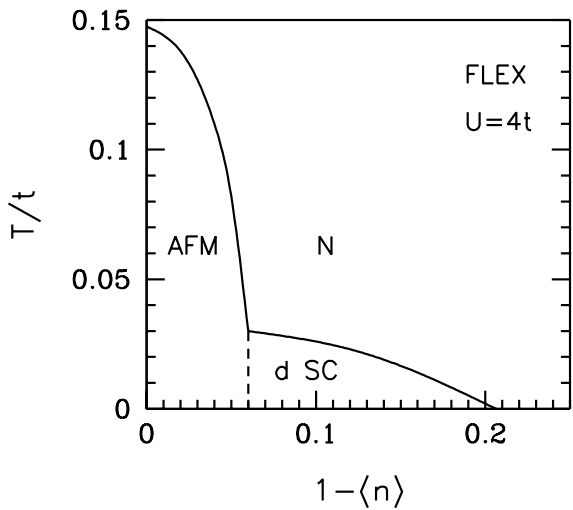}
\fi
\caption{
Sketch of the 
phase diagram of the 2D Hubbard model 
within the FLEX approximation for $U=4t$.
From [Bickers {\it et al.} 1989, Bickers and White 1991].
}
\label{8.1}
\end{figure}

In addition, using the FLEX approximation 
the effects of the nearest-neighbor hopping $t'$ 
has been investigated for the 2D Hubbard model
[Monthoux and Scalapino 1994, Dahm and Tewordt 1995].
This approach has been also used for studying a three-band 
CuO$_2$ model which has nearest neighbor copper-oxygen hopping
and an onsite Coulomb repulsion at the Cu sites
[Luo and Bickers 1993].
The solutions of the FLEX equations in the 
$d_{x^2-y^2}$-wave superconducting state were also 
obtained [Pao and Bickers 1994 and 1995, Monthoux and Scalapino 1994, 
Dahm and Tewordt 1995].

In the following, 
a comparison of the FLEX results with the QMC data will be 
carried out using results from Ref.~[Dahm and Bulut 1996].
First the single-particle properties and then the pairing correlations 
will be discussed.
Figure~8.2 compares the QMC data on 
the density of states $N(\omega)$ with the 
FLEX results for various values of $U/t$ at 
$\langle n\rangle=0.87$.
While for $U=4t$, 
the results from the two different approaches 
are similar at these temperatures, 
there are qualitative differences 
for $U=8t$ and $12t$.
The Fermi level within FLEX moves slower with the doping
at large $U/t$.
The correlated metallic band at the Fermi level
as well as the lower and the upper Hubbard bands
and the Mott-Hubbard pseudogap are not obtained within the 
FLEX approximation.
For $U=4t$ and $T=0.33t$, the FLEX and the QMC results 
have similar qualitative features.
However, at lower temperatures $N(\omega)$ could 
develop a pseudogap for $U=4t$ also.
Figure~8.3 shows the temperature evolution of $N(\omega)$ 
for $U=8t$ and $\langle n\rangle=0.87$ within the 
FLEX approximation. 
Comparing this with Fig.~5.3, 
one sees that the differences 
with the QMC data continue to exist as $T$ is lowered.
In Fig.~8.4, the QMC and the FLEX results on $N(\omega)$ 
are compared at half-filling for $U=8t$ and $T=0.5t$.
Here, it is seen that the development of the Mott-Hubbard 
gap at half-filling is not obtained within FLEX.

\begin{figure}[ht]
\centerline{
\epsfysize=6cm \epsffile[-207 164 367 598]{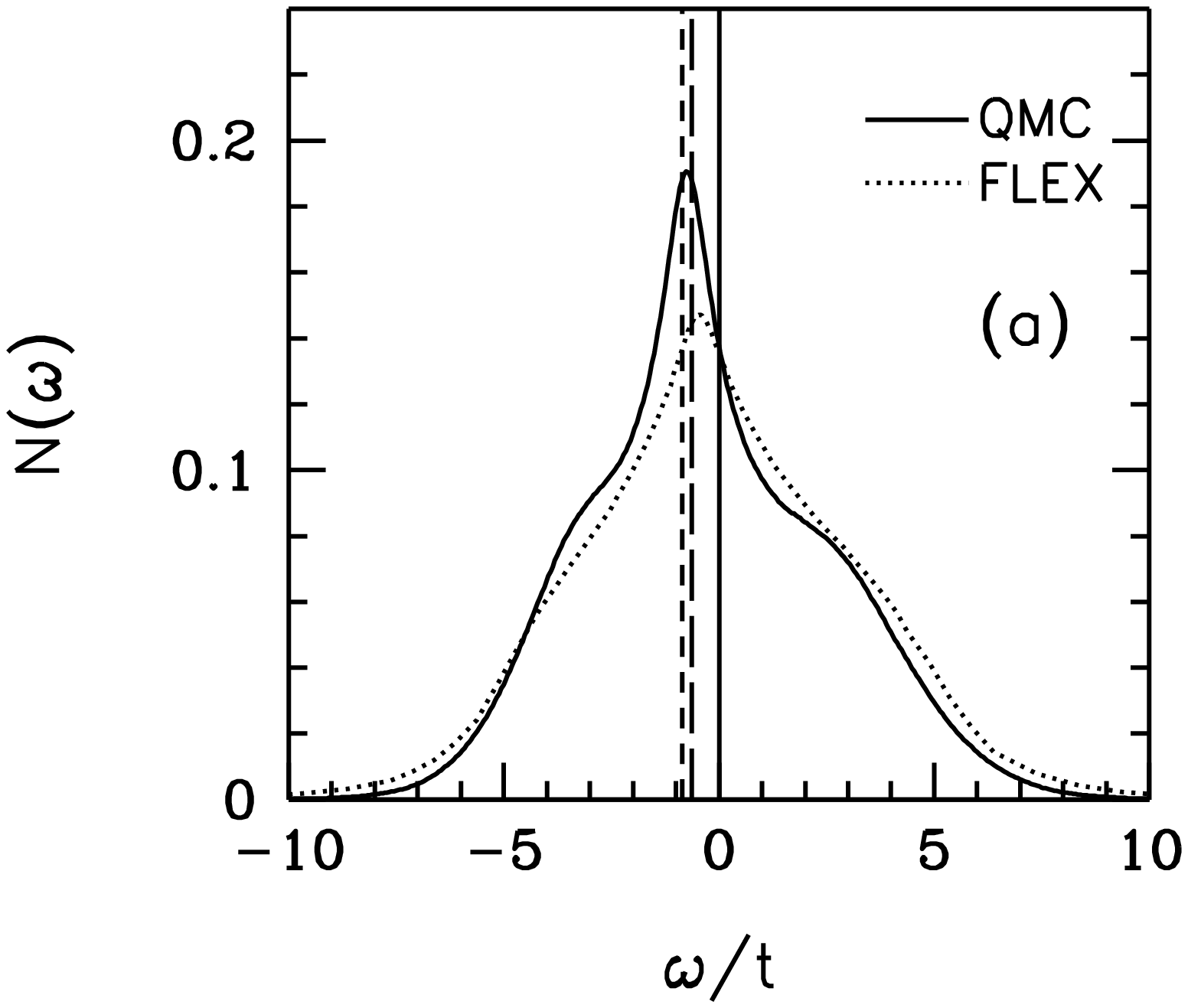}
\epsfysize=6cm \epsffile[18 164 592 598]{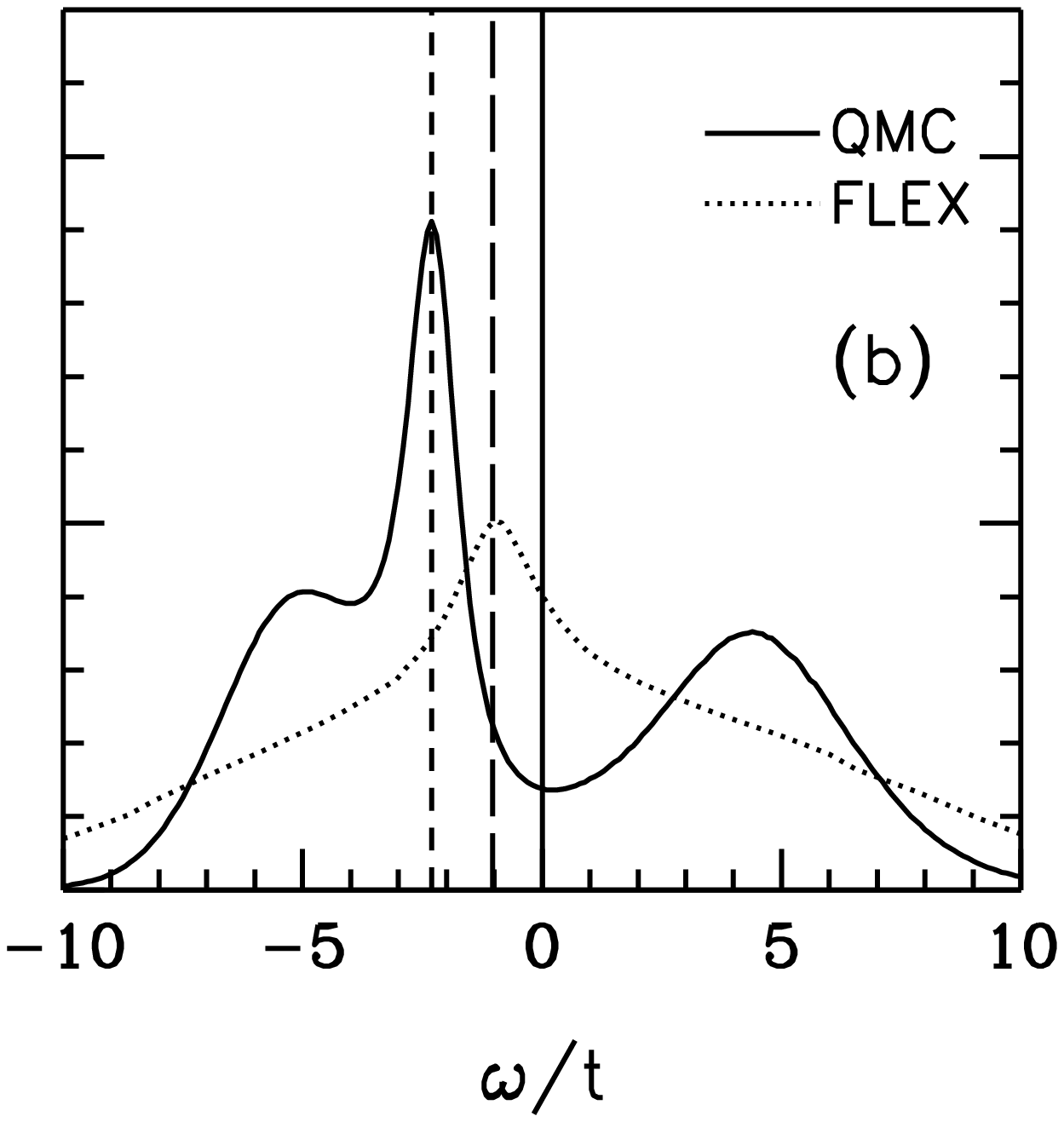}
\epsfysize=6cm \epsffile[243 164 817 598]{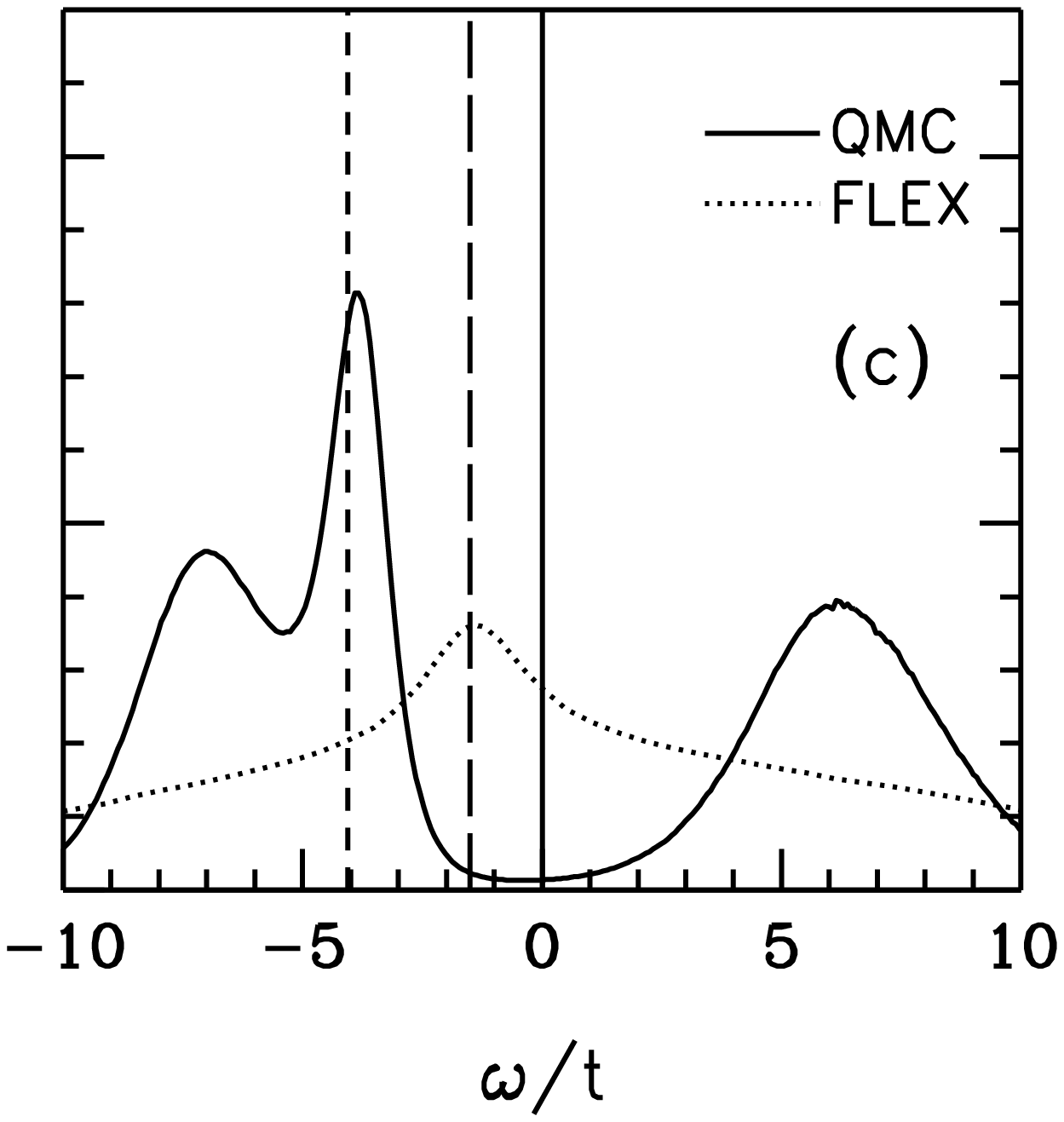}}
\caption{
Comparison of the FLEX and the QMC results on $N(\omega)$ versus 
$\omega$ at $\langle n\rangle =0.875$ 
for (a) $U=4t$, (b) $U=8t$, and (c) $U=12t$.
In (a) and (b) $T=0.33t$ was used and in (c) $T=0.5t$.
Here, the vertical long-dashed and 
the short-dashed lines denote the chemical potential
for the FLEX and the QMC calculations, respectively.
}
\label{8.2}
\end{figure}

\begin{figure}
\centering
\iffigure
\epsfysize=8cm
\epsffile[100 150 550 610]{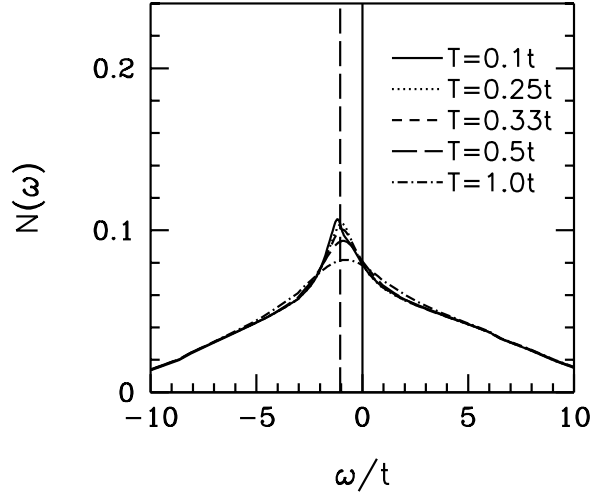}
\fi
\caption{
Temperature evolution of $N(\omega)$ versus $\omega$ 
within the FLEX approximation for $U=8t$ and 
$\langle n\rangle=0.875$.
Here, the vertical long-dashed line denotes the chemical potential.
}
\label{8.3}
\end{figure}

\begin{figure}[ht]
\centering
\iffigure
\epsfysize=8cm
\epsffile[100 150 550 610]{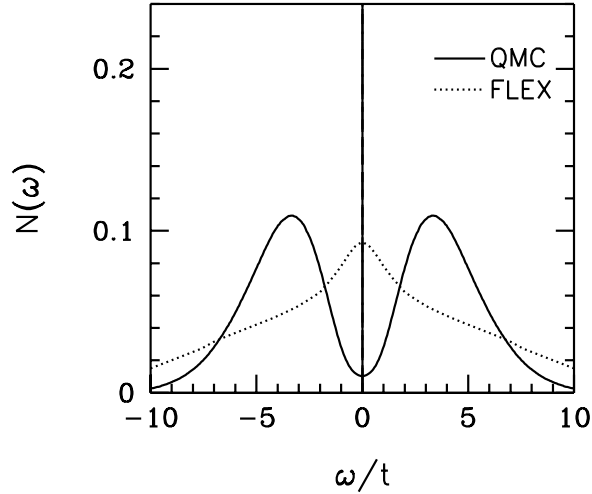}
\fi
\caption{
Comparison of the FLEX and the QMC results on $N(\omega)$ versus 
$\omega$ at half-filling 
for $U=8t$ and $T=0.5t$.
}
\label{8.4}
\end{figure}

Next, the FLEX results on the $d_{x^2-y^2}$-wave eigenvalue 
$\lambda_d$ of the Bethe-Salpeter equation in the 
particle-particle channel will be shown.
In Fig.~8.5, 
$\lambda_d$ is plotted as a function of $T/t$ for 
$U/t=4$ and 8 in the temperature regime where 
the QMC data exist.
Here, it is seen that as $U/t$ increases from 4 to 8, 
$\lambda_d$ changes by a small amount. 
For $U=4t$, the FLEX results on $\lambda_d$ versus $T/t$ are in good 
agreement with the QMC data. 
However, for $U=8t$ the FLEX approximation underestimates 
$\lambda_d$ by about a factor of two.
This is the major difference between the QMC and 
the FLEX results on $\lambda_d$.
Within FLEX the effective 
pairing interaction is also attractive in the odd-frequency 
$s$ and $p$-wave channels,
in addition to the even-frequency $d$-wave channel.

\begin{figure}[ht]
\centering
\iffigure
\epsfysize=8cm
\epsffile[100 150 550 610]{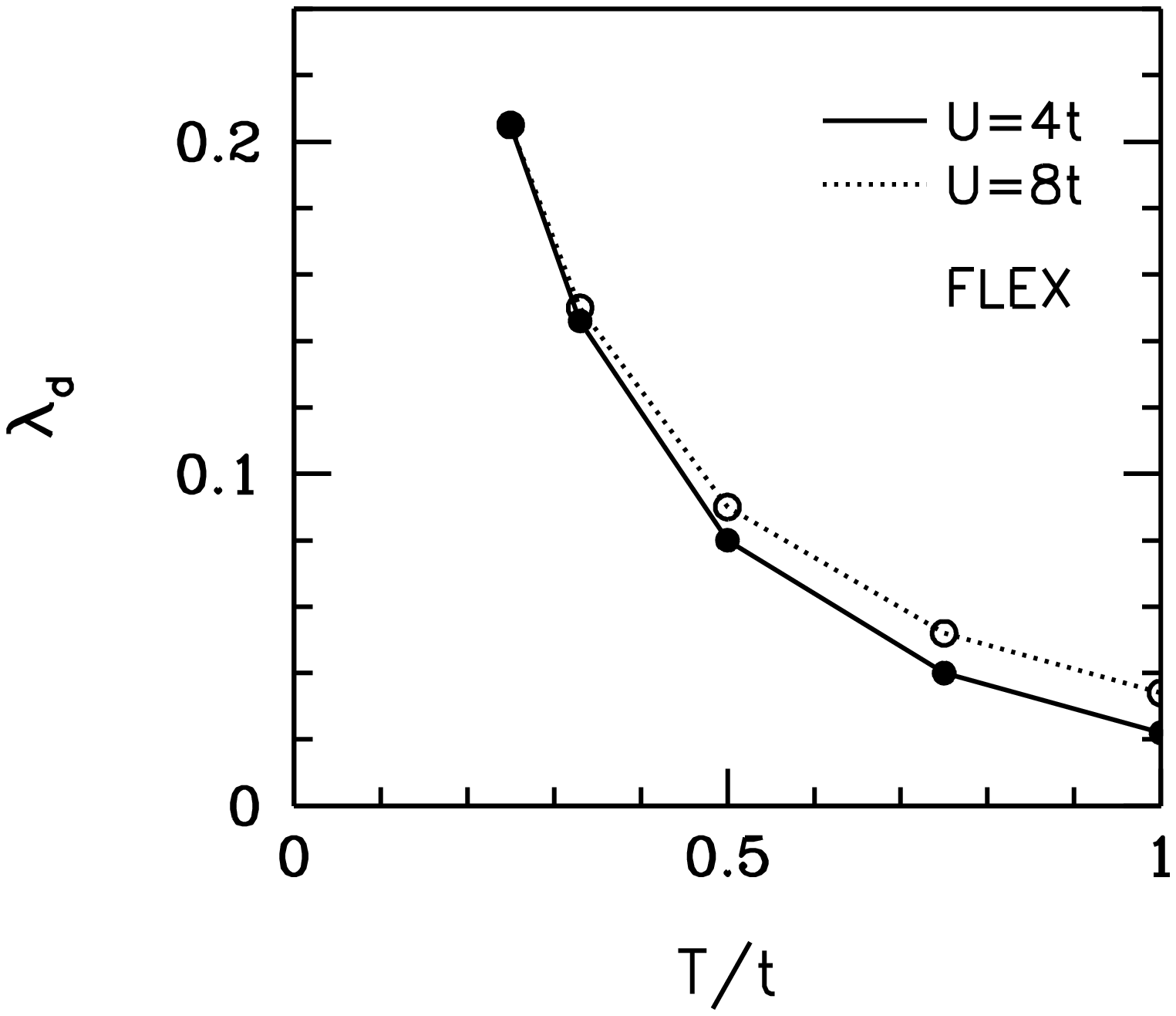}
\fi
\caption{
Temperature dependence of the $d_{x^2-y^2}$-wave 
eigenvalue within the FLEX approximation 
at $\langle n\rangle=0.875$ for $U=4t$ and $8t$.
}
\label{8.5}
\end{figure}

Comparing Fig.~5.3 with Fig.~8.3, one sees that 
the density of states at the Fermi level is  also underestimated 
by about a factor of two within FLEX.
From the simple spin-fluctuation exchange form of 
Eq.~(2.1), one expects that the $d_{x^2-y^2}$-wave 
pairing interaction increases with $U/t$.
As discussed in Section 6.2, 
the QMC data also shows that the reducible vertex $\Gamma_s$ 
gets enhanced as $U/t$ is increased from 4 to 8, even though the 
irreducible vertex $\Gamma_{Is}$ could not be obtained for 
this value of $U/t$.
Hence, it must be that the FLEX approach underestimates the 
amount of the single-particle spectral weight at the Fermi 
level, and this causes the $d_{x^2-y^2}$-wave 
eigenvalue to be smaller with respect to the QMC value.

Here, it is seen that the correlated metallic band 
which forms upon doping the AF Mott-Hubbard insulator 
is not obtained within FLEX,
which appears to cause why the FLEX 
underestimates $\lambda_d$ for $U=8t$.
Whether similar effects could take place when 
a CDW insulator is doped is an interesting problem.
 
\begin{figure}[ht]
\centering
\iffigure
\epsfysize=8cm
\epsffile[100 150 550 610]{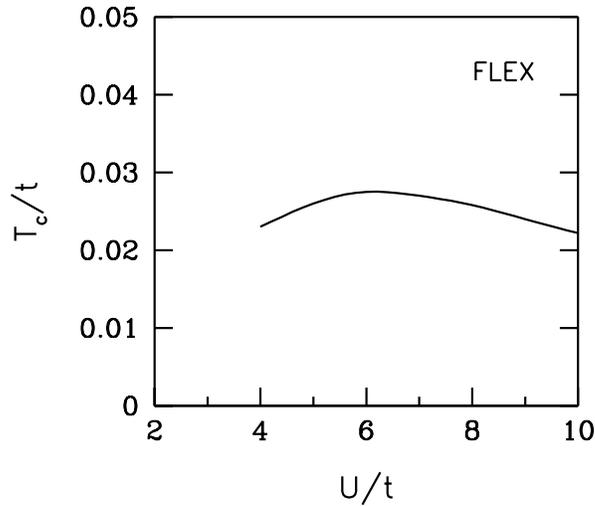}
\fi
\caption{
Superconducting transition temperature $T_c$ versus 
$U/t$ within the FLEX approximation for $\langle n\rangle=0.875$.
From [Pao and Bickers 1994].
}
\label{8.6}
\end{figure}

Above, it has been seen that within FLEX $\lambda_d$ has a weak 
dependence on $U/t$ at temperatures between $0.25t$ and $1.0t$.
Related to this is Fig.~8.6 which shows the transition
temperature $T_c$ obtained within FLEX as a function of $U/t$.
Here, it is seen that $T_c$ has a broad peak at $U\sim 6t$.
In this section, 
it has been seen that the FLEX 
underestimates the strength of the $d_{x^2-y^2}$-wave pairing 
when $U$ is of order $8t$.
An important question raised by these comparisons is 
whether for $U\sim 8t$ 
the $T_c$ should be higher than $0.025t$.
There is a chance that the maximum possible $T_c$ in the 
2D Hubbard model is higher than the FLEX estimate.
However, note that these values for $T_c$ are 
just estimates for a possible Kosterlitz-Thouless transition. 
In fact, 
the ground state of the doped 2D Hubbard model is not known, 
and it is currently beyond the reach 
of the exact many-body calculations.

\subsection{Other Monte Carlo results on the 2D Hubbard model}

The QMC data presented in the previous sections
were obtained with the determinantal 
Monte Carlo technique [White {\it et al.} 1989b].
Variational and projector Monte Carlo algorithms 
were also used for studying the 2D Hubbard model.
However,
these approaches have arrived at different conclusions
about the possibility of the $d_{x^2-y^2}$-wave pairing.
In this section, some of these calculations will be reviewed briefly. 

In order to see whether superconducting long-range 
order develops in the ground state of the doped 2D Hubbard model, 
zero temperature projector Monte Carlo calculations 
have been carried out 
[Imada 1991, Furukawa and Imada 1992].
In these calculations, 
an optimised initial state was used 
as a guiding function, mitigating the sign problem 
but making the calculation variational rather than exact.
With this technique, 
the equal-time pair-field correlation functions 
were calculated for the $s$, extended $s$ and 
the $d_{x^2-y^2}$-wave singlet pairing channels.
However, no size dependence was found in the data,
which implies that there is no long-range superconducting order
with these symmetries in the ground state.

The constrained-path Monte Carlo (CPMC) 
algorithm developed in Ref.~[Zhang {\it et al.} 1995]
is a variational method which 
starts from a trial wave function 
$| \Psi_T \rangle$ and uses the same $| \Psi_T\rangle$
as a constraining wave function as the simulation 
is carried out in the space of the Slater determinants.
In calculations on up to $16\times 16$ lattices,
free-electron and unrestricted Hartree-Fock wave functions 
were used as $|\Psi_T \rangle$, 
however no $d_{x^2-y^2}$-wave superconducting long-range 
order was found [Zhang {\it et al.} 1997].
The CPMC method has been extended by using as 
the constraining function a $d_{x^2-y^2}$-wave 
BCS wave function [Guerrero {\it et al.} 1999].
Superconducting long-range order was not found in this case either.
An important issue is how well the constraining function describes the 
correlations in the Hubbard model.
In fact, a comparison of the CPMC and the DMRG results, which will
be discussed later in Section~8.5.1, has been carried out 
for the magnetic and the density correlations in the 
$12 \times 3$ Hubbard model [Bonca {\it et al.} 2000].
This comparison has found that the CPMC results depend sensitively 
on the constraining wave function especially for large $U/t \sim 8$.
In this respect, it would be useful to test how well the CPMC 
method describes the $d_{x^2-y^2}$-like 
superconducting correlations found 
in the 2-leg Hubbard ladder. 
For instance, it is known from the DMRG calculations that 
in the 2-leg Hubbard ladder for $U/t=8$,
$t_{\perp}/t \sim 1.5$ and near half-filling the pair-field
correlation function $D(\ell)$ decays as $\ell^{-\theta}$,
where $\theta \alt 1.0$.
It would be useful to see 
whether the CPMC method describes the 
power-law pairing correlations
which exist in the 2-leg Hubbard ladder. 

Another variational Monte Carlo approach uses 
as the trial wave function 
[Nakanishi {\it et al.} 1997, Yamaji {\it et al.} 1998]
\begin{equation}
\label{PsiT}
|\Psi_T\rangle = P_N P_G |\Psi_{BCS}\rangle,
\end{equation}
where
\begin{equation}
|\Psi_{BCS}\rangle = \Pi_{\bf k} 
( u_{\bf k} + v_{\bf k} 
c^{\dagger}_{{\bf k}\uparrow}c^{\dagger}_{{-\bf k}\downarrow} )
|0 \rangle 
\end{equation}
is the usual $d_{x^2-y^2}$-wave BCS wave function, 
\begin{equation}
P_G = \Pi_i [ 1-(1-g) n_{i\uparrow}n_{i\downarrow} ],
\end{equation}
is the Gutzwiller projection operator, 
and $P_N$ is an operator which projects out states 
with electron number $N$.
In this method, the magnitude of the 
$d_{x^2-y^2}$-wave gap function entering the usual BCS coefficients
$u_{\bf k}$ and $v_{\bf k}$, the parameter $g$, and the 
chemical potential $\mu$ are used as variational parameters,
and their values are determined by minimising the ground 
state energy with a Monte Carlo simulation.
With this method it is found that,
for fillings between 0.84 and 0.68, 
the $d_{x^2-y^2}$-wave state has the lowest energy, 
and for $\langle n \rangle \agt 0.84$ an SDW state is 
favored.
In addition, 
when a next-near-neighbor hopping term $t'$ is turned on, 
the superconducting condensation energy gets enhanced.
For instance, for $U=8t$ this enhancement is maximum 
when $t' \approx -0.1t$.
It should be noted that this method was also applied
to the 2-leg Hubbard ladder [Koike {\it et al.} 2000].
In this case, 
the gap function was assumed to take momentum-independent
values $\Delta_1$ and $\Delta_2$ on the bonding and 
the antibonding bands.
Treating $\Delta_1$ and $\Delta_2$ as variational parameters,
it was found that 
a superconducting state with $d_{x^2-y^2}$-like symmetry 
is stabilized when the bottom of the antibonding band 
is near the Fermi level,
which is in agreement with the exact-diagonalization 
[Yamaji and Shimoi 1994] 
and the DMRG [Noack {\it et al.} 1995 and 1997] calculations.

Next, a projector Monte Carlo approach 
[Husslein {\it et al.} 1996]
will be discussed where the ground state wave function 
of the 2D Hubbard model is estimated from 
\begin{equation}
|\Psi_g \rangle = e^{\theta H} |\Psi_0 \rangle .
\end{equation}
Here, $|\Psi_0 \rangle$ is the ground state of the 
noninteracting electrons and 
the parameter $\theta$ is taken to be 
about 8 on a $12 \times 12$ lattice.
This approach has been used to calculate the 
$d_{x^2-y^2}$-wave pair-field correlation function 
in the weak-coupling regime, $0.5t < U < 3t$,
near half-filling.
It is found that 
the system has long-range $d_{x^2-y^2}$-wave pairing 
correlations for negative values of $t'$, 
for instance, at $t'= -0.3t$ for $U=2t$.

In this section, 
it was seen that various approaches arrive at different 
conclusions about $d_{x^2-y^2}$-wave pairing 
in the 2D Hubbard model.
This emphasises the importance of the choice of 
the trial wave functions and points out 
at the need for carrying out calculations 
without introducing approximations.

\subsection{Dynamical cluster approximation and 
RG calculations for $d_{x^2-y^2}$
pairing in the 2D Hubbard model}

Recently, 
the dynamical cluster approximation 
[Maier {\it et al.} 2000] 
and the one-loop renormalization-group method 
employing a 2D Fermi surface
[Zanchi and Schulz 2000, Halboth and Metzner 2000, 
Honerkamp {\it et al.} 2001] 
were used for studying $d_{x^2-y^2}$ pairing in the 2D Hubbard model.
In this section, 
the findings of these studies will be discussed briefly. 

When the infinite dimensions limit of the Hubbard model 
is taken with proper scaling, 
the many-body problem becomes local 
[Metzner and Vollhardt 1989, M\"uller-Hartman 1989], 
and it can be mapped to an Anderson impurity problem, 
which can then be solved with various many-body techniques 
[Pruschke {\it et al.} 1995, Georges {\it et al.} 1996].
This is called the dynamical mean-field approximation (DMFA). 
The DMFA is interesting because it is a strong coupling technique.
For instance, 
for large $U/t$ and off of half-filling, 
the DMFA yields a narrow peak in 
$N(\omega)$ near the Fermi level [Jarrell 1992], 
which is also seen in the 2D QMC data.
However, 
the DMFA does not incorporate the non-local correlations, 
and hence it is not 
possible to study $d_{x^2-y^2}$ pairing with it. 
The DCA incorporates the non-local corrections to DMFA by 
mapping the lattice problem onto an embedded cluster of size
$N_c$, rather than onto an impurity problem. 
In DCA calculations, 
the dynamical correlation length is restricted to 
$L=\sqrt{N_c}$, 
and the DCA would become exact for 
$N_c \rightarrow \infty$, 
while it reduces to DMFA for $N_c=1$.
With this approach, 
the mean-field $d_{x^2-y^2}$ superconducting $T_c$'s 
were calculated for $N_c=4$ 
by using the non-crossing approximation to 
solve the cluster problem [Maier {\it et al.} 2000].
For $U=12t$, 
it is found that  $T_c$ has the maximum value of $\approx 0.05t$ 
for about 20\% doping. 
It is also found that $T_c$ increases for positive values of 
the second-nearest-neighbour hopping $t'$, 
and decreases for negative values of $t'$.
This result agrees with the DMRG calculations 
on the $t$-$J$ model which find $d_{x^2-y^2}$ pairing 
for $t'>0$ 
[White and Scalapino 1998a].
With DCA, 
the low doping regime of the 2D Hubbard model was also studied, 
and these results will be discussed in Section 8.4.

After the discovery of the high $T_c$ cuprates, 
the one-loop RG approach was extended from 
1D to 2D in order to study this problem
[Dzyaloshinskii 1987, Schulz 1987, Lederer {\it et al.} 1987].
The RG calculations are interesting,
because with this technique the particle-particle 
and the particle-hole channels are treated on equal footing.
These studies focused on the scattering processes between 
the Fermi surface regions near the van Hove singularities.
For the 2D Hubbard model, 
the AF SDW state at half-filling and 
the $d_{x^2-y^2}$ superconducting state induced 
by AF fluctuations away from half-filling were found
[Schulz 1987, Lederer {\it et al.} 1987].
The scatterings involving the full 2D Fermi surface were taken 
into account with the work of Zanchi and Schulz, 
who studied the RG flows using 
a 32-patch discretization of the 2D Fermi surface for $t'=0$ 
[Zanchi and Schulz 2000].
These calculations found two different regimes, 
one dominated by the AF correlations 
and the other by the $d_{x^2-y^2}$ pairing. 
The 2D RG calculations were also extended to the $t' \neq 0$ case
[Halboth and Metzner 2000].
In the calculations for $t'\neq 0$ by Honerkamp {\it et al.}, 
an intermediate regime is found between 
the two regimes dominated by the 
AF correlations and $d_{x^2-y^2}$ pairing
[Honerkamp {\it et al.} 2001].
In this intermediate regime, 
there are competing AF and $d_{x^2-y^2}$ 
superconducting correlations.
This regime exists only for sizeable $t'<0$ and 
it exhibits features which are similar to those seen 
in the pseudogap regime of the cuprates,
which will be discussed in the following section.
Honerkamp {\it et al.} extracted a temperature versus 
doping phase diagram from the 
2D RG flows for the 2D Hubbard model with $t'<0$, 
which is similar to that of the cuprates. 

In spite of these interesting results 
it has to be kept in mind that 
the dynamical correlation length in DCA 
is cut-off by the size of the cluster 
and the one-loop RG is a weak-coupling approach.
In addition, 
in RG calculations the single-particle self-energy corrections 
are not included at the one-loop level.
In Section 8.1, 
it was seen that the single-particle self-energy corrections
could play an important role 
in determining the strength of pairing. 
It should also be noted that, 
while the RG finds that the regime with $t'<0$ is favored, 
in the DCA calculations the mean-field $T_c$'s 
are higher for $t'>0$.
These are some of the reasons for why exact results are necessary 
in order to reach a final conclusion. 

\subsection{Low-doping regime of the 2D Hubbard model}

In the normal state of the underdoped cuprates, 
a pseudogap is seen in the excitation spectrum 
below a temperature $T^*$ which depends on the doping.
The nature of the pseudogap is an important problem 
in this field and there exist a wide range 
of ideas about its origin and its implications
[M$^2$S-HTSC VI proceedings 2000].
In this section, 
various calculations which have been carried out for 
exploring whether there is a pseudogap regime 
in the 2D Hubbard model will be discussed. 
In Ref.~[Jaklic and Prelovsek 2000], 
a review of the numerical calculations 
on the $t$-$J$ model at finite temperatures is given 
and the results of these calculations are compared with the 
anomalous normal-state properties of the cuprates 
including the pseudogap.

The normal state pseudogap, 
while exhibiting dependence on the material properties, 
is seen in the uniform magnetic susceptibility,
the low-frequency optical conductivity, the ARPES spectrum 
and various other measurements of the electronic properties. 
An interesting feature of the normal state 
pseudogap observed in the ARPES 
spectrum is that it is anisotropic on the Fermi surface: 
the pseudogap has maximum amplitude near 
the $(\pi,0)$ and $(0,\pi)$ points of the Brillouin zone 
and it is minimum for wave vectors near $(\pi/2,\pi/2)$
[Ding {\it et al.} 1996, Ronning {\it et al.} 1998].

Kampf and Schrieffer first discussed 
the possibility of a pseudogap within 
the spin-bag approach for the 2D Hubbard model 
[Kampf and Schrieffer 1990a and 1990b]. 
They showed how a pseudogap could develop because of the 
AF fluctuations already at the one-loop level 
for the single-particle self-energy 
as the AF instability is approached. 
In the FLEX calculations, 
it was also found that a pseudogap opens in the 
density of states when the strength 
of the short-range AF fluctuations
increases [Dahm and Tewordt 1995].

In spite of the sign problem, 
there are QMC data on 
$A({\bf p},\omega)$ at temperatures as low as $0.25t$ 
in the underdoped regime
[Preuss {\it et al.} 1997].
These data show indications that a pseudogap opens 
in $A({\bf p},\omega)$ as $T$ decreases. 
In particular, 
for $\langle n\rangle \simeq 0.95$, 
it is seen that spectral weight 
above the Fermi level 
for ${\bf p}$ between $(\pi,0)$ and $(\pi,\pi)$ 
decreases gradually as $T$ decreases from $0.5t$ to $0.25t$. 
In these calculations, 
the pseudogap has been attributed to the AF spin correlations, 
which are becoming larger than a lattice spacing 
for $T < 0.3t$ in the underdoped regime. 
However, 
in order to be able to make direct comparisons with 
the experimental data on the pseudogap, 
it would be necessary to reach temperatures below $0.1t$. 

In recent one-loop RG calculations 
for the Hubbard model with $t'<0$,
a saddle-point regime is found between 
the AF ordered and the $d_{x^2-y^2}$ superconducting regimes 
[Honerkamp {\it et al.} 2001].
This regime exhibits features similar 
to those seen in the pseudogap regime of the cuprates.
Here, 
the uniform magnetic and charge susceptibilities 
flow to zero because of the 
pairing and the umklapp-scattering processes, 
respectively. 
The possible existence of an umklapp-gapped 
spin-liquid phase was suggested in earlier RG calculations 
where a two-patch discretization 
of the Brillouin zone was used  
[Furukawa {\it et al.} 1998].
The similarity of this regime to the spin-gapped phase 
in the 2-leg Hubbard ladder was also noted. 
Because of the umklapp scatterings, in this regime, 
the sections of the Fermi surface which are 
near the $(\pi,0)$ and $(0,\pi)$ points 
are truncated 
while at wave vectors near $(\pi/2,\pi/2)$ 
gapless single-particle excitations remain. 
Clearly, 
these results are useful for interpreting a number of 
experimental data on the underdoped cuprates 
and, in particular, the ARPES data.

The DCA calculations have also found interesting results 
about this subject
[Huscroft {\it et al.} 2001].
Using this method, 
the low-doping and the low-temperature regime 
of the 2D Hubbard model was studied.
Here, 
QMC simulations were used to solve 
the embedded lattice problem 
with $N_c=8$.
It was found that, as $T$ is lowered, 
an anisotropic pseudogap develops in 
$A({\bf p},\omega)$ and, in addition, 
the uniform magnetic susceptibility $\chi(T)$ gets suppressed.
In particular, 
for $\langle n\rangle =0.95$, $U=6t$ and $T=0.06t$,
a pseudogap is found in $A({\bf p},\omega)$ 
for ${\bf p}=(\pi,0)$ and $(0,\pi)$ 
and not for $(\pi/2,\pi/2)$. 
For these values of $U/t$ and $\langle n\rangle$, 
a mean-field $d_{x^2-y^2}$ superconducting transition temperature of 
$T_c\simeq 0.04t$ is obtained with the DCA. 
In addition, 
a sharp drop in $\chi(T)$ is observed near the temperature 
where the pseudogap in $A({\bf p},\omega)$ opens. 
In these calculations, 
the downturn in $\chi(T)$ has been attributed to the development 
of the AF correlations. 

These results support the idea that perhaps
the normal state pseudogap seen in the cuprates 
could be understood within a 2D Hubbard framework.
However, 
there still exist a wide range of ideas 
about the origin of the pseudogap,
and it is one of the important unresolved issues 
in this field.

\subsection{Comparisons with the $t$-$J$ model} 

In the large $U/t$ limit, the Hubbard model reduces to 
[Hirsch 1985]
\begin{eqnarray}
\label{tJ}
H = -t \sum_{\langle i,j\rangle, \sigma}
( c^{\dagger}_{i,\sigma} c_{j,\sigma} + 
c^{\dagger}_{j,\sigma} c_{i,\sigma} )
+ J \sum_{\langle i,j\rangle} 
( {\bf S_i}\cdot {\bf S_j} - {1\over 4} n_in_j ) \nonumber \\ 
-{J\over 4} \sum_{i,\sigma} 
\sum_{\delta\neq\delta'} 
(   c^{\dagger}_{i+\delta,\sigma} c^{\dagger}_{i,-\sigma} 
c_{i,-\sigma} c_{i+\delta',\sigma} 
-
c^{\dagger}_{i+\delta,-\sigma} c^{\dagger}_{i,\sigma} 
c_{i,-\sigma} c_{i+\delta',\sigma}   ),
\end{eqnarray}
where $J=4t^2/U$, and the double occupancy of a site is not 
allowed.
In Eq.~(\ref{tJ}), 
$i+\delta$ and $i+\delta'$ sum over the nearest neighbors
of site $i$.
The last sum in this expression, which involves 
operators acting at three different sites, is dropped 
and what is remaining is called the $t$-$J$ model.
Clearly, the $t$-$J$ and the Hubbard models
have differences.
The numerical studies of the $t$-$J$ model 
have produced valuable information about the magnetic, density
and the superconducting correlations in this system.
Reviews of these studies can be found in 
Refs.~[Dagotto 1994, Jaklic and Prelovsek 2000]. 
In this section, 
the density and the pairing correlations 
in the $t$-$J$ and the Hubbard models will be compared.

In Section 8.5.1, the results on the phase separation and 
the density correlations in the 2D $t$-$J$ model obtained 
with various numerical techniques will be discussed.
These data will be compared with the CPMC and the DMRG
calculations for the $12\times 3$ Hubbard lattice.
In Section 8.5.2, the nature of the pairing correlations seen 
in the 2D $t$-$J$ model will be compared with those 
in the 2D Hubbard model.
The results on the 2-leg $t$-$J$ and Hubbard ladders will be 
compared in Section 8.5.3.

\subsubsection{Comparisons with the density correlations in the 
2D $t$-$J$ model} 

The QMC simulations [Moreo and Scalapino 1991] and 
the exact-diagonalization [Dagotto {\it et al.} 1992b]
calculations on the 2D Hubbard model
did not find any indication of phase separation of the system 
into hole-rich and hole-poor regions.
The 2-leg Hubbard ladder does not show any evidence 
for phase separation either
[Noack {\it et al.} 1994].
In contrast to the Hubbard model, 
the 2D $t$-$J$ model phase 
separates at any electron filling for sufficiently large
values of the interaction strength $J/t$.
Various techniques including the variational
and the exact-diagonalization calculations [Emery {\it et al.} 1990],
the high-temperature series expansions [Putikka {\it et al.} 1992],
and the exact-diagonalization 
calculations [Dagotto 1994]
were used for studying phase separation in the 
$t$-$J$ model. 
Recently, 
the Green's function Monte Carlo (GFMC) 
[Hellberg and Manousakis 1997, 1999 and 2000, 
Calandra {\it et al.} 1998]
and 
the DMRG [White and Scalapino 1998a, 
Rommer {\it et al.} 2000]
were used for obtaining the phase-separation boundary 
in the $t$-$J$ model.
Currently, 
where the true phase-separation boundary lies is controversial,
since the approaches noted above arrive at different conclusions. 

\begin{figure}
\centering
\iffigure
\epsfysize=8cm
\epsffile[100 150 550 610]{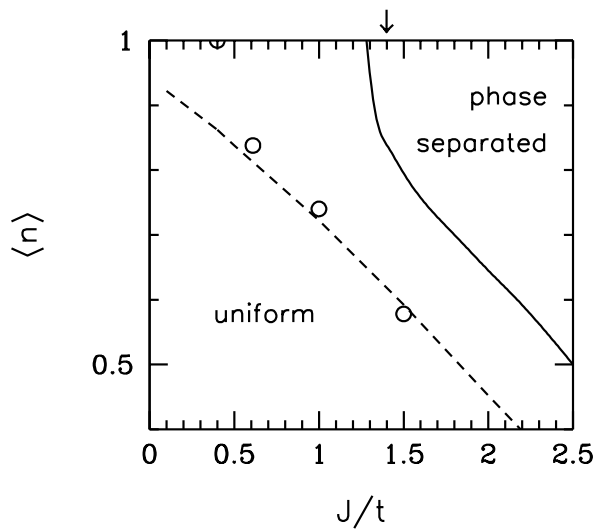}
\fi
\caption{
Sketch of the phase-separation boundary for the 2D $t$-$J$
model obtained from various calculations.
The solid line represents the phase-separation boundary 
obtained by the high-temperature series expansion 
[Putikka {\it et al.} 1992].
The open circles represent the results 
of the GFMC calculations 
from Ref.~[Calandra {\it et al.} 1998].
The dashed line is from Ref.~[Hellberg and Manousakis 1997],
which was also obtained by the GFMC.
The arrow indicates the DMRG result 
for the critical value of 
$J/t$ for phase separation on a six-leg $t$-$J$ ladder
in the limit of zero doping
[Rommer {\it et al.} 2000].
}
\label{8.7}
\end{figure}

Emery {\it et al.} suggested that the 2D $t$-$J$ model phase separates 
at all interaction strengths
[Emery {\it et al.} 1990].
Subsequent calculations found that the phase separation 
occurs only for $J \agt t$, 
as indicated by the solid curve in Fig.~8.7
[Putikka {\it et al.} 1992].
However, recently, 
the GFMC calculations [Calandra {\it et al.} 1998] 
suggested that the phase-separation boundary 
might occur at lower $J/t$ values 
near half-filling. 
This is indicated by the empty circles in Fig.~8.7.
In these calculations, 
next to the phase-separation boundary lies the 
regime where the doped holes form 
$d_{x^2-y^2}$-wave pairs. 
The recent GFMC calculations by Hellberg and Manousakis,
on the other hand,
find that the critical value of $J/t$ for phase separation 
extrapolates to zero at low dopings 
as shown by the dashed curve in Fig.~8.7
[Hellberg and Manousakis 1997, 1999 and 2000].
In these calculations, the $t$-$J$ model phase separates 
in the parameter regime appropriate for the cuprates. 
For instance, at 15\% hole doping the critical value of 
$J/t$ is about 0.4.
The DMRG calculations find that, 
in the physically relevant regime,
the ground state of the $t$-$J$ model
is striped and not phase separated
[White and Scalapino 1998a, 2000].
The DMRG calculations are carried out on large lattices
with periodic boundary conditions at the long edges and 
open boundary conditions at the short edges.
In these calculations the phase separation occurs for 
$J > t$.
For instance, in the six-leg Hubbard ladder, 
the critical value of $J/t$ is $\sim 1.4$
as the doping approaches zero,
which is indicated by the arrow in Fig.~8.7.

The stripes observed in the DMRG calculations 
are a domain wall of holes 
across which there is a $\pi$-phase shift in the AF background.
According to the DMRG calculations, 
the stripe formation represents an instability of the 
system where pairs of bound holes 
combine to form domain walls.
In particular, through the stripe correlations the system
reduces the frustration of the AF background and lowers
the kinetic energy of the holes.
However, the issue of 
the phase separation versus the stripe formation 
in the 2D $t$-$J$ model remains controversial.
The role of the open boundary conditions in producing 
the static stripes was questioned
[Hellberg and Manousakis 2000],
and the need to have more than one pair of holes in order to 
see the stripe correlations was noted
[White and Scalapino 2000].
The calculation of the dynamical spin and charge
susceptibilities for sufficiently large $t$-$J$ systems would resolve 
these issues.

The possibility of stripe formation in strongly 
correlated systems were noted in mean-field calculations of the 2D 
Hubbard model soon after the discovery of the cuprates
[Zaanen and Gunnarson 1989, Poilblanc and Rice 1989, 
Schulz 1989, Machida 1989].
There has been a surge of interest in this field 
since the observation of static stripe ordering 
in the neutron scattering experiments on 
La$_{1.6-x}$Nd$_{0.4}$Sr$_x$CuO$_4$ [Tranquada {\it et al.} 1995].
Beyond mean-field, the nature of 
the stripe correlations in the Hubbard model 
were studied with the DMRG and 
the CPMC techniques for the 3-leg Hubbard ladder 
[Bonca {\it et al.} 2000].
The findings of these calculations are important and 
they will be briefly described here.

The CPMC is an approximate method which projects 
the ground state from a trial state 
as has been discussed in Section~8.2.
Both the DMRG and the CPMC calculations find that for 
$U \agt 6t$ there are static stripes in the ground state of
a $12 \times 3$ Hubbard lattice doped with six holes 
when open boundary conditions are used.
In Figures 8.8 and 8.9, the rung magnetization 
\begin{equation}
S^z(i) = \sum_{j=1}^3 \langle S^z(i,j) \rangle
\end{equation}
and the rung hole density
\begin{equation}
\rho(i) = \sum_{j=1}^3 \langle \rho(i,j) \rangle
\end{equation}
are plotted as a function of the rung location $i$
at $U=8t$.
Here, $S^z(i,j)$ and $\rho(i,j)$ are the spin and hole-density 
operators at the $j$'th site of the $i$'th rung. 
In Fig.~8.8, it is seen that 
between rungs 3 and 4 the spins are ferromagnetically coupled,
causing a $\pi$-phase shift, and the magnetisation density is small.
The same occurs at rungs 9 and 10.
In Fig.~8.9, it is seen that at these sites the holes form domain walls.
In these figures the results of the unrestricted Hartree-Fock (UHF)
calculations are also shown.
The UHF approximation produces results in agreement with the 
CPMC and the DMRG calculations when the reduced value of 
$U=3t$ is used.
This behavior is similar in spirit to using a reduced 
Coulomb repulsion within RPA for the 
spin susceptibility as discussed in Section~3.3.
It is important to note that the structure of the stripes seen in these
calculations for $U \agt 6t$ is quite similar to those found in the 
three-leg $t$-$J$ ladder [White and Scalapino 1998b].
However, for $U< 6t$ these features disappear and no evidence 
is found for static stripes in the ground state.
In this regime, only some evidence for low-lying states 
with stripes are found.
At weak $U/t$, the density of virtual holes due 
to the double occupancy increases,
and, it has been noted that this might 
weaken the stripe correlations.
At this point, 
it is of interest to explore whether there is a 
relation between the stripe patterns 
which are observed within the 
presence of the open boundary conditions and 
any possible "4${\bf k}_F$" CDW correlations.
It should also be noted that the CPMC method does not 
show evidence for stripe correlations 
even at large $U/t$ when the free-electron 
trial wave function is used instead of the UHF wave function.
Hence, caution is necessary in choosing a trial wave function 
which is suitable for the ground state. 

\begin{figure}
\centering
\iffigure
\epsfig{file=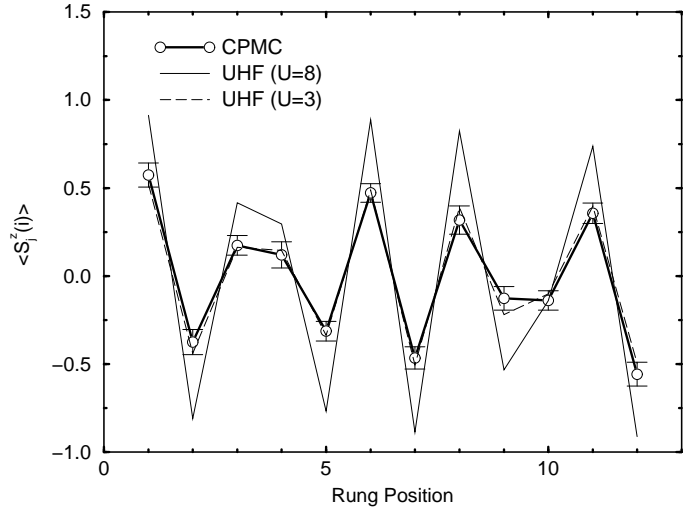,height=8cm,angle=-90}
\fi
\caption{
Rung-magnetisation versus the 
rung location for the $12 \times 3$ Hubbard lattice doped
with six holes. 
From [Bonca {\it et al.} 2000]. 
}
\label{8.8}
\end{figure}

\begin{figure}
\centering
\iffigure
\epsfig{file=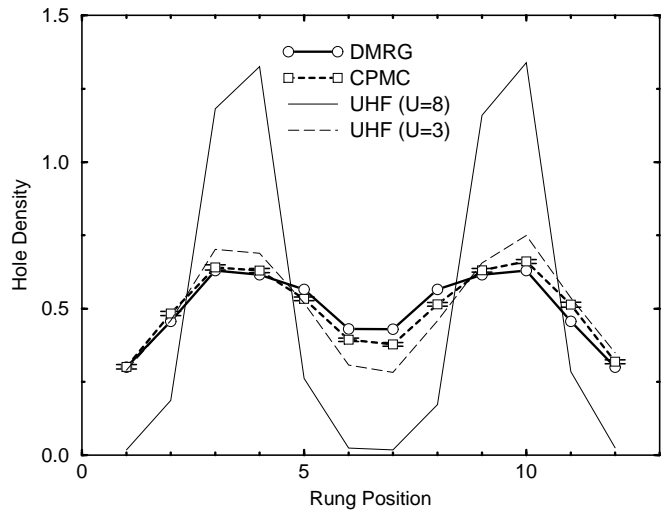,height=8cm,angle=-90}
\fi
\caption{
Rung-hole density versus the 
rung location for the $12 \times 3$ Hubbard lattice doped
with six holes.
From [Bonca {\it et al.} 2000]. 
}
\label{8.9}
\end{figure}

These results were found using open boundary conditions.
The DMRG calculations are not carried out with periodic 
boundary conditions, but the CPMC calculations can be.
When carried out with the periodic boundary 
conditions on square lattices, 
the CPMC calculations do not find 
stripes in the hole and spin densities and in the hole and 
spin correlation functions.
This indicates that the static stripe pattern seen in the 
ground state results from the open boundary conditions breaking 
the translational symmetry.
With the periodic boundary conditions, there might be low-lying 
states with stripe patterns but it might be difficult to detect them 
numerically.
Hence, whether static stripes exist in the ground state 
of the 2D Hubbard model with periodic boundary conditions 
remains as an important open question.
Here and in Section~4, it has been seen that there are similarities
in the density correlations of the $t$-$J$ model and the Hubbard 
model with large $U/t$.
But, there are differences as well.

\subsubsection{Comparisons with the superconducting correlations in the 
2D $t$-$J$ model} 

When two-holes are doped into the half-filled $t$-$J$ model, 
they form a $d_{x^2-y^2}$-wave bound pair for interaction strengths
relevant to the high-$T_c$ cuprates,
$J/t\sim 0.35$ [Poilblanc 1993, Dagotto 1994].
However, much controversy exists over what happens when more 
than one pair of holes is doped.
The GFMC calculations by Calandra {\it et al.} find that they form
a $d_{x^2-y^2}$-wave superconducting ground state 
for $J/t \sim 0.35$ and dopings relevant to the 
high-$T_c$ cuprates [Calandra {\it et al.} 1998].
However,
the GFMC calculations by Hellberg and Manousakis 
find that this parameter regime lies right at the boundary for 
the phase separation of holes
[Hellberg and Manousakis 1997 and 1999].

The DMRG calculations find that in the same regime 
the ground state has static stripes, and the system does not 
have $d_{x^2-y^2}$-wave superconducting long-range order
[White and Scalapino 1998a, 1999].
The fact that the static stripes compete with the superconductivity 
is an important result of these calculations.
However, when the static stripe patterns are broken by a 
second-near-neighbor hopping $t' > 0$, it is seen that 
the system develops long-range 
$d_{x^2-y^2}$-wave superconducting order,
and in this regime the low-lying stripe
correlations coexist with the long-range superconducting order.

Both the $t$-$J$ and the Hubbard models exhibit
$d_{x^2-y^2}$-wave pairing correlations due 
to the AF spin fluctuations.
However, there are differences between these models, 
one of them being the no-double-occupancy constraint
in the $t$-$J$ model.
It is possible that this constraint is more favorable 
of the stripe correlations 
and, because of this, the 
one-band Hubbard or a three-band CuO$_2$ model
might have weaker tendency for stripe formation
than the $t$-$J$ model
[Daul {\it et al.} 2000, Jeckelmann {\it et al.} 1998].

Whether the doped 2D Hubbard model develops 
static stripe patterns in the ground state is 
an important issue.
In the previous section, we have seen from the QMC data that 
the doped 2D Hubbard model has $d_{x^2-y^2}$-wave  
pairing correlations which develop as the temperature is lowered 
and they are not weak correlations.
However, if indeed a static striped phase 
is more favored in the ground state 
then this could suppress the 
growth of the superconducting correlations at a temperature
lower than where the QMC simulations are carried out.
In this respect, the comparison of the CPMC and 
the DMRG calculations for the $12\times 3$ Hubbard 
lattice are useful
[Bonca {\it et al.} 2000].
However, further study is necessary before reaching conclusions 
about the ground state of the doped 2D Hubbard model.

The results shown here emphasize the interplay of the 
magnetic, density and the pairing correlations 
in the Hubbard and $t$-$J$ models. 
In the cuprates, 
the substitution of nonmagnetic impurities
gives useful information about the interplay of these
correlations in these materials. 
It is known that,
when substituted in place of planar Cu, 
nonmagnetic impurities strongly suppress the superconducting $T_c$ 
[Xiao {\it et al.} 1990] 
and locally enhance the AF correlations
[Mahajan {\it et al.} 1994 and 2000].
To the extend that a nonmagnetic impurity can be 
considered as a pure potential scatterer, 
these experimental data give information about 
the magnetic and pairing response of these materials 
when perturbed in the density channel.

\subsubsection{Comparisons with the 2-leg $t$-$J$ ladder} 

The Lanczos calculations on the 2-leg $t$-$J$
ladder showed that this system, 
which has a spin gap in the insulating state, 
can exhibit superconducting correlations upon doping
[Dagotto {\it et al.} 1992a].
The mean-field calculations noted the $d_{x^2-y^2}$-like 
structure of the superconducting pairs 
in the $t$-$J$ ladder [Gopalan {\it et al.} 1994].
The DMRG calculations on long $t$-$J$ ladders 
established that in the ground state there are power-law decaying 
superconducting correlations 
[Hayward {\it et al.} 1995].
In Ref.~[Dagotto and Rice 1996],
the properties of $n$-leg spin-$1/2$ ladders are reviewed 
and comparisons are made with various ladder compounds.

There are differences in the nature of the $d_{x^2-y^2}$-like
superconducting correlations seen in the $t$-$J$ and the Hubbard 
ladders.
 For instance, 
in the Hubbard ladder for $t_{\perp}/t \sim 1.0$, 
the superconducting correlations are weak,
and the pair-field correlation function 
$D(\ell)$ decays as $\ell^{-2}$, which is like the $U=0$ case
[Noack {\it et al.} 1994].
Only when the distribution of the single-particle spectral weight 
is such that it can make use of the momentum structure 
in $\Gamma_I$, does the system 
exhibit strong pairing correlations.
On the other hand, 
in the isotropic $t$-$J$ ladder, 
the system has strong pairing correlations.
For instance,
$D(\ell)$ decays approximately as $\ell^{-1}$
for $\langle n\rangle=0.8$ and $J/t=1$
[Hayward {\it et al.} 1995].
In addition, 
Schulz has shown that $D(\ell)$ decays as 
$\ell^{-1/2}$ in the limit $\langle n\rangle \rightarrow 1$
[Schulz 1999].
Hence, for isotropic hopping the $t$-$J$ ladder has 
stronger pairing correlations compared to the 
Hubbard ladder.
Probably, this is related to the fact that 
in the $t$-$J$ model the exchange term $J$,
which is an effective attractive interaction between the 
nearest-neighbor antiparallel spins, 
is introduced by hand.
On the other hand, 
in the Hubbard model, the effective attractive interaction 
which is responsible for the pairing 
is generated by the onsite Coulomb 
repulsion through higher-order many-body processes
only when the system has the suitable conditions.

\subsection{Implications for $d_{x^2-y^2}$-wave pairing in the cuprates} 

The presence of $d_{x^2-y^2}$ pairing correlations 
in the Hubbard model and the nature of the effective 
interaction mediating it are issues 
which are of interest for studying high
temperature superconductivity in the cuprates.
This is so especially after it became clear that the superconducting 
gap function in the high $T_c$ cuprates is 
of the $d_{x^2-y^2}$-wave type.
Within this context,
enormous amount of research has been carried out on this model.

In this article, the numerical studies 
of $d_{x^2-y^2}$-wave pairing 
in the Hubbard model have been reviewed.
The Monte Carlo simulations have shown the presence of the 
$d_{x^2-y^2}$-wave pairing correlations in the Hubbard model,
even though sufficiently low temperatures, where long-range 
order could establish, have not been reached.
For $U/t=4$ and at the temperatures 
where the simulations are carried out,
the values of the $d_{x^2-y^2}$-wave eigenvalues of the 
particle-particle Bethe-Salpeter equation are in agreement 
with the FLEX calculations. 
When carried out at low temperatures,
the FLEX calculations find $T_c$'s of order $0.025t$.
Since $t$ is estimated to be about $0.45$~eV 
for a single-band model of the cuprates
[Hybertson {\it et al.} 1990], 
this value of the $T_c$ corresponds to $\sim 130$~K.
This is a high value reflecting the electronic 
energy scales of the model.
Furthermore, the comparison of the FLEX and the QMC data
suggests that the pairing could be stronger for $U=8t$.
However, these are only mean-field estimates 
for a possible Kosterlitz-Thouless transition
and, in fact, 
the ground state of the doped 2D Hubbard model is not known. 
It might be that an additional parameter needs to be tuned 
in order to create optimum conditions for $d_{x^2-y^2}$-wave  
superconductivity in two dimensions.
For instance, 
in the 2-leg ladder case, it was necessary to tune $t_{\perp}/t$.
In this respect, 
the second-nearest-neighbour hopping $t'$ may play a role
for the enhancement of the $d_{x^2-y^2}$ pairing in the 2D system
like $t_{\perp}$ does for the 2-leg ladder. 
It is also possible that a three-band CuO$_2$ model 
with an onsite Coulomb repulsion at the Cu sites offers 
additional degrees of freedom for creating more favourable 
conditions for pairing.

The QMC results reviewed here show that the short-range AF
correlations are responsible for the $d_{x^2-y^2}$ pairing
correlations in the 2D Hubbard model.
These results also suggest that effects which enhance the 
magnitude of 
$\Gamma_{Is}({\bf q}\sim (\pi,\pi),i\omega_m=0)$ 
and the 
single-particle spectral weight near the $(\pi,0)$ and the 
$(0,\pi)$ points of the Brillouin zone act to enhance 
the $d_{x^2-y^2}$-wave superconducting correlations.
Clearly, other effects which enhance 
$d_{x^2-y^2}$-wave pairing can exist.
Exploring these possibilities is an active research field.

\setcounter{equation}{0}\setcounter{figure}{0}
\section{Summary and conclusions}

In this paper, the numerical studies of the 2D and 
the 2-leg Hubbard models have been reviewed.
For the 2D Hubbard model, 
data from the QMC simulations have been shown.
These data represent what has been obtained over the years 
in the parameter regime allowed by the sign problem. 
For the 2-leg Hubbard ladder, 
the QMC results at finite temperatures 
and the DMRG data on the ground state have been shown. 

Here, the emphasis has been placed on the 
$d_{x^2-y^2}$-wave superconducting 
correlations observed in the Hubbard model. 
In order to develop an understanding of the 
origin of these correlations, results on the spin, 
charge and the single-particle excitations have been shown along 
with the data 
on the particle-particle and the particle-hole interactions. 
The observation of the short-range AF fluctuations by the 
NMR and the neutron scattering experiments, 
and the unusual single-particle spectrum 
seen in the ARPES data are properties
which support using a Hubbard framework for studying 
the pairing correlations of the cuprates.

In the 2D Hubbard model, 
upon hole doping the long-range AF order is 
destroyed and the system exhibits short-range AF correlations. 
The maximum-entropy analysis of the QMC data shows that 
the AF and the Coulomb correlations 
strongly affect the single-particle properties. 
In particular, it has been seen that, 
as the strength of $U/t$ increases, significant amount of single-particle 
weight remains pinned near the Fermi level at the $(\pi,0)$ and 
$(0,\pi)$ points of the Brillouin zone. 
These generate phase space for scatterings in the 
$d_{x^2-y^2}$-wave BCS channel.
The results on the particle-particle and the particle-hole 
irreducible interactions have been also presented. 
It has been seen that, for $U=4t$, 
a properly-renormalized single spin-fluctuation 
exchange interaction can describe the momentum, the Matsubara-frequency
and the temperature dependence of the effective-particle-particle
interaction. 
This means that, for these values of $U/t$ and $T/t$, 
the attraction in the $d_{x^2-y^2}$ channel is mediated 
by the AF spin fluctuations. 

The DMRG results on the 2-leg ladder, 
which have been presented in Section~7 are important. 
These calculations represent the first example 
where a purely repulsive
onsite interaction leads to superconducting correlations which are 
enhanced over the noninteracting ($U=0$) case in the ground state
of a bulk system.
For certain values of $t_{\perp}/t$ 
and in the intermediate coupling regime, 
this model exhibits enhanced 
power-law $d_{x^2-y^2}$-like superconducting correlations. 
The QMC simulations for this system show that 
the effective particle-particle interaction $\Gamma_I$ peaks at 
$(\pi,\pi)$ momentum transfer and, in addition, 
the $d_{x^2-y^2}$ pairing correlations are strongest when there 
is significant amount of single-particle spectral weight 
near the Fermi level at the $(\pi,0)$ and $(0,\pi)$ 
points in the Brillouin zone. 
This way the system makes optimum use of the momentum structure 
in $\Gamma_I$ for $d_{x^2-y^2}$-wave pairing. 

These data were also compared with various other approaches 
to the Hubbard model, such as the 
diagrammatic FLEX approximation, the 
variational and the projector Monte Carlo simulations,
the dynamical cluster approximation and the one-loop 
RG calculations. 
In addition, the similarities and the differences with the 
$t$-$J$ model were briefly discussed. 
The implications for the $d_{x^2-y^2}$-wave 
superconductivity seen in the cuprates were also noted.

It is difficult to reach conclusions 
about the strongly correlated systems. 
Nevertheless, here, the following conclusions are given:

(1) An onsite Coulomb repulsion can lead to 
superconducting correlations which are enhanced with respect to the 
noninteracting ($U=0$) case in the ground state
of a bulk system.
This is proven for the case of the 2-leg Hubbard ladder.

(2) The 2D Hubbard model exhibits $d_{x^2-y^2}$-wave pairing 
correlations which grow as the temperature is lowered in the parameter 
regime where the QMC simulations are carried out. 
The fastest growing pairing correlations occur in the 
singlet $d_{x^2-y^2}$-wave channel.
These correlations are not weak.

(3) The QMC simulations also find that the effective pairing interaction 
in the 2D and the 2-leg Hubbard models 
in the intermediate coupling regime and 
at temperatures greater or of order $J/2$
is consistent with the 
spin-fluctuation exchange approximation. 

(4) Two factors which create optimum conditions for 
$d_{x^2-y^2}$ pairing in the Hubbard model are enhanced 
single-particle spectral weight at the Fermi level near the 
$(\pi,0)$ and $(0,\pi)$ points in the Brillouin zone
and large weight in the effective pairing interaction
$\Gamma_I$ at large momentum transfers. 
Clearly, 
there can be other ways of enhancing 
the $d_{x^2-y^2}$  pairing correlations.

\vskip 0.5 truecm
{\bf Acknowledgments}

Most of the numerical data presented here are from
various calculations carried out with 
T. Dahm, R.M. Noack, D.J. Scalapino and S.R. White.
The QMC data shown here were obtained 
at the San Diego Supercomputer Center.

\setcounter{equation}{0}\setcounter{figure}{0}
\section{Appendix}

The results shown in Sections 3 through 7 of this review 
were obtained with the determinantal QMC and the DMRG techniques.
Here, these two approaches to the many-body problem 
will be described briefly. 

\subsection{Determinantal QMC technique}

The determinantal QMC algorithm used in obtaining 
the QMC data shown here
is described in Ref.~[White {\it et al.} 1989b].
Reviews of this technique can be found 
in Refs.~[Scalapino 1993, Muramatsu 1999].
The basic idea of this approach 
is due to Blankenbecler, Scalapino and Sugar 
[Blankenbecler {\it et al.} 1981].
Here, the purpose is to calculate the expectation value 
of an operator $O$ at finite 
temperature $T$ in the grand canonical ensemble, 
\begin{equation}
\langle O\rangle = {1\over Z} {\rm Tr}\, Oe^{-\beta H}
\end{equation}
where
\begin{equation}
Z={\rm Tr}\, e^{-\beta H}.
\end{equation}
It is convenient to take the Hubbard hamiltonian 
as $H=K+V$ where
\begin{equation}
\label{K}
K = -t \sum_{ \langle ij\rangle,\sigma} 
( c^{\dagger}_{i\sigma} c_{j\sigma} + h.c.) 
-\mu \sum_i (n_{i\uparrow} + n_{i\downarrow})
\end{equation}
and
\begin{equation}
V=U\sum_i 
(n_{i\uparrow}- {1\over 2}) (n_{i\downarrow}- {1\over 2}).
\end{equation}
After discretizing the inverse temperature into $L$ time slices, 
$\beta=L\Delta\tau$, 
the Trotter approximation is used to rewrite 
the partition function as 
\begin{equation}
Z={\rm Tr}\, e^{-L\Delta\tau H} 
\cong {\rm Tr}\, ( e^{-\Delta\tau V} e^{-\Delta\tau K})^L.
\end{equation}
Next,
the interaction term $V$, 
which is quartic in the fermion operators, 
is reduced to a bilinear form with 
the Hubbard-Stratonovich transformation 
\begin{equation}
e^{-\Delta\tau U (n_{i\uparrow}-1/2)(n_{i\downarrow}-1/2)} = 
{ e^{-\Delta\tau U/4} \over 2 }
\sum_{S_{i\ell}=\pm 1} \,
e^{-\Delta\tau S_{i\ell} 
\lambda (n_{i\uparrow} - n_{i\downarrow})}
\end{equation}
where $\cosh({\lambda\Delta\tau}) = \exp({U\Delta\tau/2})$.
Here, 
at each lattice site $i$ and 
for each time slice $\ell$, an Ising spin 
$S_{i\ell}= \pm 1$ is used for the decoupling. 
This form of the Hubbard-Stratonovich decoupling 
was introduced by Hirsch
[Hirsch 1985].
After integrating out the fermion degrees of freedom, 
the partition function is given by 
\begin{equation} 
Z = \sum_{\{S_{i\ell}\}} \, 
{\rm det}\, M^{\uparrow}(\{ S_{i\ell} \})
{\rm det}\, M^{\downarrow}(\{ S_{i\ell} \})
\end{equation}
where
\begin{equation}
M^{\sigma}(\{S_{i\ell}\}) = I + B^{\sigma}_L B^{\sigma}_{L-1}...B^{\sigma}_1
\end{equation}
and 
\begin{equation}
B^{\sigma}_{\ell} = e^{ -\sigma \Delta\tau \nu(\ell) } 
e^{-\Delta\tau k}
\end{equation}
with electron spin $\sigma=\pm 1$.
Here, 
the summation is over all configurations
$\{S_{i\ell}\}$ of the Ising fields, 
$I$ is the $N\times N$ unit matrix, 
$B_{\ell}^{\sigma}$'s are $N\times N$ matrices where
$\nu(\ell)_{ij} = \delta_{ij} S_{i\ell}$ 
and  $k$ is the matrix representation 
of the kinetic energy operator, Eq.~(\ref{K}).
Similar expressions can be obtained for the expectation value  
$\langle O \rangle$.
For instance, 
the equal-time single-particle correlation function 
$\langle c_{j\sigma} c_{j'\sigma}^{\dagger}\rangle$ 
can be expressed as 
\begin{equation} 
\label{cc}
\langle c_{j\sigma} c_{j'\sigma}^{\dagger} \rangle = 
{1\over Z}
\sum_{\{ S_{i\ell}\} } \,
G_{\sigma} (j,j'; \{S_{i\ell}\}) 
{\rm det}\,M^{\uparrow}(\{S_{i\ell}\})
{\rm det}\,M^{\downarrow}(\{S_{i\ell}\})
\end{equation}
where
\begin{equation}
G_{\sigma}(j,j';\{S_{i\ell}\}) = 
\left[ 
(I + B^{\sigma}_L B^{\sigma}_{L-1}...B^{\sigma}_1)^{-1} 
\right]_{jj'}.
\end{equation}

The summation over the Ising spin variables 
in Eq.~(\ref{cc}) is evaluated using Monte Carlo 
sampling techniques.
In this way,
Eq.~(\ref{cc}) becomes
\begin{equation}
\langle  c_{j\sigma} c^{\dagger}_{j'\sigma} \rangle = 
\langle G_{\sigma} (j,j';\{S_{i\ell}\}) 
\rangle_{P}
\end{equation}
where $\langle ...\rangle_{P}$
is evaluated by averaging over spin configurations 
$\{ S_{i\ell}\}$ 
generated with the probability distribution 
\begin{equation}
P(\{S_{i\ell}\}) = 
{1\over Z} 
{\rm det}\, M^{\uparrow}(\{S_{i\ell}\})
{\rm det}\, M^{\downarrow}(\{S_{i\ell}\}).
\end{equation}
Similarly, 
the time-dependent single-particle Green's function 
$\langle c_{j\sigma}(\tau) c_{j'\sigma}(\tau') \rangle$
is calculated by averaging 
$G_{\sigma}(j,\tau; j',\tau'; \{S_{i\ell}\})$,
which can also be expressed in terms of $B^{\sigma}_{\ell}$'s.

The Hubbard-Stratonovich transformation maps 
the interacting fermion problem on the 2D lattice 
to that of non-interacting fermions 
in a random Ising field on a 3D lattice,
the third dimension being the imaginary time axis $\tau$.
At this stage, 
for a given spin configuration $\{S_{i\ell}\}$, 
the Wick's theorem applies 
to the higher order correlation functions. 
Hence,
the higher-order correlation functions 
can be obtained by averaging over the 
products of the single-particle propagators
in random Ising fields.
For instance, 
the two-particle correlation function 
\begin{equation}
\langle 
T\,
c_{\uparrow}(i_4,\tau_4) c_{\downarrow}(i_3,\tau_3)
c^{\dagger}_{\downarrow}(i_2,\tau_2) c^{\dagger}_{\uparrow}(i_1,\tau_1)
\rangle,
\end{equation}
which is used in extracting the particle-particle 
reducible interaction $\Gamma$ in Section 6, 
can be obtained from 
\begin{equation}
\langle 
G_{\uparrow}  ( i_4,\tau_4;i_1,\tau_1;\{S_{i\ell}\} ) 
G_{\downarrow}( i_3,\tau_3;i_2,\tau_2;\{S_{i\ell}\} ) 
\rangle_P.
\end{equation}

For $\langle n\rangle=1$, 
the product of the fermion determinants,
${\rm det}\, M^{\uparrow}(\{S_{i\ell}\}) \,
 {\rm det}\, M^{\downarrow}(\{S_{i\ell}\})$,
is positive.
However, 
away from half-filling and for $U>0$, 
this product is not positive for all configurations 
$\{S_{i\ell}\}$.
In this case, 
one can use the probability distribution 
\begin{equation}
\tilde{P}(\{S_{i\ell}\}) = 
{  
|{\rm det}\, M^{\uparrow}(\{S_{i\ell}\}) \,
 {\rm det}\, M^{\downarrow}(\{S_{i\ell}\})|
\over 
\sum_{ \{S_{i\ell}\} } 
|{\rm det}\, M^{\uparrow}(\{S_{i\ell}\}) \,
 {\rm det}\, M^{\downarrow}(\{S_{i\ell}\})|
},
\end{equation}
in order to calculate the expectation value $\langle O\rangle$ with 
\begin{equation}
\label{O}
\langle O\rangle = 
{ 
\langle 
O( \{S_{i\ell}\} ) {\rm sign}( \{S_{i\ell}\} ) 
\rangle_{\tilde{P}}
\over 
\langle {\rm sign}( \{S_{i\ell}\} ) \rangle_{\tilde{P}}
}.
\end{equation}
Here,
${\rm sign}(\{S_{i\ell}\})$ is the sign of 
${\rm det}\, M^{\uparrow}(\{S_{i\ell}\}) \, 
 {\rm det}\, M^{\downarrow}(\{S_{i\ell}\})$
and the average 
$\langle ...\rangle_{\tilde{P}}$ 
is calculated with the probability distribution $\tilde{P}$. 
The denominator in Eq.~(\ref{O}),
\begin{equation}
\langle {\rm sign}\rangle \equiv 
\langle {\rm sign}(\{S_{i\ell}\})\rangle_{\tilde{P}},
\end{equation}
becomes small when the number of configurations 
with positive signs is close to that with the negative signs.
It has been shown that away from half-filling 
$\langle {\rm sign}\rangle$ 
decreases exponentially as $T$ decreases
[Loh {\it et al.} 1990].
This causes large statistical errors in 
$\langle O\rangle$ and it is the cause of the sign problem. 
The temperature and the doping regime of the 2D Hubbard model 
which cannot be studied with QMC 
because of the sign problem is shown at the end of Section 6.
In spite of the sign problem, 
useful information has been extracted from the 2D 
and the 2-leg Hubbard models with QMC simulations.
Currently, the search for ways of removing 
the sign problem is an important field of study.
Improvements in the sign problem
which decreases the temperatures accessible 
to QMC simulations by even a factor of two would be valuable. 

\subsection{DMRG technique}

The DMRG technique which was invented by White 
represents a significant development in the application 
of the renormalization group 
ideas to interacting lattice models
[White 1992 and 1993].
Currently, it is the numerical method of choice 
for studying the ground state properties of 
quasi-1D interacting systems. 
The brief discussion of this technique given 
here follows closely that of 
Ref.~[White 1993]. 
For simplicity, 
the following discussion will be for a 1D lattice.

One of the differences between the DMRG and the standard real-space
RG approach is in the treatment of the boundary conditions 
in the basic blocking procedure.
In the standard approach for a finite 1D system, 
the chain is broken into finite blocks $B_{\ell}$ 
of $\ell$ sites.  
Here, first 
the Hamiltonian for two identical blocks is diagonalized 
and then the lowest eigenstates are kept 
to form a new approximate Hamiltonian describing a
larger block $B'_{2\ell}$ with $2\ell$ sites.
This procedure is illustrated in Fig.~10.1(a).
White and Noack have shown that this was of blocking introduces 
large errors for a model of a free particle on a 1D lattice
because of the way the boundary conditions are treated
during the blocking
[White and Noack 1992].
In the standard RG scheme, 
neglecting the connection of the two blocks 
to the neighboring blocks corresponds to 
setting the wave function of the particle to 0 outside of the blocks. 
Consequently, 
the low lying states from the previous iteration 
cause a `kink` in the wave function in the middle of the enlarged 
block $B'_{2\ell}$.
Hence,
in order to accurately represent the states of block $B'_{2\ell}$, 
it is necessary to use 
almost all of the states of block $B_{\ell}$, 
and any truncation of these states introduces large errors
in the RG process.
White and Noack have noted that 
using a combination of boundary conditions 
during the blocking fixes this deficiency 
for a free particle on a 1D lattice. 
They have also shown that an alternative approach
is to use a superblock 
which is composed of more than two blocks. 
In this way,
two of the blocks in the superblock are used 
to form a larger block for the next iteration 
while the effect of the other blocks 
is to apply a variety of boundary conditions. 
The DMRG technique uses a superblock during the RG iteration,
which is illustrated in Fig.~10.1(b)
for a 1D lattice of $L$ sites.
This superblock is composed of two blocks 
with $\ell$ and $\ell'=L-\ell-2$ sites 
and two additional sites in the middle, 
which can be considered as additional blocks.
The block $B_{\ell}$ and 
its neighboring lattice site are called the `system',
while the rest of the superblock is called the `environment'.
With each iteration, 
the system block gets enlarged by one lattice site. 

\begin{figure}
\centering
\iffigure
\mbox{
\subfigure[]{
\epsfysize=5.5cm
\epsffile[150 250 470 520]{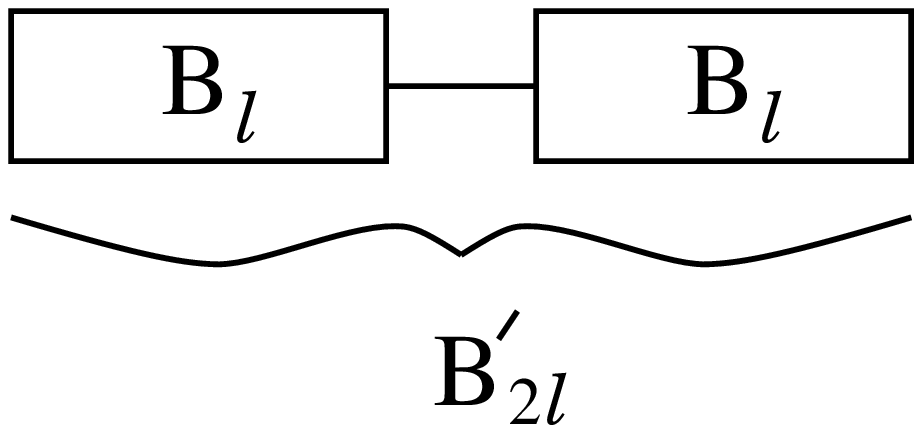}}
\quad
\subfigure[]{
\epsfysize=5.5cm
\epsffile[70 250 600 520]{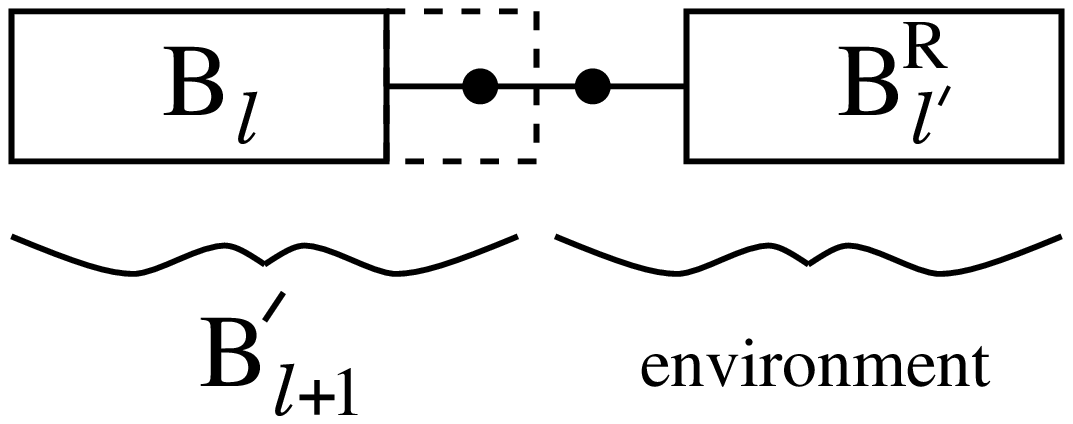}}}
\fi
\caption{
(a) Standard blocking scheme used 
for real-space RG for a 1D system.
Here, the Hamiltonian for two blocks 
$B_{\ell}$ each with $\ell$ sites is 
diagonalized and truncated to form an approximate Hamiltonian 
for the new block $B'_{2\ell}$ with $2\ell$ sites. 
(b) DMRG blocking scheme for a finite 1D system using a superblock 
which is composed of blocks $B_{\ell}$ and $B^R_{\ell'}$,
and two additional sites in between. 
Here, first the Hamiltonian for the superblock is diagonalized 
and the density matrix is formed 
for the enlarged block $B'_{\ell+1}$.
The Hamiltonian for block $B'_{\ell+1}$ is then expressed 
in a reduced basis composed of the 
leading eigenstates of this density matrix.
In the next iteration, 
the block $B'_{\ell+1}$ replaces $B_{\ell}$.
}
\label{10.1}
\end{figure}

Another novel feature of the DMRG technique is 
the use of the density matrices 
to choose the states which are to be 
kept during the iteration [White 1992 and 1993].
In the standard RG approach, 
the lowest $m$ eigenstates of the Hamiltonian for two
blocks is used in  forming the truncated Hamiltonian 
for the larger block,
as illustrated in Fig.~10.1(a).
This would be a reasonable approximation for a model 
where the coupling between the blocks is weak. 
However, 
for an interacting system such as the Hubbard model, 
each block is strongly coupled to its environment. 
In this case, 
White has shown that it is better to use during the truncation 
the eigenstates of the density matrix of the system,
and, in fact, 
the optimal states to be kept are the eigenvectors 
of the density matrix of the system with the largest eigenvalues. 
The use of the superblock 
and the density matrix formulation are the 
important new ideas used by the DMRG technique.

At this point, 
the DMRG algorithm for a finite 1D lattice can be 
summarized as follows:
(1) The first step is to diagonalize the Hamiltonian 
for the superblock, which is illustrated in Fig.~10.1(b),
using exact diagonalization techniques
in order to find the ground state $\psi_{ij}$. 
Here, 
the index $i$ refers to the system part of the lattice 
and $j$ refers to the environment. 
(2) Next, the reduced density matrix for block $B'_{\ell+1}$ 
is calculated by tracing over the environment,
\begin{equation}
\rho_{ii'}=\sum_j \, \psi_{ij}\psi^*_{i'j},
\end{equation}
and then $\rho$ is diagonalized.
(3) At this stage, the Hamiltonian for $B'_{\ell+1}$ can be 
transformed to a reduced basis which consists of the $m$
leading eigenstates of $\rho$.
Typically a few hundred eigenstates of $\rho$ are kept. 
(4) In the next iteration, this approximate Hamiltonian for 
$B'_{\ell+1}$ is used in place of the Hamiltonian for $B_{\ell}$.

With this procedure, the system block is built up 
by adding one lattice site at a time 
as the iteration continues form one end 
of the lattice to the other end. 
With each sweep through the lattice, 
a better approximation is obtained for 
the Hamiltonian of the block $B_{\ell}$. 
In addition, 
during the sweeps, 
the Hamiltonians of the system blocks from the previous sweep 
are used as the Hamiltonians for the environment blocks.
For 1D lattices, 
reflections of $B_{\ell}$ can be used in place of $B^R_{\ell'}$ 
in order to build up the superblock to the proper size 
during the first sweep.
In order to extend this algorithm to 2D, 
a row of neighboring sites can be added 
to the system block at each iteration, or
a single site can be added
using a connected 1D path through the 2D lattice. 
For higher dimensional lattices, 
an empty environment block
$B^R_{\ell'}$ is used in the first sweep. 

The DMRG technique is more accurate 
when the number of connections 
between the system and the environment blocks is minimized. 
Hence, 
usually open boundary conditions are employed.
When periodic boundary conditions are used, 
the accuracy decreases. 
For 2D lattices, the accuracy of DMRG also decreases 
as a function of the width of the lattice. 
However, 
if a larger number of the eigenstates of $\rho$ are kept 
during the truncation, the accuracy of the algorithm increases.
Finally, 
the DMRG is most accurate for large $U$ and least accurate for $U=0$.

With this technique, 
various equal-time correlation functions for the 2-leg
Hubbard ladder have been calculated 
on lattices up to $2\times 32$ in size
[Noack {\it et al.} 1996].
For the 3-leg Hubbard ladder 
there are DMRG data for a $3\times 12$ lattice
[Bonca {\it et al.} 2000]. 
For the $t$-$J$ model, 
larger systems can be studied with this technique. 
In this case, 
2D clusters up to $28\times 8$ in size have been studied 
using cylindrical boundary conditions
[White and Scalapino 1998a].
Currently,
the DMRG technique is used in a wide variety of fields, 
and the efforts to develop new algorithms 
to study larger clusters continues.
In addition, 
there is continuing work to extend the DMRG technique 
in order to calculate the dynamical correlation functions. 

\newpage

\section{References}

\begin{list}{}{\setlength{\itemindent}{-1cm}\setlength{\itemsep}{0pt}
\setlength{\parsep}{0pt}}
\item $^*$ E-mail: nbulut@ku.edu.tr

\item ABRAHAMS, E., BALATSKY, A., SCHRIEFFER, J. R., and 
ALLEN, P. B., 
1993,
{\it Phys. Rev.} B, {\bf 47}, 513.

\item ABRAHAMS, E., BALATSKY, A., SCALAPINO, D. J., and
SCHRIEFFER, J. R.,
1995,
{\it Phys. Rev.} B, {\bf 52}, 1271. 

\item AEBI, P., OSTERWALDER, J., SCHWALLER, P., SCHLAPBACH, L., 
SHIMODA, M., MOCHIKU, T., and KADOWAKI, K.,  
1994,
{\it Phys. Rev. Lett.}, {\bf 72}, 2757.

\item ANDERSON, P. W., 
1987,
{\it Science}, {\bf 235}, 1196.

\item ANDERSON, P. W., 
1997,
{\it Adv. Phys.}, {\bf 46}, 3.

\item ANDERSON, P. W., BASKARAN, G., ZOU, Z., and HSU, T.,
1987,
{\it Phys. Rev. Lett.}, {\bf 58}, 2790.

\item ANDERSON, P. W., and ZOU, Z.,
1988,
{\it Phys. Rev. Lett.}, {\bf 60}, 132.

\item BALATSKY, A. V., and ABRAHAMS, E.,
1992,
{\it Phys. Rev.} B, {\bf 45}, 13125.

\item BALENTS, L., and FISHER, M. P. A.,
1996,
{\it Phys. Rev.} B, {\bf 53}, 12133.

\item BECCA, F., CAPONE, M., and SORELLA, S., 
2000,
{\it Phys. Rev.} B, {\bf 62}, 12700.

\item BEDNORZ, J. G., and M\"ULLER, K. A., 
1986, 
{\it Z. Phys.} B, {\bf 64}, 189.

\item BEENEN, J., and EDWARDS, D. M., 
1995,
{\it J. Low Temp. Phys.}, {\bf 99}, 403;
{\it Phys. Rev.} B, {\bf 52}, 13636.

\item BEREZINSKII, V. L., 
1974,
{\it Pisma Zh. Eksp. Teor. Fiz.}, {\bf 20}, 628
[1974, {\it JETP Lett.}, {\bf 20}, 287].

\item BERK, N. F., and SCHRIEFFER, J. R., 
1966, 
{\it Phys. Rev. Lett.}, {\bf 17}, 433.

\item BICKERS, N. E., SCALAPINO, D. J., and SCALETTAR, R. T.,
1987, 
{\it Int. J. Mod. Phys.} B, {\bf 1}, 687.

\item BICKERS, N. E., SCALAPINO, D. J., and WHITE, S. R.,
1989, 
{\it Phys. Rev. Lett.}, {\bf 62}, 961.

\item BICKERS, N. E., and WHITE, S. R., 
1991, 
{\it Phys. Rev.} B, {\bf 43}, 8044.

\item BIRGENEAU, R. J.,
1990,
in {\it Physical Properties of High Temperature Superconductors II}, 
edited by D. M. Ginsberg (World Scientific, Singapore).

\item BLANKENBECLER, R., SCALAPINO, D. J., and SUGAR, R. L., 
1981,
{\it Phys. Rev.} D, {\bf 24}, 2278.

\item BONCA, J., GUBERNATIS, J. E., GUERRERO, M., 
JECKELMANN, E., and WHITE, S. R.,
2000,
{\it Phys. Rev.} B, {\bf 61}, 3251.

\item BULUT, N., HONE, D., SCALAPINO, D. J., and BICKERS, N. E., 
1990,
{\it Phys. Rev.} B, {\bf 41}, 1797.

\item BULUT, N., 
1990,
in {\it Dynamics of Magnetic Fluctuations in High-Temperature
Superconductors},
edited by G. Reiter, P. Horsch, and G.C. Psaltakis (Plenum).

\item BULUT, N., and SCALAPINO, D. J., 
1991, 
{\it Phys. Rev. Lett.}, {\bf 67}, 2898.

\item BULUT, N., and SCALAPINO, D. J., 
1992, 
{\it Phys. Rev. Lett.}, {\bf 68}, 706.

\item BULUT, N., SCALAPINO, D. J., and WHITE, S. R., 
1993, 
{\it Phys. Rev.} B, {\bf 47}, 2742;
{\it Phys. Rev.} B, {\bf 47}, 6157;
{\it Phys. Rev.} B, {\bf 47}, 14599.

\item BULUT, N., SCALAPINO, D. J., and WHITE, S. R., 
1994a, 
{\it Phys. Rev. Lett.}, {\bf 72}, 705;
{\it Phys. Rev.} B, {\bf 50}, 7215;
{\it Phys. Rev. Lett.}, {\bf 73}, 748.

\item BULUT, N., SCALAPINO, D. J., and WHITE, S. R., 
1994b, 
{\it Phys. Rev.} B, {\bf 50}, 9623.

\item BULUT, N., SCALAPINO, D. J., and WHITE, S. R., 
1995, 
{\it Physica} C, {\bf 246}, 85.

\item BULUT, N., and SCALAPINO, D. J., 
1995, 
{\it J. Phys. Chem. Solids}, {\bf 56}, 1597.

\item BULUT, N.,
1996,
{\it Tr. J. Phys.}, {\bf 20}, 548.

\item BULUT, N., and SCALAPINO, D. J., 
1996, 
{\it Phys. Rev.} B, {\bf 54}, 14971.

\item CALANDRA, M., BECCA, F., and SORELLA, S.,
1998,
{\it Phys. Rev. Lett.}, {\bf 81}, 5185.

\item CHEN, L., BOURBONNAIS, C., LI, T., and TREMBLAY, A.-M. S.
1991, 
{\it Phys. Rev. Lett.}, {\bf 66}, 369.

\item CHEN, Y. C., MOREO, A., ORTOLANI, F., DAGOTTO, E., 
and LEE, T. K.,
1994,
{\it Phys. Rev.} B, {\bf 50}, 655. 

\item CYROT, M., 
1986,
{\it Solid State Commun.}, {\bf 60}, 253.

\item DAGOTTO, E., ORTOLANI, F., and SCALAPINO, D. J., 
1991, 
{\it Phys. Rev.} B, {\bf 46}, 3183;
DAGOTTO, E., MOREO, A., ORTOLANI, F., RIERA, J., and SCALAPINO, D. J., 
1991, 
{\it Phys. Rev. Lett.}, {\bf 67}, 1918.

\item DAGOTTO, E., RIERA, J., and SCALAPINO, D. J., 
1992a, 
{\it Phys. Rev.} B, {\bf 45}, 5744.

\item DAGOTTO, E., MOREO, A., ORTOLANI, F., POILBLANC, D., and 
RIERA, J., 
1992b, 
{\it Phys. Rev.} B, {\bf 45}, 10741.

\item DAGOTTO, E., MOREO, A., ORTOLANI, F., RIERA, J., and 
SCALAPINO, D. J., 
1992c, 
{\it Phys. Rev.} B, {\bf 45}, 10107.

\item DAGOTTO, E., 
1994,
{\it Rev. Mod. Phys.}, {\bf 66}, 763.

\item DAGOTTO, E., and NAZARENKO, A., and BONINSEGNI, M.,
1994,
{\it Phys. Rev. Lett.}, {\bf 73}, 728.

\item DAGOTTO, E., and RICE, T. M., 
1996, 
{\it Science}, {\bf 271}, 618.

\item DAHM, T., and TEWORDT, L., 
1995, 
{\it Physica} C, {\bf 246}, 61;
1995, 
{\it Phys. Rev.} B, {\bf 52}, 1297;
1995,
{\it Phys. Rev. Lett.}, {\bf 74}, 793. 

\item DAHM, T., and BULUT, N.,
1996,
unpublished.

\item DAHM, T., and SCALAPINO, D. J., 
1997, 
{\it Physica} C, {\bf 288}, 33.

\item DAMASCELLI, A., LU, D. H., and SHEN, Z.-X., 
2001,
{\it J. Electron Spectr. Relat. Phenom.}, {\bf 117-118}, 165.

\item DAUL, S., SCALAPINO, D. J., and WHITE, S. R., 
2000, 
{\it Phys. Rev. Lett.}, {\bf 84}, 4188.

\item DESSAU, D., S., {\it et al.},
1993,
{\it Phys. Rev. Lett.}, {\bf 71}, 2781.

\item DING, H. {\it et al.}, 
1996,
{\it Nature}, {\bf 382}, 51.

\item DONIACH, S., and ENGELSBERG, S., 
1966, 
{\it Phys. Rev. Lett.}, {\bf 17}, 750.

\item DOPF, G., MURAMATSU, A., and HANKE, W.,
1992a,
{\it Phys. Rev. Lett.}, {\bf 68}, 353.

\item DOPF, G., WAGNER, J., DIETERICH, P., MURAMATSU, A., 
and HANKE, W., 
1992b, 
{\it Phys. Rev. Lett.}, {\bf 68}, 2082.

\item DORNEICH, A., ZACHER, M. G., GR\"OBER, C., and EDER, R., 
2000,
{\it Phys. Rev.} B, {\bf 61}, 12816.

\item DUFFY, D., and MOREO, A., 
1995,
{\it Phys. Rev.} B, {\bf 51}, 11882.

\item DZYALOSHINSKII, I., 
1987,
{\it Zh. Eksp. Teor. Fiz.} {\bf 93} 1487
[{\it Sov. Phys. JETP} {\bf 66}, 848 (1987)].

\item EMERY, V. J., 
1986, 
{\it Synth. Metals}, {\bf 13}, 21.

\item EMERY, V. J., KIVELSON, S. A., and LIN, H. Q., 
1990, 
{\it Phys. Rev. Lett.}, {\bf 64}, 475.

\item ENDRES, H., NOACK, R. M., HANKE, W., POILBLANC, D., and 
SCALAPINO, D. J., 
1996,
{\it Phys. Rev.} B, {\bf 53}, 5530.

\item FURUKAWA, N., and IMADA, M., 
1991, 
{\it J. Phys. Soc. Jpn.}, {\bf 60}, 3604.

\item FURUKAWA, N., and IMADA, M., 
1992, 
{\it J. Phys. Soc. Jpn.}, {\bf 61}, 3331.

\item FURUKAWA, N., RICE, T. M., and SALMHOFER, M., 
1998,
{\it Phys. Rev. Lett.}, {\bf 81}, 3195.

\item GEORGES, A., KOTLIAR, G., KRAUTH, W., and ROZENBERG, M.J., 
1996,
{\it Rev. Mod. Phys.} {\bf 68}, 13.

\item GOFRON, K., {\it et al.},
1993,
{\it J. Phys. Chem. Solids}, {\bf 54}, 1193. 

\item GOPALAN, S., RICE, T. M., and SIGRIST, M.,
1994,
{\it Phys. Rev.} B, {\bf 49}, 8901.

\item GR\"OBER, C., EDER, R., and HANKE, W.,
2000,
{\it Phys. Rev.} B, {\bf 62}, 4336.

\item GUERRERO, M., ORTIZ, G., and GUBERNATIS, J. E., 
1999, 
{\it Phys. Rev.} B, {\bf 59}, 1706. 

\item HALBOTH, C. J. and METZNER, W., 
2000,
{\it Phys. Rev.} B, {\bf 61}, 7364.

\item HARDY, W. N., BONN, D. A., MORGAN, D. C., LIANG, R., and ZHANG, K., 
1993,
{\it Phys. Rev. Lett.} {\bf 70}, 3999.

\item HAAS, S., MOREO, A., and DAGOTTO, E.,
1995,
{\it Phys. Rev. Lett.}, {\bf 74}, 4281.

\item HAYWARD, C. A., POILBLANC, D., NOACK, R. M., 
SCALAPINO, D. J., and HANKE, W.,
1995, 
{\it Phys. Rev. Lett.}, {\bf 75}, 926. 

\item HELLBERG, C. S., and MANOUSAKIS, E., 
1997, 
{\it Phys. Rev. Lett.}, {\bf 78}, 4609. 

\item HELLBERG, C. S., and MANOUSAKIS, E., 
1999, 
{\it Phys. Rev. Lett.}, {\bf 83}, 132. 

\item HELLBERG, C. S., and MANOUSAKIS, E., 
2000, 
{\it Phys. Rev.} B, {\bf 61}, 11787. 

\item HIRSCH, J. E., 
1985, 
{\it Phys. Rev. Lett.}, {\bf 54}, 1317;
1985, 
{\it Phys. Rev.} B, {\bf 31}, 4403. 

\item HIRSCH, J. E., and TANG, S.,
1989, 
{\it Phys. Rev. Lett.}, {\bf 62}, 591. 

\item HONERKAMP, C. SALMHOFER, M., FURUKAWA, N., and RICE, T.M., 
2001,
{\it Phys. Rev.} B, {\bf 63}, 035109.

\item HUBBARD, J.,
1963, 
{\it Proc. Roy. Soc.} A, {\bf 276}, 238.

\item HUSCROFT, C., JARRELL, M., MAIER, Th., MOUKOURI, S., 
and TAHVILDARZADEH, A. N., 
2001,
{\it Phys. Rev. Lett.}, {\bf 86}, 139.

\item HUSSLEIN, T., MORGENSTERN, I., NEWNS, D. M., 
PATTNAIK, P. C., SINGER, J. M., and MATUTTIS, H. G.,
1996,
{\it Phys. Rev.} B, {\bf 54}, 16179.

\item HYBERTSEN, M. S., STECHEL, E. B., SCHL\"UTER, M., and
JENNISON, D. R., 
1990, 
{\it Phys. Rev.} B, {\bf 41}, 11068.

\item IMADA, M., 
1991,
{\it J. Phys. Soc. Jpn.}, {\bf 60}, 2740.

\item IMADA, M., and KOHNO, M., 
2000,
{\it Phys. Rev. Lett.}, {\bf 84}, 143.

\item IMAI, T., SLICHTER, C. P., YOSHIMURA, K., and KOSUGE, K., 
1993,
{\it Phys. Rev. Lett.}, {\bf 70}, 1002.

\item ITOH, Y., YASUOKA, H., FUJIWARA, Y., UEDA, Y., 
MACHI, T., TOMENO, I., TAI, K., KOSHIZUKA, N., and TANAKA, S.,
1992, 
{\it J. Phys. Soc. Jpn.}, {\bf 61}, 1287.

\item JAKLIC, J., and PRELOVSEK, P., 
2000,
{\it Adv. Phys.}, {\bf 49}, 1.

\item JARRELL, M., 
1992,
{\it Phys. Rev. Lett.}, {\bf 69}, 168.

\item JARRELL, M., and GUBERNATIS, J. E., 
1996,
{\it Phys. Rep.}, {\bf 269}, 134.

\item JECKELMANN, E., SCALAPINO, D. J., and WHITE, S. R.,  
1998,
{\it Phys. Rev.} B, {\bf 58}, 9492.

\item KAMPF, A. P., and SCHRIEFFER, J. R., 
1990a, 
{\it Phys. Rev.} B, {\bf 41}, 6399.

\item KAMPF, A. P., and SCHRIEFFER, J. R., 
1990b, 
{\it Phys. Rev.} B, {\bf 42}, 7967.

\item KANAMORI, J.,
1963, 
{\it Prog. Theor. Phys.}, {\bf 30}, 275.

\item KIVELSON, S. A., ROKHSAR, D. S., and SETHNA, J. P., 
1987,
{\it Phys. Rev.} B, {\bf 35}, 8865.

\item KOIKE, S., YAMAJI, K., and YANAGISAWA, T.,
2000,
{\it Physica} B, {\bf 284}, 417.

\item LEDERER, P. MONTHAMBAUX, G., and POILBLANC, D.,
1987,
{\it J. Phys.} {\bf 48}, 1613. 

\item LEGGETT, A., J., 
1975,
{\it Rev. Mod. Phys.}, {\bf 47}, 331.

\item LIU, Z., and MANOUSAKIS, E., 
1992,
{\it Phys. Rev.} B, {\bf 44}, 2414.

\item LOH, E. Y, GUBERNATIS, J. E., SCALETTAR, R. T., WHITE, S. R., 
SCALAPINO, D. J., and SUGAR, R. L., 
1990,
{\it Phys. Rev.}, B, {\bf 41}, 9301.

\item LUO, J., and BICKERS, N. E.,
1993, 
{\it Phys. Rev.} B, {\bf 47}, 12153.

\item M$^2$S-HTSC VI, 
2000,
{\it Proceedings of the International Conference on Materials 
and Mechanisms of Superconductivity 
and High Temperature Superconductors},
edited by K. Salama, W.K. Chu and P.W.C. Chu,
{\it Physica} C, {\bf 341-348}.

\item MACHIDA, K.,
1989,
{\it Physica} C, {\bf 158}, 192.

\item MAHAN, G. D.,
1981,
``Many-particle Physics", (Plenum, New York).

\item MAHAJAN, A. V., ALLOUL, H., COLLIN, G., MARUCCO, J. F.,
1994,
{\it Phys. Rev. Lett.}, {\bf 72}, 3100.

\item MAHAJAN, A. V., ALLOUL, H., COLLIN, G., MARUCCO, J. F.,
2000,
{\it Euro. Phys. J.} B, {\bf 13}, 457.

\item MAIER, Th., JARRELL, M., PRUSCHKE, T., and KELLER, J.,
2000,
{\it Phys. Rev. Lett.},  {\bf 85}, 1524.

\item MARTINDALE, J. A., BARRETT, S. E., KLUG, C. A., O'HARA, K. E., 
DESOTO, S. M., SLICHTER, C. P., FRIEDMAN, T. A., and GINSBERG, D. M., 
1992, 
{\it Phys. Rev. Lett.}, {\bf 68}, 702. 

\item METZNER, W. and VOLLHARDT, D., 
1989,
{\it Phys. Rev. Lett.}, {\bf 62}, 324.

\item MILA, F., and RICE, T. M., 
1989,  
{\it Physica} C, {\bf 157}, 561.

\item MILLIS, A. J.,  MONIEN, H., and PINES, D., 
1990, 
{\it Phys. Rev.} B, {\bf 42}, 167.

\item MIYAKE, K., SCHMITT-RINK, S., and VARMA, C. M., 
1986,
{\it Phys. Rev.} B, {\bf 34}, 6554.

\item MONTHOUX, P., BALATSKY, A. V., and PINES, D., 
1991, 
{\it Phys. Rev. Lett.}, {\bf 67}, 3448;
{\it Phys. Rev.} B, {\bf 46}, 14803. 

\item MONTHOUX, P., and PINES, D., 
1992, 
{\it Phys. Rev. Lett.}, {\bf 69}, 961;
{\it Phys. Rev.} B, {\bf 50}, 16015. 

\item MONTHOUX, P., and SCALAPINO, D. J., 
1994, 
{\it Phys. Rev. Lett.}, {\bf 72}, 1874. 

\item MOREO, A., and SCALAPINO, D. J.,
1991,
{\it Phys. Rev.} B, {\bf 43}, 8211.

\item MOREO, A., and SCALAPINO, D. J., and DAGOTTO, E.,
1991,
{\it Phys. Rev.} B, {\bf 43}, 11442.

\item MOREO, A.,
1992,
{\it Phys. Rev.} B, {\bf 45}, 5059.

\item MOREO, A., HAAS, S., SANDVIK, A. W., and DAGOTTO, E., 
1995,
{\it Phys. Rev.} B, {\bf 51}, 12045.

\item MORIYA, T., TAKAHASHI, Y., and UEDA, K., 
1990,
{\it J. Phys. Soc. Jpn.}, {\bf 59}, 2905.

\item MORIYA, T., and UEDA, K., 
2000,
{\it Adv. Phys.}, {\bf 49}, 555.

\item M\"ULLER-HARTMANN, E., 
1989,
{\it Z. Phys.} B, {\bf 74}, 507.

\item MURAMATSU, A., 
1999,
in {\it Quantum Monte Carlo Methods in Physics and Chemistry},
edited by M.P. Nightingale and C.J. Umrigar (Kluwer Academic).

\item NAKANISHI, T., YAMAJI, K., and YANAGISAWA, T.,
1997, 
{\it J. Phys. Soc. Jpn.}, {\bf 66}, 294.

\item NEWNS, D. M., TSUEI, C. C., HUEBENER, R. P., 
VAN BENTUM, P. J. M., PATTNAIK, P. C., and CHI, C. C., 
1994, 
{\it Phys. Rev. Lett.}, {\bf 73}, 1695.

\item NOACK, R. M., WHITE, S. R., and SCALAPINO, D. J., 
1994, 
{\it Phys. Rev. Lett.}, {\bf 73}, 882.

\item NOACK, R. M., WHITE, S. R., and SCALAPINO, D. J., 
1995, 
{\it Europhys. Lett.}, {\bf 30}, 163.

\item NOACK, R. M., WHITE, S. R., and SCALAPINO, D. J., 
1996, 
{\it Physica} C, {\bf 270}, 281.

\item NOACK, R. M., BULUT, N., SCALAPINO, D. J., and ZACHER, M.G., 
1997, 
{\it Phys. Rev.} B, {\bf 56}, 7162.

\item PAO, C. H., and BICKERS, N. E.,
1994, 
{\it Phys. Rev. Lett.}, {\bf 72}, 1870.

\item PAO, C.-H., and BICKERS, N. E.,
1995, 
{\it Phys. Rev.} B, {\bf 51}, 16310.

\item PENNINGTON, C. H., and SLICHTER, C. P.,
1990,
in {\it Physical Properties of High Temperature Superconductors II}, 
edited by D. M. Ginsberg (World Scientific, Singapore).

\item PENNINGTON, C. H., and SLICHTER, C. P.,
1991
{\it Phys. Rev. Lett.}, {\bf 66}, 381.

\item POILBLANC, D. and RICE, T. M., 
1989,
{\it Phys. Rev.} B, {\bf 39}, 9749.

\item POILBLANC, D., 
1993,
{\it Phys. Rev.} B, {\bf 48}, 3368;
1994,
{\it Phys. Rev.} B, {\bf 49}, 1477.

\item PREUSS, R., MURAMATSU, A., von der LINDEN, W.,
DIETRICH, P., ASSAAD, F. F., and HANKE, W., 
1994,
{\it Phys. Rev.  Lett.}, {\bf 73}, 732.

\item PREUSS, R., HANKE, W., and von der LINDEN, W., 
1995,
{\it Phys. Rev.  Lett.}, {\bf 75}, 1344.

\item PREUSS, R., HANKE, W., GR\"OBER, C.,  and EVERTZ, H. G., 
1997,
{\it Phys. Rev.  Lett.}, {\bf 79}, 1122.

\item PRUSCHKE, T., JARRELL, M., and FREERICKS, J. K.,
1995,
{\it Adv. Phys.}, {\bf 42}, 187.

\item PUTIKKA, W. O., LUCHINI, M. U., and RICE, T. M., 
1992,
{\it Phys. Rev. Lett.},  {\bf 68}, 538.

\item PUTIKKA, W. O., GLENISTER, R. L., SINGH, R. R. P., and
TSUNETSUGU, H., 
1994, 
{\it Phys. Rev. Lett.}, {\bf 73}, 170.

\item ROMMER, S., WHITE, S. R., and SCALAPINO, D. J., 
2000,
{\it Phys. Rev.} B, {\bf 61}, 13424.

\item RONNING, F. {\it et al.},
1998,
{\it Science}, {\bf  282}, 2067.

\item SCALAPINO, D. J.,
1993, 
in {\it Proceedings of the Summer School on Modern Perspectives 
in Many-Body Physics},
Canberra (World Scientific, Singapore).

\item SCALAPINO, D. J.,
1995, 
{\it Phys. Rep.}, {\bf 250}, 330.

\item SCALAPINO, D. J., LOH, E. Jr., and HIRSCH, J. E., 
1986,
{\it Phys. Rev.} B, {\bf 34}, 8190.

\item SCALAPINO, D. J., WHITE, S. R., and ZHANG, S. C., 
1992,
{\it Phys. Rev. Lett.}, {\bf 68}, 2830.

\item SCALAPINO, D. J., WHITE, S. R., and ZHANG, S. C., 
1993,
{\it Phys. Rev.} B, {\bf 47}, 7995.

\item SCALETTAR, R. T., SCALAPINO, D. J., SUGAR, R. L., 
and WHITE, S. R., 
1991,
{\it Phys. Rev.} B, {\bf 44}, 770.

\item SCHRIEFFER, J. R., WEN, X. G., and ZHANG, S. C., 
1988,
{\it Phys. Rev. Lett.}, {\bf 60}, 944.

\item SCHRIEFFER, J. R., WEN, X. G., and ZHANG, S. C., 
1989,
{\it Phys. Rev.} B, {\bf 39}, 11663.

\item SCHRIEFFER, J. R., 
1994,
{\it Solid State Commun.}, {\bf 92}, 129.

\item SCHRIEFFER, J. R., 
1995,
{\it J. Low Temp. Phys.}, {\bf 99}, 397.

\item SCHULZ, H. J., 
1987,
{\it Europhys. Lett.} {\bf 4}, 609.

\item SCHULZ, H. J., 
1989,
{\it J. Physique}, {\bf 50}, 2833.

\item SCHULZ, H. J., 
1999,
{\it Phys. Rev.} B, {\bf 59}, R2471.

\item SHEN, Z.-X., DESSAU, D. S., WELLS, B. O., KING, D. M., SPEICER, W. E.,
ARKO, A. J., MARSHALL, D., LOMBARDO, L. W., KAPITULNIK, A.,
DICKINSON, P., DONIACH, S., DICARLO, J., LOESER, T., and PARK, C. H., 
1993,
{\it Phys. Rev. Lett.}, {\bf 70}, 1553.

\item SHEN, Z.-X., and DESSAU, D. S., 
1995, 
{\it Phys. Rep.}, {\bf 253}, 1.

\item SILVER, R., N., SILVIA, D., S., and GUBERNATIS, J. E.,
1990,
{\it Phys. Rev.} B, {\bf 41}, 2380.

\item TAKIGAWA, M.,
1990,
in {\it Dynamics of Magnetic Fluctuations in High-Temperature
Superconductors},
edited by G. Reiter, P. Horsch, and G.C. Psaltakis (Plenum).

\item TAKIGAWA, M., SMITH, J. L., and HULTS, W. L., 
1991,
{\it Physica} C, {\bf 185}, 1105;
{\it Phys. Rev.} B, {\bf 44}, 7764.

\item TRANQUADA,  J. M., STERNLIEB, B. J., AXE, J. D., 
NAKAMURA, Y., and UCHIDA, S., 
1995, 
{\it Nature}, {\bf 375}, 561.

\item TSUEI, C. C., CHI, C. C., NEWNS, D. M., PATTNAIK, P. C., 
and DAUMLING, M.,
1992, 
{\it Phys. Rev. Lett.}, {\bf 69}, 2134.

\item TSUEI, C. C., KIRTLEY, J. R., CHI, C. C., YU-JAHNES, L. S., 
GUPTA, A., SHAW, T., SUN, J. Z., and KETCHEN, M. B., 
1994, 
{\it Phys. Rev. Lett.}, {\bf 73}, 593.

\item TSUEI, C. C., and KIRTLEY, 
2000, 
{\it Rev. Mod. Phys.}, {\bf 72}, 969.

\item VAN HARLINGEN, D., 
1995, 
{\it Rev. Mod. Phys.}, {\bf 67}, 515.

\item VARMA, C. M., LITTLEWOOD, P. B., SCHMITT-RINK, S., 
ABRAHAMS, E., and RUCKENSTEIN, A. E., 
1989,
{\it Phys. Rev. Lett.}, {\bf 63}, 1996.

\item VEKIC, M., and WHITE, S. R., 
1993,
{\it Phys. Rev.} B, {\bf 47}, 1160.

\item WHITE,  S. R., SCALAPINO, D. J., SUGAR, R. L., 
BICKERS, N. E., and SCALETTAR, R. T., 
1989a, 
{\it Phys. Rev.} B, {\bf 39}, 839.

\item WHITE,  S. R., SCALAPINO, D. J., SUGAR, R. L., 
LOH, E. Y., GUBERNATIS, J. E., 
and SCALETTAR, R. T., 
1989b, 
{\it Phys. Rev.} B, {\bf 40}, 506.

\item WHITE, S. R., SCALAPINO, D. J., SUGAR, R. L., and 
BICKERS, N. E.,
1989c,
{\it Phys. Rev. Lett.}, {\bf 63}, 1523.

\item WHITE, S. R., 
1991,
{\it Phys. Rev.} B, {\bf 44}, 4670;
1992,
{\it Phys. Rev.} B, {\bf 46}, 5678.

\item WHITE,  S. R., 
1992, 
{\it Phys. Rev. Lett.}, {\bf 69}, 2863.

\item WHITE,  S. R., 
1993, 
{\it Phys. Rev.} B, {\bf 48}, 10345.

\item WHITE,  S. R., and NOACK, R. M.,  
1992, 
{\it Phys. Rev. Lett.}, {\bf 68}, 3487.

\item WHITE,  S. R., and SCALAPINO, D., 
1998a, 
{\it Phys. Rev. Lett.}, {\bf 80}, 1272;
{\it Phys. Rev. Lett.}, {\bf 81}, 3227.

\item WHITE,  S. R., and SCALAPINO, D., 
1998b, 
{\it Phys. Rev.} B, {\bf 57}, 3031.

\item WHITE,  S. R., and SCALAPINO, D., 
1999, 
{\it Phys. Rev.} B, {\bf 60}, 753.

\item WHITE,  S. R., and SCALAPINO, D., 
2000, 
{\it Phys. Rev.} B, {\bf 61}, 6320.

\item WOLLMAN, D. A., VAN HARLINGEN, D. J., LEE, W. C., GINSBERG, D. M., 
and LEGGETT, A. J., 
1993, 
{\it Phys. Rev. Lett.}, {\bf 71}, 2134.

\item XIAO, G., CIEPLAK, M. Z., XIAO, J. Q., CHIEN, C. L., 
1990,
{\it Phys. Rev.} B, {\bf 42}, 8752.

\item YAMAJI, K., and SHIMOI, Y., 
1994, 
{\it Physica} C, {\bf 222}, 349.

\item YAMAJI, K., and YANAGISAWA, T., NAKANISHI, T., and KOIKE, S., 
1998, 
{\it Physica} C, {\bf 340}, 225.

\item ZAANEN, J., and GUNNARSON, O., 
1989, 
{\it Phys. Rev.} B, {\bf 40}, 7391.

\item ZACHER, M. G., ARRIGONI, E. HANKE, W., and
SCHRIEFFER, J. R., 
1998,
{\it Phys. Rev.} B, {\bf 57}, 6370.

\item ZANCHI, D. and SCHULZ, H.J., 
2000,
{\it Phys. Rev.} B, {\bf 61}, 13609.

\item ZHANG, S., CARLSON, J., and GUBERNATIS, J. E.,
1995, 
{\it Phys. Rev. Lett.}, {\bf 74}, 3652.

\item ZHANG, S., CARLSON, J., and GUBERNATIS, J. E.,
1997, 
{\it Phys. Rev. Lett.}, {\bf 78}, 4486.

\end{list}
\end{document}